%% file: FittingPaper_Accepted_arxiv.tex
\documentclass[traditabstract]{aa}

\usepackage{natbib} 

\usepackage{mathrsfs}

\usepackage{amsmath}

\usepackage{txfonts}

\usepackage{multirow}

\usepackage{graphicx}

\usepackage[table]{xcolor}

\newcommand{\appropto}{\mathrel{\vcenter{
  \offinterlineskip\halign{\hfil$##$\cr
    \propto\cr\noalign{\kern2pt}\sim\cr\noalign{\kern-2pt}}}}}

\def\lsimeq
{\hbox{\raise0.5ex\hbox{$<\lower1.06ex\hbox{$\kern-1.07em{\sim}$}$}}}

\begin{document}

\title{Light-curve modelling constraints on the obliquities and aspect angles of the young \emph{Fermi} pulsars}

\author{M. Pierbattista\inst{1,2,3} 
\and A. K. Harding\inst{4} 
\and I. A. Grenier\inst{5,6} 
\and T. J. Johnson\inst{7}
\and P. A. Caraveo\inst{2,8}
\and M. Kerr\inst{9}
\and P. L. Gonthier\inst{10}
}

\institute{ 
	{Nicolaus Copernicus Astronomical Center, Rabia\'nska 8, PL-87-100 Toru\'n, Poland; \email{mpierba@gmail.com}}
\and{INAF-Istituto di Astrofisica Spaziale e Fisica Cosmica, 20133 Milano, Italy}
\and {Fran\c{c}ois Arago Centre, APC, Universit\'e Paris Diderot, CNRS/IN2P3, CEA/Irfu, Observatoire de Paris, Sorbonne Paris Cit\'e, 10 rue A. Domon et L. Duquet, 75205 Paris Cedex 13, France }
\and {Astrophysics Science Division, NASA Goddard Space Flight Center, Greenbelt, MD 20771, U.S.A.}
\and {Laboratoire AIM, Universit\'e Paris Diderot/CEA-IRFU/CNRS, Service d'Astrophysique, CEA Saclay, 91191 Gif sur Yvette, France}
\and {Institut Universitaire de France}
\and{National Research Council Research Associate, National Academy of Sciences, Washington, DC 20001, resident at Naval Research Laboratory, Washington, DC 20375, USA}
\and{Istituto Nazionale di Fisica Nucleare, Sezione di Pavia, Via Bassi 6, I-27100 Pavia, Italy}
\and {W. W. Hansen Experimental Physics Laboratory, Kavli Institute for Particle Astrophysics and Cosmology, Department of Physics and SLAC National Accelerator Laboratory, Stanford University, Stanford, CA 94305, U.S.A.}
\and {Hope College, Department of Physics, Holland MI, U.S.A. }
}

\date{}
 
\abstract
{In more than four years of observation the Large Area Telescope on board 
the \emph{Fermi} satellite has identified pulsed $\gamma$-ray emission 
from more than 80 young or middle-aged pulsars, in most cases providing 
light curves with high statistics. Fitting the observed profiles with geometrical 
models can provide estimates of the magnetic obliquity $\alpha$ and of the 
line of sight angle $\zeta$, yielding estimates of the radiation beaming 
factor and radiated luminosity.

Using different $\gamma$-ray emission geometries (Polar Cap, Slot Gap, 
Outer Gap, One Pole Caustic) and core plus cone geometries for the radio 
emission, we fit $\gamma$-ray light curves for 76 young or middle-aged 
pulsars and we jointly fit their $\gamma$-ray plus radio light curves when 
possible.

We find that a joint radio plus $\gamma$-ray fit strategy is important to obtain 
$(\alpha,\zeta)$ estimates that can explain simultaneously detectable radio and 
$\gamma$-ray emission:
when the radio emission is available, the inclusion of the radio light curve
in the fit leads to important changes in the $(\alpha,\zeta)$ solutions. The most
pronounced changes are observed for  Outer Gap and One Pole Caustic models
for which the $\gamma$-ray only fit leads to underestimated $\alpha$ or $\zeta$  
when the solution is found to the left or to the right of the main $\alpha$-$\zeta$ 
plane diagonal respectively.
The intermediate-to-high altitude magnetosphere models, Slot Gap, Outer Gap, and
One pole Caustic, are favoured in explaining the observations. We find no apparent 
evolution of $\alpha$ on a time scale of $10^6$ years.  
For all emission geometries our derived $\gamma$-ray beaming factors are generally 
less than one and do not significantly evolve with the spin-down power. A more 
pronounced beaming factor vs. spin-down power correlation is observed for Slot Gap 
model and radio-quiet pulsars and for the Outer Gap model and radio-loud pulsars.
The beaming factor distributions 
exhibit a large dispersion that is less pronounced for the Slot Gap case and that 
decreases from radio-quiet to radio-loud solutions. For all models, the correlation 
between $\gamma$-ray luminosity and spin-down power is consistent with a square 
root dependence. 
The  $\gamma$-ray luminosities obtained by using the beaming factors estimated
in the framework of each model do not exceed the spin-down power. This suggests 
that assuming a beaming factor of one for all objects, as done in other studies, likely 
overestimates the real values. 
The data show a relation between the pulsar spectral characteristics and the width 
of the accelerator gap. The relation obtained in the case of the Slot Gap model is 
consistent with the theoretical prediction.} 

\authorrunning{Pierbattista et al. 2013}
\titlerunning{Magnetic obliquity and line of sight constraints}

\maketitle

\section{Introduction} 

The advent of the Large Area Telescope \citep[LAT,][]{aaa+09a} on the \emph{Fermi} satellite has significantly increased
our understanding of the high-energy emission from pulsars.  After more than four years of observations 
the LAT has detected pulsed emission from more than 80 young or middle-aged pulsars, collecting an unprecedented amount 
of data for these sources \citep[][]{2PC}. This has allowed the study of the collective properties of the $\gamma$-ray pulsar population 
\citep{pie10,wr11,twc11,pghg12} and of the pulse profiles. The light-curve analysis can be approached by studying the number of 
peaks and morphology or by modelling the $\gamma$-ray profiles to estimate pulsar orientations and constrain the model that best describes 
the observations. The first type of analysis has been performed by \cite{wrwj09} and \cite{pie10}, who studied light-curve peak separation 
and  multiplicities in light of intermediate and high-altitude gap magnetosphere models. The second type of analysis has been 
performed for a small set of pulsars by \cite{rw10} and \cite{pie10} for young and middle-aged  pulsars, and \cite{vhg09} for millisecond pulsars.  
They used the simulated  emission patterns of proposed models to fit the observed light curves and estimate 
the magnetic obliquity angle $\alpha$ (the angle between the pulsar rotational and magnetic axes) and the observer line of sight angle $\zeta$
(the angle between the observer direction and the pulsar rotational axis),
showing that the outer magnetosphere models are favoured in 
explaining the pulsar light curves observed by \emph{Fermi}.
What these first studies suggest is that with the new high-statistics of the LAT pulsar light curves, fitting the observed 
profiles with different emission models has become a powerful tool to give estimates of the pulsar 
orientation, beaming factor, and luminosity, and to constrain the geometric emission models. 

After discovery of the pulsed high-energy emission from the Crab pulsar \citep{mbc+73}, emission gap models
were the preferred physical descriptions of magnetospheric processes that produce $\gamma$-rays. 
These models predict the existence of regions in the magnetosphere where the Goldreich 
\& Julian  force-free condition \citep{gj69} is locally violated and particles can be accelerated up to a few TeV. Three gap 
regions were identified in the pulsar magnetosphere: the Polar Cap region \citep{stu71}, 								
above the pulsar polar cap; the Slot Gap region \citep{aro83b}, 													
along the last closed magnetic field line; the Outer Gap region \citep{chr86}, 									
between the null charge surface and the light cylinder. \cite{dhr04}  calculated
the pulsar  emission patterns of each model, according to the pulsar magnetic field, spin period, $\alpha$, and gap width and 
position. The \cite{dhr04} model is based on the assumptions that the magnetic field 
of a pulsar is a vacuum dipole swept-back by the pulsar rotation \citep{deu55} and that the $\gamma$-ray 
emission is tangent to the magnetic field lines and radiated in the direction of the accelerated electron velocity in the 
co-rotating frame. The emission pattern of a pulsar is then obtained by computing the direction of $\gamma$-rays
from a gap region located at the altitude range characteristic of that model. 
Note that the number of radiated $\gamma$-rays depends only on the emission gap width and maximum emission 
radius, which are assumed parameters.

The aim of this paper is to compare the light curves of the young and middle-aged LAT pulsars listed in the second pulsar 
catalog \citep[][hereafter PSRCAT2]{2PC} with the emission patterns predicted by theoretical models. We use the \cite{dhr04} 
geometric model to  calculate the radio emission patterns according to radio core plus cone models \citep{gvh04,sgh07,hgg07,pghg12}, 
and the $\gamma$-ray emission patterns according to the Polar Cap model \citep[PC,][]{mh03}, the Slot Gap model \citep[SG,][]{mh04a}, 
the Outer Gap model \citep[OG,][]{crz00}, and an alternative formulation of the OG model that differs just in the emission gap width and
luminosity formulations, the One Pole Caustic \citep[OPC,][]{rw10,wrwj09} model. We use them to fit the observed 
light curves and obtain estimates of $\alpha$, $\zeta$, outer gap width $w_\mathrm{OG/OPC}$, and slot gap width $w_\mathrm{SG}$, 
as well as the ensuing beaming factor and luminosity. Using these estimates, we study the collective properties of some non-directly 
observable characteristics of the LAT pulsars, namely their beaming factors, $\gamma$-ray luminosity, magnetic alignment, and 
correlation between the width of the accelerator gap and the observed spectral characteristics.

For each pulsar of the sample and each model, the estimates of  $\alpha$ and $\zeta$ we obtain represent the
best-fit solution in the framework of that specific model.
We define the \emph{optimum-solution} as that solution characterised by the highest log-likelihood value among the four 
emission models, and we define the \emph{optimum-model} as the corresponding model. Hereafter we will stick to this nomenclature 
in the descriptions of the fit techniques and in the discussion of the results.

The radio and/or $\gamma$-ray nature of the pulsars of our sample have been classified according to the flux criterion
adopted in PSRCAT2: radio-quiet (RQ) pulsars, with radio flux detected at 1400 MHz  $S_{1400}<30\mu$Jy
and radio-loud (RL) pulsars with $S_{1400}>30\mu$Jy. The $30\mu$Jy  flux threshold was introduced in PSRCAT2 to 
favour observational characteristics instead of discovery history in order to have more homogeneous pulsar samples.
Yet, radio light curves  were available for 2 RQ pulsars, J0106$+$4855 and J1907$+$0602, that show a radio flux $S_{1400}<30\mu$Jy (PSRCAT2). 
We include these two radio-faint (RF) pulsars in the RQ sample and the results of their joint $\gamma$-ray plus radio analysis are given 
in Appendix \ref{JointFits_RQ2RL}. 

The outline of this paper is as follows. In Section \ref{Data} we describe the data selection criteria adopted to build the 
$\gamma$-ray and radio light curves. In Section \ref{Simulation} we describe the method we use to calculate the pulsed
emission patterns and light curves. Sections \ref{Individual gamma-ray fit} and \ref{Fitting both the gamma-ray and radio emission} 
describe the fitting techniques used for the RQ and RL pulsars, respectively. The results are discussed in Section \ref{Results}.

In Appendix \ref{GoodFitMethod} we describe the method used to give an estimate of the relative goodness of the fit solutions.
In Appendix \ref{PopDis} we show further results obtained from the pulsar population synthesis study of  \cite{pghg12} that
we will compare with results obtained in Sections \ref{A-Z best solutions plane} and \ref{HighECut}. Appendices \ref{GammaFitRes}, 
\ref{JointFitRes}, and \ref{JointFits_RQ2RL} show, for each model, the best-fit $\gamma$-ray light curves for RQ LAT pulsars, the 
best-fit $\gamma$-ray and radio light curves for RL LAT pulsars, and the best-fit $\gamma$-ray and radio light curves of two 
RQ-classified  LAT pulsars for which a radio light curve exists.

\section{Data selection and LAT pulsar light curves}
\label{Data}
We have analysed the 35 RQ and 41 RL young or middle-aged pulsars listed in Tables \ref{A_Z_gamma1} and \ref{JointAlpZetFit}, respectively. 
Their $\gamma$-ray and radio light curves have been published  in PSRCAT2. 
For a spin period $P$ and spin period first time derivative $\dot{P}$ , their characteristic age spans the interval 
$10^{3.1} < \tau_{ch} = P/2\dot{P}  < 10^{6.5}$ years, assuming a negligible spin period at birth and a spin-down rate due 
to magnetic dipole radiation.

We have performed $\gamma$-ray only fits for all RQ objects and joint $\gamma$-ray plus radio fits for all RL objects. 
The $\gamma$-ray  light curve of the RL pulsar J1531$-$5610 has a very low number of counts 
(PSRCAT2) so we have not attempted to fit its $\gamma$-ray  profile and it is not included in our analysis.

The Crab (J0534+2200) is the only RL pulsar of our sample that shows aligned $\gamma$-ray and radio peaks. As stated 
by \cite{vjh12}, this could be explained by assuming a wide radio beam that originates at higher altitude \citep{man05} in the 
same magnetospheric region as the $\gamma$-rays, and possibly of caustic nature \citep{rwh+12}.
This interpretation is not compatible with the radio emission site near the magnetic poles assumed in this paper since it 
does not predict aligned radio and $\gamma$-ray peaks as observed in the Crab pulsar.
The joint radio plus $\gamma$-ray fits and the $\gamma$-ray only fit yield the same pulsar orientations that can explain the 
$\gamma$-ray light curve, but largely fails to reproduce the radio light curve at 1400 MHz. 
We decided to show the joint fit results for the Crab pulsar to show how the radio emission 
model used in this paper fails to explains the Crab radio light curve.

For each analysed pulsar, the selected dataset spans 3 years of LAT observation, from 2008 August 4 to 2011 August 4. 
In order to have high background rejection only photons with energy $E_{ph}>100$ MeV and belonging to the \emph{source} 
event class, as defined in the P7\_V6 instrument response function, have been used. To avoid spurious detection 
due to the $\gamma$-rays scattered from the Earth atmosphere,  
events with zenith angle  $\ge100^{\circ}$ have been excluded. A detailed description of the criteria adopted in the data selection 
can be found in PSRCAT2.

The photon rotational phases have been computed by using the TEMPO 2 software \citep{hem06} with a \emph{Fermi} LAT
plug-in\footnote{http://fermi.gsfc.nasa.gov/ssc/data/analysis/user/Fermi\_plug\_doc.pdf} written by Lucas Guillemot \citep{rkp+11}.
The pulsar ephemerides have been generated by the \emph{Fermi} Pulsar Search Consortium  \cite[PSC, ][]{rap+12} and by the 
\emph{Fermi} Pulsar Timing Consortium \cite[PTC, ][]{sgc+08}. The PTC is an international collaboration of radio astronomers and  
\emph{Fermi} collaboration members with the purpose of timing radio pulsars and pulsar candidates discovered by the PSC to provide 
the most up to date radio ephemerides and light curves. 
\begin{figure}[t]
\centering
\includegraphics[width=0.49\textwidth]{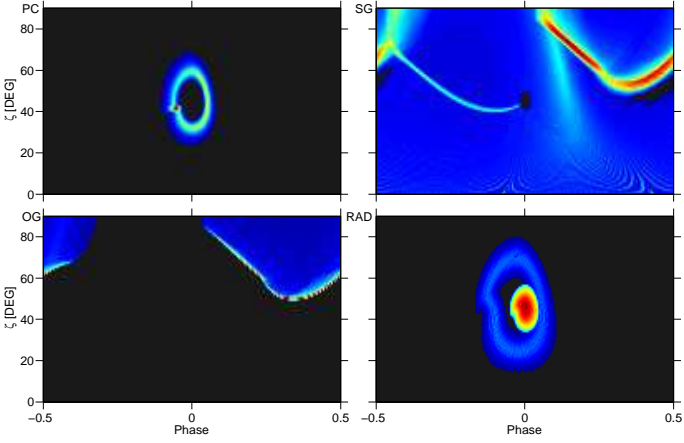}
\caption{The top left to bottom right panels illustrate phase-plots obtained for the PC, SG, OG/OPC, 
and radio (core plus cone) models respectively, with a magnetic field strength of 
$B_\mathrm{G}$=10$^8$ Tesla and spin period of 30 ms for the PC and radio cases, and gap widths 
of 0.04 and 0.01 for the SG and OG/OPC cases, respectively. All the plots are given for an obliquity 
$\alpha = 45^\circ$. The emission flux increases from black to red.}
\label{phase-plots}
\end{figure}

The $\gamma$-ray light curves used in this paper are those published in PSRCAT2. They have been obtained by a photon weighting 
technique that uses a pulsar spectral model, the instrument point spread function, and a model for the $\gamma$-ray emission from the 
observed region to evaluate the probability that each photon originates from the pulsar of interest or from the diffuse background or nearby 
sources  \citep{ker11}. A binned light curve is then 
obtained by summing the probabilities of all the photons within the phase bin edges. This method gives a high background rejection and 
increases the sensitivity to pulsed emission by more than 50\% compared to the standard non-weighted version of the of $H$-test \citep[][]{ker11}. 
The higher signal-to-noise ratio in the resulting light curves allows tighter fits in our analyses.
The complete description of the LAT pulsar light-curves generation procedure can be found in  \cite{ker11} 
and PSRCAT2.

According to the probability distribution of the weighted photons, the pulsar light-curve background is computed as
\begin{equation}
\label{background}
B= \left( \sum_{i=1}^{n_{ph}} w_{i} -  \sum_{i=1}^{n_{ph}} w_{i}^{2}\right) n_{bin}^{-1}
\end{equation}
where $w_i$ is the weight (probability) associated with the \emph{i-th} photon, $n_{ph}$ is the total number of photons in the
light curve, and $n_{bin}$ is the number of light-curve bins. The pulsar light-curve background represents the DC light-curve 
emission component that does not originate from the pulsars.
The error associated with the \emph{j-th} phase bin of the light curve, 
corresponding to the standard deviation of the photon weights in that bin, is
\begin{equation}
\label{noise}
\sigma_{j} = \left(\sum_{i=1}^{N_j} w_{i}^2\right)^{0.5}
\end{equation}
where $N_j$ is the number of photon weights in the \emph{j-th} bin. More details can be found in PSRCAT2.

The radio profiles of the RL LAT pulsars have been obtained in collaboration with the PSC and PTC. 
They have been built from observations mainly performed at 1400 MHz from 
Green Bank Telescope (GBT), Parkes Telescope, Nan\c cay Radio Telescope  (NRT), Arecibo Telescope, 
the Lovell Telescope at Jodrell Bank, and the Westerbork Synthesis Radio Telescope  \citep{sgc+08}.

\section{Simulation of the LAT pulsars emission patterns and light curves}
\label{Simulation}

\subsection{Phase-plots}
\label{Phase-plotsSection}
\begin{figure}[b]
\centering
\includegraphics[width=0.49\textwidth]{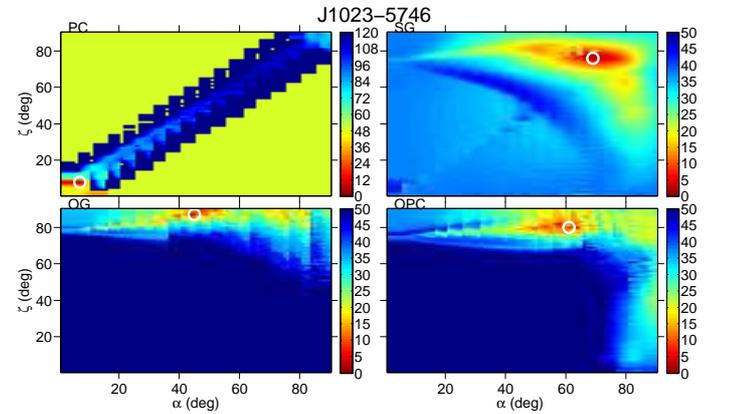}
\caption{$\alpha$-$\zeta$ log-likelihood maps obtained by fitting the $\gamma$-ray light curve of pulsar J1023$-$5746
	       with each $\gamma$-ray model phase plot. The fit has been performed with $\chi^2$ estimator and FCBin light curves. 
	       A white circle shows the position of the best-fit solutions. The colour-bar is in effective $\sigma=(| \ln L -\ln L_{max}|)^{0.5}$ 
	       units, zero corresponds to the best-fit solution. The diagonal band present in the PC panel is due to the fact that the emission 
	       region is located close to the polar cap and shines mainly when $\zeta_{obs}\cong\alpha$. Elsewhere, for  $|\zeta_{obs}-\alpha|>\rho /2$
	       with $\rho$ the opening angle of the PC emission cone, no PC emission is visible from the pulsars and the simulated light curves for those 
	       angles are fitted as flat background emission. This generates the observed yellow flat field in the log-likelihood map.}
\label{RQMapEx}
\end{figure}

A pulsar \emph{phase-plot} as a two-dimensional matrix, containing the pulsar emission at all 
rotational phases (light curve), for all the possible values of $\zeta$, and obtained for the specific set of pulsar 
parameters: period $P$, surface magnetic field $B_\mathrm{G}$, gap width $w$, and $\alpha$. For each of the 
LAT pulsars the pulsar $B_\mathrm{G}$ and $w$ of the various models have been computed as described  in 
\cite{pghg12}.

\begin{table*}
\def\arraystretch{1.2}
\centering
\begin{tabular}{| c || c | c | c | c | c || c | c | c | c | c |}
\hline
& $ \alpha_{PC}$ & $ \alpha_{SG}$  & $ \alpha_{OG}$  & $ \alpha_{OPC}$ & $\alpha_\mathrm{others}$ & $ \zeta_{PC}$  &  $ \zeta_{SG}$  &  $ \zeta_{OG}$  &  $ \zeta_{OPC}$ & $\zeta_\mathrm{others}$\\
& $^\circ$ & $^\circ$  & $^\circ$  & $^\circ$ & $^\circ$& $^\circ$  &  $^\circ$  &  $^\circ$  &  $^\circ$ &  $^\circ$ \\
\hline
\hline
J0007$+$7303 & $ 4^{6}_{4}$  & $ 31^{1}_{1}$  & $ 19^{1}_{1}$  & $ 12^{1}_{1}$  &   & $ 3^{1}_{1}$  & $ 72^{1}_{1}$  & $ 87^{1}_{1}$  & $ 74^{1}_{1}$   & \\
\hline
J0106$+$4855 & $ 88^{1}_{1}$  & $ 90^{1}_{4}$  & $ 20^{2}_{2}$  & $ 4^{3}_{4}$  &   & $ 84^{1}_{1}$  & $ 90^{1}_{2}$  & $ 90^{1}_{1}$  & $ 90^{1}_{1}$   & \\
\hline
J0357$+$3205 & $ 10^{1}_{10}$  & $ 71^{1}_{1}$  & $ 88^{1}_{1}$  & $ 70^{1}_{1}$  &   & $ 3^{1}_{1}$  & $ 26^{2}_{1}$  & $ 81^{1}_{1}$  & $ 71^{1}_{1}$   & \\
\hline
J0622$+$3749 & $ 9^{1}_{9}$  & $ 36^{2}_{5}$  & $ 24^{1}_{2}$  & $ 9^{2}_{1}$  &   & $ 3^{1}_{1}$  & $ 51^{7}_{4}$  & $ 90^{1}_{1}$  & $ 89^{1}_{1}$   & \\
\hline
J0633$+$0632 & $ 25^{1}_{1}$  & $ 69^{3}_{3}$  & $ 82^{1}_{1}$  & $ 73^{1}_{1}$  &   & $ 17^{1}_{1}$  & $ 84^{2}_{2}$  & $ 56^{1}_{1}$  & $ 32^{1}_{1}$   & \\
\hline
J0633$+$1746 & $ 10^{1}_{1}$  & $ 42^{1}_{1}$  & $ 66^{1}_{1}$  & $ 4^{2}_{4}$  &   & $ 4^{1}_{1}$  & $ 51^{1}_{1}$  & $ 90^{1}_{1}$  & $ 84^{1}_{1}$   & $60.0-90.0^{(1)}$\\
\hline
J0734$-$1559 & $ 12^{4}_{1}$  & $ 35^{1}_{2}$  & $ 7^{1}_{1}$  & $ 88^{2}_{1}$  &   & $ 3^{1}_{1}$  & $ 57^{1}_{4}$  & $ 89^{1}_{1}$  & $ 18^{1}_{2}$   & \\
\hline
J1023$-$5746 & $ 7^{1}_{1}$  & $ 69^{2}_{2}$  & $ 45^{1}_{1}$  & $ 61^{1}_{1}$  &   & $ 8^{1}_{1}$  & $ 76^{1}_{1}$  & $ 87^{1}_{1}$  & $ 80^{1}_{1}$   & \\
\hline
J1044$-$5737 & $ 4^{2}_{4}$  & $ 64^{1}_{1}$  & $ 70^{1}_{1}$  & $ 10^{1}_{1}$  &   & $ 9^{1}_{1}$  & $ 51^{1}_{1}$  & $ 81^{1}_{1}$  & $ 76^{1}_{1}$   & \\
\hline
J1135$-$6055 & $ 14^{2}_{3}$  & $ 31^{2}_{7}$  & $ 6^{1}_{6}$  & $ 75^{1}_{1}$  &   & $ 3^{1}_{1}$  & $ 70^{1}_{2}$  & $ 80^{1}_{1}$  & $ 12^{1}_{1}$   & \\
\hline
J1413$-$6205 & $ 8^{1}_{1}$  & $ 61^{1}_{1}$  & $ 18^{1}_{1}$  & $ 49^{1}_{2}$  &   & $ 9^{1}_{1}$  & $ 53^{1}_{1}$  & $ 81^{1}_{1}$  & $ 75^{1}_{1}$   & \\
\hline
J1418$-$6058 & $ 7^{1}_{1}$  & $ 62^{2}_{1}$  & $ 44^{1}_{1}$  & $ 60^{1}_{2}$  &   & $ 8^{1}_{1}$  & $ 77^{1}_{1}$  & $ 83^{1}_{1}$  & $ 84^{1}_{1}$   & \\
\hline
J1429$-$5911 & $ 4^{2}_{4}$  & $ 66^{3}_{3}$  & $ 77^{1}_{1}$  & $ 67^{1}_{1}$  &   & $ 7^{1}_{1}$  & $ 82^{2}_{2}$  & $ 42^{1}_{1}$  & $ 21^{1}_{1}$   & \\
\hline
J1459$-$6053 & $ 14^{1}_{3}$  & $ 36^{1}_{1}$  & $ 79^{1}_{1}$  & $ 78^{1}_{1}$  &   & $ 3^{1}_{1}$  & $ 70^{1}_{1}$  & $ 48^{1}_{1}$  & $ 12^{1}_{1}$   & \\
\hline
J1620$-$4927 & $ 9^{1}_{1}$  & $ 74^{2}_{3}$  & $ 7^{1}_{1}$  & $ 4^{2}_{4}$  &   & $ 7^{1}_{1}$  & $ 18^{2}_{3}$  & $ 89^{1}_{1}$  & $ 81^{1}_{1}$   & \\
\hline
J1732$-$3131 & $ 8^{1}_{1}$  & $ 46^{1}_{1}$  & $ 31^{1}_{1}$  & $ 75^{1}_{1}$  &   & $ 7^{1}_{1}$  & $ 48^{1}_{1}$  & $ 86^{1}_{1}$  & $ 75^{1}_{2}$   & \\
\hline
J1746$-$3239 & $ 10^{1}_{1}$  & $ 76^{2}_{3}$  & $ 27^{2}_{1}$  & $ 8^{1}_{1}$  &   & $ 4^{1}_{1}$  & $ 21^{1}_{4}$  & $ 89^{1}_{1}$  & $ 89^{1}_{1}$   & \\
\hline
J1803$-$2149 & $ 8^{1}_{1}$  & $ 60^{1}_{2}$  & $ 89^{1}_{1}$  & $ 48^{1}_{1}$  &   & $ 9^{1}_{1}$  & $ 61^{1}_{1}$  & $ 81^{1}_{1}$  & $ 77^{1}_{1}$   & $88.0-92.0^{(2)}$\\
\hline
J1809$-$2332 & $ 4^{2}_{4}$  & $ 62^{1}_{1}$  & $ 70^{1}_{1}$  & $ 39^{1}_{1}$  &   & $ 9^{1}_{1}$  & $ 54^{1}_{1}$  & $ 78^{1}_{1}$  & $ 72^{1}_{1}$   & \\
\hline
J1813$-$1246 & $ 4^{2}_{4}$  & $ 40^{3}_{2}$  & $ 8^{1}_{1}$  & $ 7^{1}_{1}$  &   & $ 10^{1}_{1}$  & $ 87^{2}_{1}$  & $ 78^{1}_{1}$  & $ 75^{1}_{1}$   & \\
\hline
J1826$-$1256 & $ 4^{2}_{4}$  & $ 70^{2}_{2}$  & $ 45^{1}_{1}$  & $ 61^{1}_{1}$  &   & $ 7^{1}_{1}$  & $ 82^{1}_{1}$  & $ 89^{1}_{1}$  & $ 84^{1}_{1}$   & \\
\hline
J1836$+$5925 & $ 9^{1}_{1}$  & $ 89^{1}_{1}$  & $ 81^{1}_{1}$  & $ 85^{1}_{1}$  &   & $ 2^{1}_{2}$  & $ 22^{1}_{1}$  & $ 90^{1}_{1}$  & $ 88^{1}_{1}$   & \\
\hline
J1838$-$0537 & $ 10^{1}_{1}$  & $ 59^{1}_{1}$  & $ 2^{5}_{2}$  & $ 8^{1}_{1}$  &   & $ 3^{1}_{1}$  & $ 46^{1}_{1}$  & $ 80^{1}_{1}$  & $ 77^{1}_{1}$   & \\
\hline
J1846$+$0919 & $ 3^{2}_{3}$  & $ 46^{1}_{1}$  & $ 27^{2}_{1}$  & $ 18^{1}_{5}$  &   & $ 10^{1}_{1}$  & $ 45^{1}_{1}$  & $ 90^{1}_{1}$  & $ 87^{2}_{1}$   & \\
\hline
J1907$+$0602 & $ 7^{1}_{1}$  & $ 64^{1}_{1}$  & $ 29^{1}_{1}$  & $ 17^{1}_{1}$  &   & $ 9^{1}_{1}$  & $ 53^{1}_{1}$  & $ 81^{1}_{1}$  & $ 73^{1}_{1}$   & \\
\hline
J1954$+$2836 & $ 7^{1}_{1}$  & $ 60^{1}_{4}$  & $ 40^{1}_{1}$  & $ 20^{1}_{1}$  &   & $ 8^{1}_{1}$  & $ 79^{1}_{1}$  & $ 87^{1}_{1}$  & $ 74^{1}_{1}$   & \\
\hline
J1957$+$5033 & $ 3^{7}_{3}$  & $ 66^{5}_{1}$  & $ 89^{1}_{1}$  & $ 76^{1}_{1}$  &   & $ 4^{1}_{1}$  & $ 24^{4}_{1}$  & $ 84^{1}_{1}$  & $ 71^{1}_{1}$   & \\
\hline
J1958$+$2846 & $ 13^{1}_{1}$  & $ 41^{1}_{1}$  & $ 64^{1}_{1}$  & $ 49^{1}_{2}$  &   & $ 5^{1}_{1}$  & $ 53^{1}_{1}$  & $ 90^{1}_{1}$  & $ 85^{1}_{1}$   & \\
\hline
J2021$+$4026 & $ 15^{1}_{4}$  & $ 89^{1}_{1}$  & $ 2^{4}_{2}$  & $ 7^{1}_{1}$  &   & $ 1^{1}_{1}$  & $ 19^{1}_{1}$  & $ 86^{1}_{1}$  & $ 82^{1}_{1}$   & \\
\hline
J2028$+$3332 & $ 7^{1}_{1}$  & $ 46^{1}_{2}$  & $ 48^{1}_{1}$  & $ 90^{1}_{1}$  &   & $ 7^{1}_{1}$  & $ 51^{1}_{1}$  & $ 89^{1}_{1}$  & $ 85^{3}_{1}$   & \\
\hline
J2030$+$4415 & $ 90^{1}_{1}$  & $ 90^{1}_{2}$  & $ 22^{1}_{1}$  & $ 8^{1}_{1}$  &   & $ 90^{1}_{1}$  & $ 90^{1}_{1}$  & $ 90^{1}_{1}$  & $ 90^{1}_{1}$   & \\
\hline
J2055$+$2539 & $ 9^{1}_{9}$  & $ 70^{2}_{2}$  & $ 89^{1}_{1}$  & $ 89^{1}_{1}$  &   & $ 3^{1}_{1}$  & $ 28^{1}_{2}$  & $ 89^{1}_{1}$  & $ 66^{1}_{1}$   & \\
\hline
J2111$+$4606 & $ 4^{2}_{4}$  & $ 61^{1}_{1}$  & $ 6^{1}_{6}$  & $ 21^{1}_{2}$  &   & $ 9^{1}_{1}$  & $ 51^{1}_{1}$  & $ 81^{1}_{1}$  & $ 72^{1}_{1}$   & \\
\hline
J2139$+$4716 & $ 11^{1}_{1}$  & $ 37^{4}_{3}$  & $ 71^{1}_{1}$  & $ 79^{1}_{1}$  &   & $ 20^{1}_{1}$  & $ 52^{5}_{4}$  & $ 87^{1}_{1}$  & $ 70^{1}_{1}$   & \\
\hline
J2238$+$5903 & $ 90^{1}_{1}$  & $ 90^{1}_{1}$  & $ 86^{1}_{1}$  & $ 75^{1}_{1}$  &   & $ 88^{1}_{1}$  & $ 90^{1}_{1}$  & $ 48^{1}_{1}$  & $ 90^{1}_{1}$   & \\
\hline
\end{tabular}
\centering
\caption{$\alpha$ and $\zeta$ best-fit solutions resulting from the $\gamma$-ray fit of the 33 RQ plus 2 RF pulsars. The last column lists independent $\zeta$ 
estimates found in the literature. Superscript and subscript refer to upper and lower errors, respectively. The errors bigger than 1 correspond to the $3\sigma$ 
statistical error. $^{(1)}$ \cite{cbd+03}; $^{(2)}$ \cite{nr08}}
\label{A_Z_gamma1}
\end{table*}

Let us define the instantaneous co-rotating frame (ICF) as the inertial reference frame instantaneously 
co-rotating with the magnetospheric emission point. The direction of the photon generated at the emission  
point in the pulsar magnetosphere as seen from an observer frame (OF) has been computed according to
\cite{bs10a}, as it follows: 
\emph{(i)} the magnetic field in the OF has been computed as given by the retarded vacuum dipole formula;
\emph{(ii)} the magnetic field in the ICF has been computed by Lorentz transformation of the OF magnetic field;
\emph{(iii)} the direction of the $\gamma$-ray photons in the ICF, $\boldsymbol \eta_\mathrm{ICF}$, has been assumed 
parallel to {\bf B}$_\mathrm{ICF}$;
\emph{(iv)} the direction of the $\gamma$-ray photons in the OF, $\boldsymbol \eta_\mathrm{OF}$ has been computed 
by correcting $\boldsymbol \eta_\mathrm{ICF}$ for the light aberration effect.

We computed the $\gamma$-ray and radio phase-plots of each pulsar for the PC, SG, OG, and OPC 
$\gamma$-ray models and a radio core plus cone model. OPC and OG emission geometries are 
described by the same phase-plot. 
Examples of phase-plots are shown, for all the models, in Figure~\ref{phase-plots}. 

In our computation, each phase-plot has been sampled in $45\times90$ steps in phase and $\zeta$ angle, respectively. 
Phase-plots were produced for every degree in $\alpha$, from 1$^\circ$ to 90$^\circ$.
Given a pulsar phase-plot evaluated for a specific $\alpha$, the light curve observed at a particular $\zeta_{LTC}$ is 
obtained by cutting horizontally across the phase-plot at constant $\zeta_{LTC}$.

A detailed description of  $\gamma$-ray models, radio model, and of the phase-plot generation strategy used in this paper 
can be found in \cite{pghg12}. 

\subsection{Light-curves binning and normalisation}
\label{BinningNorm}
\begin{figure*}[tbp!]
\centering
\includegraphics[width=1\textwidth]{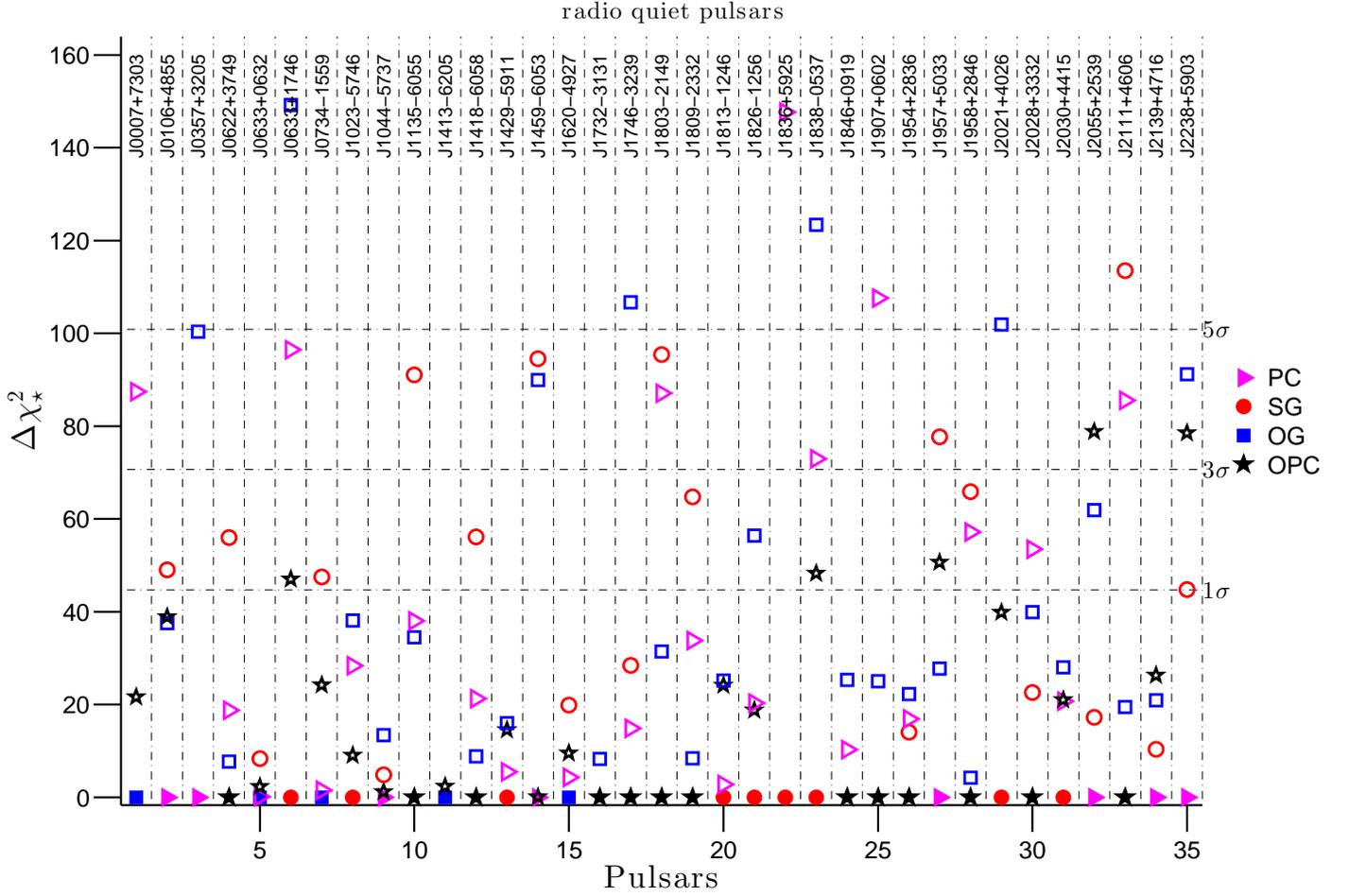}
\caption{Comparison of the relative goodness of the fit solutions obtained for the RQ LAT pulsars between the optimum-model and alternative models. 
The comparison is expressed as the $\Delta \chi^2_\star$ difference between the optimum and alternative model. The horizontal dash-dot lines indicate 
the confidence levels at which to reject a model solution compared to the optimum-solution. Triangles, circles, squares, and stars refer to the PC, SG, OG, 
and OPC models, respectively. Hereafter in all the figures of the paper, filled and empty symbols refer to the solutions of the optimum and alternative 
models, respectively.}
\label{SigGam}
\end{figure*}

The simulated pulsar $\gamma$-ray light curves, generated as described in section \ref{Phase-plotsSection}, are first computed in Regular Binning 
(RBin) where the phase interval 0 to 1 is divided into $N_{bin}$ equal intervals
and the light curve is built counting the photons in each bin.  By fitting between RBin light curves, all the 
phase regions (peak or valleys) have the same statistical weight: in the case of significant pulsed emission over very few bins, 
the fit solution will be strongly dominated by the off-peak level and not by the pulsed emission. 
Since most of the observed LAT light curves exhibit emission concentrated in narrow peaks 
and a wide off-peak or bridge region, we increased the statistical weight of the peak regions by using 
Fixed Count Binning (FCBin) light curves.
In FCBin the size of each phase bin is re-defined in order to contain the same sum of weights per bin, obtained
by dividing the total sum of weights by the total number of bins. 

The simulated $\gamma$-ray light curves, obtained as described in section \ref{Phase-plotsSection}, are computed in arbitrary intensity 
units and do not include background emission modelling. This means that before they are used to fit the LAT profiles, they must be 
scaled to the observed light curves, and a value for the background emission must be added.

Using the FCBin light curves helps the fit to converge to a solution making use of the main morphological information at its disposal: 
the level of pulsed to flat DC emission from the pulsar model and the level of flat background B from Equation~\ref{background}.

Let us define $C$ as the normalisation constant of the simulated light curve. Imposing equality between the total photon count in the observed 
and modelled light curves yields an average constant $C_\mathrm{bar}$ near which the fit solution should converge:
\begin{equation}
\label{normalisation_condition}
\sum_j N_\mathrm{obs,j} = \sum_\mathrm{j} \left( C_\mathrm{bar} \times N_\mathrm{mod,j} + B^{'}_\mathrm{j} \right)
\end{equation}
where $N_\mathrm{mod,j}$ and $N_\mathrm{obs,j}$ are the j-bin values of the simulated and observed FCBin light curves respectively and 
$B^{'}_\mathrm{j}$ is the background emission obtained from the constant background emission $B$ computed in Equation~\ref{background}. 
Prior to being used in Equation \ref{normalisation_condition} both simulated light curve and background emission 
have been re-binned by applying the same binning technique as was used to obtain the observed light curve.

\section{Radio-quiet pulsar $(\alpha,\zeta)$ estimates: $\gamma$-ray fit only}
\label{Individual gamma-ray fit}

\begin{table}[hbp!]
\def\arraystretch{1.2}
\centering
\begin{tabular}{| c | c | c | c | c | c |  }
\hline
 & & PC & SG & OG & OPC \\
\hline
\hline
\multirow{2}{*}{$1 \sigma$}  & $|\Delta \alpha|^{\circ}$ &$1$ & $1$ & $1$ & $1$\\
& $|\Delta\zeta|^{\circ}$   & $1$ & $1$ & $1$ & $1$\\
\multirow{2}{*}{$2 \sigma$}  & $|\Delta \alpha|^{\circ}$ &$2$ & $1$ & $9$ & $28$\\
& $|\Delta\zeta|^{\circ}$   & $1$ & $1$ & $2$ & $9$\\
\hline
\end{tabular}
\caption{Estimate of the systematic errors on $\alpha$ and $\zeta$ obtained from the comparison of the FCBin and
RBin fits.}
\label{SisTab}
\end{table}
\begin{table*}
\def\arraystretch{1.2}
\centering
\begin{tabular}{| c || c | c | c | c | c || c | c | c | c | c |}
\hline
& $ \alpha_{PC}$ & $ \alpha_{SG}$  & $ \alpha_{OG}$  & $ \alpha_{OPC}$ & $\alpha_\mathrm{others}$ & $ \zeta_{PC}$  &  $ \zeta_{SG}$  &  $ \zeta_{OG}$  &  $ \zeta_{OPC}$ & $\zeta_\mathrm{others}$\\
& $^\circ$ & $^\circ$  & $^\circ$  & $^\circ$ & $^\circ$& $^\circ$  &  $^\circ$  &  $^\circ$  &  $^\circ$ & $^\circ$\\
\hline
\hline
J0205$+$6449 & $ 79^{2}_{2}$  & $ 75^{2}_{2}$  & $ 73^{2}_{2}$  & $ 80^{2}_{2}$  &   & $ 89^{2}_{2}$  & $ 86^{2}_{2}$  & $ 90^{2}_{2}$  & $ 89^{2}_{2}$  & $85.7-90^{(2)}$\\
\hline
J0248$+$6021 & $ 6^{2}_{2}$  & $ 60^{2}_{2}$  & $ 56^{2}_{2}$  & $ 55^{2}_{2}$  &   & $ 3^{2}_{2}$  & $ 53^{2}_{2}$  & $ 65^{2}_{2}$  & $ 56^{2}_{2}$  & \\
\hline
J0534$+$2200 & $ 12^{2}_{2}$  & $ 53^{2}_{2}$  & $ 50^{2}_{2}$  & $ 50^{2}_{2}$  &   & $ 14^{2}_{2}$  & $ 74^{2}_{2}$  & $ 74^{2}_{2}$  & $ 73^{2}_{2}$  & $60.1-64.35^{(2)}$\\
\hline
J0631$+$1036 & $ 6^{2}_{2}$  & $ 48^{2}_{2}$  & $ 87^{2}_{2}$  & $ 75^{2}_{2}$  &   & $ 3^{2}_{2}$  & $ 67^{2}_{2}$  & $ 72^{2}_{2}$  & $ 64^{2}_{2}$  & \\
\hline
J0659$+$1414 & $ 9^{2}_{2}$  & $ 30^{2}_{2}$  & $ 78^{2}_{2}$  & $ 66^{2}_{2}$  &   & $ 4^{2}_{2}$  & $ 32^{2}_{2}$  & $ 73^{2}_{2}$  & $ 75^{2}_{2}$  & \\
\hline
J0729$-$1448 & $ 42^{2}_{2}$  & $ 67^{2}_{2}$  & $ 79^{2}_{2}$  & $ 86^{4}_{2}$  &   & $ 42^{2}_{2}$  & $ 78^{2}_{2}$  & $ 84^{2}_{2}$  & $ 71^{2}_{2}$  & \\
\hline
J0742$-$2822 & $ 6^{2}_{2}$  & $ 63^{2}_{2}$  & $ 76^{2}_{3}$  & $ 85^{2}_{2}$  &   & $ 10^{2}_{2}$  & $ 77^{2}_{2}$  & $ 86^{3}_{2}$  & $ 68^{2}_{2}$  & \\
\hline
J0835$-$4510 & $ 3^{2}_{3}$  & $ 45^{2}_{2}$  & $ 71^{2}_{2}$  & $ 56^{2}_{2}$  &  $43^{(1)}$/$70^{(3)}$ & $ 4^{2}_{2}$  & $ 69^{2}_{2}$  & $ 83^{2}_{2}$  & $ 77^{2}_{2}$  & $62.95-64.27^{(2)}$\\
\hline
J0908$-$4913 & $ 7^{2}_{2}$  & $ 70^{2}_{2}$  & $ 75^{2}_{2}$  & $ 65^{2}_{2}$  &   & $ 6^{2}_{2}$  & $ 90^{2}_{2}$  & $ 90^{2}_{2}$  & $ 88^{2}_{2}$  & \\
\hline
J0940$-$5428 & $ 6^{2}_{2}$  & $ 55^{4}_{3}$  & $ 52^{3}_{2}$  & $ 49^{2}_{3}$  &   & $ 12^{3}_{2}$  & $ 56^{4}_{6}$  & $ 62^{2}_{2}$  & $ 55^{2}_{2}$  & \\
\hline
J1016$-$5857 & $ 7^{2}_{7}$  & $ 57^{2}_{2}$  & $ 69^{2}_{2}$  & $ 65^{2}_{2}$  &   & $ 9^{2}_{2}$  & $ 70^{2}_{2}$  & $ 82^{2}_{2}$  & $ 56^{2}_{2}$  & \\
\hline
J1019$-$5749 & $ 6^{2}_{2}$  & $ 14^{2}_{4}$  & $ 83^{2}_{2}$  & $ 83^{2}_{2}$  &   & $ 4^{2}_{2}$  & $ 6^{3}_{3}$  & $ 86^{2}_{2}$  & $ 86^{2}_{2}$  & \\
\hline
J1028$-$5819 & $ 7^{2}_{2}$  & $ 73^{2}_{2}$  & $ 82^{2}_{2}$  & $ 90^{2}_{2}$  &   & $ 7^{2}_{2}$  & $ 83^{2}_{2}$  & $ 87^{2}_{2}$  & $ 89^{2}_{2}$  & \\
\hline
J1048$-$5832 & $ 6^{2}_{2}$  & $ 62^{2}_{2}$  & $ 87^{2}_{2}$  & $ 87^{2}_{2}$  &   & $ 8^{2}_{2}$  & $ 74^{2}_{2}$  & $ 76^{2}_{2}$  & $ 73^{2}_{2}$  & \\
\hline
J1057$-$5226 & $ 10^{2}_{2}$  & $ 46^{2}_{2}$  & $ 77^{2}_{2}$  & $ 73^{2}_{2}$  &   & $ 7^{2}_{2}$  & $ 45^{2}_{2}$  & $ 87^{2}_{2}$  & $ 73^{2}_{2}$  & \\
\hline
J1105$-$6107 & $ 26^{2}_{2}$  & $ 71^{2}_{2}$  & $ 66^{2}_{2}$  & $ 65^{2}_{2}$  &   & $ 39^{2}_{2}$  & $ 85^{2}_{2}$  & $ 81^{2}_{2}$  & $ 82^{2}_{2}$  & \\
\hline
J1112$-$6103 & $ 15^{2}_{2}$  & $ 45^{2}_{2}$  & $ 64^{2}_{2}$  & $ 62^{2}_{2}$  &   & $ 5^{2}_{2}$  & $ 38^{2}_{3}$  & $ 77^{2}_{2}$  & $ 77^{2}_{2}$  & \\
\hline
J1119$-$6127 & $ 9^{2}_{2}$  & $ 55^{2}_{2}$  & $ 74^{2}_{2}$  & $ 61^{2}_{2}$  &   & $ 7^{2}_{2}$  & $ 52^{2}_{2}$  & $ 68^{2}_{2}$  & $ 53^{2}_{2}$  & \\
\hline
J1124$-$5916 & $ 90^{2}_{2}$  & $ 84^{2}_{2}$  & $ 83^{2}_{2}$  & $ 84^{2}_{2}$  &   & $ 89^{2}_{2}$  & $ 89^{2}_{2}$  & $ 88^{2}_{2}$  & $ 89^{2}_{2}$  & $68.0-82.0^{(2)}$\\
\hline
J1357$-$6429 & $ 3^{2}_{3}$  & $ 50^{2}_{2}$  & $ 55^{2}_{2}$  & $ 49^{2}_{2}$  &   & $ 8^{2}_{2}$  & $ 54^{2}_{2}$  & $ 60^{2}_{2}$  & $ 54^{2}_{2}$  & \\
\hline
J1410$-$6132 & $ 7^{2}_{7}$  & $ 19^{2}_{4}$  & $ 87^{2}_{2}$  & $ 75^{2}_{2}$  &   & $ 9^{2}_{2}$  & $ 6^{2}_{2}$  & $ 76^{2}_{2}$  & $ 86^{2}_{2}$  & \\
\hline
J1420$-$6048 & $ 11^{2}_{2}$  & $ 52^{2}_{2}$  & $ 55^{2}_{2}$  & $ 55^{2}_{2}$  &   & $ 5^{2}_{2}$  & $ 53^{2}_{2}$  & $ 57^{2}_{2}$  & $ 52^{2}_{2}$  & \\
\hline
J1509$-$5850 & $ 10^{2}_{2}$  & $ 46^{2}_{2}$  & $ 85^{2}_{2}$  & $ 56^{2}_{2}$  &   & $ 6^{2}_{2}$  & $ 66^{2}_{2}$  & $ 76^{2}_{2}$  & $ 65^{2}_{2}$  & \\
\hline
J1513$-$5908 & $ 30^{2}_{2}$  & $ 50^{2}_{2}$  & $ 60^{2}_{2}$  & $ 45^{2}_{2}$  &   & $ 26^{2}_{2}$  & $ 54^{2}_{2}$  & $ 59^{2}_{2}$  & $ 55^{2}_{2}$  & \\
\hline
J1648$-$4611 & $ 15^{2}_{2}$  & $ 60^{2}_{2}$  & $ 69^{2}_{2}$  & $ 69^{2}_{2}$  &   & $ 11^{2}_{2}$  & $ 56^{2}_{2}$  & $ 67^{2}_{2}$  & $ 67^{2}_{2}$  & \\
\hline
J1702$-$4128 & $ 8^{2}_{2}$  & $ 56^{2}_{2}$  & $ 63^{2}_{2}$  & $ 56^{2}_{2}$  &   & $ 6^{2}_{2}$  & $ 59^{2}_{2}$  & $ 62^{2}_{2}$  & $ 59^{2}_{2}$  & \\
\hline
J1709$-$4429 & $ 11^{2}_{2}$  & $ 42^{2}_{2}$  & $ 73^{2}_{2}$  & $ 46^{2}_{2}$  &   & $ 3^{2}_{2}$  & $ 63^{2}_{2}$  & $ 72^{2}_{2}$  & $ 63^{2}_{2}$  & $49.0-57.8^{(2)}$\\
\hline
J1718$-$3825 & $ 16^{4}_{2}$  & $ 45^{2}_{2}$  & $ 80^{2}_{2}$  & $ 49^{2}_{2}$  &   & $ 3^{2}_{2}$  & $ 65^{2}_{2}$  & $ 55^{2}_{2}$  & $ 61^{2}_{2}$  & \\
\hline
J1730$-$3350 & $ 16^{2}_{2}$  & $ 60^{2}_{2}$  & $ 79^{2}_{2}$  & $ 60^{2}_{2}$  &   & $ 11^{2}_{2}$  & $ 63^{2}_{2}$  & $ 68^{2}_{2}$  & $ 62^{2}_{2}$  & \\
\hline
J1741$-$2054 & $ 3^{2}_{3}$  & $ 31^{2}_{2}$  & $ 84^{2}_{2}$  & $ 72^{2}_{2}$  &   & $ 4^{2}_{2}$  & $ 26^{2}_{2}$  & $ 90^{2}_{2}$  & $ 76^{2}_{2}$  & \\
\hline
J1747$-$2958 & $ 8^{2}_{2}$  & $ 56^{2}_{2}$  & $ 87^{2}_{2}$  & $ 56^{2}_{2}$  &   & $ 7^{2}_{2}$  & $ 77^{2}_{2}$  & $ 79^{2}_{2}$  & $ 77^{2}_{2}$  & \\
\hline
J1801$-$2451 & $ 16^{2}_{2}$  & $ 81^{2}_{2}$  & $ 74^{2}_{2}$  & $ 74^{2}_{2}$  &   & $ 11^{2}_{2}$  & $ 74^{2}_{2}$  & $ 85^{2}_{2}$  & $ 78^{2}_{2}$  & \\
\hline
J1833$-$1034 & $ 86^{2}_{2}$  & $ 55^{2}_{2}$  & $ 65^{2}_{2}$  & $ 89^{2}_{2}$  &   & $ 81^{2}_{2}$  & $ 75^{2}_{2}$  & $ 87^{2}_{2}$  & $ 66^{2}_{2}$  &  $85.1-85.6^{(2)}$\\
\hline
J1835$-$1106 & $ 7^{2}_{2}$  & $ 67^{2}_{2}$  & $ 74^{2}_{6}$  & $ 86^{4}_{2}$  &   & $ 6^{2}_{2}$  & $ 61^{2}_{2}$  & $ 89^{2}_{2}$  & $ 72^{3}_{2}$  & \\
\hline
J1952$+$3252 & $ 11^{2}_{2}$  & $ 51^{2}_{2}$  & $ 65^{2}_{2}$  & $ 65^{2}_{2}$  &   & $ 9^{2}_{2}$  & $ 80^{2}_{2}$  & $ 86^{2}_{2}$  & $ 83^{2}_{2}$  & \\
\hline
J2021$+$3651 & $ 7^{2}_{2}$  & $ 73^{2}_{2}$  & $ 68^{2}_{2}$  & $ 84^{2}_{2}$  &   & $ 7^{2}_{2}$  & $ 83^{2}_{2}$  & $ 90^{2}_{2}$  & $ 88^{2}_{2}$  & $76.0-82.0^{(2)}$\\
\hline
J2030$+$3641 & $ 8^{2}_{2}$  & $ 60^{2}_{2}$  & $ 87^{2}_{2}$  & $ 67^{2}_{2}$  &   & $ 8^{2}_{2}$  & $ 65^{2}_{2}$  & $ 77^{2}_{2}$  & $ 68^{2}_{2}$  & \\
\hline
J2032$+$4127 & $ 16^{2}_{2}$  & $ 41^{2}_{2}$  & $ 59^{2}_{2}$  & $ 65^{2}_{2}$  &   & $ 7^{2}_{2}$  & $ 54^{2}_{2}$  & $ 60^{2}_{2}$  & $ 89^{2}_{2}$  & \\
\hline
J2043$+$2740 & $ 6^{2}_{6}$  & $ 59^{6}_{5}$  & $ 76^{2}_{3}$  & $ 66^{2}_{2}$  &   & $ 9^{2}_{2}$  & $ 79^{4}_{3}$  & $ 88^{2}_{2}$  & $ 87^{2}_{2}$  & \\
\hline
J2229$+$6114 & $ 4^{2}_{4}$  & $ 42^{2}_{2}$  & $ 75^{2}_{2}$  & $ 65^{2}_{2}$  &   & $ 3^{2}_{2}$  & $ 62^{2}_{2}$  & $ 55^{2}_{2}$  & $ 55^{2}_{2}$  & $38.0-54.0^{(2)}$\\
\hline
J2240$+$5832 & $ 13^{3}_{2}$  & $ 67^{5}_{5}$  & $ 70^{3}_{3}$  & $ 71^{2}_{2}$  &   & $ 4^{2}_{2}$  & $ 88^{2}_{3}$  & $ 89^{2}_{3}$  & $ 88^{2}_{4}$  & \\
\hline
\end{tabular}
\centering
\caption{$\alpha$ and $\zeta$ best-fit solutions resulting from the joint radio plus $\gamma$-ray fit of the 41 RL pulsars. The central and last columns list independent 
$\alpha$ and $\zeta$ estimates, found in the literature, respectively. Superscript and subscript refer to upper and lower errors, respectively.
The errors bigger than 2 correspond to  the $3\sigma$ statistical error. 
$^{(1)}$ \cite{jhv+05}; $^{(2)}$ \cite{nr08}; $^{(3)}$ $\alpha =\zeta+6.5$ found by \cite{jhv+05} with $\zeta\sim63.5$ from \cite{nr08}.}
\label{JointAlpZetFit}
\end{table*}
We have used the PC, SG, and OG/OPC phase-plots and a $\chi^2$ estimator to fit the LAT pulsar $\gamma$-ray light curves sampled with RBin 
and FCBin in phase. The free parameters of the fits are: the $\alpha$ and $\zeta$ angles, both sampled every 
degree in the interval 1$^\circ$ to 90$^\circ$; the final light-curve normalisation factor $C$ sampled every $0.1C_\mathrm{bar}$ in the interval 
$0.5C_\mathrm{bar}$ to $1.5C_\mathrm{bar}$ with $C_\mathrm{bar}$ from Equation \ref{normalisation_condition}; the light-curve phase shift $\phi$, sampled in 45 steps between 0 and 1. 

For each type of light-curve binning, we have obtained a log-likelihood matrix of dimension $90_\alpha \times 90_\zeta \times 45_{\phi} \times 11_{norm}$. 
Maximising the matrix over $\phi$ and $C$ yields the $\alpha$-$\zeta$ log-likelihood map.
The location and shape of the maximum in this map give the best-fit estimates on $\alpha$ and $\zeta$ and their errors.
An example of  $\alpha$-$\zeta$ log-likelihood map is given in Figure \ref{RQMapEx} for the pulsar J1023$-$5746. The corresponding best-fit 
$\gamma$-ray light curve is shown in Figure \ref{fitGm7}.
The comparison of the set of solutions obtained with the two light-curve binning modes shows that FCBin best matches the sharp peak structures of the 
observed profiles because of the higher density of bins across the peaks. Hereafter the $\alpha$ and $\zeta$ estimates given for RQ pulsars are those obtained 
with FCBin light curves. They are listed with their respective statistical errors in Table \ref{A_Z_gamma1}.

In order to estimate the systematic errors on the derivation of $(\alpha,\zeta)$ due to  
the choice of fitting method, we have compared the sets of solutions obtained with the FCBin and RBin light curves. 
Their cumulative distributions give the errors at the 1$\sigma$ and 2$\sigma$ confidence levels displayed in Table \ref{SisTab}.
Because OG and OPC models predict sharp peaks and no off-pulse emission, we expect the differences between $\alpha$ and $\zeta$ 
obtained with RBin and FCBin light curves to be the largest with these models. It explains their large 2$\sigma$ values in Table \ref{SisTab}. 
The results in Table \ref{SisTab} most importantly show that the fitting method itself yields an uncertainty of a few 
degrees at least on $\alpha$ and $\zeta$. It is generally much larger than the statistical errors derived from the log-likelihood map. For this 
reason we have set a minimum error of 1$^{\circ}$ in Table \ref{A_Z_gamma1}.

Figure \ref{SigGam} compares, for each pulsar, the relative goodness of the fit solutions obtained with the different models. The light curves from the modelled phase-plots 
can reproduce the bulk shape of the observed light curves, but not the fine details. Furthermore, the observed light curves having a large number of counts have very small 
errors. So the reduced $\chi^2$ values of the best fits remain 
large because the errors on the data are small compared to the model variance. On the other hand, the figures in Appendix \ref{GammaFitRes} show that the optimum-models 
reasonably describe the light-curve patterns in most cases. To quantify the relative merits of the models, we have therefore set the model variance in order 
to achieve a reduced $\chi^2_\star$ of 1 for the optimum-model. This variance has then been used to calculate the $\chi^2_\star$ value of other model solutions 
and to derive the $\Delta \chi^2_\star$ difference between the optimum-model and any other model. In Appendix \ref{GoodFitMethod}, we show how to relate the 
original log-likelihood values obtained for each fit, given in Table \ref{Like_gamma1}, and the $\Delta \chi^2_\star$ differences between models.

The $\Delta \chi^2_\star$ difference is plotted in Figure \ref{SigGam} for each pulsar and each non-optimum-model. The $\chi^2$ probability density function for 
the 41 degrees of freedom of the fits gives us the confidence levels above which the alternative models are significantly worse. The levels are labelled on the plot. 
The results indicate that one or two models can be rejected for nearly half the pulsars, but we see no systematic trend against a particular model. We also note 
that the geometrically similar OG and OPC models give significantly different solutions in several instances. This is because the gap width evolves differently in 
the two models.

\begin{figure*}
\centering
\includegraphics[width=0.52\textwidth]{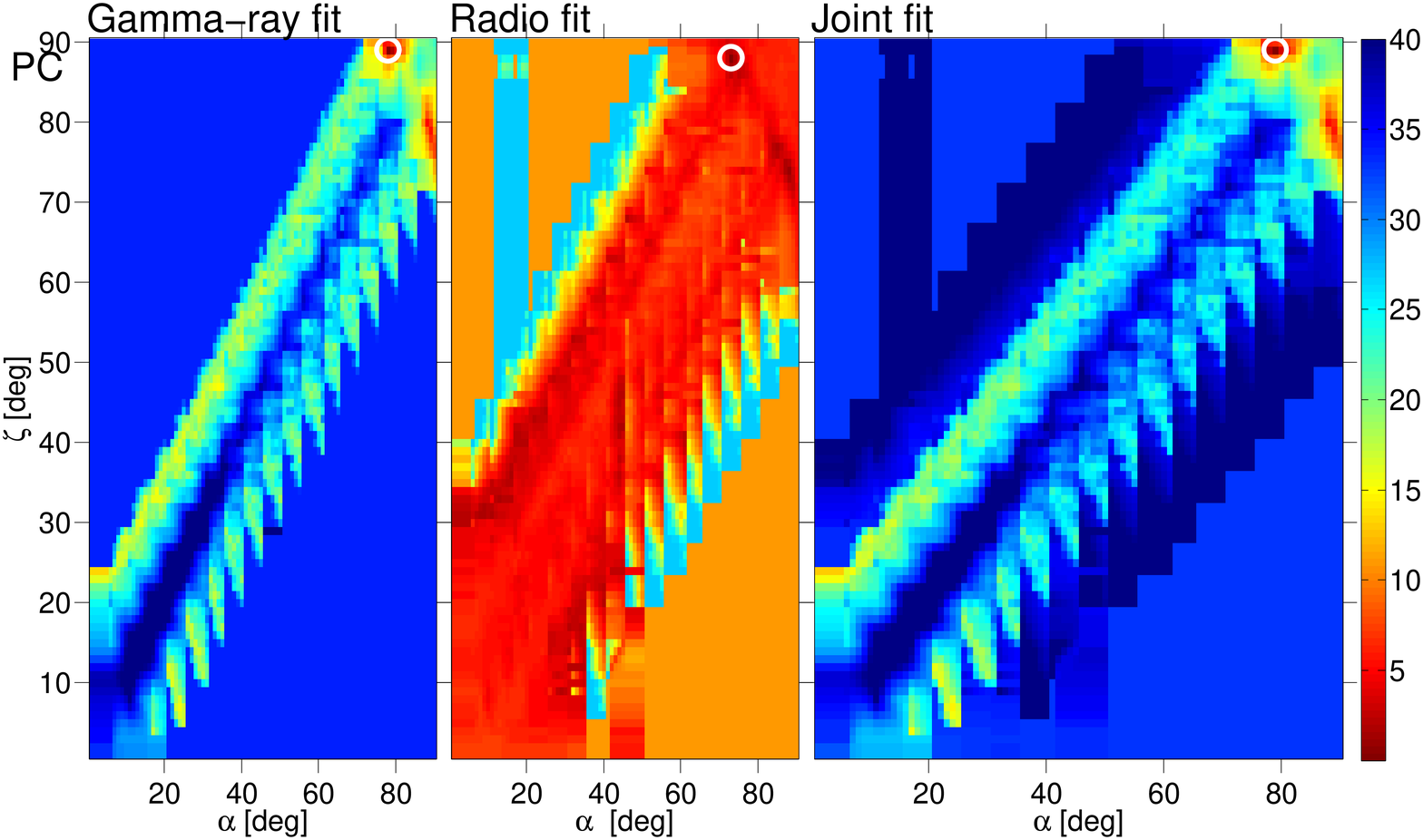}
\includegraphics[width=0.52\textwidth]{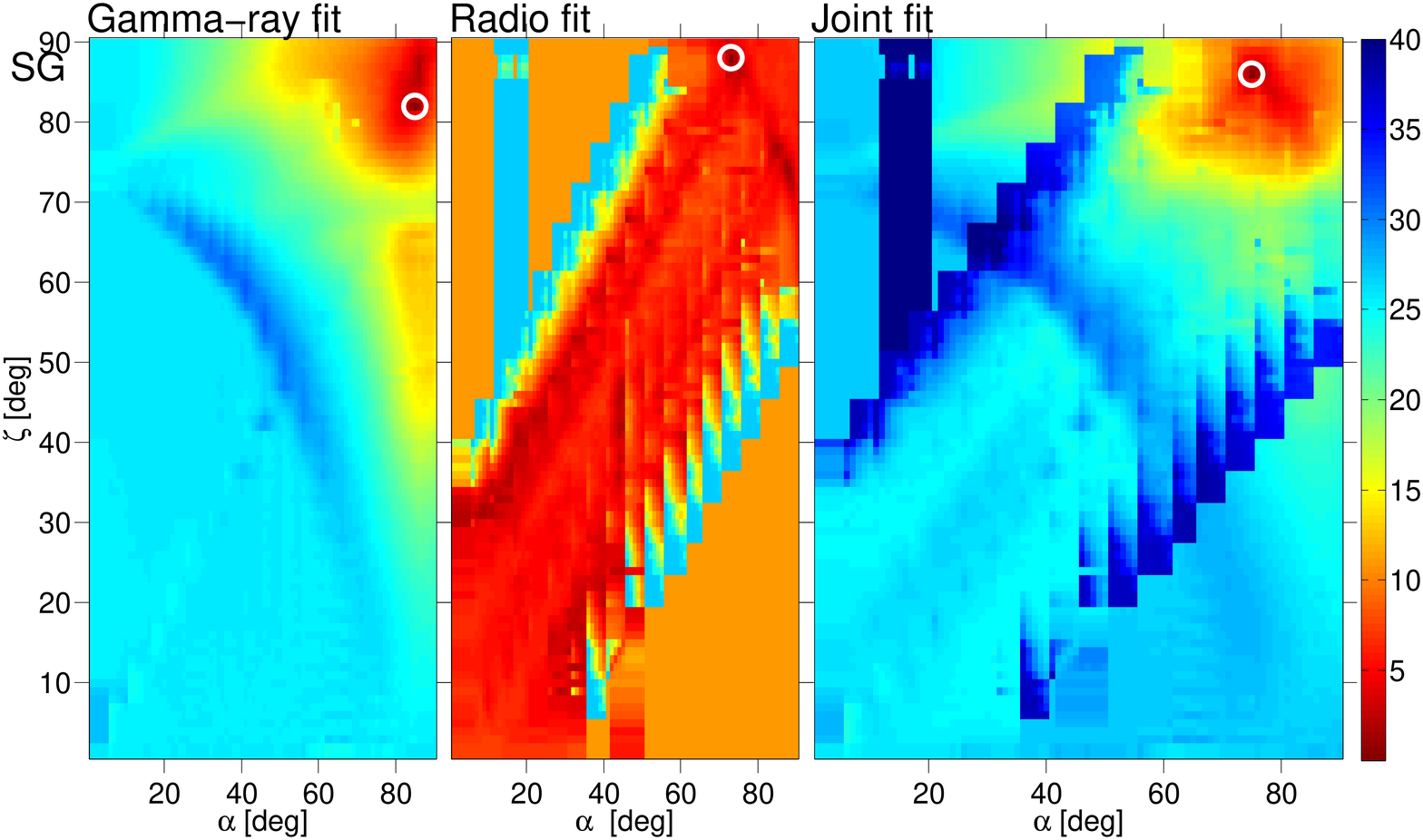}
\includegraphics[width=0.52\textwidth]{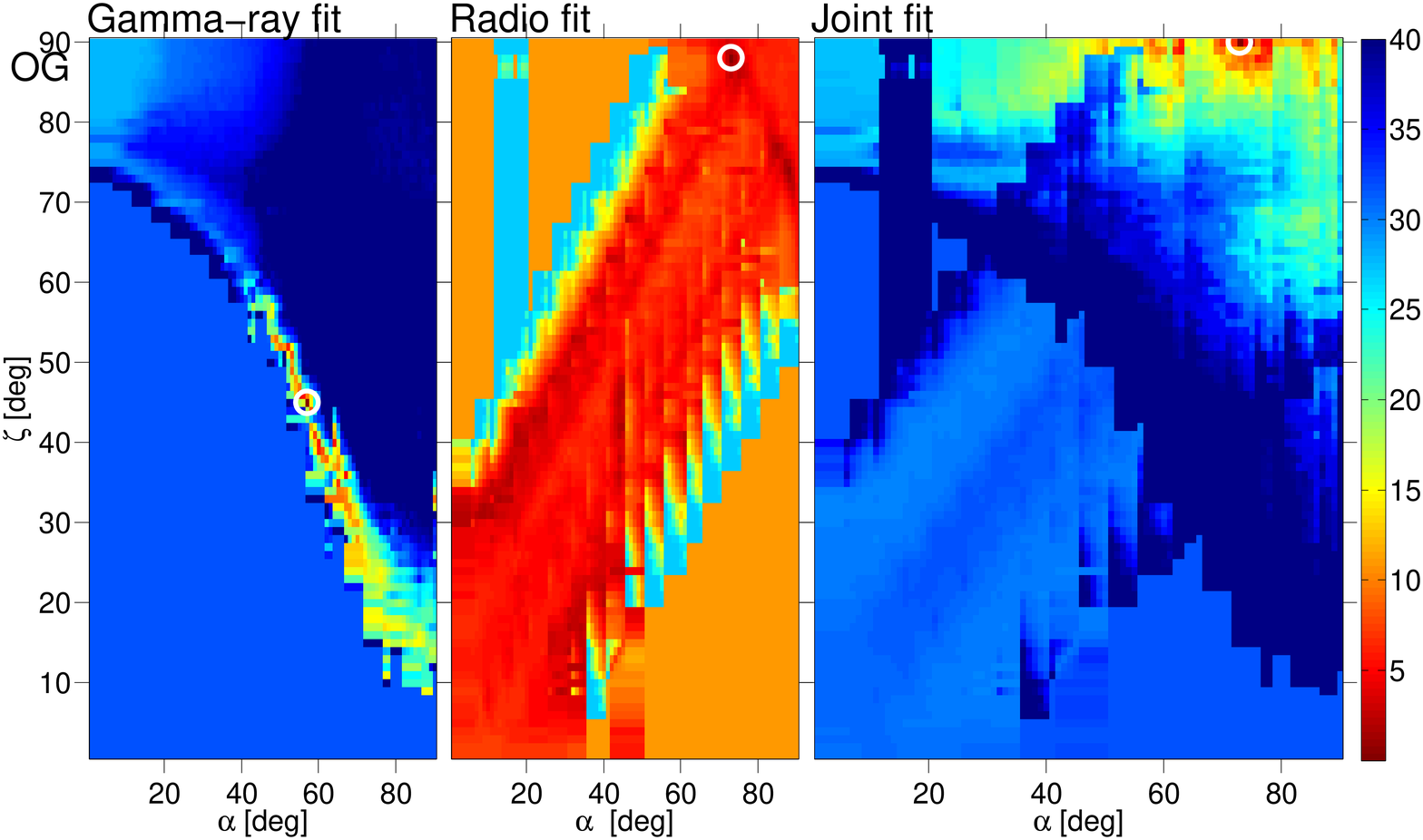}
\includegraphics[width=0.52\textwidth]{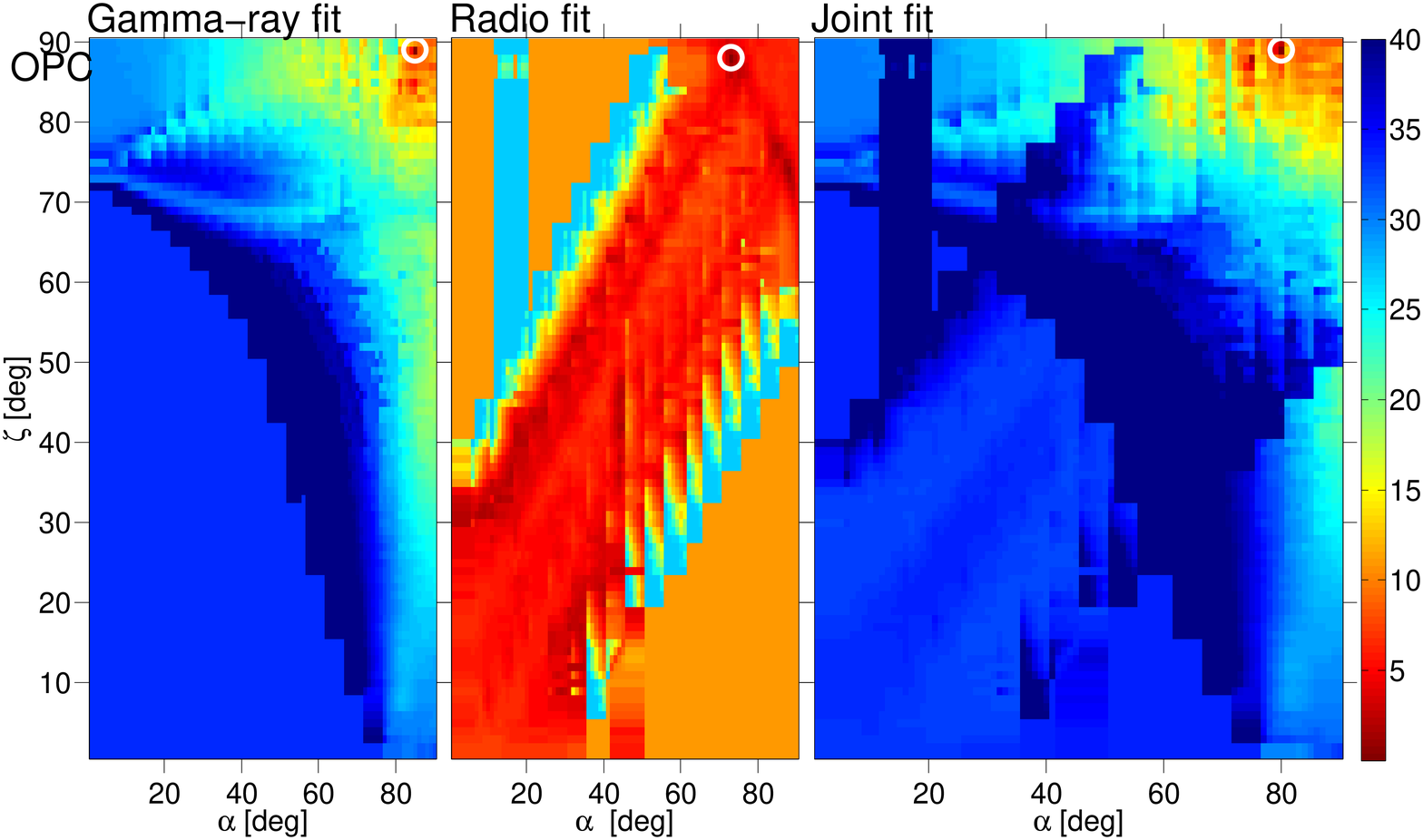}
\caption{For each model the $(\alpha$,$\zeta)$ log-likelihood map for the $\gamma$-ray  
fit, the radio  fit, and the sum of these two maps for pulsar  J0205$+$6449 is shown. A white circle shows the position of the 
best fit solution for each log-likelihood map. The colour-bar is in effective $\sigma=(| \ln L -\ln L_{max}|)^{0.5}$  units, zero corresponds to the 
best-fit solution.}
\label{RLMapEx}
\end{figure*}
\begin{figure*}
\centering
\includegraphics[width=1\textwidth]{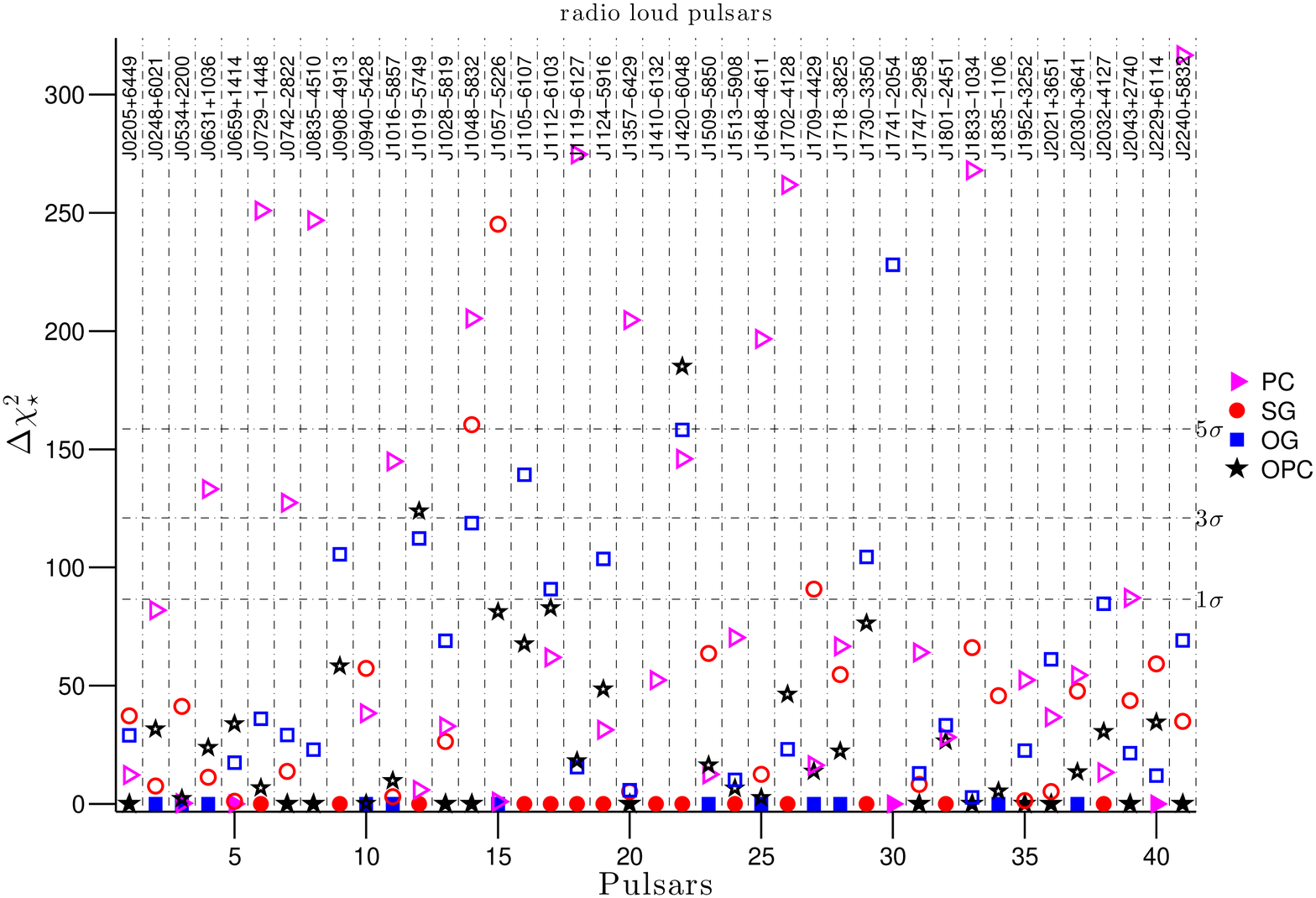}
\caption{Comparison of the relative goodness of the fit solutions obtained for the RL LAT pulsars between the optimum-model and alternative models. The comparison is 
expressed as the $\Delta \chi^2_\star$ difference between the optimum and alternative model. The horizontal dash-dot lines indicate the confidence levels at which to 
reject a model solution compared to the optimum-solution. Triangles, circles, squares, and stars refer to the PC, SG, OG, and OPC models, respectively.}
\label{SigJoint}
\end{figure*}

\section{Radio-loud pulsar $(\alpha,\zeta)$ estimates: fitting both the $\gamma$-ray and radio emission}
\label{Fitting both the gamma-ray and radio emission}

The strategy we have adopted to jointly fit radio and $\gamma$-ray profiles consists of summing the log-likelihood maps obtained
by fitting the radio and $\gamma$-ray light curves individually. Because of the much larger signal-to-noise ratio in the radio than 
in $\gamma$-rays, and since the $\gamma$-ray and radio models are equally uncertain, the radio log-likelihood map is more 
constraining and the joint fit is largely dominated by the radio-only solution. 
To lower the weight of the radio fit and make it comparable with the $\gamma$-ray fit we have implemented a two-step strategy: 
we have first fitted the radio profiles by using a standard deviation evaluated from the relative uncertainty in the $\gamma$-ray light curve. 
We have then used the best-fit light curves of this first fit to evaluate an optimised standard deviation in the radio and use it to fit again the radio light curves. 
A detailed description of the joint fit technique is given in Sections \ref{Individual radio fit} and \ref{JointFit}.

\subsection{Radio fit only}
\label{Individual radio fit}

We have implemented a fit of the RBin radio profiles using 5 free parameters, the same four defined in Section 
\ref{Individual gamma-ray fit}, $\alpha$, $\zeta$, phase shift $\phi$, and normalisation factor, equally stepped 
in the same intervals, plus a  flat background emission level sampled in 16 steps over an interval that includes 
the averaged minimum of the observed light curve.

The first fit is done with the standard deviation $\sigma_{peak}$ evaluated as the average relative $\gamma$-ray 
uncertainty in the on-peak region times the maximum radio intensity value  \citep{jvhg11,vjh12}. Hereafter we will 
refer to this first fit as the $\sigma_{peak}$ radio fit.
The second fit is implemented by using a standard deviation value evaluated from the best-fit results of the first fit, on the 
basis of a reduced $\chi^2=1$ criterion. 

Let us define $N^{*}_\mathrm{mod}$ the best fit light curve obtained in the first step which yields a maximum log-likelihood:
\begin{equation}
\label{opt1}
\ln L_{max}=-\frac{1}{2\sigma^2_{\gamma-peak}}\sum_j [N_\mathrm{obs,j}-N^{*}_\mathrm{mod,j}]^2
\end{equation}
{with $L_{max}$ function of the best-fit $\alpha$ and $\zeta$ obtained from the first fit.}
By making use of the reduced $\chi^2$=1 criterion, Equation \ref{opt1} gives
\begin{equation}
\label{RedChi2}
\frac{1}{n_{free}}\sum_j \frac{[N_\mathrm{obs,j}-N^{*}_\mathrm{mod,j}]^2}{\sigma_*^2}=1,
\end{equation}
where $n_{free}=(n_{bin}-5)$ is the number of the free parameters and $\sigma_{*}$ is the newly optimised value for 
the standard deviation. Combining Equations \ref{opt1} and \ref{RedChi2} yields
\begin{equation}
\label{SigmaOptRadio}
\sigma_{*}^2=-\frac{2\ln L_{max}}{n_{free}}\sigma_{peak}^2.
\end{equation}
The new optimised $\sigma_{*}$ is a function of the $\alpha$ and $\zeta$ solutions obtained in the first step. It has been used to implement a 
new fit of the radio light curves, hereafter the $\sigma_{*}$ radio fit.

\subsection{Joint $\gamma$-ray plus radio estimate of the LAT pulsar orientations}
\label{JointFit}

Since the radio and $\gamma$-ray emissions occur simultaneously and independently, and since the 
$\gamma$-ray and radio log-likelihood maps have been evaluated in a logarithmic scale for the same free parameters, the joint
$(\alpha$,$\zeta)$ log-likelihood map is obtained by summing the $\gamma$-ray and radio maps.  

We have summed the $\gamma$-ray log-likelihood maps, evaluated by fitting FCBin light curves 
(Section \ref{Individual gamma-ray fit}), with the radio log-likelihood maps, evaluated by fitting RBin 
light curves (Section \ref{Individual radio fit}) with either $\sigma_{peak}$ or $\sigma_{*}$.
Among the two sets of solutions obtained for each pulsar, we have selected the solution characterised by the highest final log-likelihood value.
An example of  a joint $\gamma$-ray plus radio $\alpha$-$\zeta$ estimate is given in Figure \ref{RLMapEx} for the pulsar J0205$+$6449.
The corresponding best-fit light curves are shown in Figure \ref{fitJoint_GmR1}. The log-likelihood values $L$ of the final results are listed in Table \ref{JointLike}.

Because of statistical fluctuations, and/or the difference in the radio and $\gamma$-ray profile accuracy, and/or the inadequacy of the 
assumed emission geometries to describe the data, the $(\alpha$,$\zeta)$ solutions obtained from the joint fit did not always supply 
both radio and $\gamma$-ray emission at those angles. In those cases, the next highest log-likelihood $(\alpha$,$\zeta)$ 
solution with non-zero radio and $\gamma$-ray pulsed emission was chosen. For some light curves with low statistics and/or signal-to-noise ratio, 
the joint fit method found a flat light curve as the best solution for the SG model. This is the case for pulsars J0729$-$1448, 
J1112$-$6103, J1801$-$2451, and J1835$-$1106. For those, we have selected the non-flat light curve with the highest log-likelihood value as the SG solution.  

Table \ref{JointAlpZetFit} lists the ($\alpha,\zeta$) estimates obtained for the RL pulsars from the optimised $\sigma_*$ fit. 
Since the estimates are obtained by merging two 1$^{\circ}$ resolution log-likelihood maps, 
we conservatively assign a minimum statistical error of 2$^{\circ}$. As for RQ pulsars in section \ref{Individual gamma-ray fit}, we compare in 
Figure \ref{SigJoint} the relative goodness of the fits obtained between the optimum-solution and alternative models for the RL pulsars. We have 
derived the $\Delta \chi^2_\star$ difference between two models according to Appendix \ref{GoodFitMethod}, by making use of the log-likelihood obtained
for each fit and listed in Table \ref{JointLike} and for 81 degrees of freedom. 
It shows that the tight additional constraint provided by the radio data forces the fits to converge to rather comparable light-curve shapes, so that the solutions 
often gather within 1$\sigma$ from the optimum-solution. It also shows that the PC model is more often significantly rejected than the other, more widely beamed, 
models.

In order to estimate the systematic errors on the derivation of $\alpha$ and $\zeta$, we have studied how the sets of solutions obtained with the two joint-fit 
methods ($\gamma$-ray fit plus $\sigma_{peak}$ radio fit and $\gamma$-ray fit plus $\sigma_{*}$ radio fit) depart from each other.
Table \ref{SisTabJ} lists the 1$\sigma$ and 2 $\sigma$ systematic errors on $\alpha$ and $\zeta$ for each model. It shows how the joint-fit strategy yields 
uncertainties of few a degrees at least in $\alpha$ and $\zeta$. They largely exceed the statistical errors shown in Table \ref{JointAlpZetFit}. 
 
\begin{table}[h!]
\def\arraystretch{1.2}
\centering
\begin{tabular}{| c | c | c | c | c | c |  }
\hline
 & & PC & SG & OG & OPC \\
\hline
\hline
\multirow{2}{*}{$1 \sigma$}  & $|\Delta \alpha|^{\circ}$ &$4$ & $18$ & $3$ & $5$\\
& $|\Delta\zeta|^{\circ}$   & $12$ & $30$ & $3$ & $1$\\
\multirow{2}{*}{$2 \sigma$}  & $|\Delta \alpha|^{\circ}$ &$49$ & $46$ & $14$ & $18$\\
& $|\Delta\zeta|^{\circ}$   & $51$ & $50$ & $14$ & $9$\\
\hline
\end{tabular}
\caption{Estimate of the systematic errors on $\alpha$ and $\zeta$ obtained from the comparison 
of the two joint fit methods $\gamma$-ray fit plus $\sigma_{peak}$ radio fit and $\gamma$-ray fit plus  $\sigma_{*}$ radio fit.}
\label{SisTabJ}
\end{table}

\section{Results}
\label{Results}
\begin{figure*}[t!]
\begin{center}
\includegraphics[width=1\textwidth]{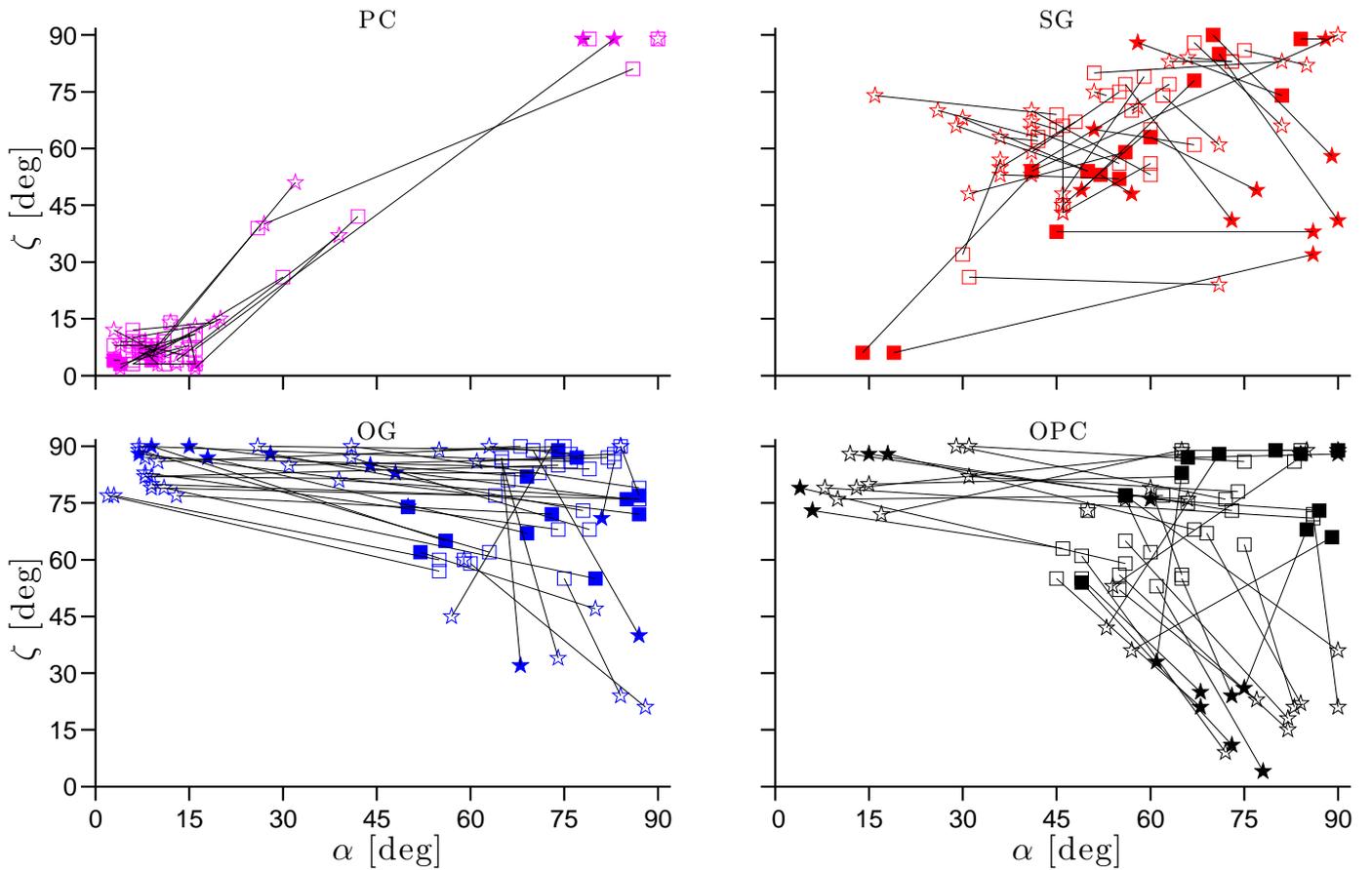}
\caption{Distribution of the $\alpha$-$\zeta$ best-fit solutions obtained, for the RL pulsars sample and in the framework of each model, 
by fitting the $\gamma$-ray light curves alone (stars) and by jointly fitting the $\gamma$-ray and the radio light curves (squares). Recall that filled and empty 
symbols refer to the solutions of the optimum and alternative models, respectively.}
\label{SolutionMigration}
\end{center}
\end{figure*}

For the RQ LAT pulsars, the best-fit light curves obtained by the fits in FCBin mode are shown in Figures \ref{fitGm1} to \ref{fitGm35}. 
while Figures \ref{fitJoint_GmR1}  to \ref{fitJoint_GmR41} show the radio and $\gamma$-ray best-fit light curves obtained from the 
joint $\gamma$-ray plus radio fits for the RL LAT pulsars. 
In Figures \ref{fitJoint_GmR_2_1} and \ref{fitJoint_GmR_2_2} we give the joint radio plus $\gamma$-ray fit results for the RF pulsars 
J0106$+$4855 and J1907$+$0602.
All radio light curves shown in Appendices have been plotted with the errors (optimised standard deviations $\sigma_*$) evaluated 
as described in section \ref{Individual radio fit}. The $\alpha$ and $\zeta$ estimates for RQ and RL pulsars are indicated in Tables 
\ref{A_Z_gamma1} and \ref{JointAlpZetFit} respectively.

In addition to the $\chi ^2$ fits to the FCBin and RBin $\gamma$-ray light curves described above, we have also tested maximum log-likelihood fits with 
Poisson statistics. We have checked that while the individual pulsar $(\alpha,\zeta)$ estimates can change according to the method used,
the collective properties of the LAT pulsar population discussed below, such as the correlation between luminosity and beaming factor with 
$\dot{E}$, are robust and not strongly dependent on the fitting strategy. 

\subsection{Comparison of the $\gamma$-ray geometrical models}
\label{CompModels}

We can compare the merits of the models in terms of frequency of achieving the optimum-model in the sample of LAT pulsar light 
curves. Table \ref{discriminationTab} shows, for each model, the number of optimum-solutions that are better than the other 
models by at least 1$\sigma$ (left) and the number of non-optimum-solutions that are rejected at more than 3$\sigma$ (right). 
We give those counts for the RQ, RL, and all pulsars of the sample. 
Table \ref{discriminationTab} shows that, in the majority of cases, there is no statistically best optimum-solution. In the few cases where 
there are, most are SG and PC and only one is OPC. The PC emission geometry, in general, most poorly describe the observations; the 
PC model is rejected at more than 3$\sigma$ confidence level for almost the 60\% of the RL pulsars and for almost half of the total pulsars 
of the sample. The SG and PC models are 
rejected at more than 3$\sigma$ nearly equally for RQ pulsars. Thus, the outer magnetosphere models, SG, OG and OPC,  overall seem to best 
describe the observed LAT pulsar light curves.
This geometrical trend concurs with the absence of a super-exponential cut-off in the recorded $\gamma$-ray spectra (PSRCAT2) 
to rule out a PC origin of the $\gamma$-ray beam in most of the LAT pulsars, but not all. We note that the RF and RQ pulsars J0106$+$4855 
and J2238$+$5903 respectively, have a PC optimum-solution and that the other models are very strongly rejected. On the other hand, the 
PC optimum-solution obtained for pulsar J2238$+$5903 has $\alpha$ and $\zeta$ angles so close that it should be observed as RL or RF
object and so it is likely to be incorrect, unless the radio emitting zone actually lies at higher altitude than in our present model. In any case, 
$\gamma$-ray beams originating at medium to high altitude in the magnetosphere largely dominate the LAT sample.

\begin{table*}
\def\arraystretch{1.2}
\centering
\begin{tabular}{| c | c | c || c | c || c | c |  }
\hline
\multicolumn{7}{|c|}{optimum-solutions by at least 1$\sigma$} \\
\hline
& \multicolumn{2}{c||}{RQ} & \multicolumn{2}{c||}{RL} & \multicolumn{2}{c|}{RL+RQ}\\
\hline
         & no. & $\%$ &  no. & $\%$  &  no. & $\%$ \\
\hline
PC   & 2               & 40 				& 1              &  33.3     	        & 3            & 37.5 \\
\hline
SG   & 3              & 60  				& 1             &  33.3   		        & 4            &  50.0 \\
\hline
OG   & 0              & 0    				& 0             & 0		       		& 0            &  0   \\
\hline
OPC& 0               & 0	 	 			& 1             &  33.3   			& 1            &  12.5 \\
\hline
\hline
Total& 5            &  	          	                &  3            &     		         	& 8          &   \\
\hline
\end{tabular}
~~~~~~~~~~~~~~~~~~~~~~~~
\centering
\begin{tabular}{| c | c | c || c | c || c | c |  }
\hline
\multicolumn{7}{|c|}{Solutions rejected by more than 3$\sigma$} \\
\hline
& \multicolumn{2}{c||}{RQ} & \multicolumn{2}{c||}{RL} & \multicolumn{2}{c|}{RL+RQ}\\
\hline
         & no. & $\%$ &  no. & $\%$  &  no. & $\%$ \\
\hline
PC   & 10               & 30.3				& 17              &  58     	        & 27            & 44 \\
\hline
SG   & 11              & 33.4  				& 4             &  14   		        & 15            &  24 \\
\hline
OG   & 8              & 24.2   				& 4             & 14	       		& 12            &  19   \\
\hline
OPC& 4               & 12.1 	 			& 4             &  14   			& 8            &  13 \\
\hline
\hline
Total& 33            &  	          	                &  29            &     		         & 62          &   \\
\hline
\end{tabular}

\caption{Left: for each model, the number (and frequency in the sample) of optimum-solutions that yield a better fit than the other models by at least 1$\sigma$. 
Right: for each model, the number and frequency of solutions that are rejected by more than 3$\sigma$ compared to the optimum-model. 
The values are given for the RQ, RL, and total LAT pulsar samples.}
\label{discriminationTab}
\end{table*}

\begin{figure*}
\begin{center}
\includegraphics[width=0.65\textwidth]{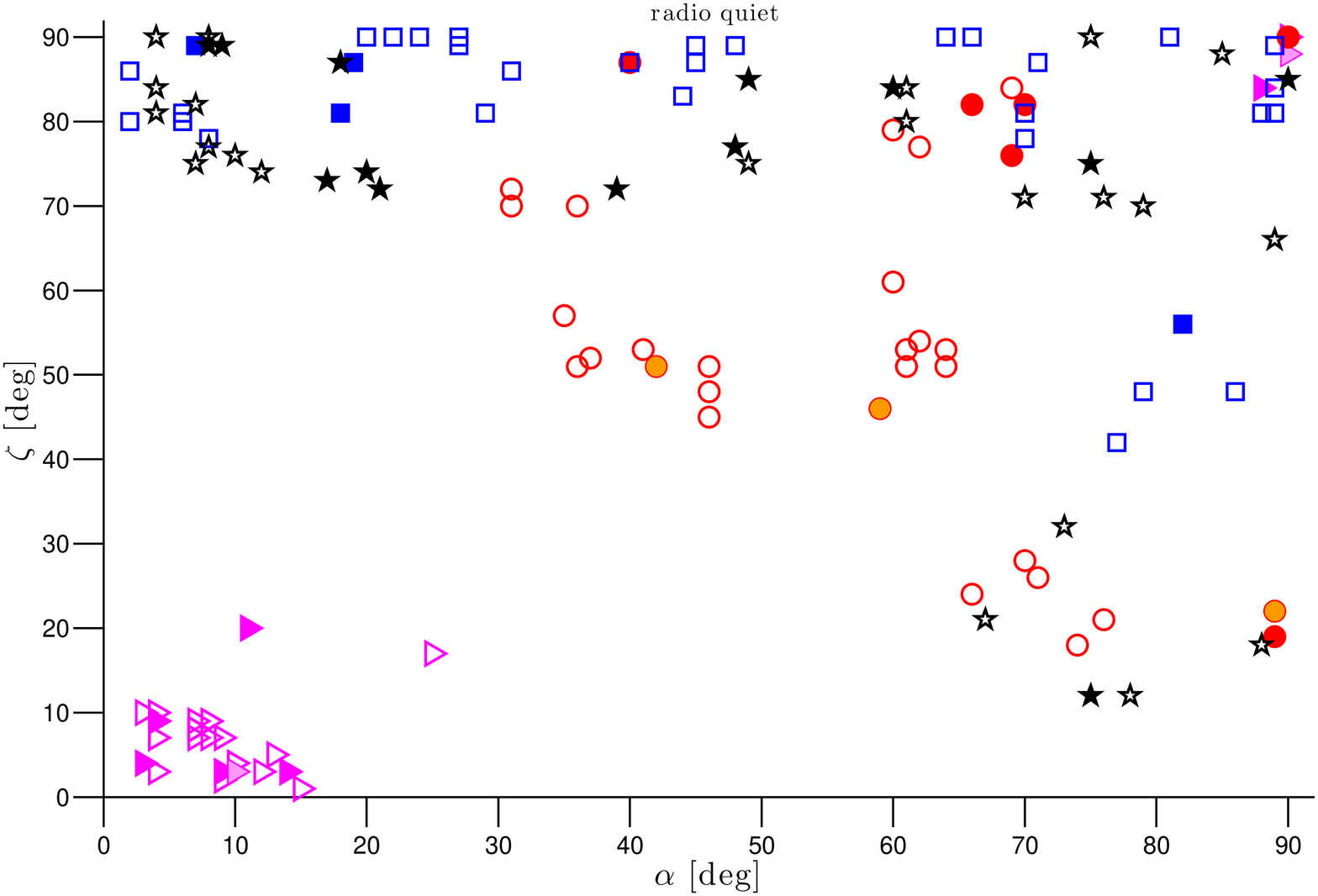}
\includegraphics[width=0.65\textwidth]{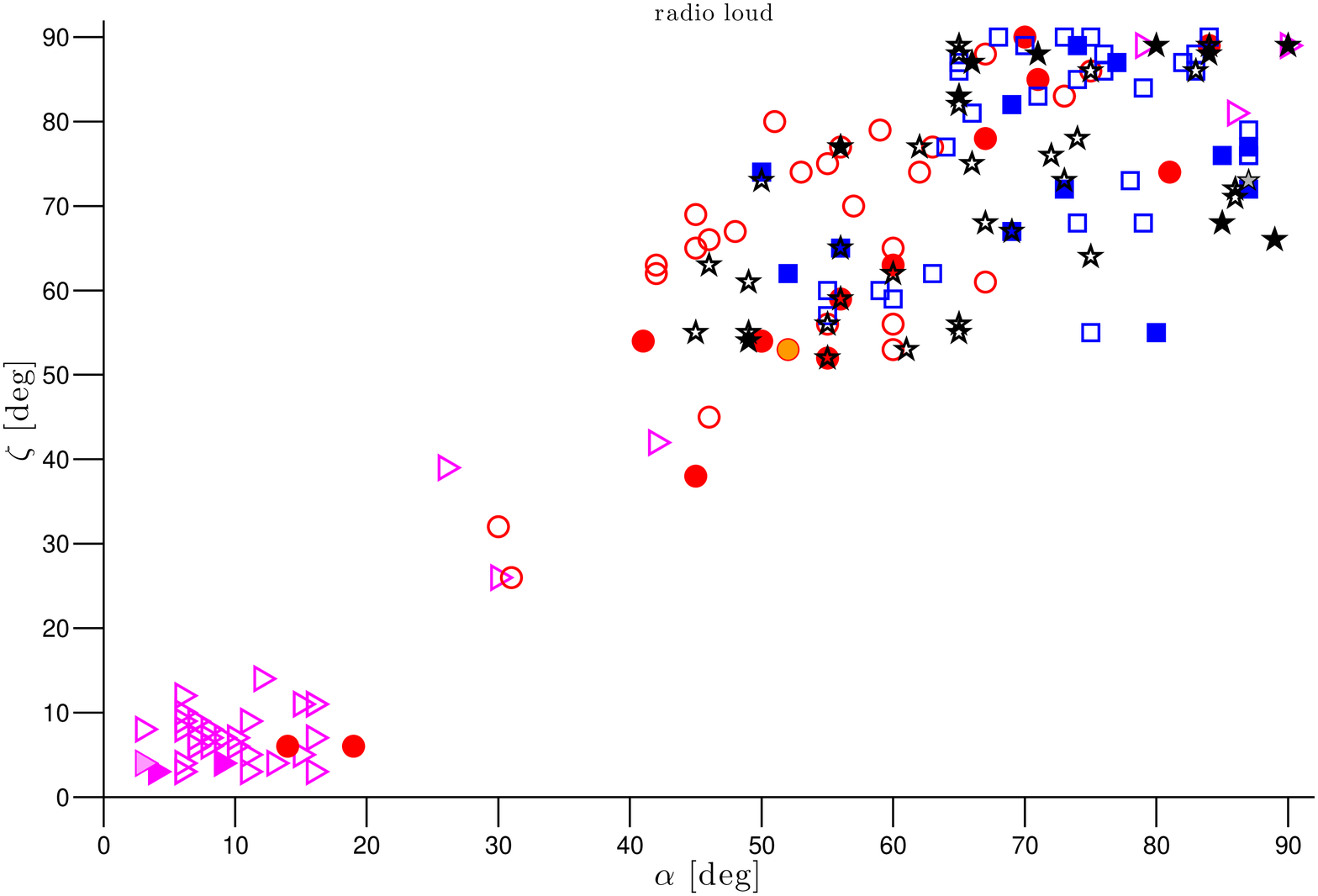}
\caption{$\alpha$-$\zeta$ plane distribution of RQ (top panel) and RL (bottom panel) fit solutions for the PC (magenta triangles), SG 
(red circles), OG (blue squares), and OPC (black stars) models. Recall that filled and empty symbols refer to best-fit solutions of the 
optimum and alternative models, respectively. The optimum-solutions that are better than the other models by more than 1$\sigma$ 
are plotted as light-colour-filled symbols.}
\label{AlpaZetaDistG}
\end{center}
\end{figure*}
The fit results can point to which model best explains the emission from each pulsar but they do not single out a model  that is 
able to explain all the observed light curves. This suggests that none of the assumed emission geometries can explain 
the variety of the LAT sample.

\subsection{Impact of the radio emission geometry on the pulsar orientation estimate}
\begin{table*}
\def\arraystretch{1.2}
\centering
\begin{tabular}{| c || c | c | c | c | c || c | c | c | c | c |}
\hline
& $ \alpha_{PC}$ & $ \alpha_{SG}$  & $ \alpha_{OG}$  & $ \alpha_{OPC}$ & $\alpha_\mathrm{others}$ & $ \zeta_{PC}$  &  $ \zeta_{SG}$  &  $ \zeta_{OG}$  &  $ \zeta_{OPC}$ & $\zeta_\mathrm{others}$\\
& $^\circ$ & $^\circ$  & $^\circ$  & $^\circ$ & $^\circ$& $^\circ$  &  $^\circ$  &  $^\circ$  &  $^\circ$ & $^\circ$\\
\hline
\hline
J0205$+$6449 & \cellcolor{gray!30} $ 78^{1}_{1}$  &   $ 85^{2}_{2}$  &   $ 57^{1}_{1}$  &   $ 85^{1}_{1}$  &   & \cellcolor{gray!30} $ 89^{1}_{1}$  &   $ 82^{1}_{1}$  &   $ 45^{1}_{1}$  & \cellcolor{gray!30} $ 89^{1}_{1}$  & $88.9-94.3^{(2)}$ \\
\hline
J0248$+$6021 &   $ 10^{6}_{1}$  &   $ 41^{1}_{2}$  &   $ 9^{1}_{1}$  &   $ 82^{1}_{1}$  &   &   $ 7^{1}_{1}$  &   $ 70^{1}_{1}$  &   $ 90^{1}_{1}$  &   $ 15^{1}_{1}$  & \\
\hline
J0534$+$2200 & \cellcolor{gray!30} $ 12^{1}_{1}$  & \cellcolor{gray!30} $ 51^{1}_{1}$  & \cellcolor{gray!30} $ 50^{1}_{1}$  & \cellcolor{gray!30} $ 50^{1}_{1}$  &   & \cellcolor{gray!30} $ 14^{1}_{1}$  & \cellcolor{gray!30} $ 75^{1}_{1}$  & \cellcolor{gray!30} $ 74^{1}_{1}$  & \cellcolor{gray!30} $ 73^{1}_{1}$  & $60.10-64.35^{(2)}$\\
\hline
J0631$+$1036 &   $ 16^{1}_{1}$  &   $ 36^{2}_{2}$  &   $ 7^{1}_{1}$  &   $ 83^{2}_{1}$  &   & \cellcolor{gray!30} $ 3^{1}_{1}$  &   $ 55^{2}_{5}$  &   $ 88^{1}_{1}$  &   $ 21^{1}_{2}$  & \\
\hline
J0659$+$1414 & \cellcolor{gray!30} $ 10^{1}_{1}$  &   $ 36^{1}_{1}$  &   $ 28^{1}_{1}$  &   $ 12^{1}_{1}$  &   &   $ 9^{1}_{1}$  &   $ 57^{1}_{3}$  &   $ 88^{1}_{1}$  &   $ 88^{1}_{1}$  & \\
\hline
J0729$-$1448 &   $ 16^{1}_{16}$  &   $ 49^{15}_{3}$  &   $ 8^{2}_{1}$  &   $ 8^{2}_{8}$  &   &   $ 2^{5}_{2}$  &   $ 49^{8}_{20}$  & \cellcolor{gray!30} $ 87^{3}_{2}$  &   $ 79^{1}_{1}$  & \\
\hline
J0742$-$2822 &   $ 16^{1}_{2}$  &   $ 41^{2}_{3}$  &   $ 7^{5}_{7}$  &   $ 75^{2}_{3}$  &   & \cellcolor{gray!30} $ 13^{1}_{7}$  &   $ 55^{1}_{3}$  & \cellcolor{gray!30} $ 90^{1}_{1}$  &   $ 26^{1}_{1}$  & \\
\hline
J0835$-$4510 & \cellcolor{gray!30} $ 4^{2}_{4}$  &   $ 16^{1}_{1}$  &   $ 48^{1}_{1}$  &   $ 66^{1}_{1}$  &  $43^{(1)}$/$70^{(2)}$ & \cellcolor{gray!30} $ 4^{1}_{1}$  &   $ 74^{1}_{1}$  & \cellcolor{gray!30} $ 83^{1}_{1}$  & \cellcolor{gray!30} $ 76^{1}_{1}$  & $62.95-64.27^{(2)}$\\
\hline
J0908$-$4913 & \cellcolor{gray!30} $ 7^{1}_{1}$  &   $ 89^{1}_{3}$  &   $ 10^{1}_{1}$  &   $ 13^{1}_{1}$  &   & \cellcolor{gray!30} $ 6^{1}_{1}$  &   $ 58^{3}_{4}$  &   $ 86^{1}_{1}$  &   $ 79^{1}_{1}$  & \\
\hline
J0940$-$5428 &   $ 19^{1}_{1}$  & \cellcolor{gray!30} $ 41^{16}_{26}$  &   $ 80^{1}_{2}$  &   $ 68^{3}_{5}$  &   & \cellcolor{gray!30} $ 14^{6}_{3}$  & \cellcolor{gray!30} $ 65^{10}_{11}$  &   $ 47^{2}_{1}$  &   $ 25^{8}_{5}$  & \\
\hline
J1016$-$5857 & \cellcolor{gray!30} $ 7^{1}_{1}$  & \cellcolor{gray!30} $ 58^{7}_{6}$  &   $ 44^{1}_{3}$  &   $ 60^{1}_{1}$  &   & \cellcolor{gray!30} $ 9^{1}_{1}$  & \cellcolor{gray!30} $ 71^{2}_{2}$  & \cellcolor{gray!30} $ 85^{3}_{1}$  &   $ 79^{1}_{1}$  & \\
\hline
J1019$-$5749 & \cellcolor{gray!30} $ 20^{21}_{20}$  &   $ 41^{6}_{3}$  & \cellcolor{gray!30} $ 81^{5}_{3}$  &   $ 54^{26}_{11}$  &   & \cellcolor{gray!30} $ 15^{24}_{11}$  &   $ 53^{2}_{6}$  &   $ 71^{7}_{22}$  &   $ 53^{24}_{29}$  & \\
\hline
J1028$-$5819 & \cellcolor{gray!30} $ 7^{1}_{1}$  &   $ 66^{2}_{1}$  &   $ 31^{1}_{1}$  & \cellcolor{gray!30} $ 90^{1}_{1}$  &   & \cellcolor{gray!30} $ 7^{1}_{1}$  & \cellcolor{gray!30} $ 84^{1}_{1}$  & \cellcolor{gray!30} $ 85^{1}_{1}$  & \cellcolor{gray!30} $ 89^{1}_{1}$  & \\
\hline
J1048$-$5832 & \cellcolor{gray!30} $ 4^{2}_{4}$  &   $ 71^{1}_{1}$  &   $ 39^{1}_{1}$  &   $ 60^{1}_{1}$  &   & \cellcolor{gray!30} $ 8^{1}_{1}$  &   $ 61^{1}_{1}$  &   $ 81^{1}_{1}$  & \cellcolor{gray!30} $ 76^{1}_{1}$  & \\
\hline
J1057$-$5226 & \cellcolor{gray!30} $ 10^{1}_{1}$  & \cellcolor{gray!30} $ 46^{1}_{1}$  & \cellcolor{gray!30} $ 77^{1}_{1}$  &   $ 15^{1}_{1}$  &   & \cellcolor{gray!30} $ 7^{1}_{1}$  & \cellcolor{gray!30} $ 45^{1}_{1}$  & \cellcolor{gray!30} $ 87^{1}_{1}$  &   $ 88^{1}_{1}$  & \\
\hline
J1105$-$6107 &   $ 9^{1}_{1}$  &   $ 90^{1}_{2}$  &   $ 9^{1}_{1}$  &   $ 15^{4}_{2}$  &   &   $ 6^{1}_{1}$  &   $ 41^{8}_{5}$  & \cellcolor{gray!30} $ 82^{1}_{1}$  & \cellcolor{gray!30} $ 80^{2}_{1}$  & \\
\hline
J1112$-$6103 &   $ 8^{1}_{8}$  &   $ 86^{2}_{2}$  &   $ 9^{4}_{2}$  &   $ 10^{1}_{1}$  &   & \cellcolor{gray!30} $ 8^{1}_{1}$  & \cellcolor{gray!30} $ 38^{4}_{3}$  & \cellcolor{gray!30} $ 80^{3}_{1}$  & \cellcolor{gray!30} $ 76^{1}_{1}$  & \\
\hline
J1119$-$6127 & \cellcolor{gray!30} $ 4^{6}_{4}$  &   $ 36^{2}_{3}$  &   $ 8^{2}_{1}$  &   $ 78^{1}_{1}$  &   &   $ 2^{1}_{2}$  & \cellcolor{gray!30} $ 53^{4}_{4}$  &   $ 83^{2}_{1}$  &   $ 4^{1}_{1}$  & \\
\hline
J1124$-$5916 & \cellcolor{gray!30} $ 90^{1}_{1}$  & \cellcolor{gray!30} $ 88^{2}_{3}$  &   $ 61^{1}_{1}$  &   $ 65^{1}_{1}$  &   & \cellcolor{gray!30} $ 89^{1}_{1}$  & \cellcolor{gray!30} $ 89^{1}_{2}$  & \cellcolor{gray!30} $ 86^{1}_{1}$  & \cellcolor{gray!30} $ 89^{1}_{1}$  & $68.0-82.0^{(2)}$\\
\hline
J1357$-$6429 &   $ 10^{1}_{1}$  &   $ 30^{3}_{6}$  &   $ 3^{2}_{3}$  &   $ 73^{1}_{1}$  &   & \cellcolor{gray!30} $ 8^{1}_{1}$  &   $ 68^{2}_{2}$  &   $ 77^{1}_{1}$  &   $ 11^{2}_{1}$  & \\
\hline
J1410$-$6132 & \cellcolor{gray!30} $ 8^{1}_{1}$  &   $ 86^{4}_{5}$  &   $ 9^{1}_{2}$  &   $ 31^{1}_{2}$  &   & \cellcolor{gray!30} $ 9^{1}_{1}$  &   $ 32^{11}_{6}$  & \cellcolor{gray!30} $ 79^{1}_{1}$  & \cellcolor{gray!30} $ 90^{1}_{5}$  & \\
\hline
J1420$-$6048 & \cellcolor{gray!30} $ 14^{1}_{1}$  &   $ 57^{1}_{1}$  &   $ 2^{4}_{2}$  &   $ 77^{1}_{1}$  &   & \cellcolor{gray!30} $ 7^{1}_{1}$  &   $ 48^{1}_{1}$  &   $ 77^{1}_{1}$  &   $ 23^{1}_{1}$  & \\
\hline
J1509$-$5850 &   $ 3^{2}_{3}$  & \cellcolor{gray!30} $ 46^{1}_{1}$  &   $ 15^{1}_{1}$  &   $ 82^{1}_{1}$  &   &   $ 12^{1}_{1}$  &   $ 43^{1}_{1}$  &   $ 90^{1}_{1}$  &   $ 18^{1}_{1}$  & \\
\hline
J1513$-$5908 &   $ 10^{1}_{1}$  &   $ 29^{9}_{10}$  &   $ 88^{2}_{1}$  &   $ 68^{2}_{1}$  &   &   $ 6^{1}_{1}$  &   $ 66^{4}_{3}$  &   $ 21^{5}_{3}$  &   $ 21^{2}_{3}$  & \\
\hline
J1648$-$4611 & \cellcolor{gray!30} $ 16^{1}_{5}$  &   $ 46^{3}_{1}$  &   $ 18^{1}_{2}$  &   $ 84^{3}_{1}$  &   &   $ 3^{1}_{1}$  &   $ 43^{2}_{1}$  &   $ 87^{1}_{1}$  &   $ 22^{1}_{1}$  & \\
\hline
J1702$-$4128 & \cellcolor{gray!30} $ 10^{5}_{2}$  &   $ 31^{6}_{22}$  &   $ 7^{1}_{1}$  &   $ 4^{3}_{4}$  &   & \cellcolor{gray!30} $ 3^{1}_{3}$  & \cellcolor{gray!30} $ 48^{15}_{11}$  &   $ 89^{1}_{1}$  &   $ 79^{1}_{1}$  & \\
\hline
J1709$-$4429 & \cellcolor{gray!30} $ 13^{2}_{2}$  &   $ 26^{1}_{1}$  &   $ 13^{1}_{1}$  &   $ 6^{1}_{1}$  &   & \cellcolor{gray!30} $ 3^{1}_{1}$  &   $ 70^{1}_{1}$  &   $ 77^{1}_{1}$  &   $ 73^{1}_{1}$  & $49.0-57.8^{(2)}$\\
\hline
J1718$-$3825 & \cellcolor{gray!30} $ 16^{4}_{1}$  &   $ 41^{1}_{1}$  &   $ 11^{1}_{1}$  &   $ 72^{1}_{1}$  &   & \cellcolor{gray!30} $ 3^{1}_{1}$  & \cellcolor{gray!30} $ 67^{1}_{1}$  &   $ 79^{1}_{1}$  &   $ 9^{1}_{1}$  & \\
\hline
J1730$-$3350 &   $ 7^{1}_{1}$  &   $ 77^{3}_{4}$  &   $ 41^{1}_{3}$  & \cellcolor{gray!30} $ 56^{2}_{1}$  &   &   $ 7^{1}_{1}$  &   $ 49^{6}_{4}$  &   $ 87^{3}_{2}$  &   $ 76^{2}_{1}$  & \\
\hline
J1741$-$2054 & \cellcolor{gray!30} $ 3^{2}_{3}$  &   $ 71^{1}_{1}$  & \cellcolor{gray!30} $ 84^{1}_{1}$  &   $ 29^{1}_{1}$  &   & \cellcolor{gray!30} $ 4^{1}_{1}$  & \cellcolor{gray!30} $ 24^{1}_{1}$  & \cellcolor{gray!30} $ 90^{1}_{1}$  &   $ 90^{1}_{1}$  & \\
\hline
J1747$-$2958 & \cellcolor{gray!30} $ 9^{1}_{1}$  &   $ 73^{1}_{1}$  &   $ 41^{1}_{1}$  &   $ 90^{1}_{1}$  &   & \cellcolor{gray!30} $ 7^{1}_{1}$  &   $ 41^{1}_{1}$  &   $ 90^{1}_{1}$  &   $ 36^{1}_{1}$  & \\
\hline
J1801$-$2451 &   $ 8^{1}_{1}$  &   $ 58^{1}_{6}$  &   $ 8^{1}_{1}$  &   $ 31^{1}_{1}$  &   &   $ 6^{1}_{1}$  &   $ 88^{2}_{2}$  & \cellcolor{gray!30} $ 82^{1}_{1}$  & \cellcolor{gray!30} $ 82^{2}_{2}$  & \\
\hline
J1833$-$1034 &   $ 27^{2}_{1}$  &   $ 41^{1}_{1}$  & \cellcolor{gray!30} $ 68^{1}_{1}$  &   $ 57^{1}_{1}$  &   &   $ 40^{1}_{1}$  &   $ 59^{1}_{1}$  &   $ 32^{1}_{1}$  &   $ 36^{1}_{1}$  & $85.1-85.6^{(2)}$\\
\hline
J1835$-$1106 & \cellcolor{gray!30} $ 10^{1}_{2}$  &   $ 51^{9}_{5}$  &   $ 26^{4}_{10}$  & \cellcolor{gray!30} $ 90^{1}_{2}$  &   & \cellcolor{gray!30} $ 5^{4}_{1}$  & \cellcolor{gray!30} $ 65^{2}_{7}$  & \cellcolor{gray!30} $ 90^{1}_{3}$  &   $ 21^{6}_{3}$  & \\
\hline
J1952$+$3252 &   $ 32^{1}_{1}$  &   $ 81^{1}_{1}$  &   $ 74^{1}_{1}$  &   $ 61^{1}_{1}$  &   &   $ 51^{1}_{1}$  & \cellcolor{gray!30} $ 83^{1}_{1}$  &   $ 34^{1}_{1}$  &   $ 33^{1}_{1}$  & \\
\hline
J2021$+$3651 & \cellcolor{gray!30} $ 7^{1}_{1}$  &   $ 63^{1}_{1}$  &   $ 55^{1}_{1}$  & \cellcolor{gray!30} $ 84^{1}_{1}$  &   & \cellcolor{gray!30} $ 7^{1}_{1}$  & \cellcolor{gray!30} $ 83^{1}_{1}$  & \cellcolor{gray!30} $ 89^{1}_{1}$  & \cellcolor{gray!30} $ 88^{1}_{1}$  & $76.0-82.0^{(2)}$\\
\hline
J2030$+$3641 & \cellcolor{gray!30} $ 8^{1}_{1}$  &   $ 46^{1}_{1}$  & \cellcolor{gray!30} $ 84^{1}_{1}$  &   $ 18^{1}_{1}$  &   & \cellcolor{gray!30} $ 8^{1}_{1}$  &   $ 45^{1}_{1}$  &   $ 90^{1}_{1}$  &   $ 88^{1}_{1}$  & \\
\hline
J2032$+$4127 &   $ 83^{1}_{1}$  &   $ 90^{1}_{1}$  & \cellcolor{gray!30} $ 59^{1}_{1}$  &   $ 17^{1}_{1}$  &   &   $ 89^{1}_{1}$  &   $ 90^{1}_{1}$  & \cellcolor{gray!30} $ 60^{1}_{1}$  &   $ 72^{1}_{1}$  & \\
\hline
J2043$+$2740 & \cellcolor{gray!30} $ 4^{4}_{4}$  &   $ 46^{1}_{1}$  &   $ 63^{1}_{4}$  &   $ 90^{1}_{2}$  &   & \cellcolor{gray!30} $ 9^{1}_{1}$  &   $ 48^{1}_{1}$  & \cellcolor{gray!30} $ 90^{1}_{2}$  & \cellcolor{gray!30} $ 88^{2}_{2}$  & \\
\hline
J2229$+$6114 &   $ 15^{1}_{1}$  &   $ 36^{1}_{1}$  &   $ 84^{1}_{1}$  &   $ 73^{1}_{1}$  &   &   $ 8^{1}_{1}$  & \cellcolor{gray!30} $ 63^{1}_{1}$  &   $ 24^{1}_{1}$  &   $ 24^{1}_{1}$  & $38.0-54.0^{(2)}$\\
\hline
J2240$+$5832 &   $ 39^{4}_{16}$  &   $ 81^{5}_{8}$  &   $ 87^{2}_{2}$  &   $ 53^{4}_{1}$  &   &   $ 37^{2}_{11}$  &   $ 66^{3}_{11}$  &   $ 40^{9}_{4}$  &   $ 42^{1}_{1}$  & \\
\hline
\end{tabular}
\centering
\caption{$\alpha$ and $\zeta$ best-fit solution resulting from the $\gamma$-ray only fit of the 41 RL pulsars. The central and last columns list independent 
$\alpha$ and $\zeta$ estimates, found in the literature, respectively. Superscript and subscript refer to upper and lower errors, respectively.
The errors bigger than 1 correspond to  the $3\sigma$ statistical error. 
The solutions compatible, within the errors, with the solutions obtained by fitting jointly radio and $\gamma$-ray light curves and listed in 
Table \ref{JointAlpZetFit} are highlighted in grey cells. 
$^{(1)}$ \cite{jhv+05}; $^{(2)}$ \cite{nr08}; $^{(3)}$ $\alpha =\zeta+6.5$ found by \cite{jhv+05} with $\zeta\sim63.5$ from \cite{nr08}}
\label{TabGammaOnlyRL}
\end{table*}

Figure \ref{SolutionMigration} shows how the $(\alpha,\zeta)$ solutions obtained for the RL sample migrate, from the $\gamma$-only
solutions when we take into account the radio emission.
We have used the $\chi^2$ fit and FCBin light curves to give an $(\alpha,\zeta)$ estimate for RL \emph{Fermi} pulsars based on the $\gamma$-ray 
emission only. They are listed in Table \ref{TabGammaOnlyRL}. We have plotted those solutions as stars in 
Figure \ref{SolutionMigration}. To study how they change by including the radio emission in the fit, we have plotted as squares the 
solutions obtained with the joint $\sigma_{peak}$ radio fit and we have connected with a line the solutions of the two methods for each pulsar.

In many cases the $\gamma$-only solutions for RL pulsars
are found far away from the diagonal $(0,0)$  to $(90,90)$ in the $\alpha-\zeta$ plane where radio emission 
is more likely. Hereafter we will refer to this diagonal as the \emph{radio diagonal}. For all models 
except the PC, the introduction of the radio component in the fit  causes the $(\alpha,\zeta)$ solution to migrate from 
orientations where radio emission is unlikely toward the radio diagonal.
This suggests that a $\gamma$-ray only fit estimate of $\alpha$ and $\zeta$ for RL pulsars may give results far 
away from the radio diagonal and should be used with caution. 

In the PC model, the inclusion of the radio component in the fit produces a migration of the solutions along the \emph{radio diagonal}.
In the SG model, the extent of the migration is somewhat larger than in the PC case and it does not follow any trend (Figure \ref{SolutionMigration}). 
In the OG and OPC models the $\gamma$-ray only solutions migrate the furthest to the joint solutions in Figure \ref{SolutionMigration}.
In the outer magnetosphere models, both the $\alpha$ and $\zeta$ angles can be underestimated according to the position
of the $\gamma$-only solution with respect to the radio diagonal. When the $\gamma$-only solution is to the right of the radio diagonal, $\zeta$ migrates 
toward higher values while $\alpha$ keeps quite stable and \emph{vice versa} when the $\gamma$-only solution is to the left of the radio diagonal.

\subsection{$\alpha$-$\zeta$ plane}
\label{A-Z best solutions plane}

\begin{figure*}
\centering
\includegraphics[width=0.75\textwidth]{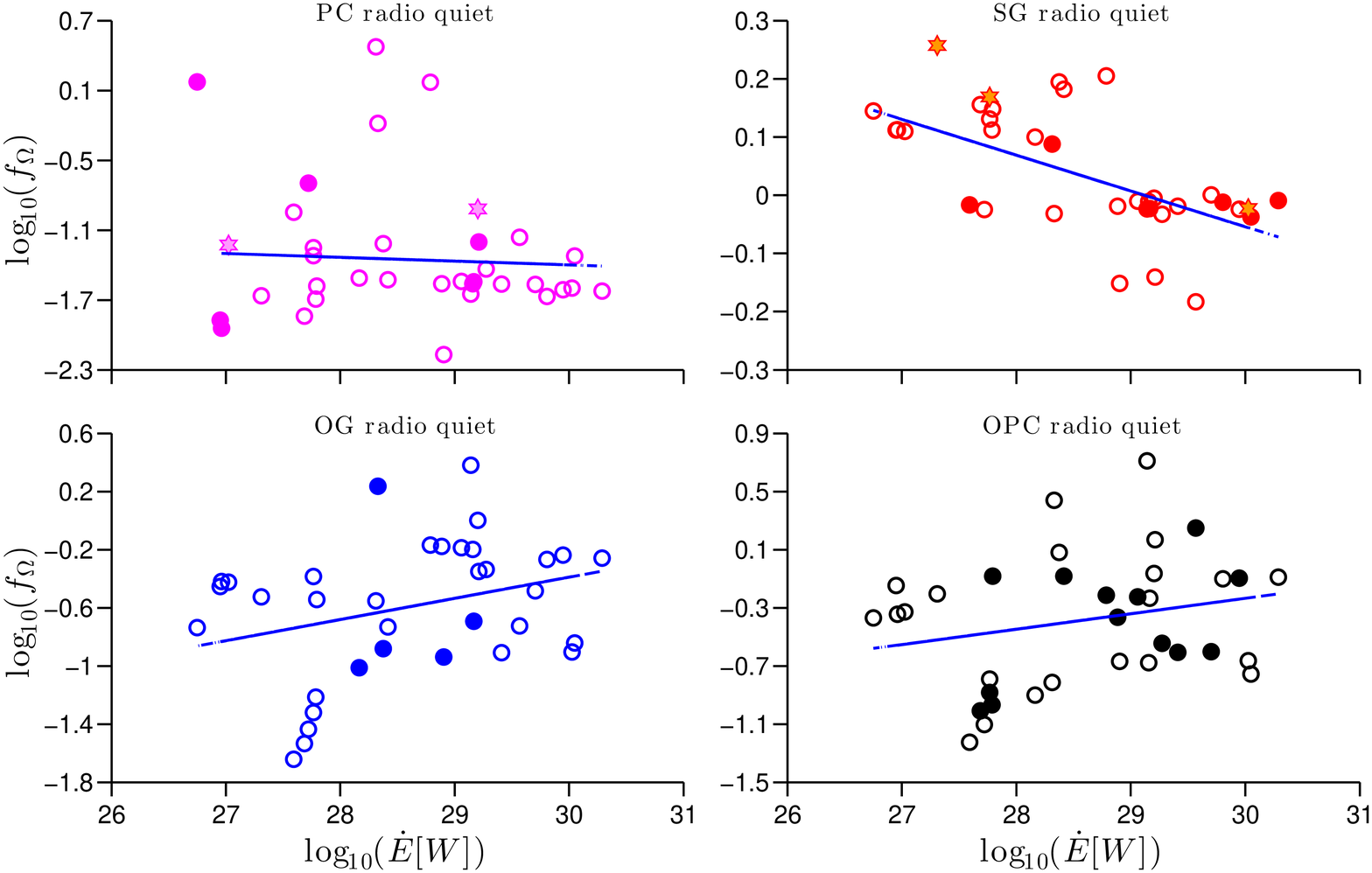}
\includegraphics[width=0.75\textwidth]{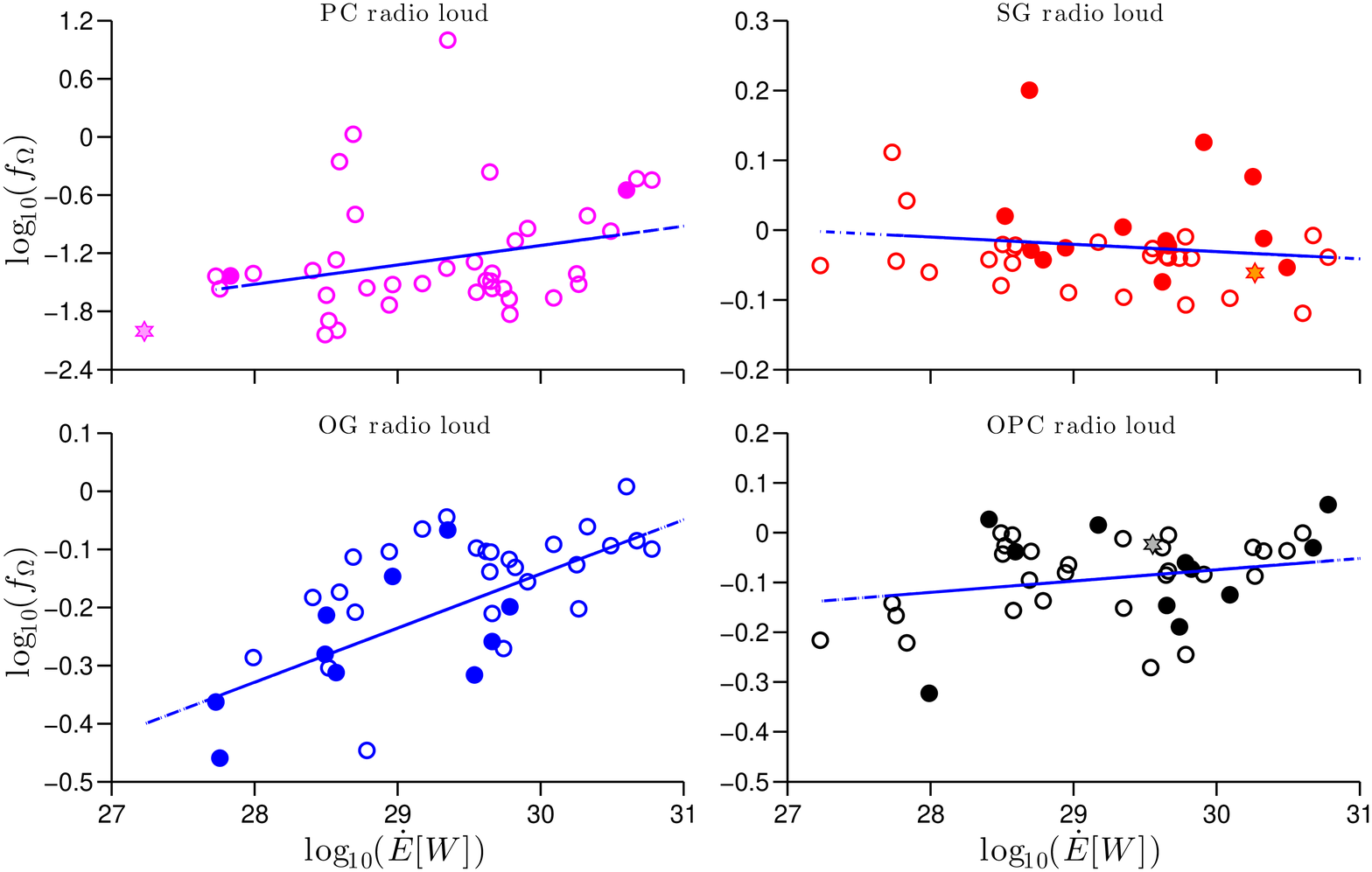}
\caption{ Beaming factor $f_{\Omega}$ versus the pulsar spin-down power $\dot{E}$ evaluated for RQ 
(top panel) and RL (bottom panel) pulsars . The lines represent the best power-law fits to the data 
points; the best fit power-law parameters with relative $1\sigma$ errors, are listed in Table \ref{TabBFfit}.
Hereafter the optimum-solutions that are better than the other models by more than 1$\sigma$ will be 
plotted as light-colour-filled hexagrams.}
\label{fOmegaEdot1}
\end{figure*}

Figure \ref{AlpaZetaDistG} shows the solutions in the $\alpha$-$\zeta$  plane for the RQ and RL pulsars in the top and bottom panels respectively.
A comparison of the $\alpha$ and $\zeta$ estimates with the values obtained from observations at 
other wavelengths show good consistency in all the reported cases (Tables \ref{A_Z_gamma1} and \ref{JointAlpZetFit}). 
Our $\zeta$ estimates are consistent with the values predicted by \cite{cbd+03} for PSR J0633$+$1746 OG 
and OPC models, and with the values predicted by \cite{nr08} for pulsars J0205$+$6449  OG/SG/OPC  models,  
J1709$-$4429 OG model, J1833$-$1034 OG model, J2021$+$3651 SG model, and J2229$+$6114 OG model. 
For PSRs J1803$-$2149, Crab, and J1124$-$5916, none of our $\zeta$ estimates is included in the 
interval predicted by other authors. For those pulsars, the values closest to the predictions 
made by \cite{nr08} are obtained by OG for J1803$-$2149, SG/OG/OPC for the Crab, and by all models for J1124$-$5916.  
In the case of the Vela pulsar, our SG model predictions $\alpha=45^\circ\pm2^\circ$ and $\zeta=69^\circ\pm2^\circ$  
are both consistent with $\alpha=43^\circ$ by \cite{jhv+05} and 
$63^\circ~\lsimeq~\zeta~\lsimeq~64^\circ$  by \cite{nr08}.

Since the radio and PC emissions are generated in the same region of the magnetosphere in narrow conical beams, coaxial with 
the magnetic axis, all the PC solutions are found along the radio diagonal.  The concentration of solutions at low $\alpha$ and 
$\zeta$ for both RQ and RL pulsars is due to the PC emission geometry, for which low $\alpha$ and $\zeta$ angles predict the highest
variety of light-curve shapes.

The majority of SG solutions, both for RQ and RL objects, are concentrated in the central-upper part of the radio diagonal. 
The paucity of low $\alpha$ and $\zeta$ solutions is due to SG geometry: 
the SG bright caustics shine generally at high $\zeta$ and tend to concentrate toward the neutron star spin equator as $\alpha$ decreases.
\begin{table*}
\def\arraystretch{1.2}
\centering
\begin{tabular}{| c || c | c | c | c |}
\hline
RQ & $ f_{\Omega,PC}$ & $ f_{\Omega,SG}$  & $ f_{\Omega,OG}$  & $ f_{\Omega,OPC}$ \\
\hline
\hline
J0007$+$7303 & 0.01 & 0.71 & 0.12  &0.21\\
\hline
J0106$+$4855 & 0.2 & 0.94 & 0.04  &0.08\\
\hline
J0357$+$3205 & 0.06 & 1.29 & 0.38  &0.47\\
\hline
J0622$+$3749 & 0.01 & 1.43 & 0.03  &0.1\\
\hline
J0633$+$0632 & 0.66 & 0.93 & 1.72  &2.75\\
\hline
J0633$+$1746 & 0.06 & 1.47 & 0.41  &0.16\\
\hline
J0734$-$1559 & 0.06 & 1.57 & 0.13  &1.21\\
\hline
J1023$-$5746 & 0.02 & 0.98 & 0.55  &0.81\\
\hline
J1044$-$5737 & 0.03 & 0.97 & 0.63  &0.21\\
\hline
J1135$-$6055 & 0.07 & 0.66 & 0.19  &1.77\\
\hline
J1413$-$6205 & 0.03 & 0.95 & 0.2  &0.58\\
\hline
J1418$-$6058 & 0.02 & 0.95 & 0.58  &0.8\\
\hline
J1429$-$5911 & 0.02 & 0.95 & 2.4  &5.14\\
\hline
J1459$-$6053 & 0.06 & 0.72 & 0.45  &1.47\\
\hline
J1620$-$4927 & 0.03 & 1.26 & 0.1  &0.13\\
\hline
J1732$-$3131 & 0.03 & 1.52 & 0.19  &0.83\\
\hline
J1746$-$3239 & 0.05 & 1.35 & 0.05  &0.13\\
\hline
J1803$-$2149 & 0.03 & 0.98 & 0.65  &0.6\\
\hline
J1809$-$2332 & 0.03 & 0.96 & 0.66  &0.43\\
\hline
J1813$-$1246 & 0.05 & 0.92 & 0.14  &0.18\\
\hline
J1826$-$1256 & 0.02 & 0.97 & 0.54  &0.79\\
\hline
J1836$+$5925 & 0.02 & 1.81 & 0.3  &0.62\\
\hline
J1838$-$0537 & 0.03 & 0.95 & 0.12  &0.22\\
\hline
J1846$+$0919 & 0.02 & 1.29 & 0.06  &0.11\\
\hline
J1907$+$0602 & 0.03 & 1 & 0.33  &0.25\\
\hline
J1954$+$2836 & 0.04 & 0.93 & 0.46  &0.28\\
\hline
J1957$+$5033 & 0.01 & 1.3 & 0.38  &0.45\\
\hline
J1958$+$2846 & 1.48 & 1.6 & 0.68  &0.61\\
\hline
J2021$+$4026 & 2.98 & 1.22 & 0.28  &0.15\\
\hline
J2028$+$3332 & 0.03 & 1.41 & 0.29  &0.83\\
\hline
J2030$+$4415 & 0.11 & 0.96 & 0.02  &0.06\\
\hline
J2055$+$2539 & 0.01 & 1.29 & 0.35  &0.71\\
\hline
J2111$+$4606 & 0.03 & 0.96 & 0.12  &0.25\\
\hline
J2139$+$4716 & 1.49 & 1.4 & 0.18  &0.43\\
\hline
J2238$+$5903 & 0.12 & 0.99 & 1  &0.86\\
\hline
\end{tabular}
~~~~~~~~~~~~~~~~~~~~~~~~
\centering
\begin{tabular}{| c || c | c | c | c |}
\hline
RL & $ f_{\Omega,PC}$ & $ f_{\Omega,SG}$  & $ f_{\Omega,OG}$  & $ f_{\Omega,OPC}$ \\
\hline
\hline
J0205$+$6449 & 0.37 & 0.98 & 0.82  &0.93\\
\hline
J0248$+$6021 & 0.01 & 0.94 & 0.24  &0.7\\
\hline
J0534$+$2200 & 0.13 & 0.9 & 0.64  &0.67\\
\hline
J0631$+$1036 & 0.01 & 0.83 & 0.52  &1\\
\hline
J0659$+$1414 & 0.04 & 1.1 & 0.31  &0.6\\
\hline
J0729$-$1448 & 0.16 & 0.94 & 0.62  &0.92\\
\hline
J0742$-$2822 & 0.04 & 0.91 & 0.66  &1.06\\
\hline
J0835$-$4510 & 0.02 & 0.8 & 0.81  &0.75\\
\hline
J0908$-$4913 & 0.02 & 0.94 & 0.79  &0.83\\
\hline
J0940$-$5428 & 0.05 & 0.92 & 0.48  &0.54\\
\hline
J1016$-$5857 & 0.03 & 0.91 & 0.55  &0.99\\
\hline
J1019$-$5749 & 0.01 & 1.05 & 0.5  &0.94\\
\hline
J1028$-$5819 & 0.03 & 0.96 & 0.86  &1.04\\
\hline
J1048$-$5832 & 0.02 & 0.94 & 0.8  &0.95\\
\hline
J1057$-$5226 & 0.04 & 1.29 & 0.43  &0.72\\
\hline
J1105$-$6107 & 0.43 & 0.97 & 0.73  &0.82\\
\hline
J1112$-$6103 & 0.11 & 1.34 & 0.7  &0.82\\
\hline
J1119$-$6127 & 0.03 & 0.84 & 0.79  &0.93\\
\hline
J1124$-$5916 & 0.15 & 0.97 & 0.87  &0.92\\
\hline
J1357$-$6429 & 0.03 & 0.91 & 0.54  &0.65\\
\hline
J1410$-$6132 & 0.04 & 1.19 & 0.75  &0.93\\
\hline
J1420$-$6048 & 0.03 & 0.87 & 0.63  &0.82\\
\hline
J1509$-$5850 & 0.03 & 0.81 & 0.71  &0.86\\
\hline
J1513$-$5908 & 0.11 & 0.88 & 0.81  &0.92\\
\hline
J1648$-$4611 & 0.05 & 0.9 & 0.49  &0.99\\
\hline
J1702$-$4128 & 0.03 & 0.91 & 0.36  &0.73\\
\hline
J1709$-$4429 & 0.01 & 0.78 & 0.63  &0.57\\
\hline
J1718$-$3825 & 9.96 & 0.8 & 0.86  &0.71\\
\hline
J1730$-$3350 & 0.04 & 1.01 & 0.9  &0.97\\
\hline
J1741$-$2054 & 0.01 & 0.89 & 0.3  &0.61\\
\hline
J1747$-$2958 & 0.03 & 0.92 & 0.79  &0.71\\
\hline
J1801$-$2451 & 0.04 & 0.95 & 0.62  &0.84\\
\hline
J1833$-$1034 & 0.36 & 0.91 & 0.79  &1.14\\
\hline
J1835$-$1106 & 0.02 & 0.95 & 0.61  &0.91\\
\hline
J1952$+$3252 & 0.08 & 0.91 & 0.74  &0.84\\
\hline
J2021$+$3651 & 0.02 & 0.98 & 0.76  &0.87\\
\hline
J2030$+$3641 & 0.03 & 0.9 & 0.35  &0.68\\
\hline
J2032$+$4127 & 1.07 & 1.59 & 0.77  &0.8\\
\hline
J2043$+$2740 & 0.04 & 0.87 & 0.52  &0.48\\
\hline
J2229$+$6114 & 0.28 & 0.76 & 1.02  &1\\
\hline
J2240$+$5832 & 0.55 & 0.95 & 0.67  &0.91\\
\hline
\end{tabular}
\caption{ Beaming factors $f_{\Omega}$ evaluated for the RQ (\emph{left}) and RL (\emph{right}) pulsars in the 
framework of each model.}
\label{FOmega1}
\end{table*}

In agreement with \cite{twc11} we show that OG and OPC $\alpha$ and $\zeta$ estimates for both RQ and RL pulsars 
are mainly observed at high $\alpha$ and $\zeta$ angles, preferably at high $\zeta$ for all obliquities for the RQ pulsars. 
Only a handful of OPC pulsars are potentially seen at  $\zeta <30^{\circ}$.
The comparison of OG and OPC solutions shows that the two different prescriptions for the gap width evolution 
do not much affect the estimation of $\alpha$ and $\zeta$.
The fact that RQ SG solutions are closer to the radio diagonal than RQ OG solutions is due to their different 
emission geometry: two-pole emission geometry \citep[emission from both poles, e.g. Two Pole Caustic model,][]{dr03} 
and one-pole emission geometry \citep[emission from just one pole, Outer Gap model, ][]{crz00} respectively. 
It follows that for lower $\alpha$ angles ($\lesssim45^{\circ}$), OG emission can be observed with large 
enough peak separation only at high $\zeta$ angles whereas in the SG geometry large peak 
separations can be observed at lower $\zeta$ angles and from both poles.

We show in figure \ref{NvisA_large_effi1p012p01p00p5_alphazeta_histo} the $\alpha$-$\zeta$ plane distribution obtained for the $\gamma$-ray visible 
pulsars from the population synthesis described in \cite{pghg12}. The comparison with the RQ and RL pulsars of Figure \ref{AlpaZetaDistG}
shows consistency between the LAT pulsars and the prediction from the Galactic population for the SG, OG, and OPC models. The PC predictions show 
an abundance of solutions at intermediate $(\alpha,\zeta)$ that are not observed in the LAT sample.

We will now use the $(\alpha,\zeta)$ solutions to study various collective properties of the LAT pulsar sample.

\subsection{Beaming factor $f_\Omega$} 
\label{Beaming factor} 
\begin{figure*}[tb!]
\centering
\includegraphics[width=0.65\textwidth]{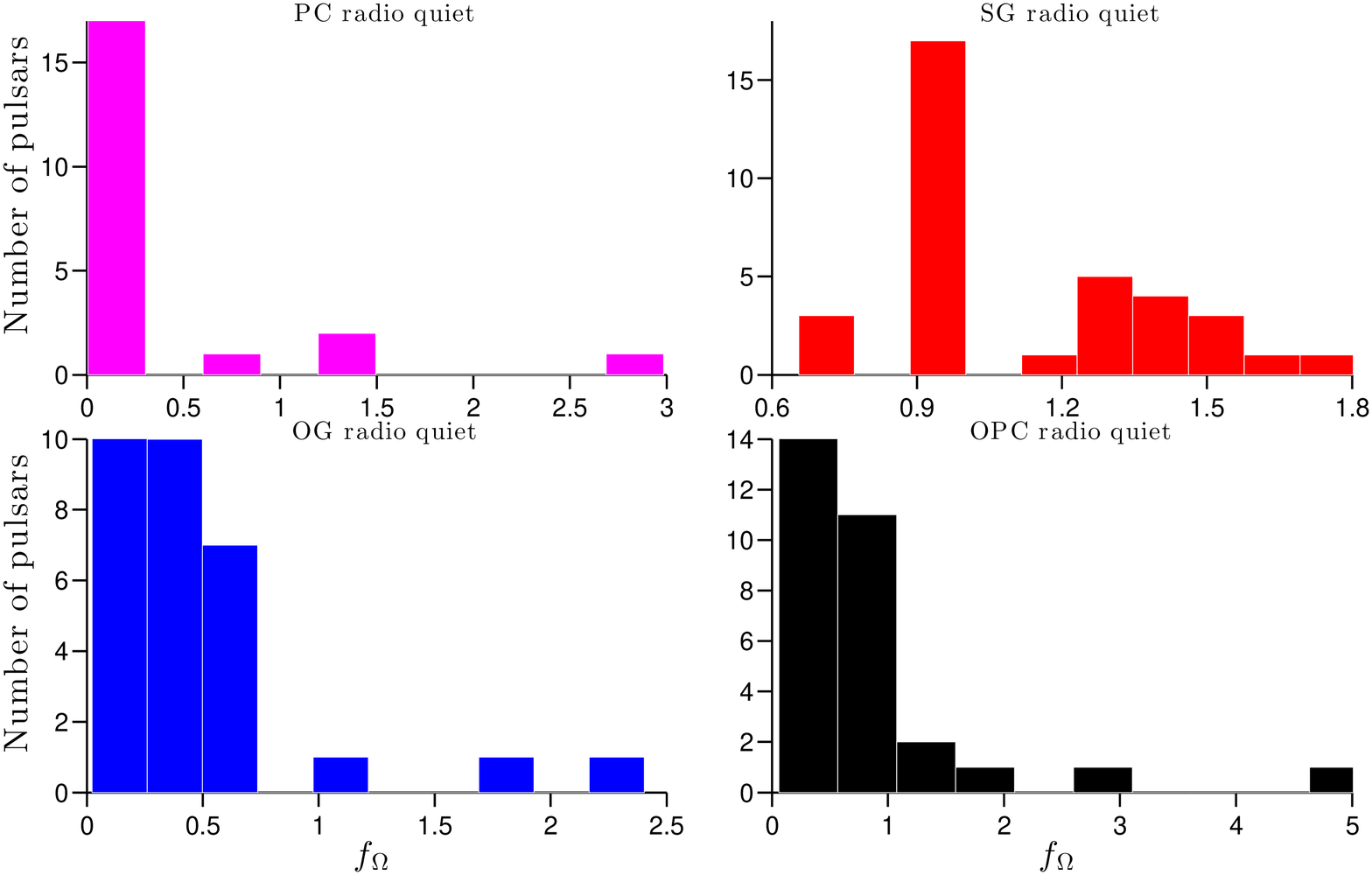}
\includegraphics[width=0.65\textwidth]{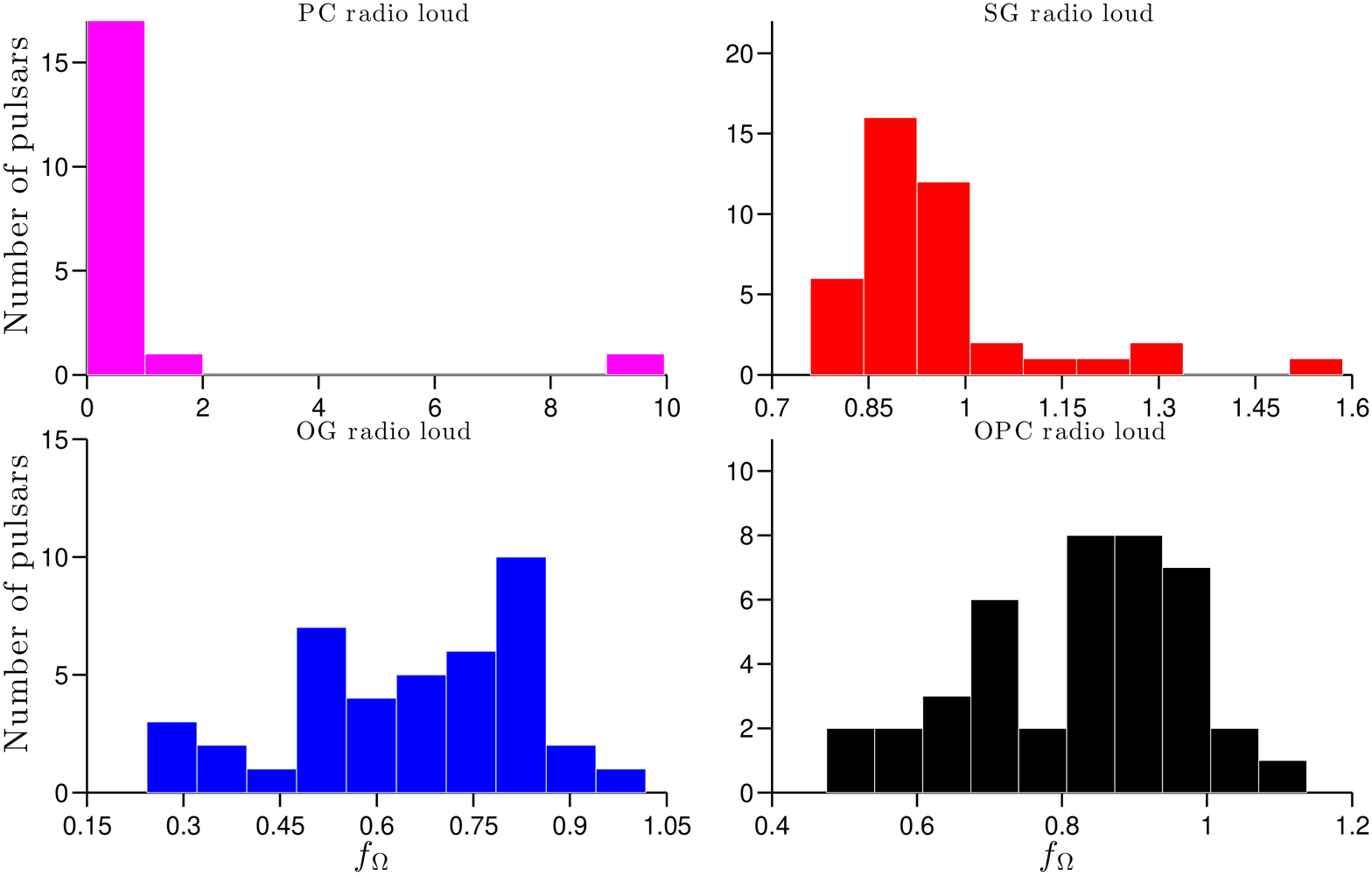}
\caption{Beaming factor $f_{\Omega}$ distribution for the RQ  (top panel) and RL (bottom panel) pulsars and all models.}
\label{FomDist}
\end{figure*}

The pulsar beaming factor $f_\Omega$ is the ratio of the total luminosity radiated over a 4$\pi$ sr solid angle
to the observed phase-averaged energy flux,
\begin{equation}
\label{LumBeam}
L_{\gamma}=4\pi f_{\Omega} F_{\mathrm{obs}} D^2,
\end{equation}
where $D$ is the pulsar distance and $F_{\mathrm{obs}}$ is the observed pulsar flux.
The LAT pulsar beaming factors $f_{\Omega}$ have been 
evaluated from each of the $(\alpha,\zeta)$ solutions and the corresponding phase-plots according to:
\begin{equation}
\label{beamFrac}
f_\mathrm{\Omega}=\frac{\int_0^\pi \sin \zeta  \int_0^{2\pi} n(\phi,\alpha_\mathrm{obs},\zeta)d\phi d\zeta}{2\int_0^{2\pi} n(\phi,\alpha_\mathrm{obs},\zeta_\mathrm{obs})d\phi}
\end{equation}
where the numerator is the integrated luminosity radiated by the pulsar in all directions for the $\alpha_\mathrm{obs}$ obliquity 
and the denominator integrates the energy flux intercepted for the observer line of sight $\zeta=\zeta_\mathrm{obs}$ \citep{wrwj09}.

Figure \ref{fOmegaEdot1} shows the beaming factor as a function of the pulsar spin-down power. The beaming factors have been derived from the 
best-fit RQ and RL $(\alpha,\zeta)$ solutions for each model. The LAT pulsar spin-down powers $\dot{E}$ have been evaluated from the periods and 
period first time derivatives given in PSRCAT2, as described in \cite{pghg12} ({with a different choice of pulsar moment of inertia, mass, and radius 
than in PSRCAT2}). The dependence of the beaming factors on $\dot{E}$ have been fitted,
using a nonlinear regression algorithm, with power laws, the indices of which are given in Table \ref{TabBFfit}. 
The goodness of each fit shown in Table \ref{TabBFfit} has been estimated by computing the coefficient of determination
$R^2$ that compares the sum of the squares of residuals  and the dataset variability (proportional to the sample variance). It is computed as
\begin{equation}
\label{DetCoeff}
R^2 = 1 - \frac{\sum_i^n (y_i - x_i)^2}{\sum_i^n (y_i - \langle y \rangle)^2} = 1 - \frac{\sum_i^n y_{r,i} ^2}{\sigma_y^2 (n-1)}
\end{equation}
where $y_i$ are the data, $x_i$ are the fit predictions, $y_{r,i}$ are the fit residuals, $\sigma_y^2$ is the data sample variance, $\langle y \rangle$ 
is the average value of the data sample, and $n$ is the
number of data points in the fit. $R^2$ ranges between 0 and 1 and a value close to 1 indicates a good correlation between data and fit predictions.
\begin{figure*}[htpp!]
\centering
\includegraphics[width=0.7\textwidth]{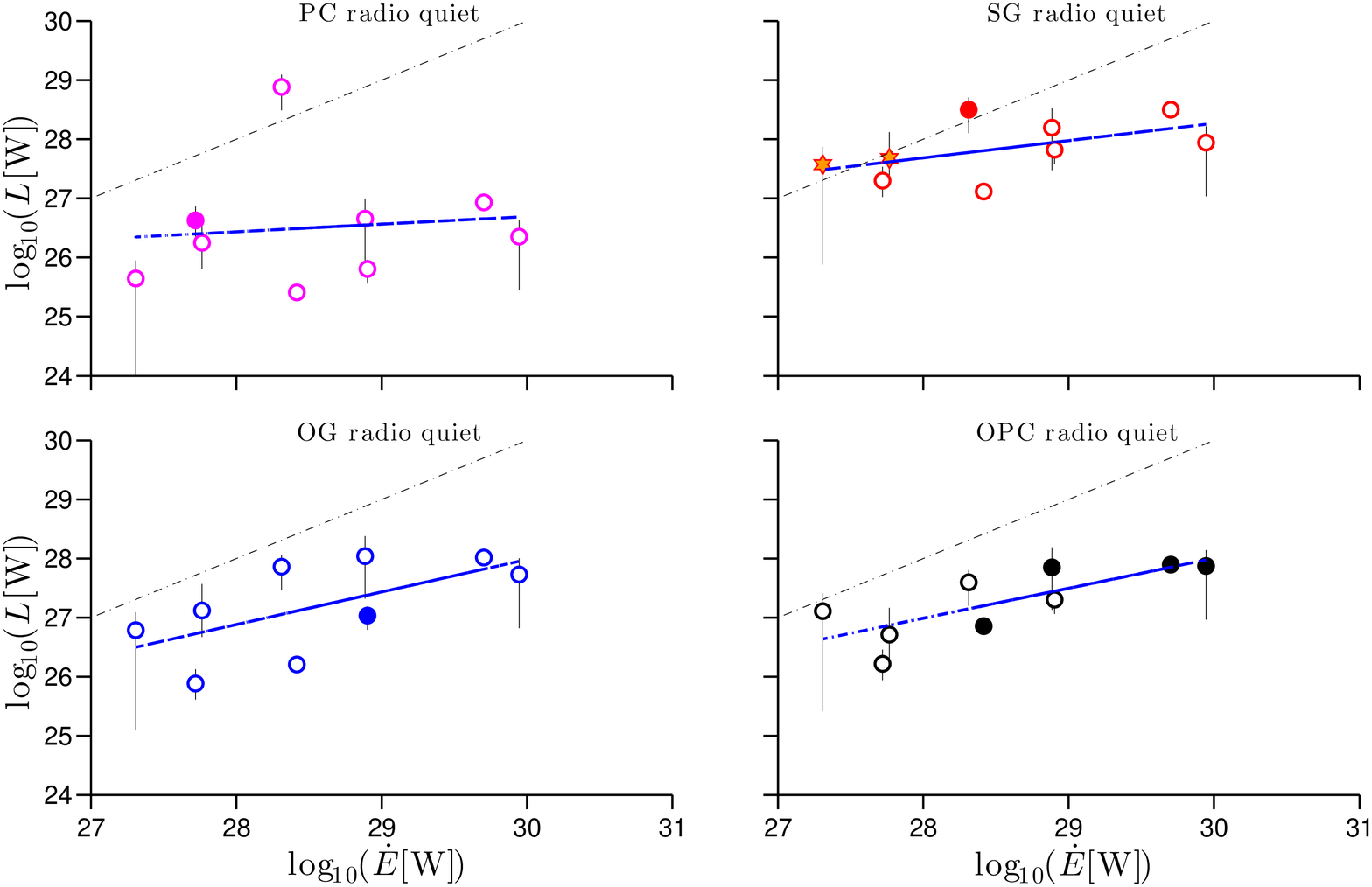}
\includegraphics[width=0.7\textwidth]{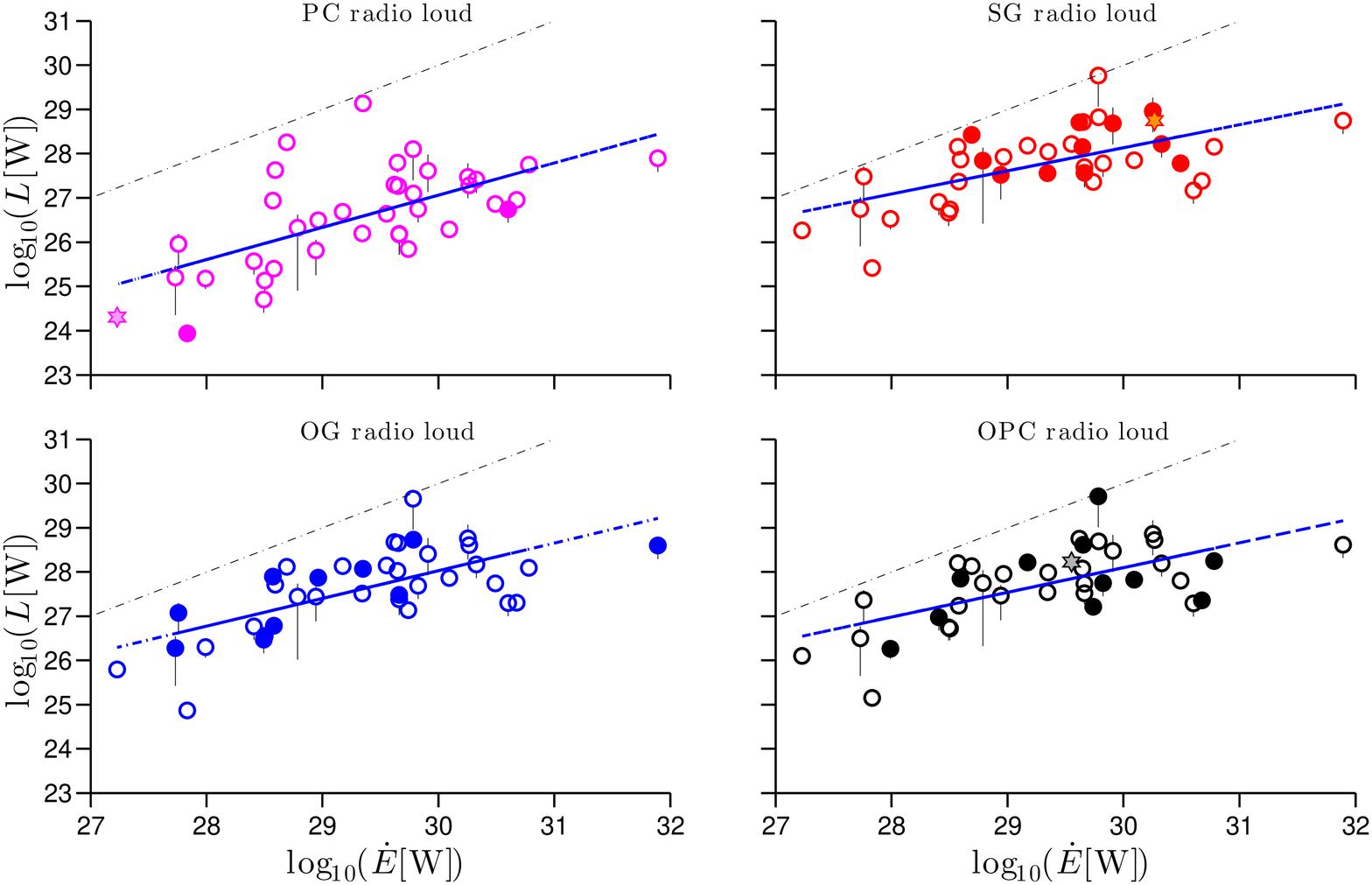}
\caption{$\gamma$-ray luminosity versus $\dot{E}$ for RQ (top panel) and RL (bottom panel) \emph{Fermi} pulsars.
The thick lines represent the best power-law fits to the data points; their parameters and $1\sigma$ errors are 
listed in Table \ref{TabBFfit}. The thin dot-dashed line
indicates 100\% conversion of $\dot{E}$ into $\gamma$-rays.}
\label{LumEdot1}
\end{figure*}

In the PC case  $f_{\Omega}$ is low as expected from the small hollow cone beam produced above the polar caps 
(Figure \ref{phase-plots}). The  $f_{\Omega}$ distribution is centred around 0.05 and 0.07 for RQ and RL objects, respectively. 
Since the PC beam size scales with the polar cap size, we expect $f_{\Omega}$ to decrease as the period increases, thus as 
$\dot{E}$ decreases. Because of the high dispersion in the sample, no trend is apparent.
In the SG case, the beaming factor of both RL and RQ pulsars remains rather stable and well constrained around $f_{\Omega}\sim1$.
A more pronounced $f_{\Omega}$-$\dot{E}$ correlation, characterised by a higher index of determination $R^2$ (Table \ref{TabBFfit}), 
is observed for the RQ pulsars. The absence of an evident correlation between $f_{\Omega}$ and $\dot{E}$ is due to the 
less strongly beamed nature of the SG emission, to the high level of off pulse emission predicted, and on the fact that, contrary to the OG, 
the bright caustics do not quickly shrink toward the pulsar equator as the pulsar ages, but they span a wider range of $\zeta$ values.
In the OG and OPC cases, the $f_{\Omega}$ values are much less dispersed for the RL pulsars than for the RQ pulsars as
indicated in \cite{pghg12}. Both OG and OPC do not show any significant $f_{\Omega}$ variation for RQ pulsars with $\dot{E}$ 
and are characterised by distributions centred around $\sim0.25$ and $\sim0.64$ for RQ and RL OG objects respectively, and $\sim0.41$ and $\sim0.85$ for 
RQ and RL OPC objects respectively. The OG model exhibits a more pronounced $f_{\Omega}$-$\dot{E}$ correlation, characterised by 
a higher index of determination $R^2$ (Table \ref{TabBFfit}), for RL pulsars.
The distribution of the beaming factor values in the framework of each model is shown in Figure \ref{FomDist}.
In all models other than the SG, the beaming factors calculated for the RQ population are numerically smaller than those calculated for the RL 
population. This is consistent with the fact that the wide SG $\gamma$-ray beams of the RL pulsars have higher probability to overlap the radio beams.
The beaming factors for RQ and RL LAT pulsars computed in the framework of each model are given in Table \ref{FOmega1}.
The $f_{\Omega}$ values are generally lower than one for all models and this suggests that to assign a beaming factor 
of one to all the pulsars (as done in PSRCAT2) is likely to represent an overestimation of the real values.

\subsection{Luminosity}
\label{LumSec}
Figure \ref{LumEdot1} shows the $\gamma$-ray luminosities versus $\dot{E}$ for RQ and RL pulsars in the upper and lower panel 
respectively. The $\gamma$-ray luminosities of the LAT pulsars have been computed with equation \ref{LumBeam} by using
the pulsar fluxes detected by the LAT above 100MeV (PSRCAT2), and the beaming factor $f_{\Omega}$ computed from the simulated
phase plot with Equation \ref{beamFrac}. The error on the LAT luminosities include the errors on the LAT fluxes 
and distances as listed in PSRCAT2.
The correlations between $\gamma$-ray luminosities and $\dot{E}$ have been fitted, using a nonlinear regression algorithm, with power laws, 
the indices and coefficient of determination $R^2$ of which are given in Table \ref{TabBFfit}. 
\begin{table*}
\def\arraystretch{1.2}
\centering
\begin{tabular}{| c  c | c | c | c || c | c | c|   }
\hline
\multicolumn{2}{|c|}{} & \multicolumn{3}{c||}{RQ}   & \multicolumn{3}{c|}{RL}  \\
\hline
\multicolumn{2}{|c|}{} & power-law index & intercept & R$^2$ & power-law index & intercept & R$^2$ \\
\hline
\multirow{2}{*}{PC}  & $f_{\Omega}$    & -0.11$\pm$0.11    &1.73$\pm$3.02    &   0.03   &           0.2$\pm$0.1     &  -7.0$\pm$2.9     & 0.10\\
                                &  $L_{\gamma}$  &  0.13$\pm$0.46   &22.82$\pm$13.02  &  0.01    &           0.73$\pm$0.15 & 5.24$\pm$4.38   & 0.40\\
\hline
\multirow{2}{*}{SG}  &  $f_{\Omega}$   & -0.07$\pm$0.01   & 1.97$\pm$ 0.41  &   0.41   &           -0.01$\pm$0.01&  0.3$\pm$ 0.3     &  0.03\\
    			       &  $L_{\gamma}$  &  0.29$\pm$0.19   &19.51$\pm$5.38  &   0.28   &           0.52$\pm$0.11  & 12.49$\pm$3.38   & 0.37\\
\hline
\multirow{2}{*}{OG}  &  $f_{\Omega}$   & 0.15$\pm$0.08    & -4.76$\pm$ 2.26 &   0.10   &         0.09$\pm$0.02   & -2.94$\pm$ 0.54    &  0.47 \\
    			        &  $L_{\gamma}$ &  0.55$\pm$0.28   &11.44$\pm$8.00   &   0.39   &         0.63$\pm$0.12   & 9.23$\pm$3.48   &  0.44\\
\hline
\multirow{2}{*}{OPC} & $f_{\Omega}$   & 0.11$\pm$0.08    & -3.42$\pm$ 2.19 &   0.06   &          0.02$\pm$0.01   &  -0.76$\pm$ 0.41    & 0.10\\
     			        &  $L_{\gamma}$  &  0.51$\pm$0.17  &12.8$\pm$4.95    &   0.59    &         0.56$\pm$0.11   & 11.33$\pm$3.37  & 0.40\\
\hline
\end{tabular}
\caption{Best power-law fits to the distribution of $f_{\Omega}$ and $L_{\gamma}$ as functions of $\dot{E}$ for each 
model and RL or RQ pulsars. The coefficient of determination R$^2$ relative to each fit is reported.}
\label{TabBFfit}
\end{table*}

For RQ and RL objects of all models, the trend $L_{\gamma}\appropto \dot{E}^{0.5}$, observed in the first LAT pulsar catalog 
\citep{aaa+10} and confirmed in PSRCAT2, is observed within the errors. The luminosity excess ($L_{\gamma} > \dot E$) 
observed in PSRCAT2 for some pulsars is solved here by computing each pulsar beaming factor from its best-fit light curve and 
emission pattern phase-plot (Equation \ref{beamFrac}). The only exception is noted for the PC luminosity of PSR J2021$+$4026 
but this results is likely incorrect since this pulsar appears to have a low $|\alpha - \zeta|$ and should be observed as RL or RF object. 
Moreover, the $\gamma$-ray luminosity distribution as a function of $\dot{E}$, evaluated in the framework of each model, appears 
much less dispersed than in the catalog. The lack of objects with $L_{\gamma} > \dot E$, using our $f_{\Omega}$ estimate,
supports the conclusion that to assign a beaming factor of 1 to all the pulsars represents an overestimate of the real value, particularly 
for low $\dot{E}$ pulsars.
The distributions observed in Figure \ref{LumEdot1} for RL pulsars are consistent with the model prediction shown in \cite{pghg12}, with 
the PC model providing the lowest luminosity values and SG and OG distributions characterised by the same dispersion. 

\begin{figure*}[htpp!]
\centering
\includegraphics[width=0.7\textwidth]{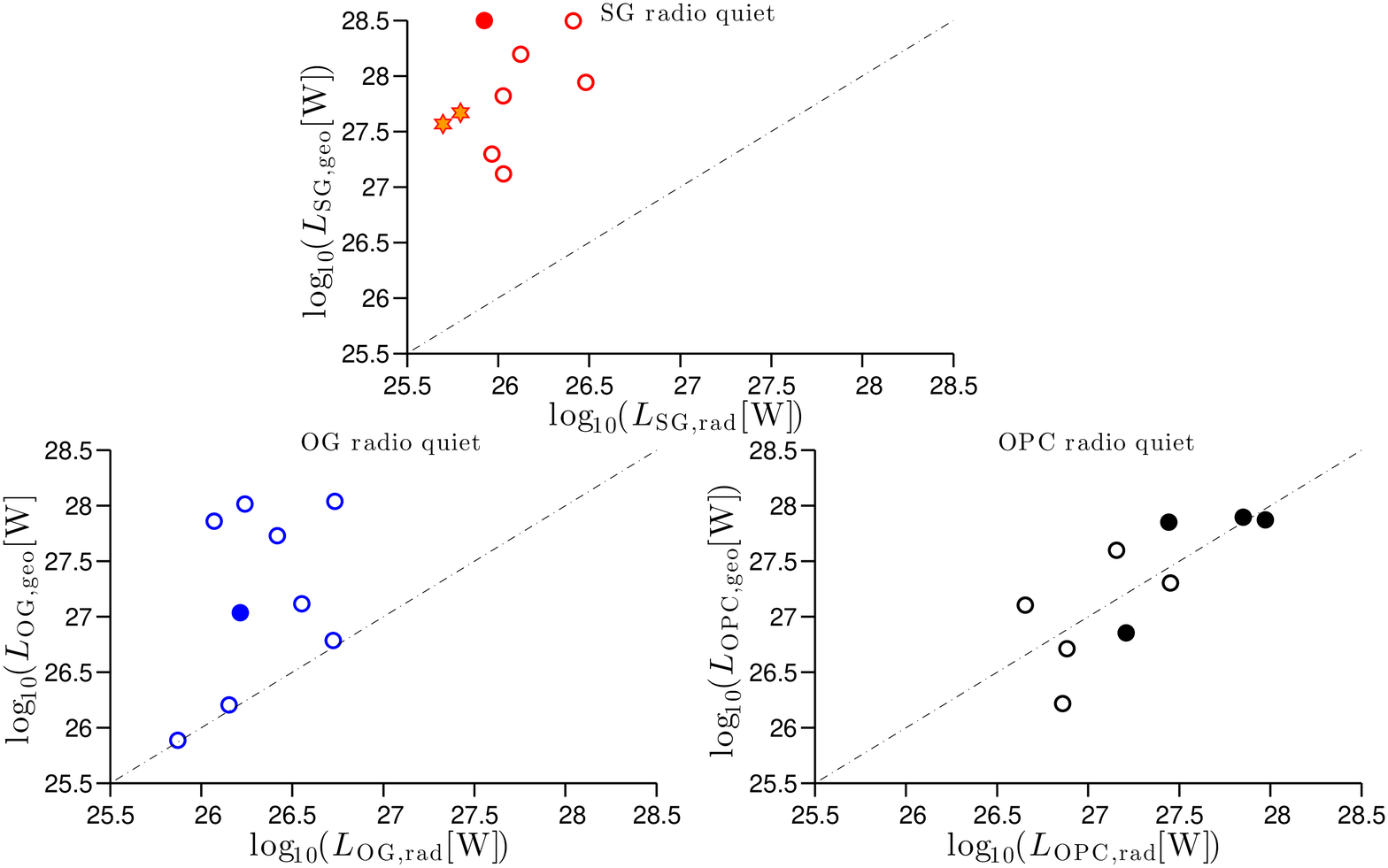}
\includegraphics[width=0.7\textwidth]{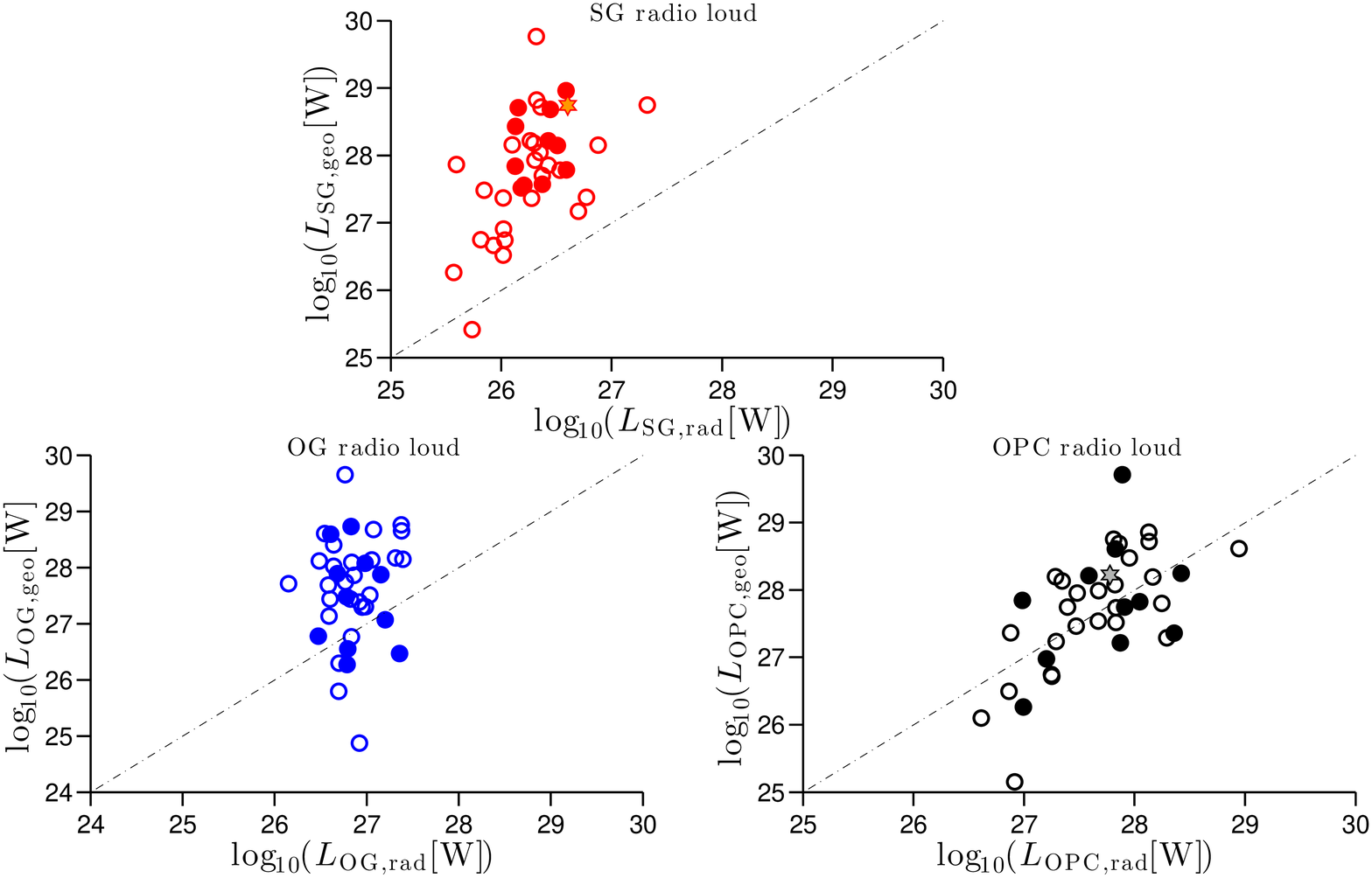}
\caption{Geometric $\gamma$-ray luminosity, $L_\mathrm{geo}$ versus the standard gap-model $\gamma$-ray luminosity $L_\mathrm{rad}$ for 
RQ (top panel) and RL (bottom panel) \emph{Fermi} pulsars and each model. The dot-dashed lines indicates $L_\mathrm{geo}=L_\mathrm{rad}$.}
\label{LumComp}
\end{figure*}

Figure \ref{LumComp} shows the geometric $\gamma$-ray luminosity of the LAT pulsars computed with Equations \ref{LumBeam} 
and \ref{beamFrac}, $L_\mathrm{geo}$, as a function of the standard gap-model $\gamma$-ray luminosity computed as $L_\mathrm{rad}= W^3 \dot{E}$.  
In some cases $L_\mathrm{geo}$ overestimates $L_\mathrm{rad}$ by more than 2 orders of magnitude for RQ pulsars and 3 orders of magnitude 
for RL pulsars. This is mainly the case for small gap-width pulsars, $W<0.1$, that are expected to shine with $L_\mathrm{rad}<0.001\dot{E}$ but 
that show larger $\gamma$-ray luminosities $L_\mathrm{geo}$. 
This inconsistency reflects the difficulties in defining a unique gap width that could simultaneously explain the light-curve shape and the observed 
pulsar flux in the framework of the same radiative-geometrical model: the observed $\gamma$-ray pulsar light-curve shapes are well explained 
by thin gaps that yet do not provide enough luminosity to predict the observed $\gamma$-ray flux. The radiative-geometrical luminosity discrepancy appears 
more pronounced for the SG pulsars, where the gap-width computation critically depends on the assumed shape of the pair formation 
front (PFF) \citep[see description of the $\lambda$ parameter in][ Section 5.2]{pghg12}. In the OG model, $L_\mathrm{geo}$ overestimates
$L_\mathrm{rad}$ just for RQ pulsars while the $L_\mathrm{geo}$ of RL objects are more distributed around 100\% of $L_\mathrm{rad}$ but showing
a large dispersion above $L_\mathrm{rad}$. The OPC is the model that shows the highest agreement between geometrical and radiative luminosity 
estimates with both RQ and RL $L_\mathrm{geo}$ homogeneously distributed around 100\% of $L_\mathrm{rad}$. This is expected since the OPC 
luminosity law is artificially designed to match observed luminosities.

\cite{pghg12} reduced the lack of $L_\mathrm{rad}$ discrepancy by choosing the highest possible $\gamma$-ray efficiency, 100\%, for the OG model and by choosing an 
appropriate PFF shape \citep[see Section 5.2 of][]{pghg12} and by setting the $\gamma$-ray efficiency to 1200\% for the SG model. The high SG 
efficiency is possibly justified by the enhanced accelerating electric field expected in case of offset polar caps \citep{hm11}. 

The geometrical approach adopted in this paper avoid the lack of $L_\mathrm{rad}$ obtained for SG and OG models \citep{pghg12} when one tries to 
simultaneously explain light-curve shape and luminosity and does not require ad-hoc $\gamma$-ray efficiency assumptions.
On the other hand our geometrical approach highlights an intrinsic inconsistency between geometric and radiative models in describing the pulsar 
magnetosphere. The geometrical model used in this paper is based on simple assumptions that do not account for the complex 
electrodynamics at the base of the radiative gap-models. This is true for both OG and SG models and cause the radiative-geometrical luminosity 
inconsistencies discussed above. The OG model requires large gap widths to produce the observed luminosities, and these gaps do not produce the 
observed thin light-curve peaks. This is suggested by the higher consistency between radiative and geometrical luminosities obtained by the OPC model 
that differs from the OG just in the gap-width formulation. 
In the SG model, radiative-geometrical luminosity inconsistencies are due to two factors: thin slot gaps required to explain the light-curve shapes do not 
produce enough luminosity to explain the observed fluxes; the electrodynamics of the low-altitude slot-gap region is not implemented in the adopted 
geometrical model.
The assumptions on the SG high-altitude emission and the inconsistencies between radiative and geometrical SG emission at low-altitude will be 
discussed in Section \ref{SGlow}.

In the current formulation of SG and OG geometrical models, both SG and OG model acceleration and emission regions are restricted to inside the light cylinder.  
In more recent and realistic global dissipative pulsar magnetosphere models, acceleration and emission also outside the light cylinder may be able to solve this 
radiative-geometrical luminosity discrepancy \citep[in preparation]{khk14,BraPrep14}.

\subsection{Magnetic alignment and Pulsar orientation}
\begin{figure}
\centering
\includegraphics[width=0.49\textwidth]{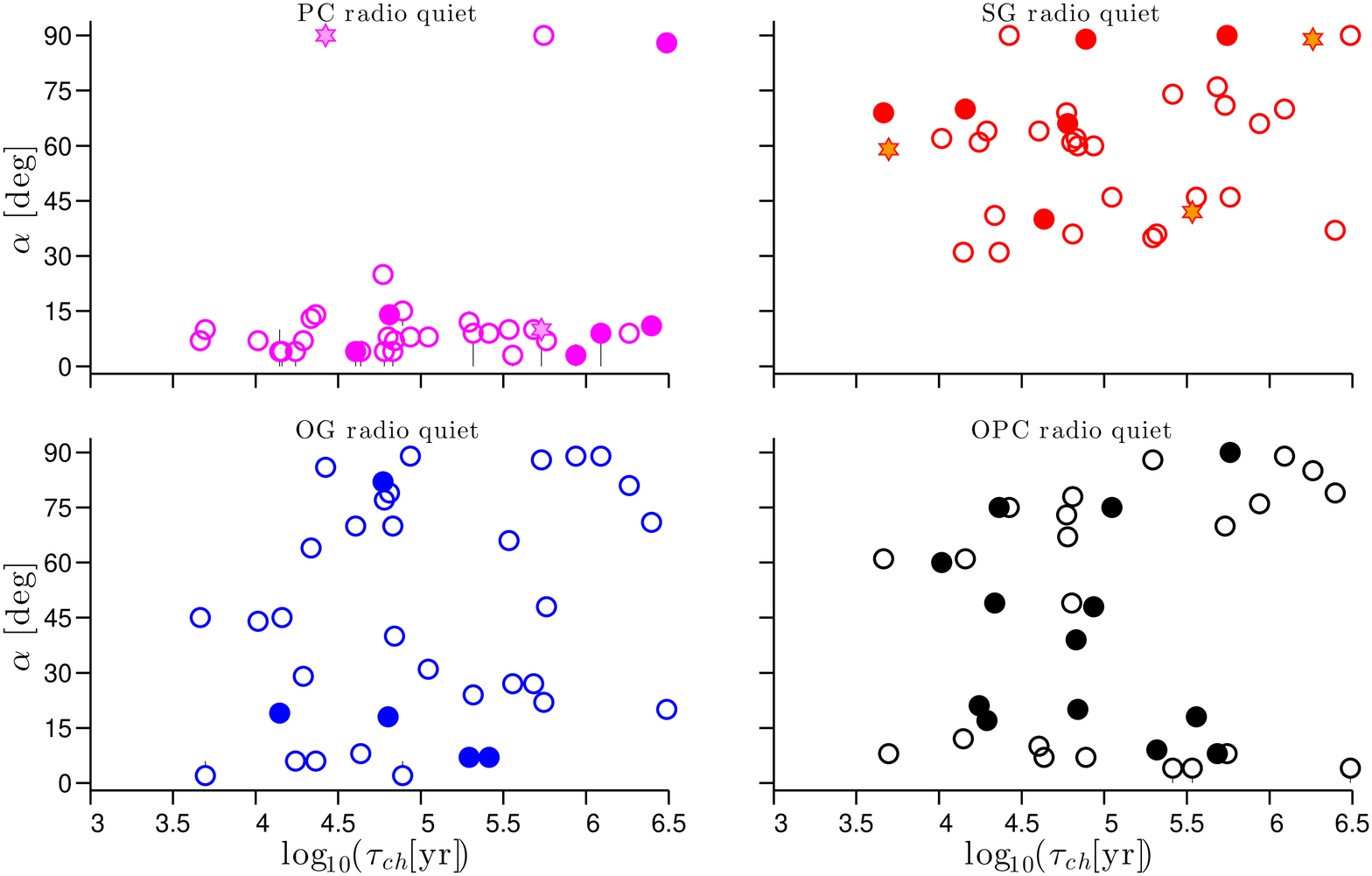}
\includegraphics[width=0.49\textwidth]{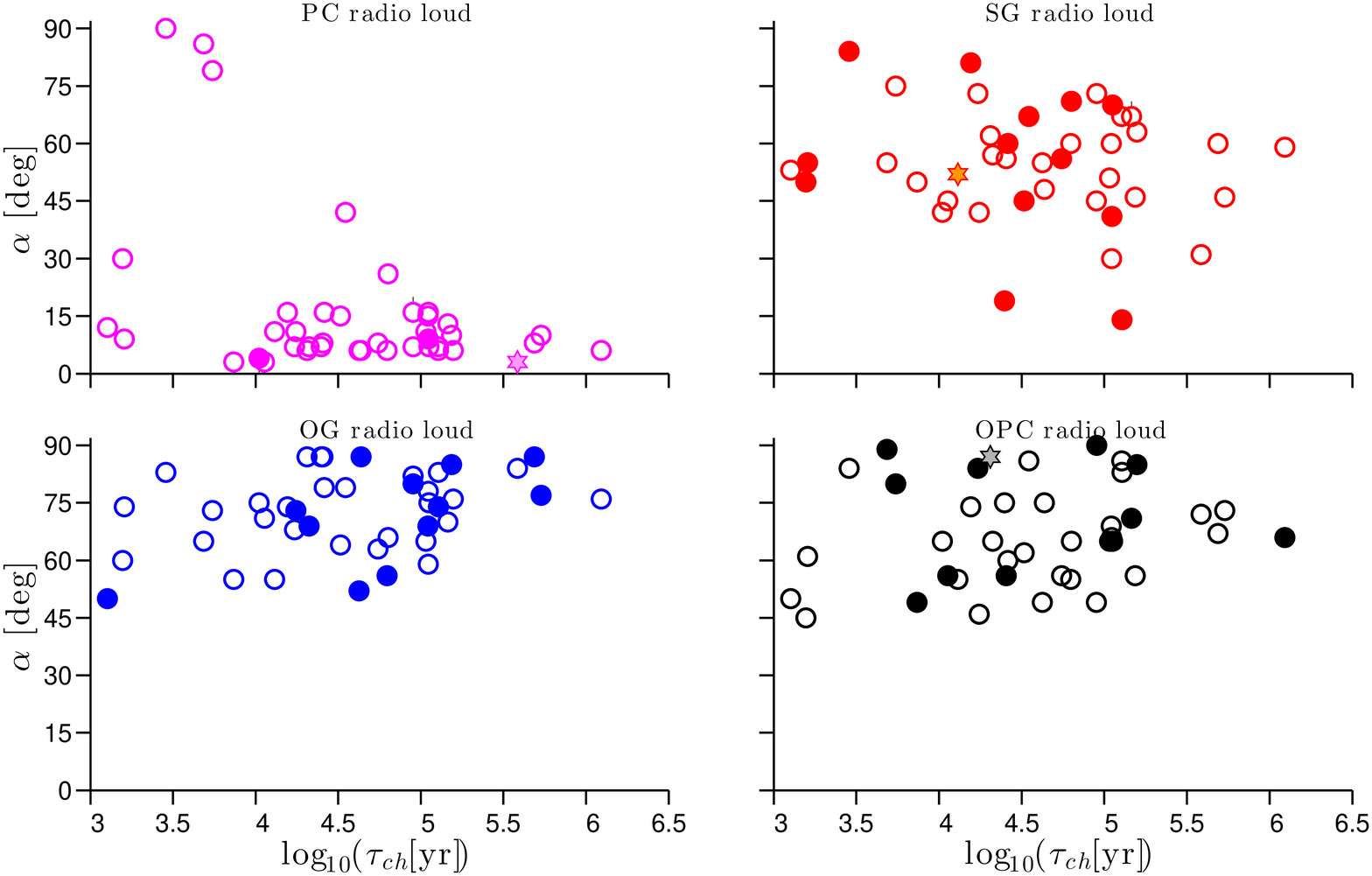}
\caption{Magnetic obliquity $\alpha$  versus characteristic age for RQ (top panel) and RL (bottom panel) \emph{Fermi} pulsars
and each model.}
\label{AgeAlph1}
\end{figure}

Figure \ref{AgeAlph1} shows $\alpha$ versus the characteristic age $\tau_{ch}$ 
for each model and pulsar type. We have tried to verify if the LAT sample shows any evidence of an alignment or misalignment of the 
magnetic and rotational axes with age. 
The possibility that magnetic and rotational axes of a pulsar could become aligned with time has been suggested by 
\cite{ycbb10} on the basis of a pulsar evolution model including two distinct effects: an exponential magnetic alignment as indicated by \cite{jon76}
and a progressive narrowing of the emission cone as the pulsar ages. The alignment of magnetic and rotational axes of a pulsar should occur 
on a timescale of $\sim 10^{6}$ yr.

\begin{figure}
\centering
\includegraphics[width=0.49\textwidth]{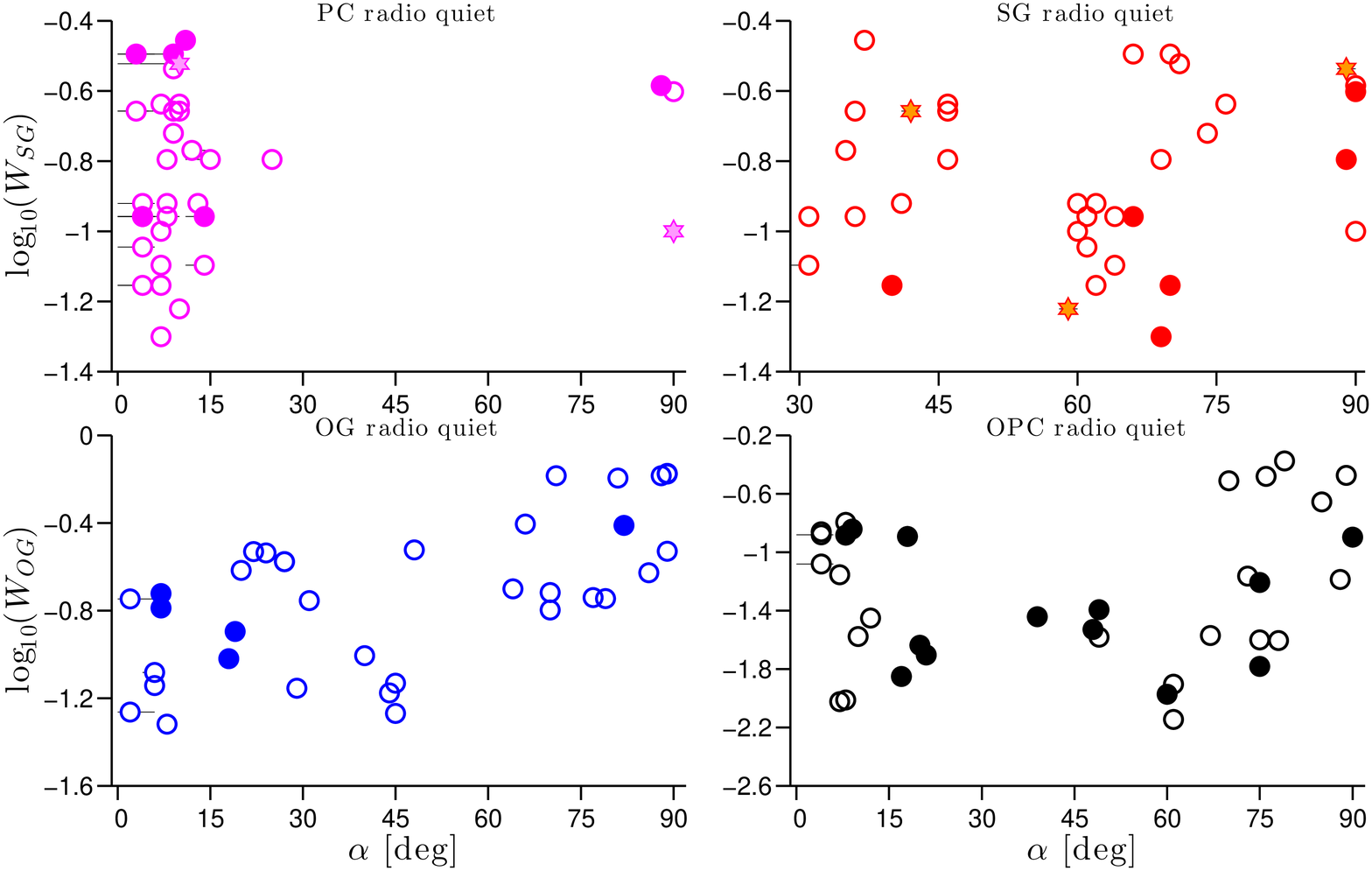}
\includegraphics[width=0.49\textwidth]{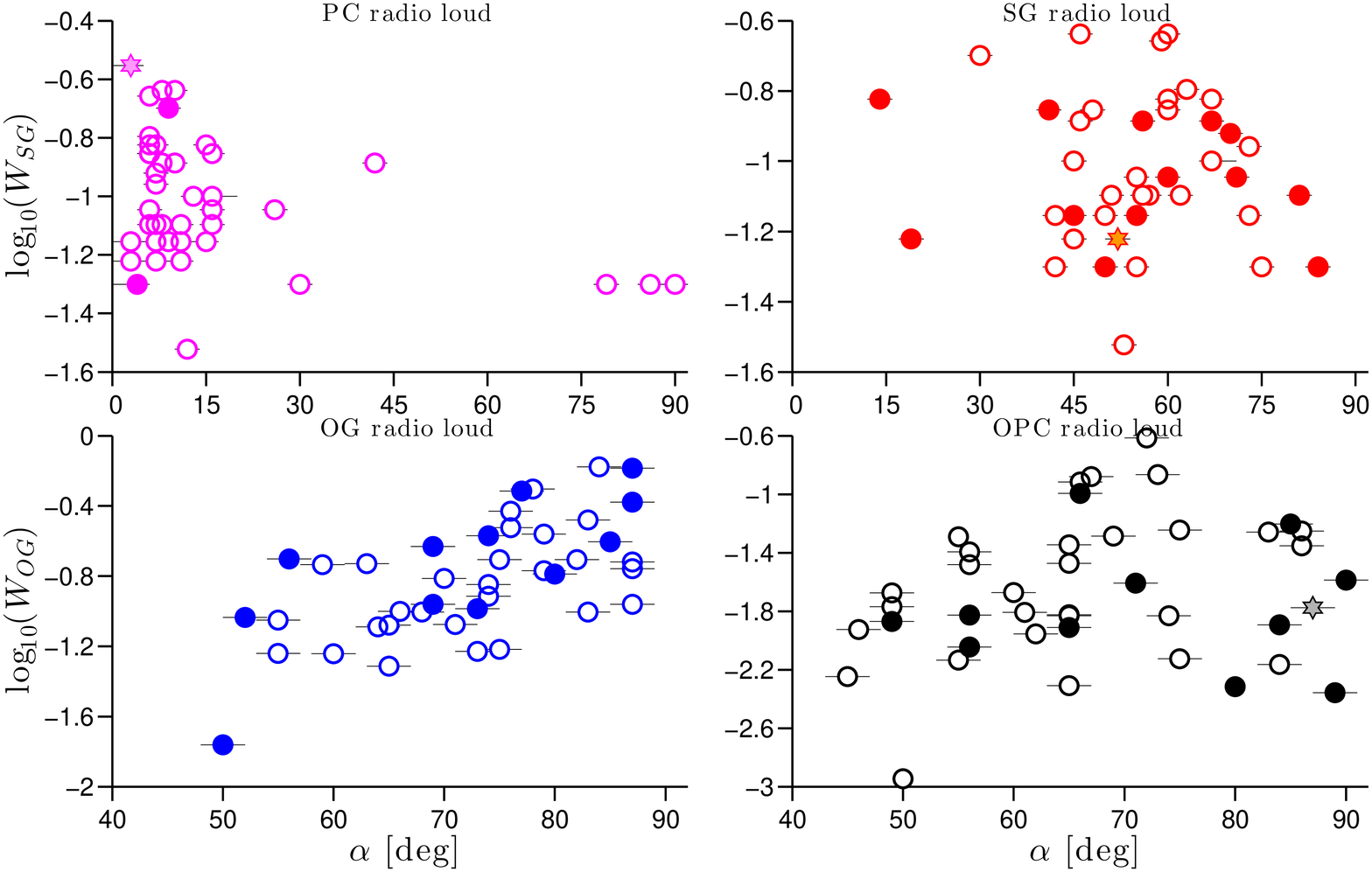}
\caption{Gap width  as a function of $\alpha$ for RQ (top panel) and RL (bottom panel) \emph{Fermi} pulsars
and each model.}
\label{FCBXSquareJAlpSGOGRGLogLog}
\end{figure}

Both RQ and RL solutions for all the models are highly dispersed and show no evidence of changes in $\alpha$ with age. 
In Figure \ref{FCBXSquareJAlpSGOGRGLogLog} we show gap width as a function of $\alpha$ for RQ 
and RL pulsars for all models. A mild dependence between gap width and $\alpha$ is present  just for the OG model
and is due to the fact that in the OG model the gap width $w_\mathrm{OG}$ is a function of $\alpha$.

\begin{figure}
\centering
\includegraphics[width=0.49\textwidth]{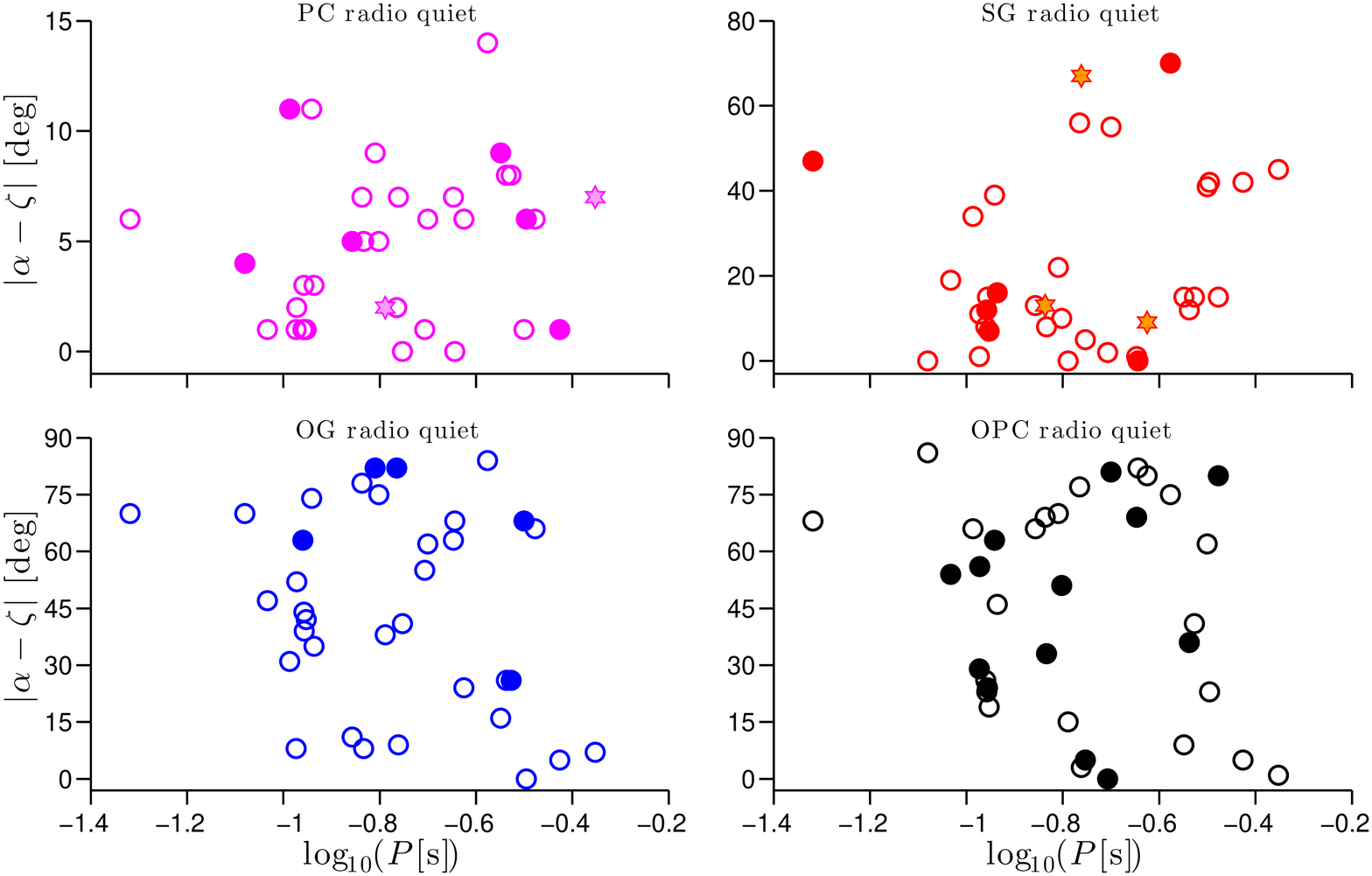}
\includegraphics[width=0.49\textwidth]{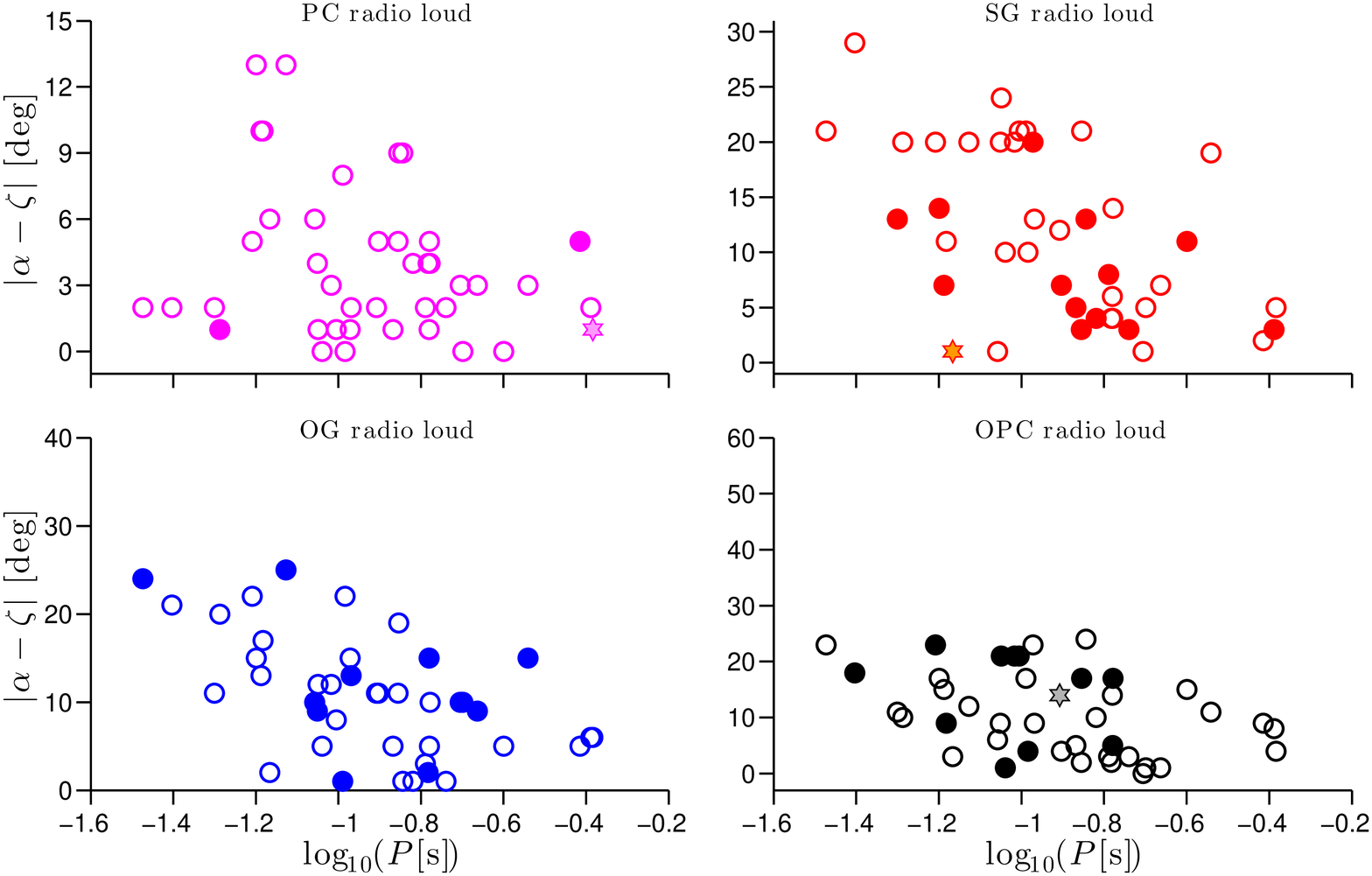}
\caption{ For each model the $\beta=|\alpha-\zeta|$ angles a function of the spin period for RQ 
(top panel) and RL (bottom panel) \emph{Fermi} pulsars is shown.}
\label{AlpZetG}
\end{figure}

Figure \ref{AlpZetG} shows the quantity $|\alpha-\zeta|$ plotted as a function of the pulsar period, for RQ and RL pulsars in all models.
It is evident how the solutions change from RQ to RL objects, appearing much less dispersed and showing slight decreasing trends with 
the spin period. This trend is due to a selection effect for which young and rapidly spinning pulsars have a wider radio beam that can 
overlap the $\gamma$-ray beam up to high $|\alpha-\zeta|$ values. As a pulsar ages, its spin period increases while polar cap size and 
radio beam size decrease and the radio beam will overlap the $\gamma$-ray beam only for  smaller $|\alpha-\zeta|$.
This trend is consistent with changes of $|\alpha-\zeta|$ as a function of the spin period, obtained, for each emission 
model, in the population synthesis study described in \cite{pie10} and shown in Figure 6.84 of that paper.

\begin{table*}
\def\arraystretch{1.2}
\centering
\begin{tabular}{| c  c | c | c | c || c | c | c|   }
\hline
\multicolumn{2}{|c|}{} & \multicolumn{3}{c||}{RQ}   & \multicolumn{3}{c|}{RL}  \\
\hline
\multicolumn{2}{|c|}{} & power-law index & intercept & R$^2$ & power-law index & intercept & R$^2$ \\
\hline
\multirow{2}{*}{SG}      & $E_\mathrm{cut}$     & -0.59$\pm$0.12   &-0.14$\pm$0.11  &  0.42    &         -0.46$\pm$0.23  &  -0.14$\pm$0.24 & 0.11\\
                                    &  $\Gamma$  &  -0.30$\pm$0.07  &-0.11$\pm$0.06  &   0.39   &          -0.13$\pm$0.07 &   0.05$\pm$0.07 & 0.11\\
\hline
\multirow{2}{*}{OG}     & $E_\mathrm{cut}$      & -0.41$\pm$0.09   &0.07$\pm$0.07  &   0.42   &          -0.25$\pm$0.15 &   0.11$\pm$0.14 & 0.08\\
                                    &  $\Gamma$  &  -0.19$\pm$0.05  &0.01$\pm$0.04  &    0.31  &          -0.06$\pm$0.05 &   0.13$\pm$0.04 & 0.04\\
\hline
\multirow{2}{*}{OPC}   & $E_\mathrm{cut}$     & -0.29$\pm$0.05    &-0.01$\pm$0.08 &   0.47   &          -0.21$\pm$0.10 &   -0.03$\pm$0.18  & 0.11\\
                                    &  $\Gamma$  &  -0.15$\pm$0.03  &-0.04$\pm$0.04  &   0.42  &          -0.09$\pm$0.03 &    0.03$\pm$0.06 & 0.26\\
\hline
\end{tabular}
\caption{Best power-law fits to the distribution of $E_\mathrm{cut}$  and $\Gamma$ as functions of the width of the acceleration gap for each 
model and RL or RQ pulsars. The coefficient of determination R$^2$ relative to each fit is reported.}
\label{SPECTRALTAB}
\end{table*}

\subsection{High-energy cutoff and spectral index versus gap width}

\label{HighECut}
\begin{figure}[htbp!]
\centering
\includegraphics[width=0.49\textwidth]{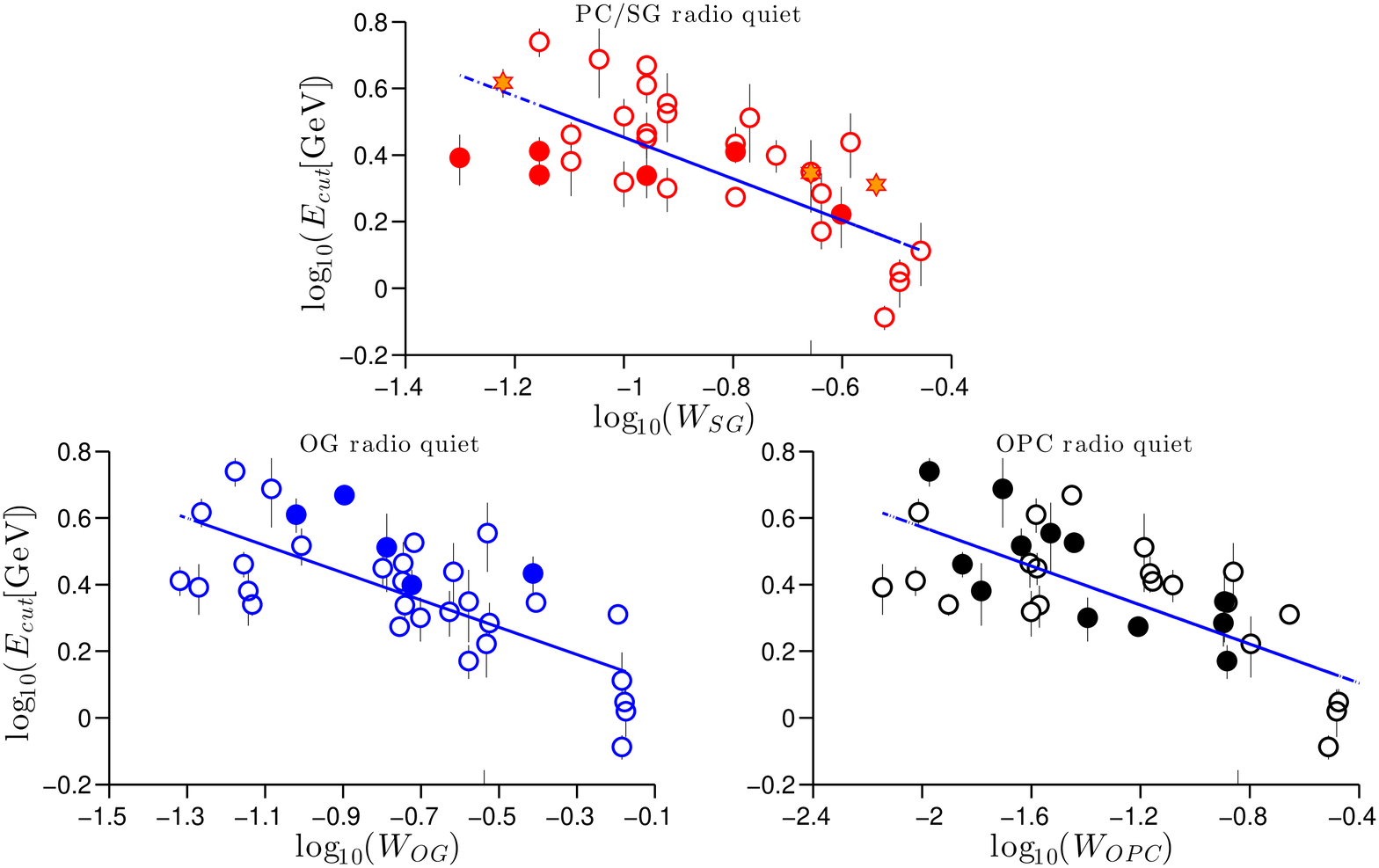}
\includegraphics[width=0.49\textwidth]{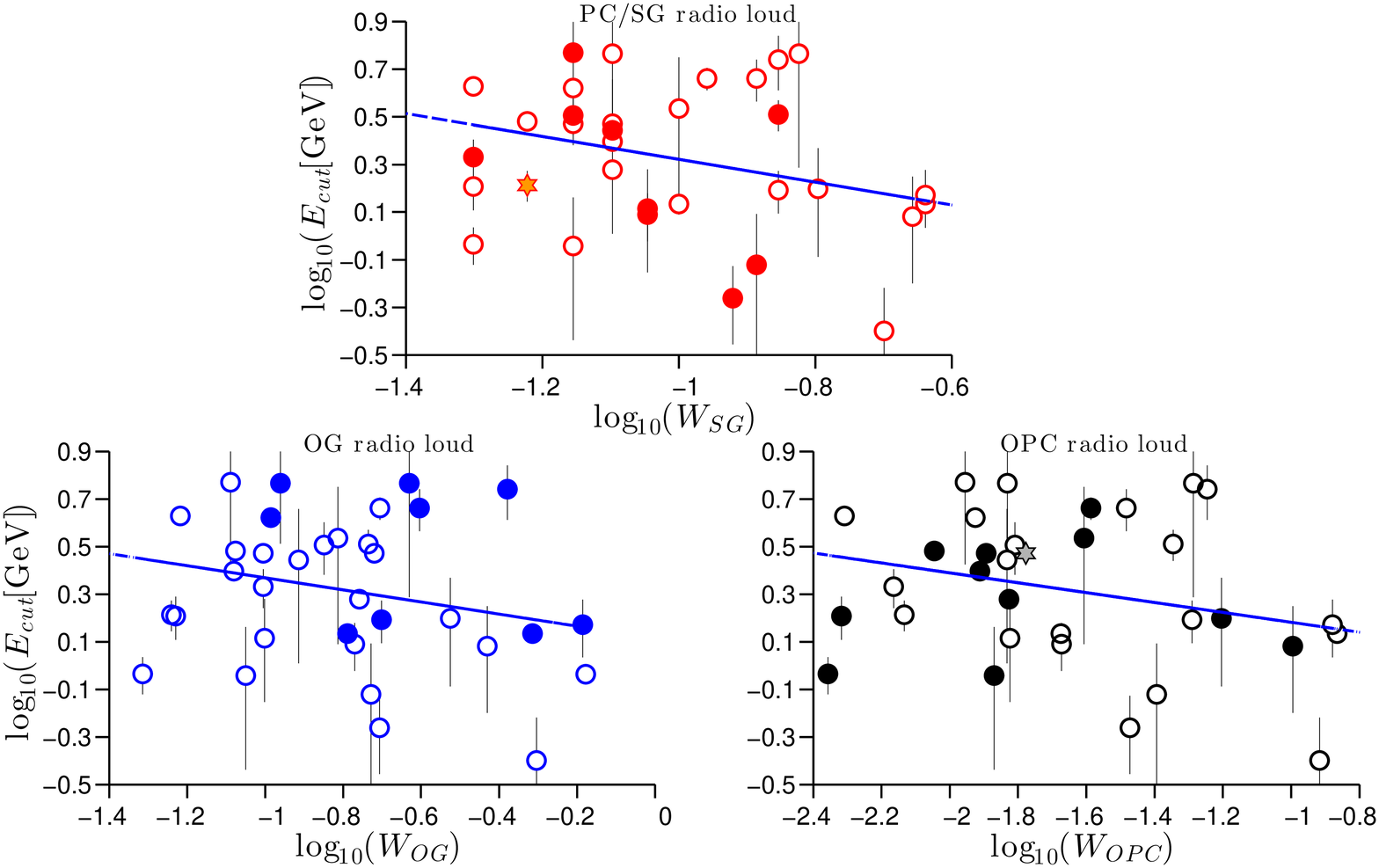}
\caption{Energy cutoff versus gap width for RQ (top panel) and RL (bottom panel) \emph{Fermi} pulsars
for each model. The best fit power law trends are given in each figure. PC and SG results are characterised by the same gap width $w_\mathrm{SG}$
and have been plotted together.}
\label{EcutGapW1}
\end{figure}

\begin{figure}[t!]
\centering
\includegraphics[width=0.49\textwidth]{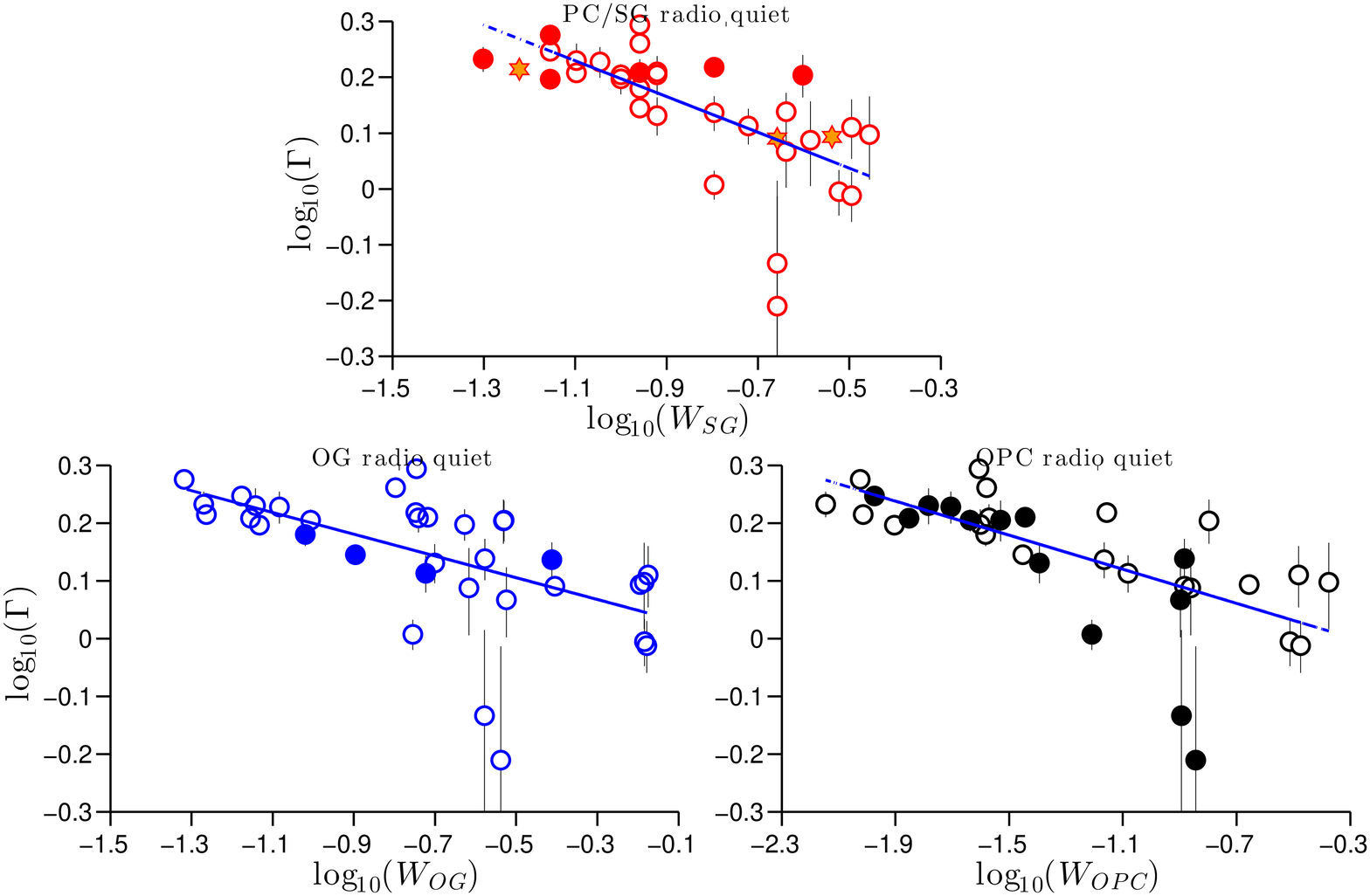}
\includegraphics[width=0.49\textwidth]{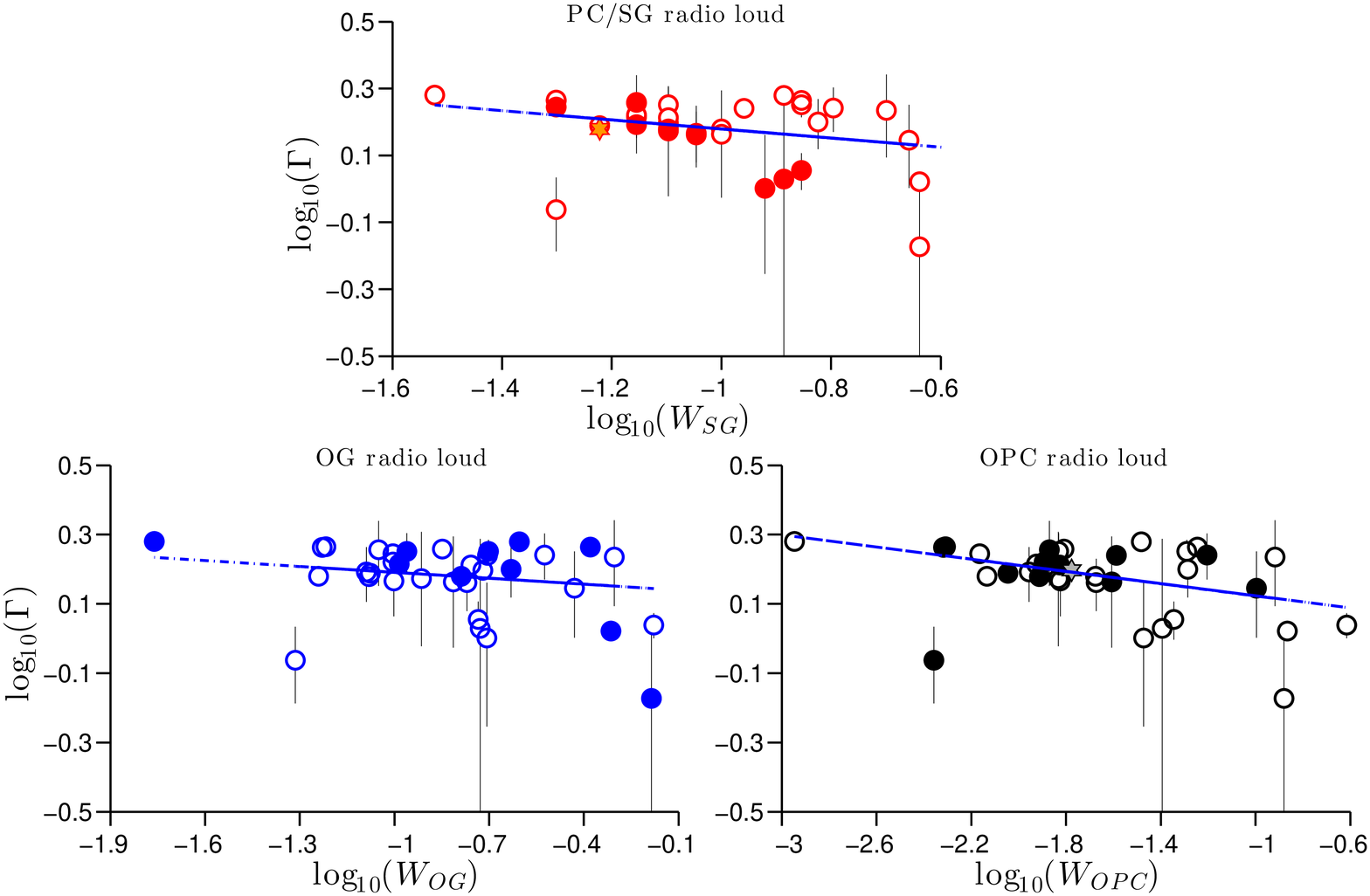}
\caption{Spectral index versus gap width for RQ (top panel) and RL (bottom panel) \emph{Fermi} pulsars
for each model. The best fit power law trends are given in each figure. PC and SG results are characterised by the same gap width $w_\mathrm{SG}$
and have been plotted together.}
\label{SpecGapW1}
\end{figure}

Figures \ref{EcutGapW1} and \ref{SpecGapW1} show the relation between observable spectral characteristics, namely the high-energy 
cutoff $E_\mathrm{cut}$ and spectral index $\Gamma$, and the width of the emission gap evaluated in the framework of each emission 
model. $\Gamma$ and $E_\mathrm{cut}$ are taken from PSRCAT2. The  SG, OG, and OPC gap widths have been calculated for each 
pulsar according to its spin characteristics as described in \cite{pghg12}. 

The spectral fits for the RL pulsars J1410$-$6132, J1513$-$5908, and J1835$-$1106 were noted as unreliable in PSRCAT2. These pulsars 
are not included in Figures \ref{EcutGapW1} and \ref{SpecGapW1}. We find a tendency for $E_\mathrm{cut}$ and $\Gamma$ to decrease 
when the gaps widens. This dependence is particularly important because it relates the spectral characteristics and the intrinsic, non-directly 
observable, gap width that controls the  acceleration and cascade electrodynamics.

A power law dependence between $E_\mathrm{cut}$ and SG, OG, and OPC gap widths can be theoretically obtained as it follows 
(see Figure \ref{EcutGapW1} and Table \ref{SPECTRALTAB} for comparison). From \cite{aaa+10b}, 
the $E_\mathrm{cut}$ dependence is defined as
\begin{equation}
E_\mathrm{cut}\propto E_{\|}^{3/4}\rho_{c}^{1/2}
\end{equation}
where $E_{\|}$ is the electric field parallel to the magnetic field $B$ lines, and $\rho_c$ is the radius of curvature of the magnetic field lines.
Since for all the implemented emission models $E_{\|}$ scales as $E_{\|}\propto w^2 B_\mathrm{LC}$, we have
\begin{equation}
E_\mathrm{cut}\propto [w^2 B_\mathrm{LC}]^{3/4} \rho_{c}^{1/2}
\end{equation}
where $w$ is the width of the emission gap. 
The light cylinder magnetic field dependence can be written as
\begin{equation}
B_\mathrm{LC}=B_\mathrm{G}\left(\frac{\Omega R}{c}\right)^3 \propto B_\mathrm{G} P^{-3}
\end{equation}
where $R$ the pulsar radius.
Since, for SG, OG, and OPC the $\gamma$-ray emission occurs mainly at high altitude, close to the light cylinder,
$\rho_{c}\propto R_{LC}\propto P$, and the $E_\mathrm{cut}$ proportionality can be expressed as
\begin{equation}
E_\mathrm{cut}\propto w^{3/2} [P B_\mathrm{G}^{-3/7}]^{-7/4}. 
\label{EcutRef}
\end{equation}
Since the slot gap width dependence follows approximately
\begin{alignat}{2}
w_\mathrm{SG}\appropto PB_\mathrm{G}^{-3/7},~~B_\mathrm{G} > 0.1 \times 10^{12}~~\mathrm{G}\\
w_\mathrm{SG}\appropto PB_\mathrm{G}^{-4/7},~~B_\mathrm{G} < 0.1 \times 10^{12}~~\mathrm{G}
\end{alignat}
the final approximate $E_\mathrm{cut,SG}=f(w_\mathrm{SG})$ dependence is
\begin{equation}
E_\mathrm{cut,SG} \appropto w_\mathrm{SG}^{3/2} w_\mathrm{SG}^{-7/4} = w_\mathrm{SG}^{-0.25}.
\end{equation}
More approximated power law dependences between $E_\mathrm{cut}$ and the OG and OPC gap widths can also be obtained from Equation
\ref{EcutRef} and from the $w_\mathrm{OG}$ and $w_\mathrm{OPC}$ dependences. From \cite{pghg12} we have that $w_\mathrm{OG}$
can be written as
\begin{alignat}{2}
w_\mathrm{OG}\propto B_\mathrm{G}^{-4/7}P^{26/21}=[B_\mathrm{G}^{-3/7}P^{13/14}]^{4/3}\approx [B_\mathrm{G}^{-3/7}P]^{4/3}\label{OGnew}\\
w_\mathrm{OPC}\propto \dot{E}^{-0.5}=B_\mathrm{G}^{-1}P^{2}=[B_\mathrm{G}^{-3/7}P^{6/7}]^{7/3}\approx [B_\mathrm{G}^{-3/7}P]^{7/3}
\label{OPCnew}
\end{alignat}
where the right-hand member of Equation \ref{OGnew} has been obtained under the assumption $P^{13/14}\approx P$, while the right-hand member of 
Equation \ref{OPCnew} has been obtained by making use of the relations $\dot{E}\propto\dot{P}P^{-3}$ and $\dot{P}P\propto B_\mathrm{G}^2$, and by 
assuming $P^{6/7}\approx P$. By solving Equations \ref{OGnew} and \ref{OPCnew} for $[B_\mathrm{G}^{-3/7}P]$ and substituting in Equation \ref{EcutRef} 
we obtain the final approximate $E_\mathrm{cut,OG}=f(w_\mathrm{OG})$ and $E_\mathrm{cut,OPC}=f(w_\mathrm{OPC})$ dependences
\begin{alignat}{2}
E_\mathrm{cut,OG} \appropto w_\mathrm{OG}^{3/2} w_\mathrm{OG}^{-21/16} \sim w_\mathrm{OG}^{0.19}\\
E_\mathrm{cut,OPC} \appropto w_\mathrm{OPC}^{3/2} w_\mathrm{OPC}^{-3/4} = w_\mathrm{OPC}^{0.75}
\end{alignat}

In Figures \ref{EcutGapW1} and \ref{SpecGapW1}, nonlinear regression power-law fits to all the data points are given for 
both pulsar types and all models. The fit indices and coefficients of determination $R^2$ are given in Table \ref{SPECTRALTAB}.

Figure \ref{NvisA_large_effi1p012p01p00p5_Ecutgap_histo} shows the behaviour of $E_\mathrm{cut}$ and $\Gamma$ with respect to the SG, OG, and OPC gap widths 
for the population synthesis results in \cite{pghg12}. The fact that no trend is apparent is due to the choice of 
spectral characteristics that have been randomly assigned from the double gaussian distribution that statistically describes the 
observed values in the LAT catalogue.
The fact that the results in Figures \ref{EcutGapW1} and \ref{SpecGapW1} show a trend that can be 
predicted theoretically encourages future efforts to confirm the trend and to improve the implemented fit strategy. Since in the phase-plot 
modelling there is no relation between $E_\mathrm{cut}$ and gap width, our results 
suggest a real physical relation between the $\gamma$-ray spectrum and gap width that can be used to discriminate 
between the proposed models. Moreover, the lack of trend in the simulation data for both $E_\mathrm{cut}$ and $\Gamma$ 
(Figures \ref{NvisA_large_effi1p012p01p00p5_Ecutgap_histo})
demonstrates that the decline observed in the present LAT sample is not due to an observation bias. A more precise $E_\mathrm{cut}=f(w)$ 
relation drawn from the analysis of a larger LAT sample should be tested in the future for both young and millisecond pulsars.
\begin{figure}[t!]
\centering
\includegraphics[width=0.49\textwidth]{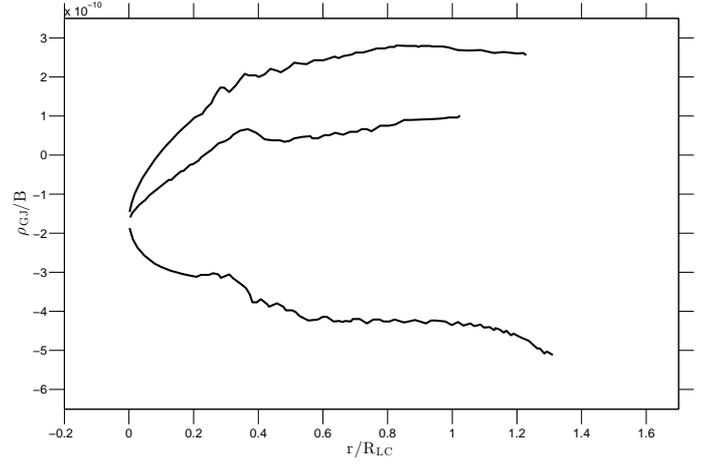}
\caption{Variation, in a force-free magnetosphere, of the ratio Goldreich-Julian charge density over the magnetic field, $\rho_\mathrm{GJ}/B$, with the 
distance from the pulsar expressed in unit of the light-cylinder radius, $r/R_\mathrm{LC}$.}
\label{HighAltSG}
\end{figure}

\subsubsection{The SG $\gamma$-ray emission}
\label{SGlow}
The SG width computation implemented in this paper follows the prescription by \cite{mh04a}. Those authors assumed that the Goldreich-Julian 
charge density, $\rho_\mathrm{GJ}$ \citep{gj69}, does not grow monotonically up to the light cylinder, as it would happen in the case of a dipolar 
magnetic field, but it levels off at high altitudes. The growing of $\rho_\mathrm{GJ}$ depends on the field line curvature that in a force free magnetosphere 
decreases toward the light cylinder (the poloidal magnetic field lines tend to get straighter) so causing the levelling off of $\rho_\mathrm{GJ}$. 
Recent implementations of force free magnetosphere pulsar models show that, at high altitudes, the variation of $\rho_\mathrm{GJ}$ with the 
distance from the pulsar is consistent with the assumption from \cite{mh04a}. In Figure \ref{HighAltSG} the variation of the quantity $\rho_\mathrm{GJ}/B$
with $B$ the pulsar magnetic field,  as a function of the distance from the pulsar in units of $R_\mathrm{LC}$ is shown. It shows how the quantity
$\rho_\mathrm{GJ}/B$ levels-off at distances larger than 0.4 $R_\mathrm{LC}$. 

At low altitudes, typically $<0.4 R_\mathrm{LC}$, the physical SG model predicts a reversal of the sign of $E_{\parallel}$ on some magnetic field lines and 
for some $\alpha$ values and no straightening of the low altitude magnetic field lines is assumed. In the current implementation of the SG emission 
geometry no reversal of the sign of $E_{\parallel}$ and no straightening of the magnetic field lines at low altitude are implemented: our modelling of the
SG geometry assumes a simplified low-altitude slot-gap region and emission is assumed from all field lines in the gap. The impact of our simplified 
prescription for the SG structure in the current paper may be an overestimation of the geometric $\gamma$-ray luminosity, $L_\mathrm{geo}$, for those 
pulsars with very high $\alpha$. However the actual impact of our assumption on the estimate of $L_\mathrm{geo}$ could be quantified just through the 
future implementation of a geometric model that accounts for the reversal of the sign of $E_{\parallel}$ in the low-altitude slot gap.

\section{Summary}

We have selected a sample of young and middle-aged  pulsars observed by the LAT during three years and described in 
PSRCAT2. We have fitted their $\gamma$-ray and radio light curves with simulated $\gamma$-ray and radio emission 
patterns. We have computed the radio emission beam according to \cite{sgh07} and we have used the geometrical 
model of \cite{dhr04} to simulate the $\gamma$-ray emission according to four gap models, 
PC, \citep{mh03}, SG, \citep{mh04a}, OG, \citep{crz00} and OPC \citep{rw10,wrwj09}. Each emission pattern 
has been described by a series of phase-plots, evaluated for 
the pulsar period, magnetic field, and gap width, and for the whole $\alpha$ interval sampled every degree. 
These phase-plots predict the pulsar light curve as a function of $\zeta$.

The simulated phase-plots have been used to fit the observed radio and $\gamma$-ray light curves according to two different
schemes: a single fit to the $\gamma$-ray profiles of RF and RQ objects and a joint fit to the $\gamma$-ray and radio light curves 
of RL pulsars.

The individual fit to the $\gamma$-ray profiles has been implemented using a $\chi^2$ estimator
and light curves binned both in FCBin  and RBin. The comparison of the results obtained with the 
two methods shows that the $\chi^2$ fit with FCBin light curves yields the closest match between the observations and
modelled profiles. We use the latter to give $\alpha$ and $\zeta$ estimates for the RQ and RF LAT pulsars and we use the RBin fit to evaluate 
the systematic uncertainties induced by the fitting method.

The joint $\gamma$-ray plus radio fit of RL pulsars uses RBin radio light curves and FCBin $\gamma$-ray light curves with a $\chi^2$ estimator. 
The log-likelihood maps in $\alpha$ and $\zeta$ obtained from the radio-only and $\gamma$-ray-only fits were summed to produce the joint solution. 
Two options were considered to couple the high signal-to-noise ratio of the radio data to the much lower signal-to-noise ratio of the $\gamma$-ray profiles
and the solution characterised by the highest log-likelihood value was selected. The systematic errors on $(\alpha,\zeta)$ for the RL pulsars  have 
been obtained by studying the difference between the solutions obtained with the two joint fit coupling schemes.

We have obtained new constraints on $\alpha$ and $\zeta$ for 33 RQ, 2 RF, and 41 RL $\gamma$-ray pulsars. We have studied how the 
$(\alpha,\zeta)$ solutions of RL pulsars obtained by fitting only the $\gamma$-ray light curves change by including the radio emission 
in the fit. We have used the $\alpha$ and $\zeta$ solutions to estimate several important pulsar parameters:  gap width, beaming factor, 
and luminosity. We have also investigated some relations between observable characteristics and intrinsic pulsar parameters, such as 
$\alpha$ as a function of age and the spectral energy cut-off and index in $\gamma$-rays as a function of the gap width.
We find no evidence for an evolution of the magnetic obliquity over the $\sim10^6$ yr of age span in the sample, but we find an 
interesting apparent change in the $\gamma$-ray spectral index $\Gamma$ and high-energy cutoff $E_\mathrm{cut}$ associated with changes
in the gap widths.

We have found that a multi-wavelength fit of $\gamma$-ray and radio light curves is important in giving a pulsar 
orientation estimate that can explain both radio and $\gamma$-ray emission.
The PC emission geometry explains only a small fraction of the observed profiles, in particular for the RL pulsars, while the intermediate to high SG 
and OG/OPC models are favoured in explaining the pulsar emission pattern of both RQ and RL LAT pulsars.
The fact that none of the assumed emission geometries is able to explain all the observed LAT light curves 
suggests that the true $\gamma$-ray emission geometry may be a 
combination of SG and OG and that we detect the respective light curves for different observer viewing angles.

Comparison of the $\alpha$ and $\zeta$ solutions obtained by fitting only the $\gamma$-ray profiles of RL pulsars 
and both their $\gamma$-ray and radio profiles suggests that in the OG and OPC models, $\alpha$ or $\zeta$ are 
underestimated when one does not account for radio emission. When the $\gamma$-only solution is to the right of the 
radio diagonal in the $\alpha$-$\zeta$ plane, $\zeta$ migrates toward higher values while $\alpha$ keeps quite stable 
and \emph{vice versa} when the $\gamma$-only solution is to the left of the radio diagonal.

The beaming factors found for the RQ and RL objects are consistent with the distributions obtained in the population 
study of \cite{pghg12}. For all the models we observe a large scatter of the beaming factors with $\dot{E}$, 
which is reduced for RL pulsars compared to RQ pulsars, except for the SG. This is because RQ pulsars are viewed at 
lower $\alpha$ and $\zeta$, and OG and OPC beams shrink towards the spin equator with decreasing $\dot{E}$ while SG 
beams do not. The low $f_{\Omega}$ values found for the PC reflect the narrow geometry of the PC beams. The $f_{\Omega}$ 
values for the SG appear to be fairly stable around 1 over 4 decades in $\dot{E}$. We find also little evolution for the OG 
and OPC beaming factors of RQ objects which gather around 0.25 and 0.39, respectively. Larger averages are obtained 
for the RL objects (0,68 for OG and 0,86 for OPC) with no evolution with $\dot{E}$ for the OPC case and some hint of an 
increase with $\dot{E}$ in the OG case. The fact that the majority of the pulsars exhibit an $f_{\Omega}$ estimate less than 
unity in all models suggests that the isotropic luminosities ($f_{\Omega}$ = 1) often quoted in other studies are likely to 
overestimate the real values. 

For all the models a power law relation consistent with $L_{\gamma}~\appropto~\dot{E}^{0.5}$ is observed for both 
RQ and RL pulsars. In contrast with PSRCAT2 we do not obtain any $\gamma$-ray luminosities significantly
higher than $\dot{E}$. Since the only difference between the luminosity computation here and that of PSRCAT2 is in the 
$f_{\Omega}$ value (assumed equal to one in the catalog), the excessively high luminosities obtained in the catalog probably 
result from a too high beaming factor. We have studied the consistency of the geometric $\gamma$-ray luminosity, 
$L_\mathrm{geo}$, obtained in this paper and the $\gamma$-ray luminosity computed in the framework of radiative 
gap-models, $L_\mathrm{rad}$. We found that $L_\mathrm{geo}$ overestimate $L_\mathrm{rad}$ of 2-3 order of magnitude
for the RQ and RL SG pulsar and for RQ OG pulsars while the $L_\mathrm{geo}$ of RL OG objects are more consistent with
their $L_\mathrm{rad}$ values while showing higher dispersion in $L_\mathrm{geo}$. For both RQ and RL OPC objects, $L_\mathrm{rad}$ 
is consistent with the $L_\mathrm{rad}$ estimates. These OG and SG geometric-radiative luminosity disagreements are due to 
inconsistencies in the formulation of the geometrical and radiative aspects of the $\gamma$-ray pulsar emission, rise the problem 
of formulating geometrical models more based on the actual pulsar electrodynamics in the framework of each gap model, and points 
to fundamental shortcomings of these electrodynamic gap models.

We find a correlation between $E_\mathrm{cut}$ and $\Gamma$ of the $\gamma$-rays 
and the accelerator gap width in the magnetosphere. The relation is consistent with the SG prediction 
$E_\mathrm{cut}\appropto w_\mathrm{SG}^{-0.25}$ just for the RL objects while the more approximated predictions formulated 
for OG and OPC models are not consistent with the observations. This $E_\mathrm{cut}$ and $\Gamma$ versus gap width proportionality 
is important because it connects the observed spectral information and the non observable size of the gap region 
on the basis of the light-curve morphology alone.

\begin{acknowledgements}
The \textit{Fermi} LAT Collaboration acknowledges generous ongoing 
support from a number of agencies and institutes that have supported 
both the development and the operation of the LAT as well as scientific 
data analysis. These include the National Aeronautics and Space 
Administration and the Department of Energy in the United States, the 
Commissariat \`a l'Energie Atomique and the Centre National de la 
Recherche Scientifique / Institut National de Physique Nucl\'eaire et de 
Physique des Particules in France, the Agenzia Spaziale Italiana and the 
Istituto Nazionale di Fisica Nucleare in Italy, the Ministry of Education, 
Culture, Sports, Science and Technology (MEXT), High Energy Accelerator 
Research Organization (KEK) and Japan Aerospace Exploration Agency 
(JAXA) in Japan, and the K.~A.~Wallenberg Foundation, the Swedish 
Research Council and the Swedish National Space Board in Sweden.

Additional support for science analysis during the operations phase is 
gratefully acknowledged from the Istituto Nazionale di Astrofisica in Italy 
and the Centre National d'\'Etudes Spatiales in France.

MP acknowledges IASF-INAF in Milan for fundamental support during 
the realisation of this project, the Nicolaus Copernicus Astronomical 
Center, grant DEC-2011/02/A/ST9/00256, for providing software and 
computer facilities needed for the conclusion of this work, and Sacha Hony
for the precious help.
And a very special thanks to Isabel Caballero, Isa, for supporting me so 
many years, for teaching me many things, and for being there with me in 
daily life.

AKH acknowledges support from the NASA Astrophysics Theory and 
Fermi GI Programs.

The authors wish to acknowledge the anonymous referee for the helpful 
suggestions and comments that enriched the paper and helped to highlight 
some of its results. 

The authors gratefully acknowledge the Pulsar Search and Timing Consortia,
all the radio scientists who contributed in providing the radio light curves used 
in this paper, and the radio observatories that generated the radio profiles used
in this paper:
the Parkes Radio Telescope is part of the Australia Telescope which is funded 
by the Commonwealth Government for operation as a National Facility managed 
by CSIRO; the Green Bank Telescope is operated by the National Radio Astronomy 
Observatory, a facility of the National Science Foundation operated under cooperative 
agreement by Associated Universities, Inc; the Arecibo Observatory is part of the National 
Astronomy and Ionosphere Center (NAIC), a national research center operated by Cornell 
University under a cooperative agreement with the National Science Foundation; the 
Nan\c cay Radio Observatory is operated by the Paris Observatory, associated with the 
French Centre National de la Recherche Scientifique (CNRS); the Lovell Telescope is 
owned and operated by the University of Manchester as part of the Jodrell Bank Centre 
for Astrophysics with support from the Science and Technology Facilities Council of the 
United Kingdom; the Westerbork Synthesis Radio Telescope is operated by Netherlands 
Foundation for Radio Astronomy, ASTRON.

\end{acknowledgements}

\bibliographystyle{aa}
\bibliography{journals,psrrefs,crossrefs,modrefs,pierba}


\appendix
\begin{onecolumn}

\section{Estimate of the goodness of the fit for each model solution}
\label{GoodFitMethod}
\input{GoodFitMethod.tex}

\clearpage

\section{Population synthesis results from \cite{pghg12}}
\label{PopDis}
\input{PopulationStudy.tex} 
\clearpage

\section{The LAT pulsar $\gamma$-ray fit light-curve results}
\label{GammaFitRes}
\input{GammaFitsRes.tex} 
\clearpage

\section{The LAT pulsar Joint fit light-curve results}
\label{JointFitRes}
\input{JointFitsRes.tex}

 \clearpage
 
\section{Joint fit of radio and $\gamma$-ray light curves of the radio-faint pulsars J0106$+$4855 and J1907$+$0602}
\label{JointFits_RQ2RL}
\input{JointFits_RQ2RL.tex}

\end{onecolumn}

\end{document}

%% file: GoodFitMethod.tex
In this Appendix we describe the calculations used to quantify the relative goodness of the fit solutions 
obtained between the optimum-model and another model. The method assumes that the 
optimum-model light curve describes reasonably well the observations and it is based on the evaluation 
of the standard deviation of all the models, $\sigma_\star$, by imposing that the reduced $\chi^2_\star$ of 
the optimum-solution is equal to unity. The difference between the $\chi^2_\star$ values reached for the 
optimum-model and the other models then provides a measure of the relative goodness of the two 
solutions.

The $\chi^2$ of the optimum-model and of another model, $\chi^2_{opt}$ and $\chi^2_{mod}$ respectively, 
are defined as
\begin{equation}
\chi^2_{opt} = \frac{\sum_j (N_\mathrm{obs,j}-N_\mathrm{opt,j})^2}{\sigma^2}
\label{1}
\end{equation}
\begin{equation}
\chi^2_{mod} = \frac{\sum_j (N_\mathrm{obs,j}-N_\mathrm{mod,j})^2}{\sigma^2}
\label{2}
\end{equation}
where $N_\mathrm{obs,j}$ and $N_\mathrm{mod,j}$ are the observed and modelled light curves respectively, 
and $\sigma$ the standard deviation of the observed light curve. The difference between these two $\chi^2$ 
can be evaluated from the log-likelihood values given in Tables C.1 and D.1 as $\Delta \chi^2 = - 2 [ln(L_{opt}) - ln(L_{mod})]$.

With the reduced $\chi^2$ of the optimum model set to 1, the standard deviation of the models, $\sigma_\star$, is 
\begin{equation}
\sigma^2_\star= \frac{\sum_j (N_\mathrm{obs,j}-N_\mathrm{opt,j})^2}{N_{dof}},
\label{3}
\end{equation}
where $N_{dof}$ is the number of degrees of freedom of each type of fit (41 for RL pulsars and 81 for RQ ones). 
With the model variance, the $\chi^2_\star$ of the optimum and other models become: 
\begin{equation}
\chi^2_{opt,\star} = N_{dof} 
\label{4}
\end{equation}
\begin{equation}
\chi^2_{mod,\star} = 
N_{dof}\frac{\sum_j (N_\mathrm{obs,j}-N_\mathrm{mod,j})^2}{\sum_j (N_\mathrm{obs,j}-N_\mathrm{opt,j})^2}.
\label{5}
\end{equation}
and their difference $\Delta\chi^2_\star$ is 
\begin{align}
\Delta\chi^2_\star = \chi^2_{mod,\star} - \chi^2_{opt,\star} = 
{1\over \sigma^2_\star}\left[\sum_j (N_\mathrm{obs,j}-N_\mathrm{mod,j})^2 - \sum_j (N_\mathrm{obs,j}-N_\mathrm{opt,j})^2 \right]\notag \\ 
= N_{dof} \frac{\sum_j (N_\mathrm{obs,j}-N_\mathrm{mod,j})^2 - \sum_j (N_\mathrm{obs,j}-N_\mathrm{opt,j})^2}{\sum_j (N_\mathrm{obs,j}-N_\mathrm{opt,j})^2} 
=  N_{dof} \left(\frac{\chi^2_{mod}-\chi^2_{opt}}{\chi^2_{opt}}\right) .
\label{6}
\end{align}
We have plotted the resulting $\Delta \chi^2_\star$ values in Figures \ref{SigGam} and \ref{SigJoint}. 
The 1$\sigma$, 3$\sigma$, and 5$\sigma$ confidence levels plotted in these figures have been obtained 
from the $\chi^2$ probability density function for the appropriate number of degrees of freedom.

%% file: PopulationStudy.tex
By synthesising a pulsar population we compared theoretical and observed distributions of observable quantities between 
the \emph{Fermi} pulsars and the predictions of different $\gamma$-ray models. We have assumed low/intermediate and high 
altitude magnetosphere emission models PC and SG, OG and OPC respectively, and core plus cone radio emission model.
Full details on the population synthesis study can be found in 
\cite{pghg12}. The plots shown in this Appendix have been obtained as additional results to the population study in \cite{pghg12} 
by using the original data at our disposal.

\subsection{$\alpha$-$\zeta$ plane}
The $\alpha$ and $\zeta$ distributions of the visible component of the simulated population for PC, SG, OG, and OPC models are shown. 
\begin{figure}[htbp!]
\begin{center}
\includegraphics[width=0.49\textwidth]{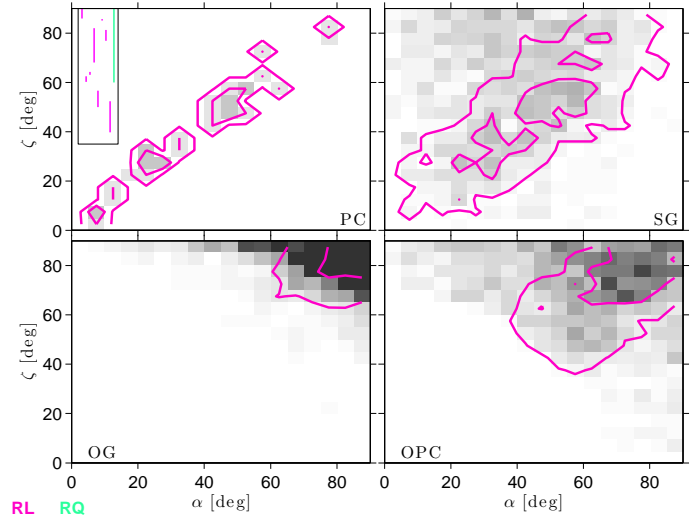} 
\caption{Number density of the visible $\gamma$-ray pulsars obtained for each model as a function of $\alpha$)and 
$\zeta$ in the population synthesis of \cite{pghg12}. The linear gray scale saturates at 1.5 star/bin. The pink 
contours outline the density obtained for the 
radio-loud $\gamma$-ray sub-sample (at 5\% and 50\% of the maximum density). The insert gives the set of $\zeta$ values measured by 
\citep{nr08} from the orientation of the wind torus seen in X rays (pink lines) and by \cite{cbd+03} from the orientation of the 
Geminga X-ray tails (green line). The separation in $\alpha$ in the insert is meaningless.}
\label{NvisA_large_effi1p012p01p00p5_alphazeta_histo}
\end{center}
\end{figure}

\subsection{High-energy cutoff and spectral index as a function of the gap width}
High energy cutoff and spectral index as a function of the width of the accelerator gap of the visible component of the simulated population for PC, SG, OG, and OPC.
In disagreement with figure \ref{EcutGapW1}, no $E_{cut}$-gap width dependence is predicted from the simulations.
\begin{figure}[htbp!]
\begin{center}
\includegraphics[width=0.49\textwidth]{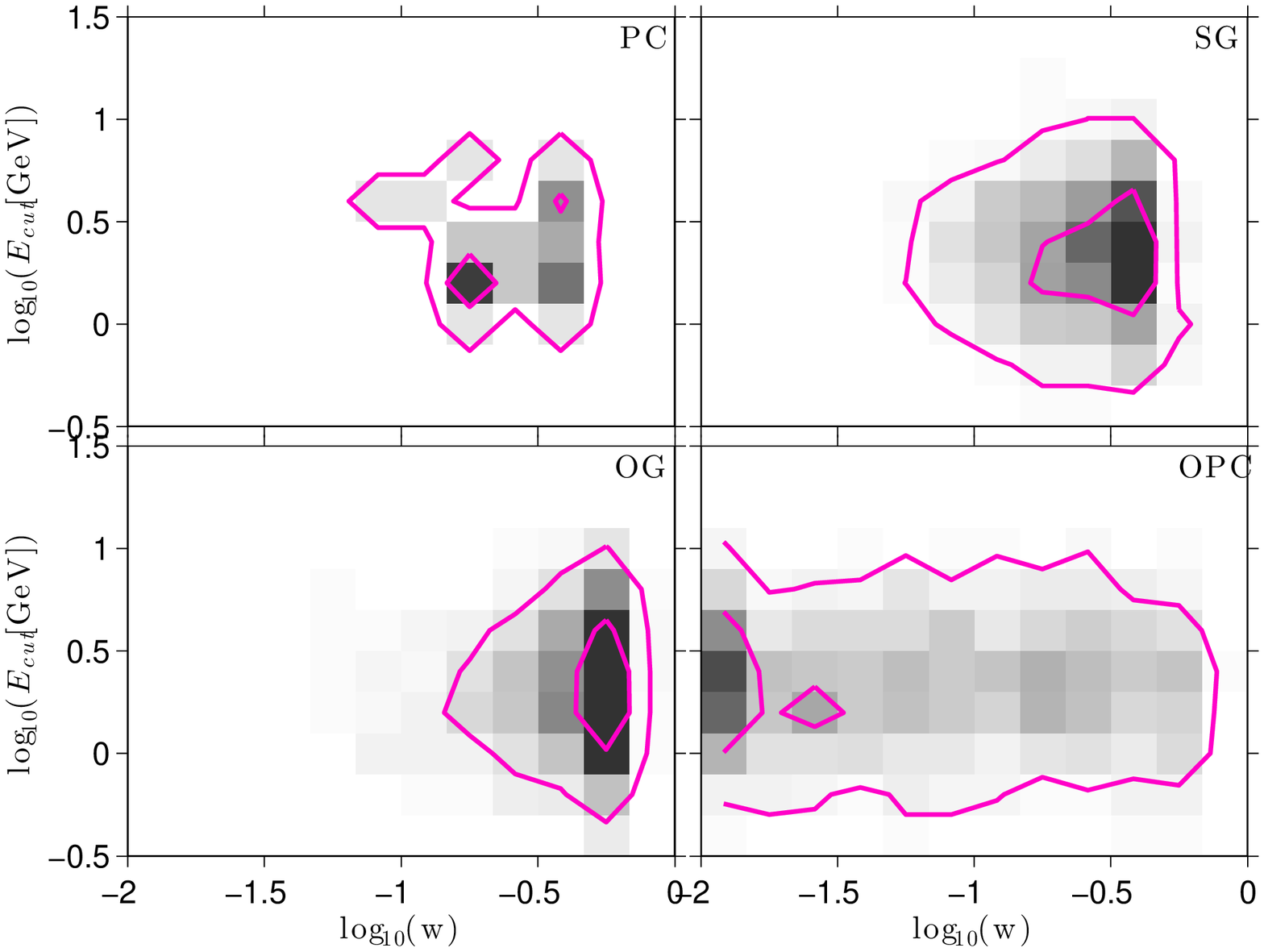}
\includegraphics[width=0.49\textwidth]{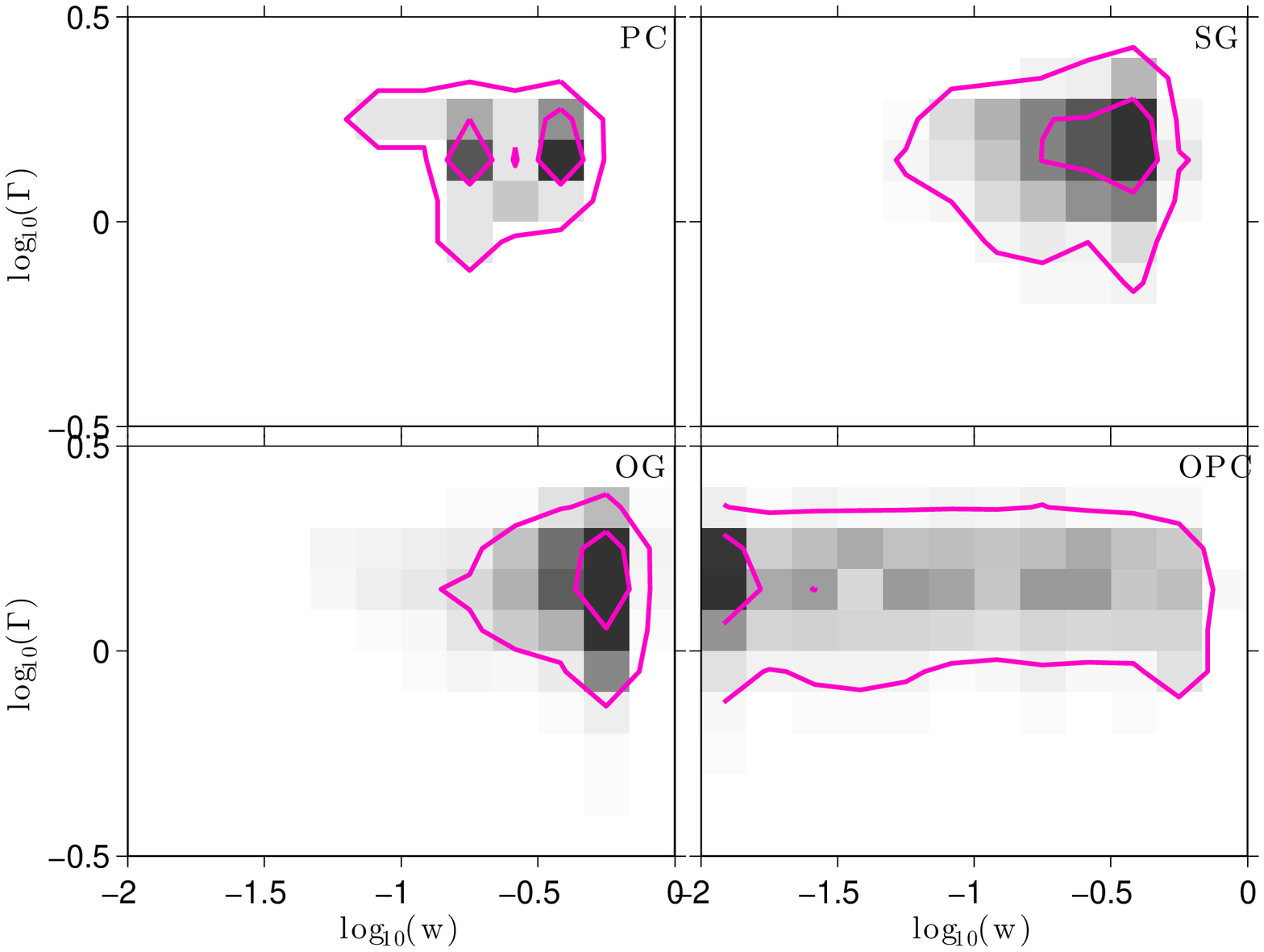}
\caption{Number density of the visible $\gamma$-ray pulsars obtained for each model as a function of gap width and high-energy cutoff (left) and of 
gap width and spectral index (right). The linear grey 
scale saturates at 8 star/bin. The pink contours outline the density obtained for the radio-loud $\gamma$-ray sub-sample (at 5\% and 50\% of the 
maximum density).}
\label{NvisA_large_effi1p012p01p00p5_Ecutgap_histo}
\end{center}
\end{figure}

%% file: GammaFitsRes.tex
\begin{table*}[htbp!]
\centering
\begin{tabular}{| c || c | c | c | c |}
\hline
& $ \ln L_{PC}$ & $ \ln L_{SG}$  & $ \ln L_{OG}$  & $ \ln L_{OPC}$ \\
\hline
\hline
J0007+7303 & $ -1855 $  & $ -7780 $  & $ -592 $  & $ -905 $  \\
\hline
J0106+4855 & $ -81 $  & $ -177 $  & $ -155 $  & $ -157 $  \\
\hline
J0357+3205 & $ -144 $  & $ -989 $  & $ -495 $  & $ -725 $  \\
\hline
J0622+3749 & $ -63 $  & $ -102 $  & $ -51 $  & $ -43 $  \\
\hline
J0633+0632 & $ -722 $  & $ -867 $  & $ -720 $  & $ -760 $  \\
\hline
J0633+1746 & $ -60831 $  & $ -18144 $  & $ -84189 $  & $ -38960 $  \\
\hline
J0734$-$1559 & $ -57 $  & $ -119 $  & $ -55 $  & $ -88 $  \\
\hline
J1023$-$5746 & $ -400 $  & $ -236 $  & $ -456 $  & $ -289 $  \\
\hline
J1044$-$5737 & $ -293 $  & $ -327 $  & $ -388 $  & $ -301 $  \\
\hline
J1135$-$6055 & $ -71 $  & $ -118 $  & $ -68 $  & $ -37 $  \\
\hline
J1413$-$6205 & $ -417 $  & $ -730 $  & $ -70 $  & $ -74 $  \\
\hline
J1418$-$6058 & $ -503 $  & $ -785 $  & $ -403 $  & $ -331 $  \\
\hline
J1429$-$5911 & $ -299 $  & $ -263 $  & $ -366 $  & $ -357 $  \\
\hline
J1459$-$6053 & $ -118 $  & $ -391 $  & $ -378 $  & $ -119 $  \\
\hline
J1620$-$4927 & $ -134 $  & $ -180 $  & $ -121 $  & $ -150 $  \\
\hline
J1732$-$3131 & $ -1057 $  & $ -1075 $  & $ -212 $  & $ -177 $  \\
\hline
J1746$-$3239 & $ -76 $  & $ -94 $  & $ -200 $  & $ -56 $  \\
\hline
J1803$-$2149 & $ -114 $  & $ -122 $  & $ -65 $  & $ -37 $  \\
\hline
J1809$-$2332 & $ -2228 $  & $ -3149 $  & $ -1472 $  & $ -1221 $  \\
\hline
J1813$-$1246 & $ -238 $  & $ -223 $  & $ -359 $  & $ -354 $  \\
\hline
J1826$-$1256 & $ -1339 $  & $ -896 $  & $ -2127 $  & $ -1306 $  \\
\hline
J1836+5925 & $ -1828 $  & $ -397 $  & $ -19937 $  & $ -17849 $  \\
\hline
J1838$-$0537 & $ -151 $  & $ -54 $  & $ -219 $  & $ -119 $  \\
\hline
J1846+0919 & $ -62 $  & $ -267 $  & $ -80 $  & $ -49 $  \\
\hline
J1907+0602 & $ -641 $  & $ -1195 $  & $ -285 $  & $ -177 $  \\
\hline
J1954+2836 & $ -240 $  & $ -228 $  & $ -262 $  & $ -170 $  \\
\hline
J1957+5033 & $ -88 $  & $ -256 $  & $ -148 $  & $ -198 $  \\
\hline
J1958+2846 & $ -468 $  & $ -509 $  & $ -215 $  & $ -195 $  \\
\hline
J2021+4026 & $ -2720 $  & $ -350 $  & $ -1220 $  & $ -690 $  \\
\hline
J2028+3332 & $ -303 $  & $ -204 $  & $ -260 $  & $ -132 $  \\
\hline
J2030+4415 & $ -239 $  & $ -159 $  & $ -268 $  & $ -240 $  \\
\hline
J2055+2539 & $ -129 $  & $ -183 $  & $ -324 $  & $ -377 $  \\
\hline
J2111+4606 & $ -162 $  & $ -198 $  & $ -77 $  & $ -52 $  \\
\hline
J2139+4716 & $ -61 $  & $ -77 $  & $ -93 $  & $ -101 $  \\
\hline
J2238+5903 & $ -212 $  & $ -443 $  & $ -683 $  & $ -618 $  \\
\hline
\end{tabular}
\caption{Best fit log-likelihood values resulting from the $\gamma$-ray fit of the 35 RQ pulsars of the analysed sample.}
\label{Like_gamma1}
\end{table*}
\clearpage

\begin{figure}[htbp!]
\centering
\includegraphics[width=0.9\textwidth]{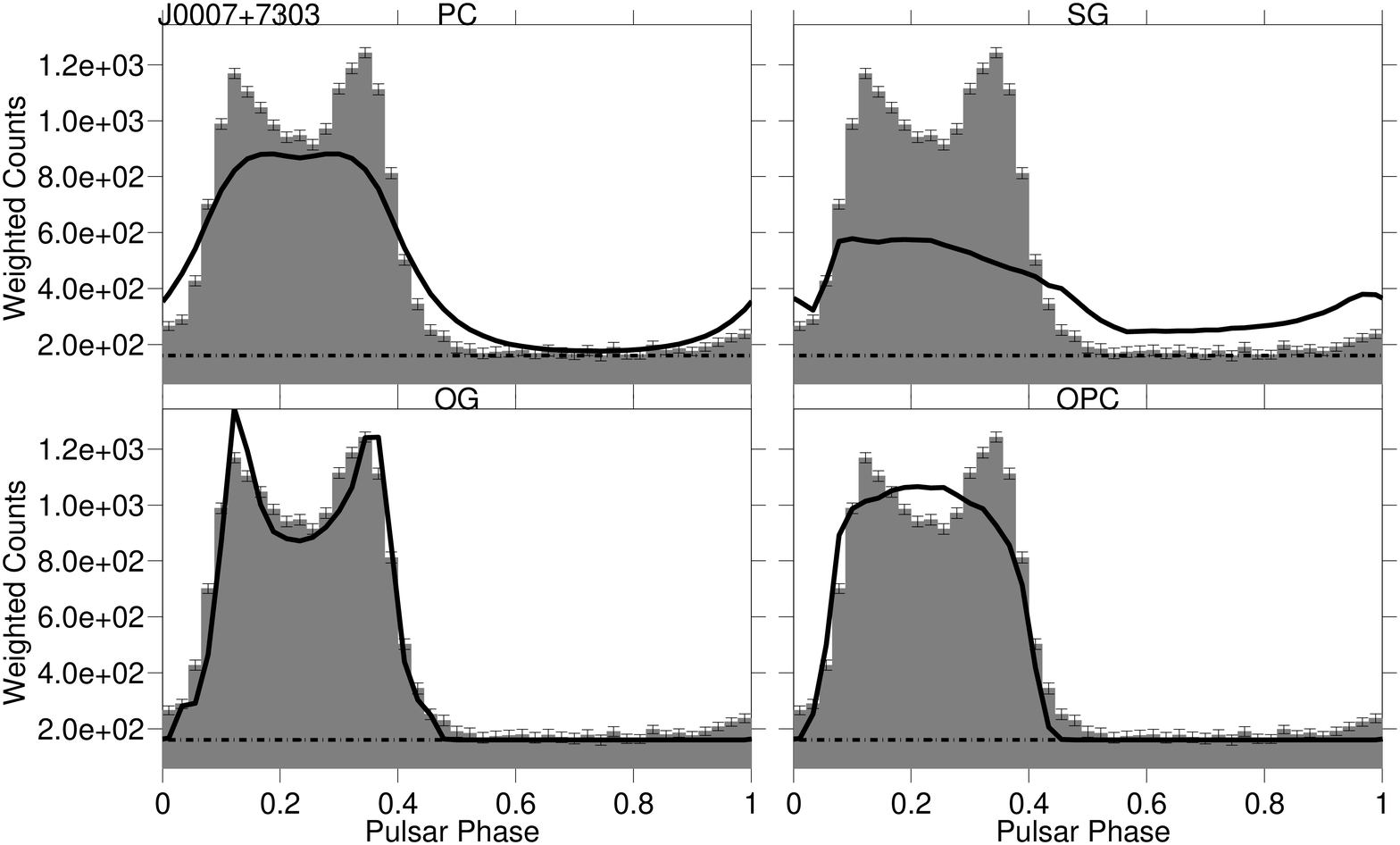}
\includegraphics[width=0.9\textwidth]{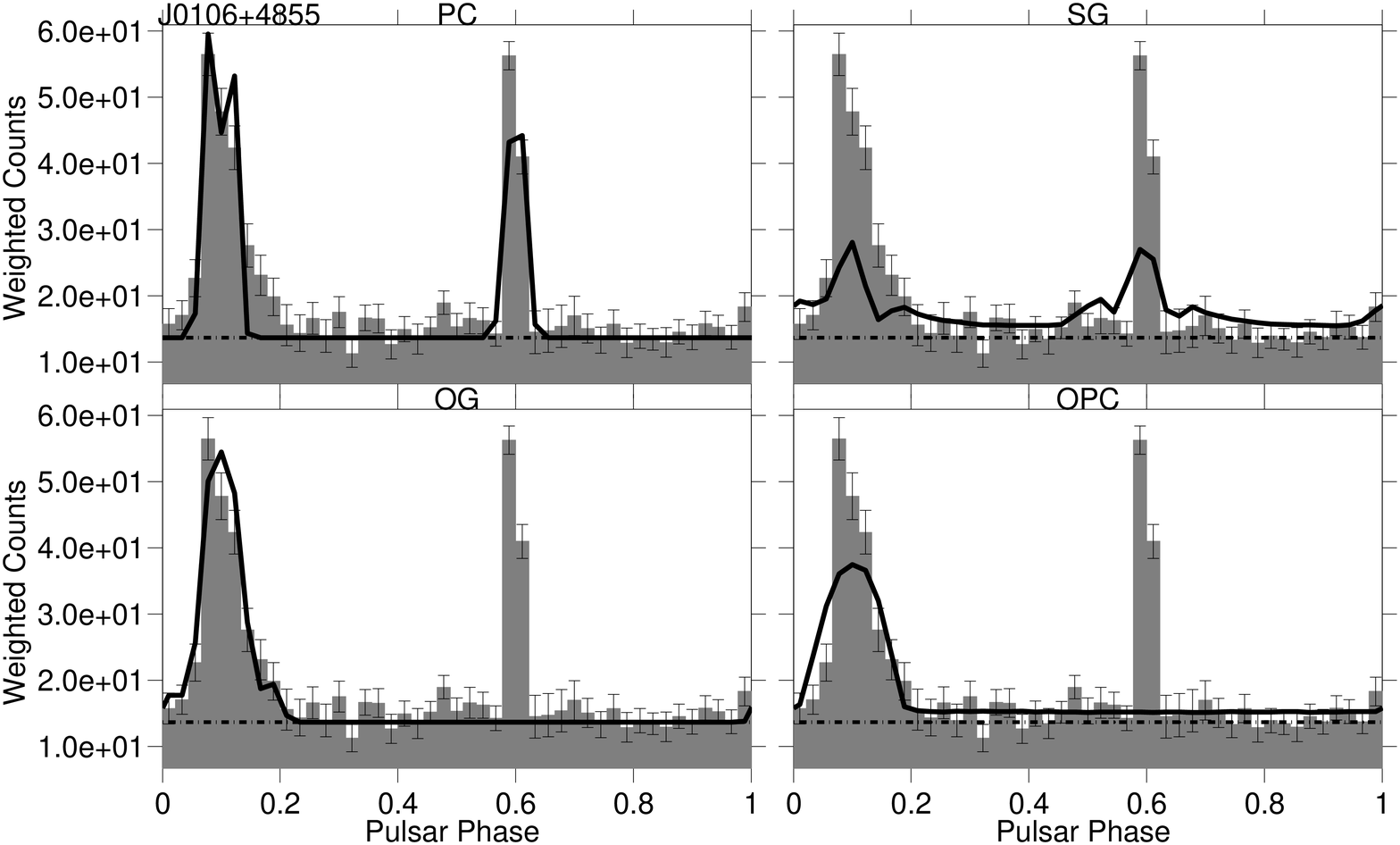}
\caption{\emph{Top:} PSR J0007+7303; \emph{bottom:} PSR J0106+4855. For each model the best $\gamma$-ray light-curve (thick black line) is superimposed on the LAT  pulsar light-curve (shaded histogram). The estimated background is indicated by the dash-dot line.}
\label{fitGm1}
\end{figure}
  
\clearpage
\begin{figure}[htbp!]
\centering
\includegraphics[width=0.9\textwidth]{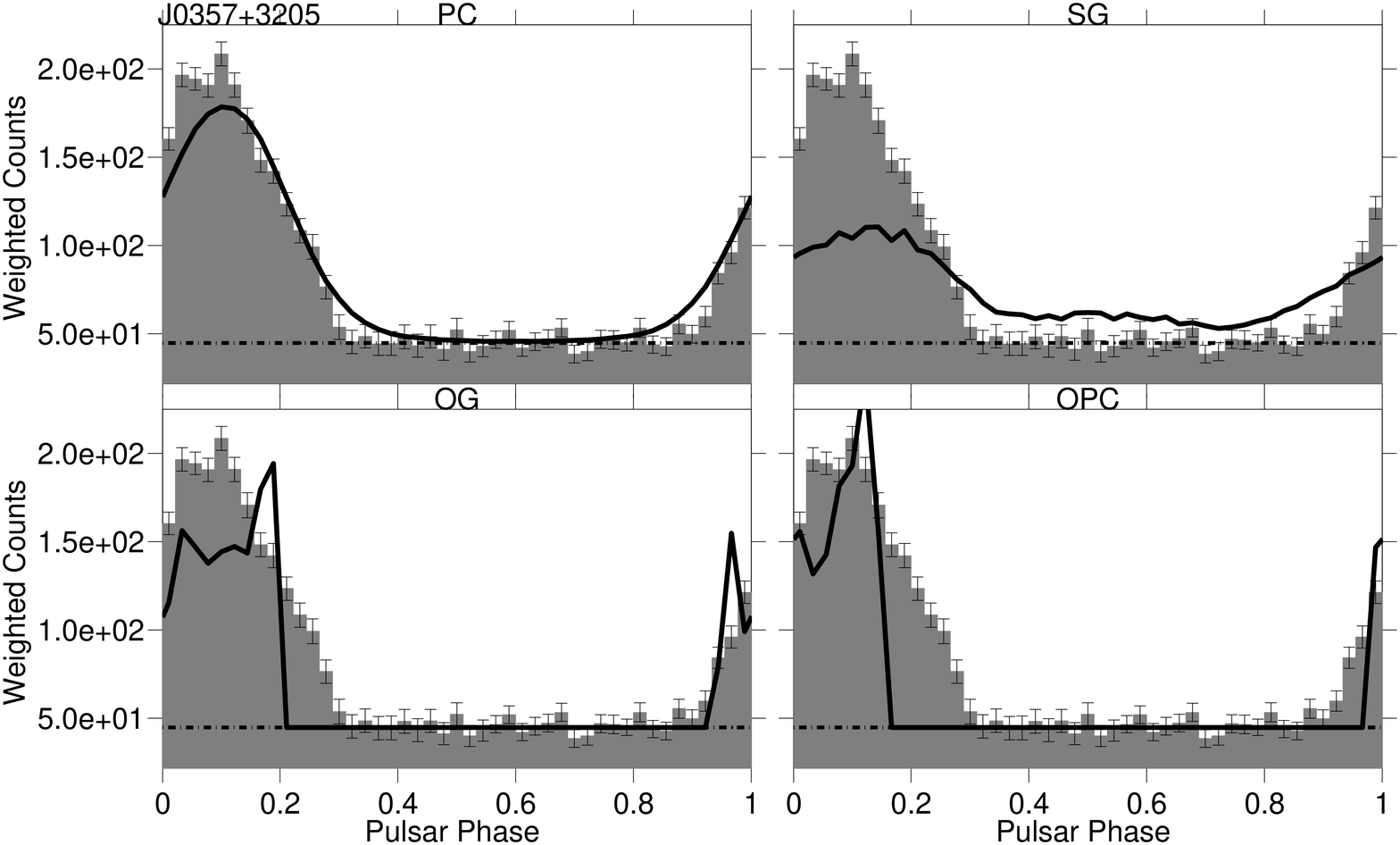}
\includegraphics[width=0.9\textwidth]{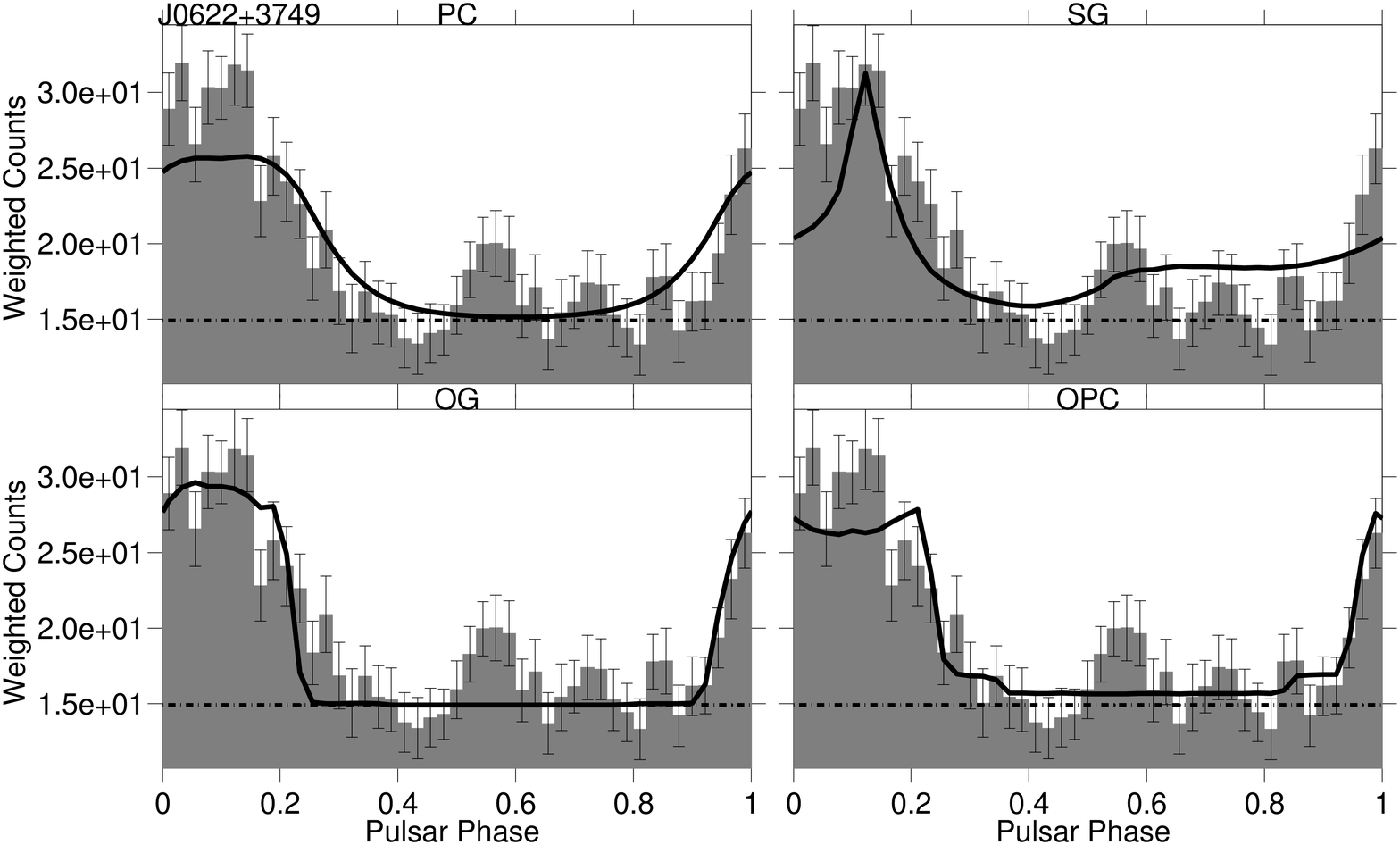}
\caption{\emph{Top:} PSR J0357+3205; \emph{bottom:} PSR J0622+3749. For each model the best $\gamma$-ray light-curve (thick black line) is superimposed on the LAT  pulsar light-curve (shaded histogram). The estimated background is indicated by the dash-dot line.}
\label{fitGm3}
\end{figure}
  
\clearpage
\begin{figure}[htbp!]
\centering
\includegraphics[width=0.9\textwidth]{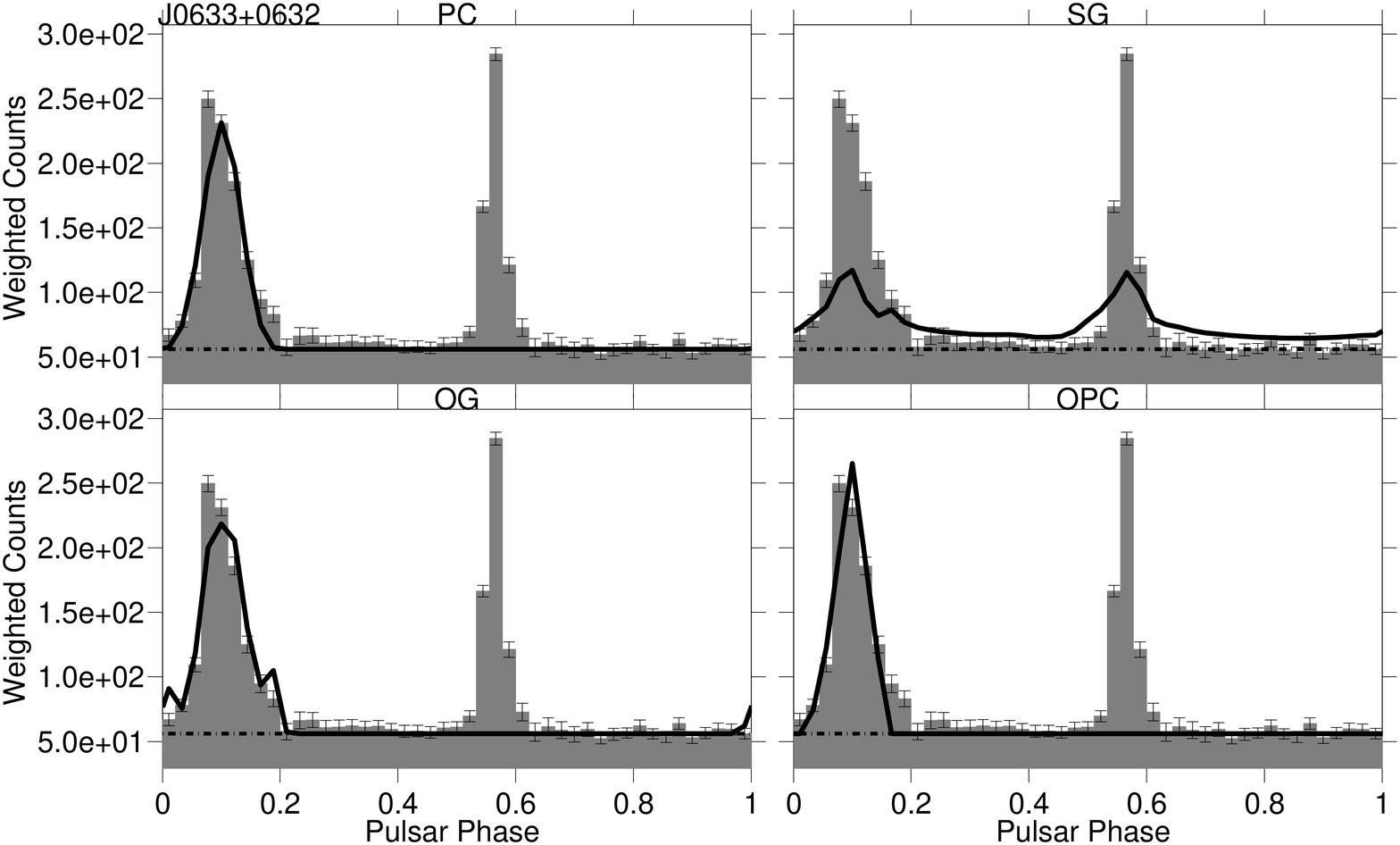}
\includegraphics[width=0.9\textwidth]{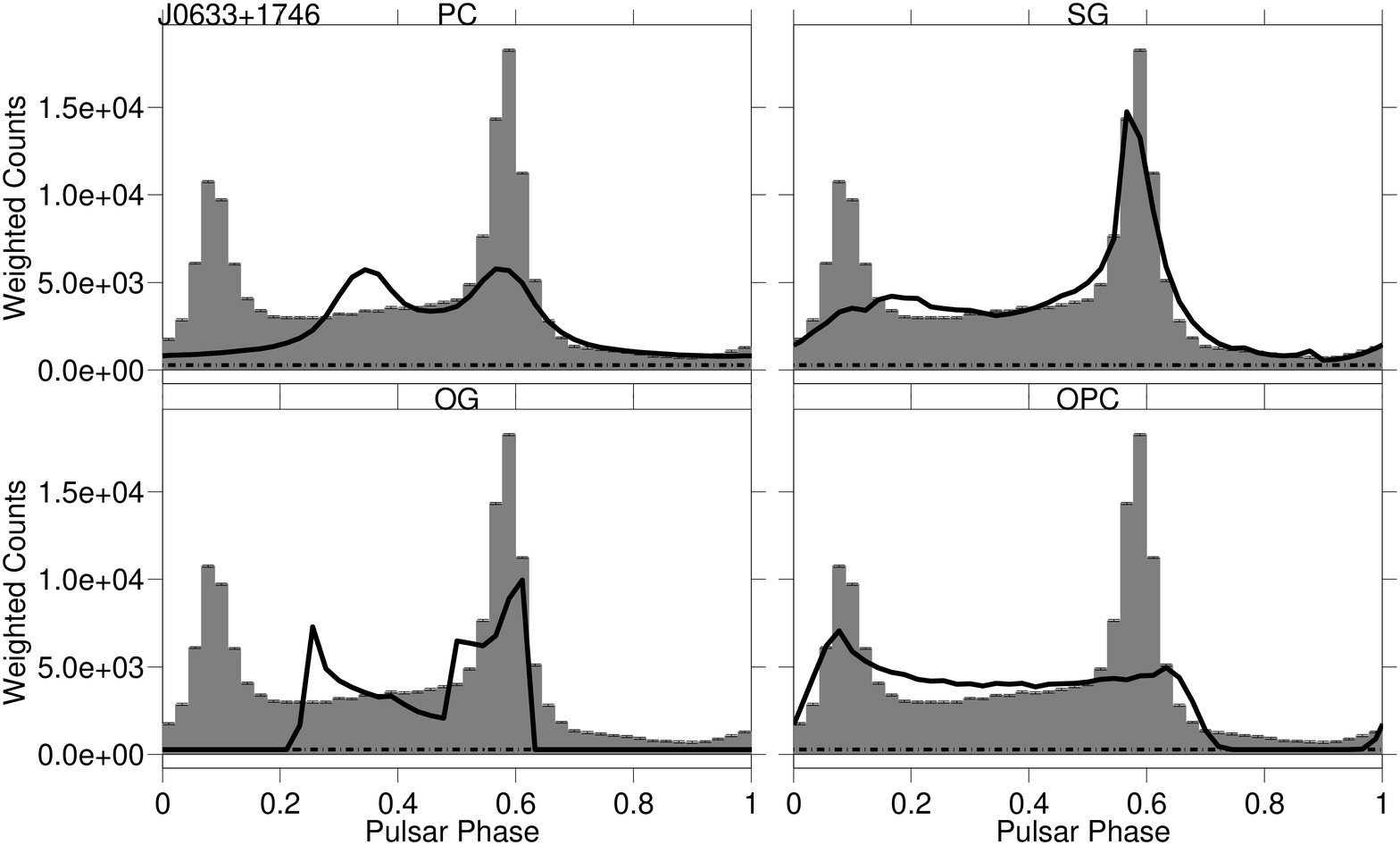}
\caption{\emph{Top:} PSR J0633+0632; \emph{bottom:} PSR J0633+1746. For each model the best $\gamma$-ray light-curve (thick black line) is superimposed on the LAT  pulsar light-curve (shaded histogram). The estimated background is indicated by the dash-dot line.}
\label{fitGm5}
\end{figure}
  
\clearpage
\begin{figure}[htbp!]
\centering
\includegraphics[width=0.9\textwidth]{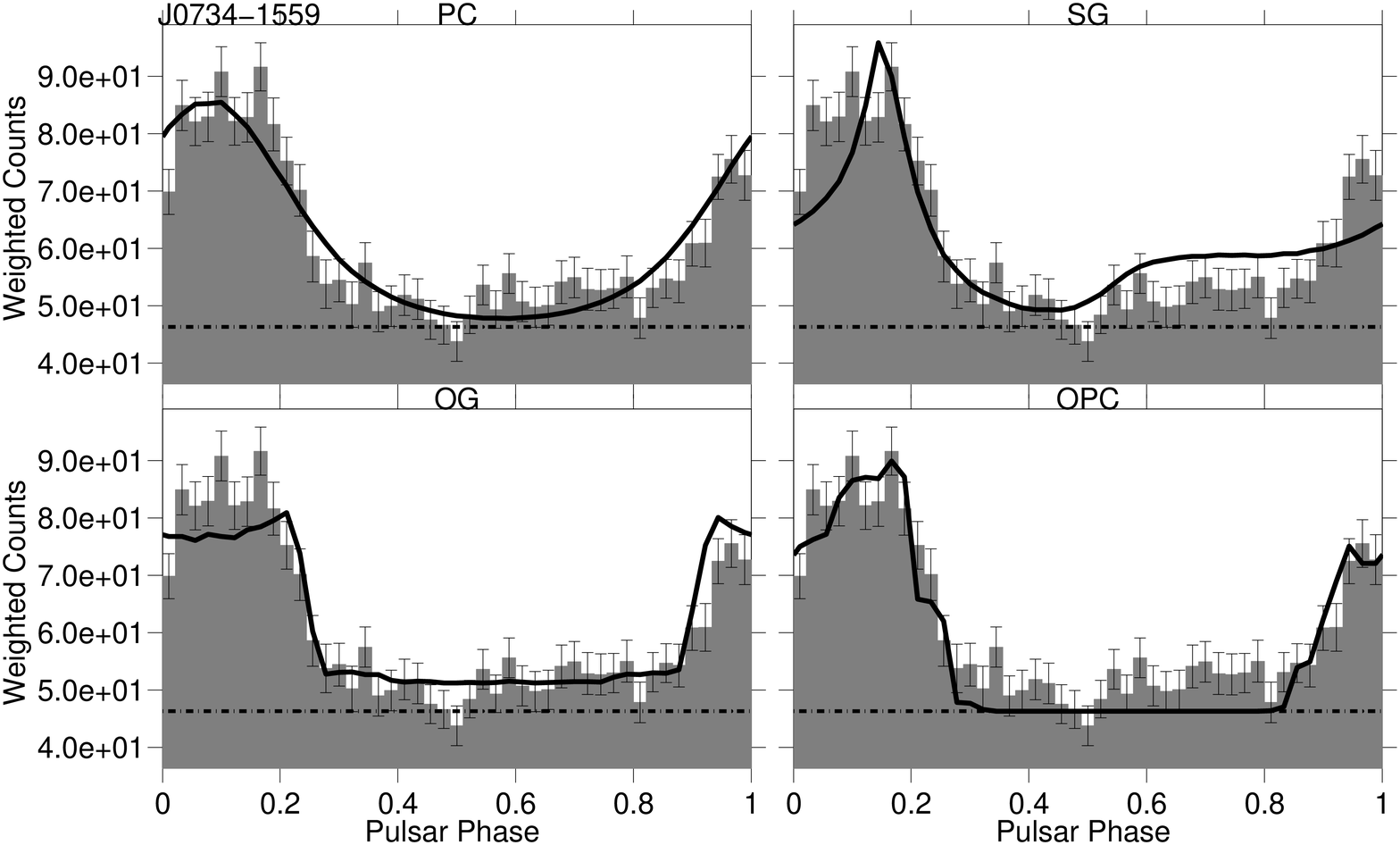}
\includegraphics[width=0.9\textwidth]{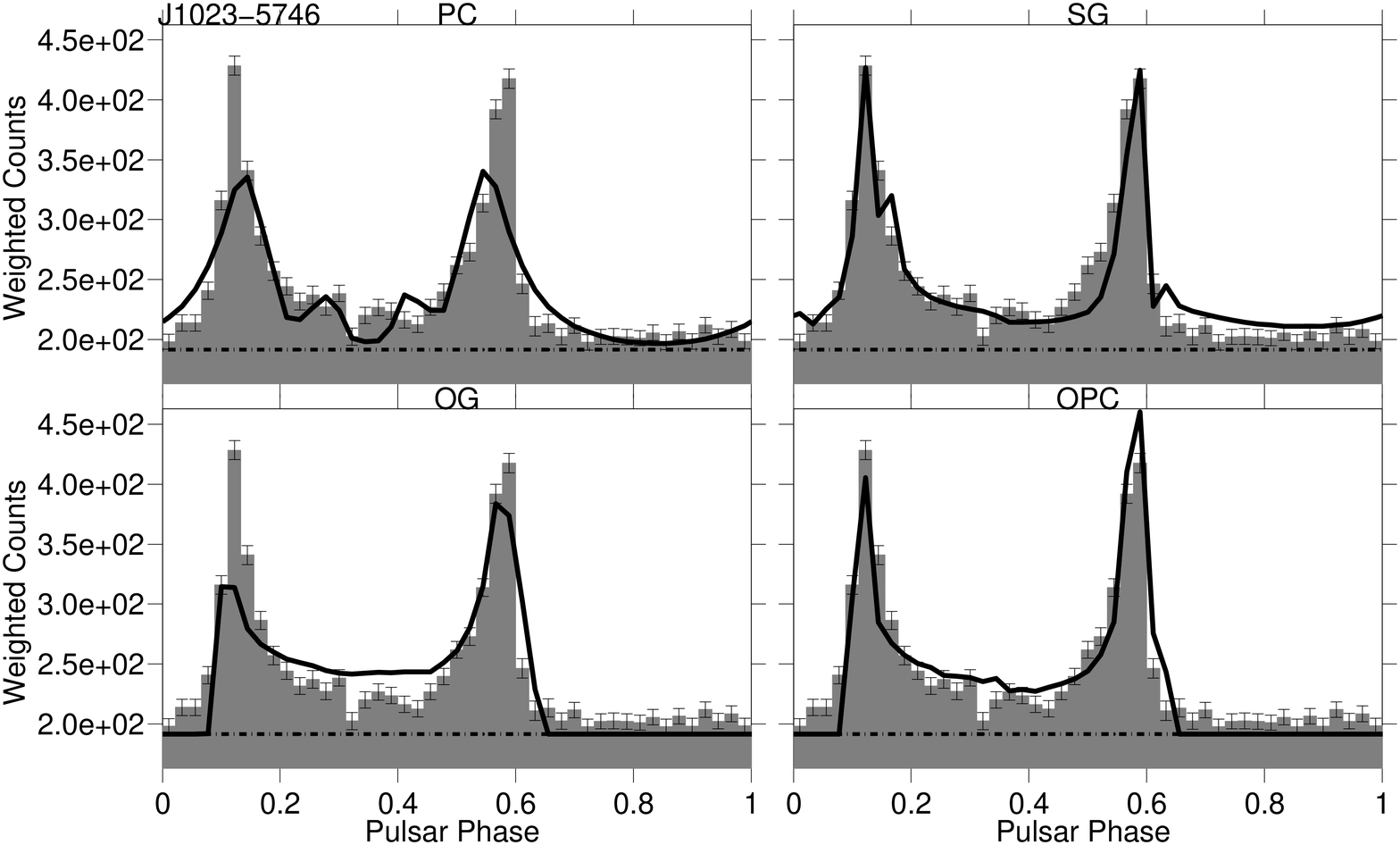}
\caption{\emph{Top:} PSR J0734-1559; \emph{bottom:} PSR J1023-5746. For each model the best $\gamma$-ray light-curve (thick black line) is superimposed on the LAT  pulsar light-curve (shaded histogram). The estimated background is indicated by the dash-dot line.}
\label{fitGm7}
\end{figure}
  
\clearpage
\begin{figure}[htbp!]
\centering
\includegraphics[width=0.9\textwidth]{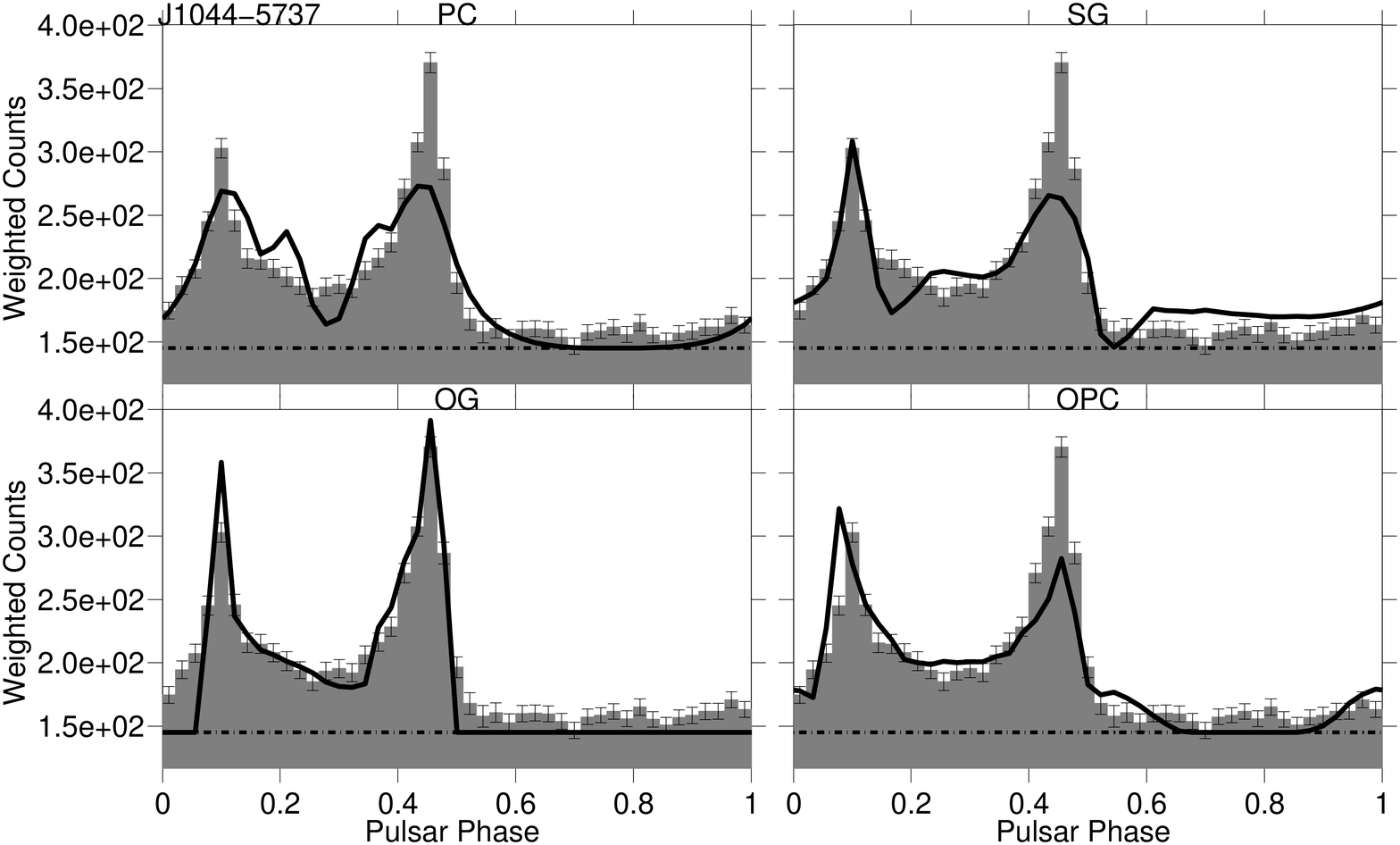}
\includegraphics[width=0.9\textwidth]{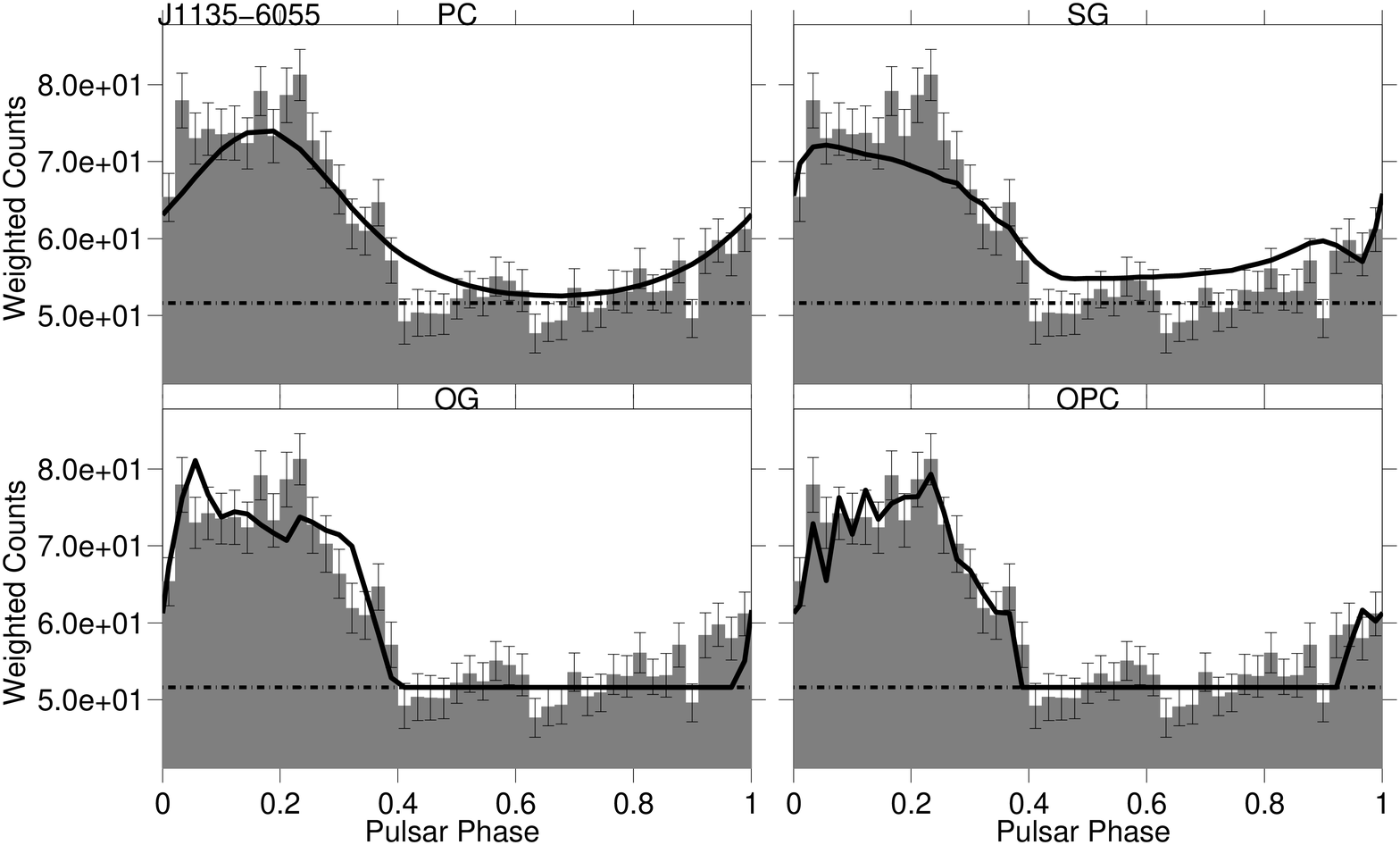}
\caption{\emph{Top:} PSR J1044-5737; \emph{bottom:} PSR J1135-6055. For each model the best $\gamma$-ray light-curve (thick black line) is superimposed on the LAT  pulsar light-curve (shaded histogram). The estimated background is indicated by the dash-dot line.}
\label{fitGm9}
\end{figure}
  
\clearpage
\begin{figure}[htbp!]
\centering
\includegraphics[width=0.9\textwidth]{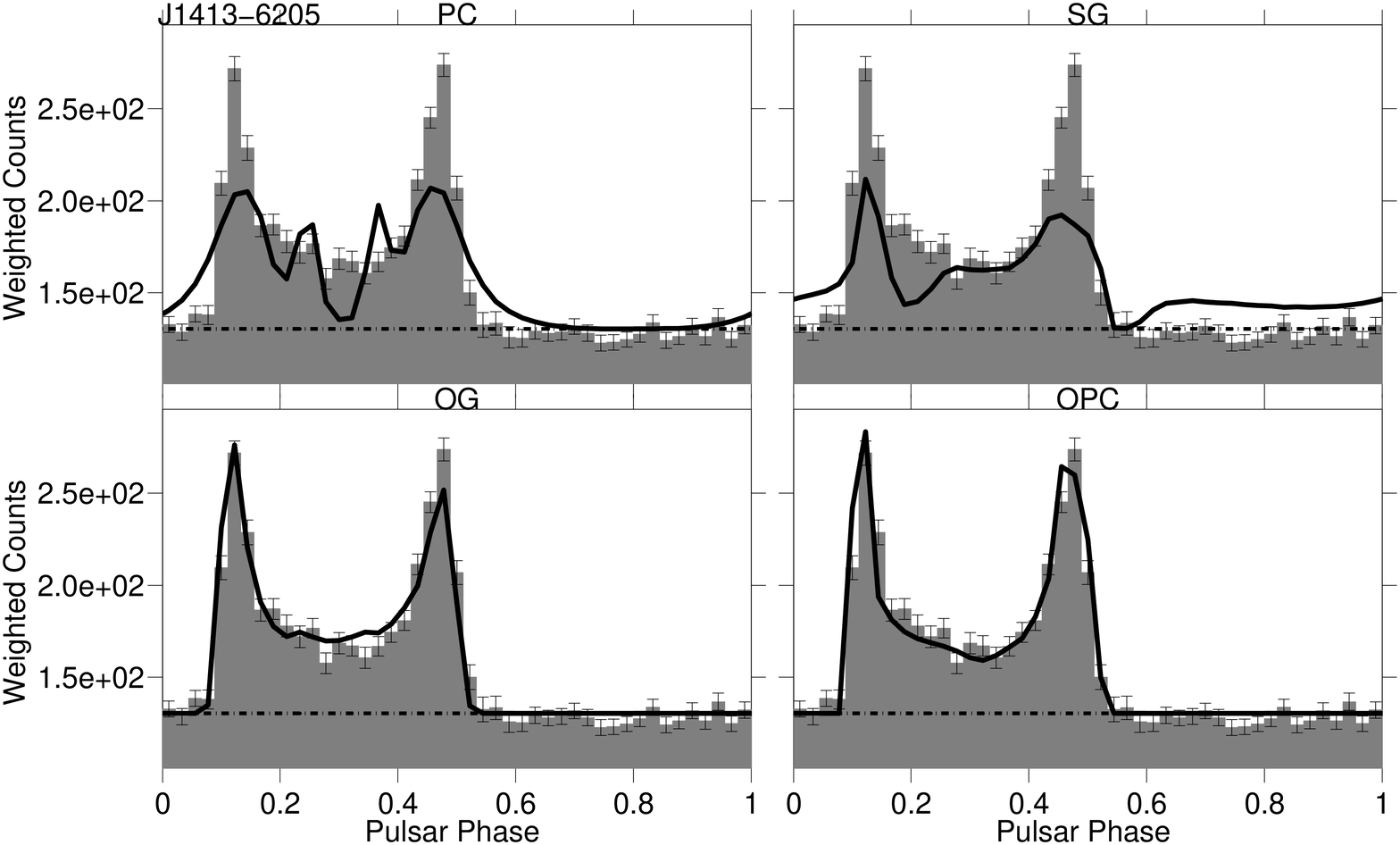}
\includegraphics[width=0.9\textwidth]{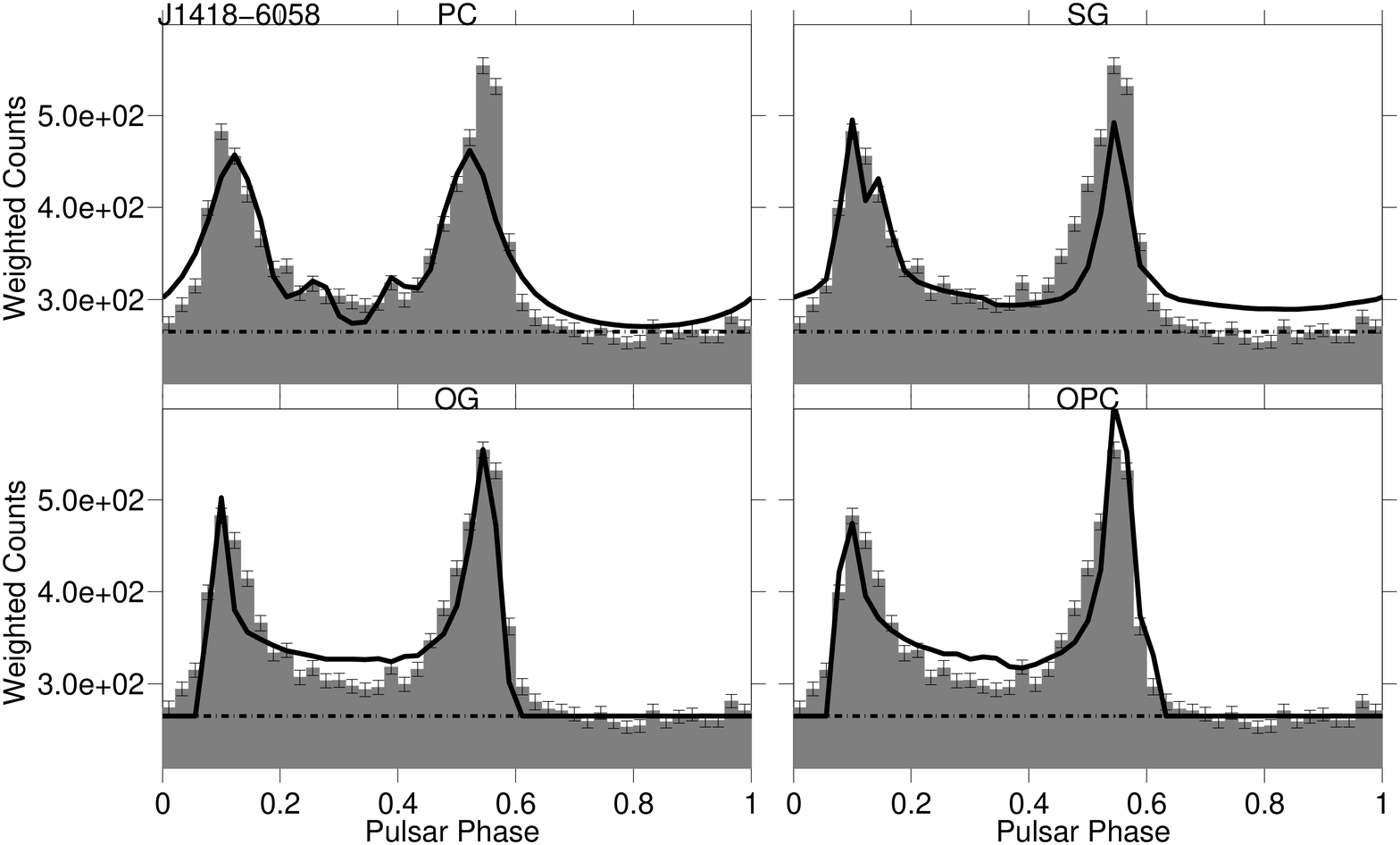}
\caption{\emph{Top:} PSR J1413-6205; \emph{bottom:} PSR J1418-6058. For each model the best $\gamma$-ray light-curve (thick black line) is superimposed on the LAT  pulsar light-curve (shaded histogram). The estimated background is indicated by the dash-dot line.}
\label{fitGm11}
\end{figure}
  
\clearpage
\begin{figure}[htbp!]
\centering
\includegraphics[width=0.9\textwidth]{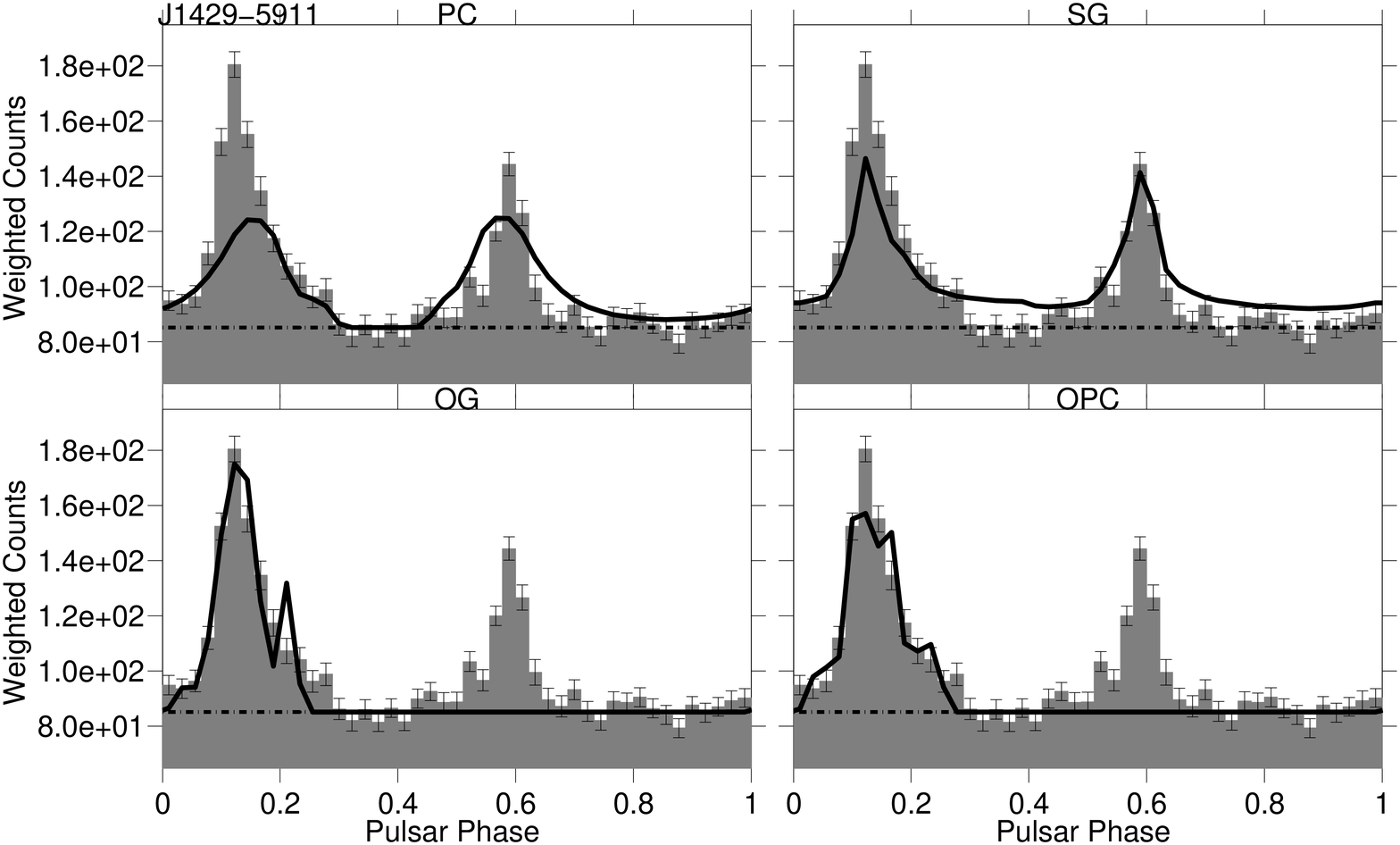}
\includegraphics[width=0.9\textwidth]{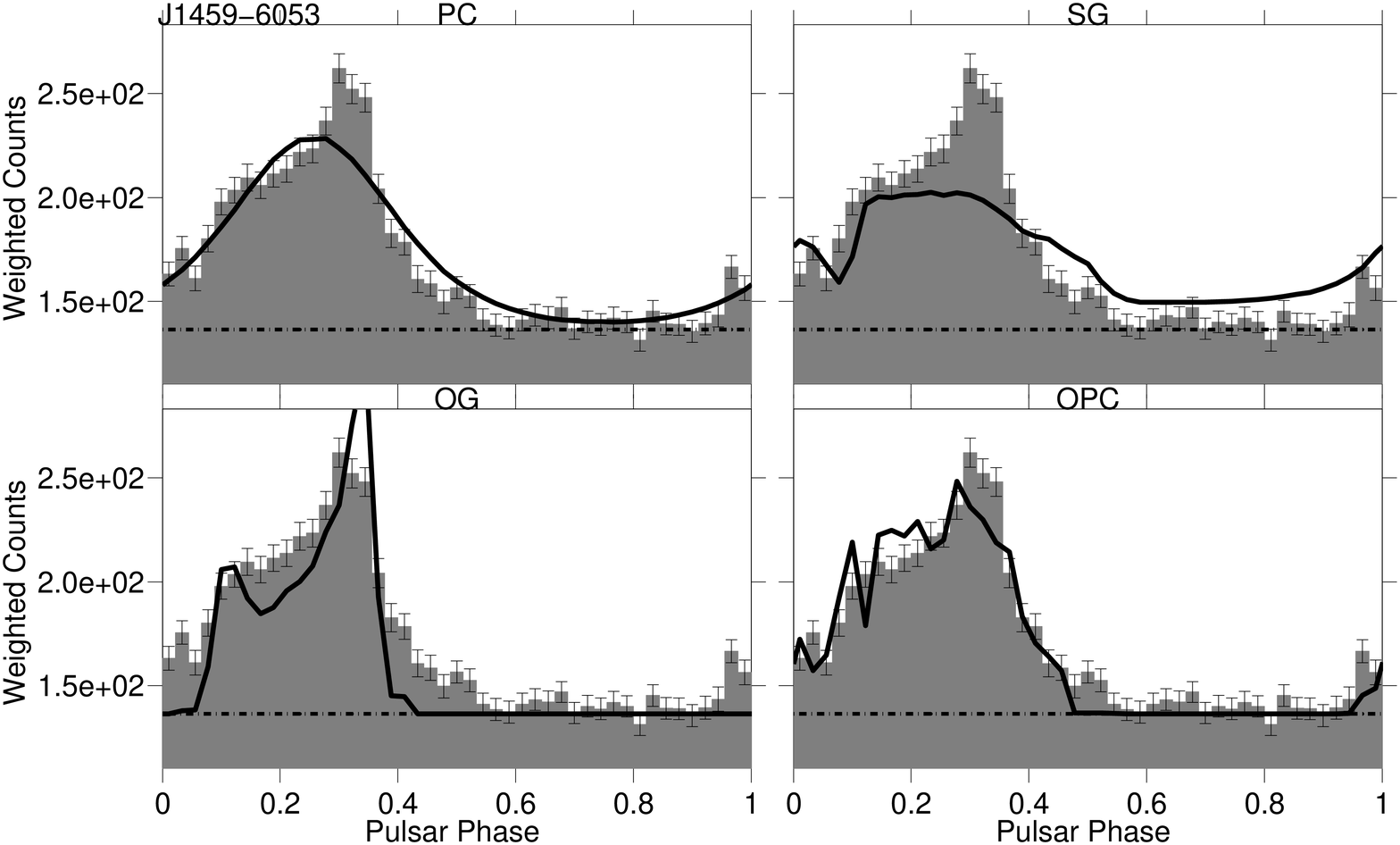}
\caption{\emph{Top:} PSR J1429-5911; \emph{bottom:} PSR J1459-6053. For each model the best $\gamma$-ray light-curve (thick black line) is superimposed on the LAT  pulsar light-curve (shaded histogram). The estimated background is indicated by the dash-dot line.}
\label{fitGm13}
\end{figure}
  
\clearpage
\begin{figure}[htbp!]
\centering
\includegraphics[width=0.9\textwidth]{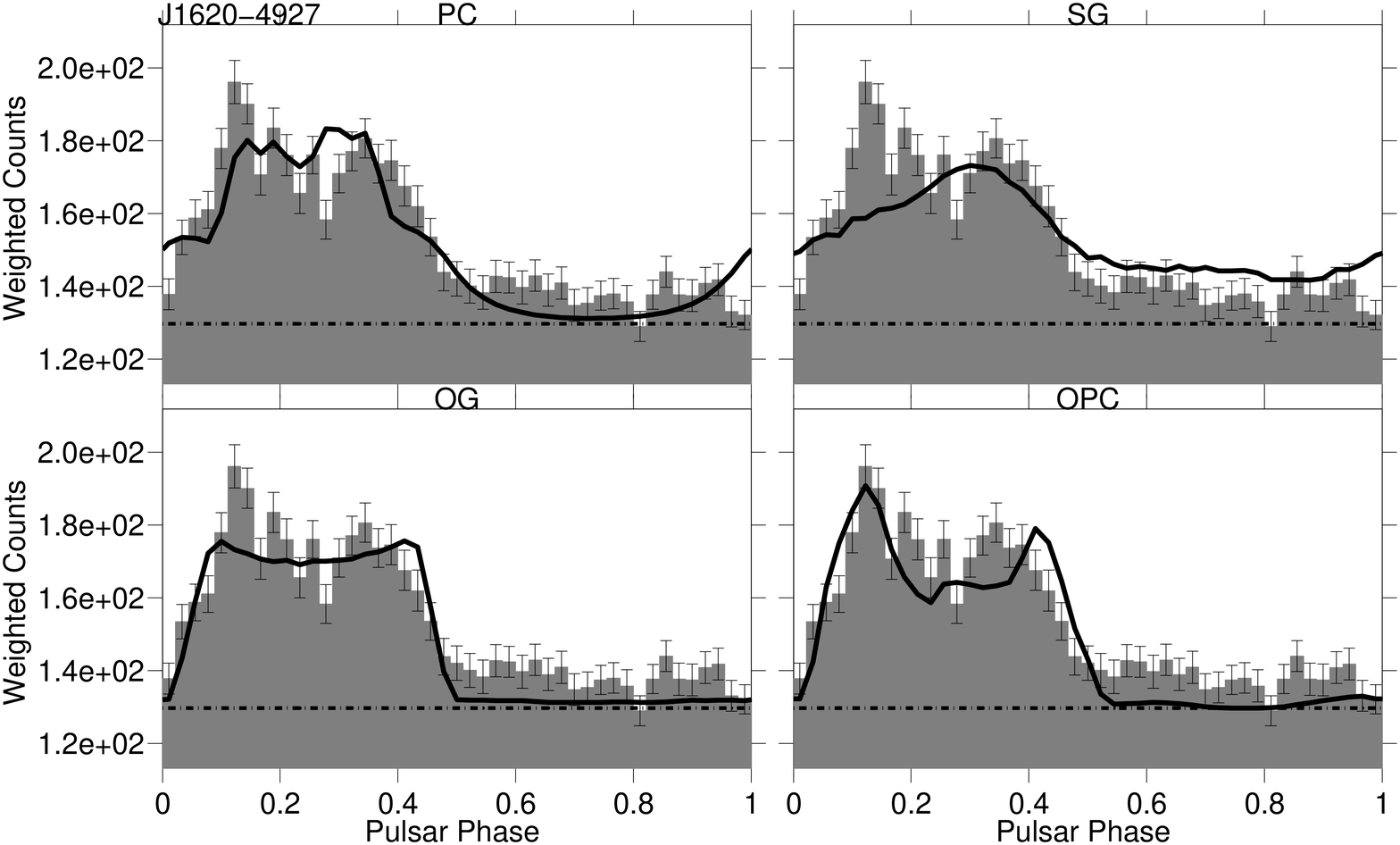}
\includegraphics[width=0.9\textwidth]{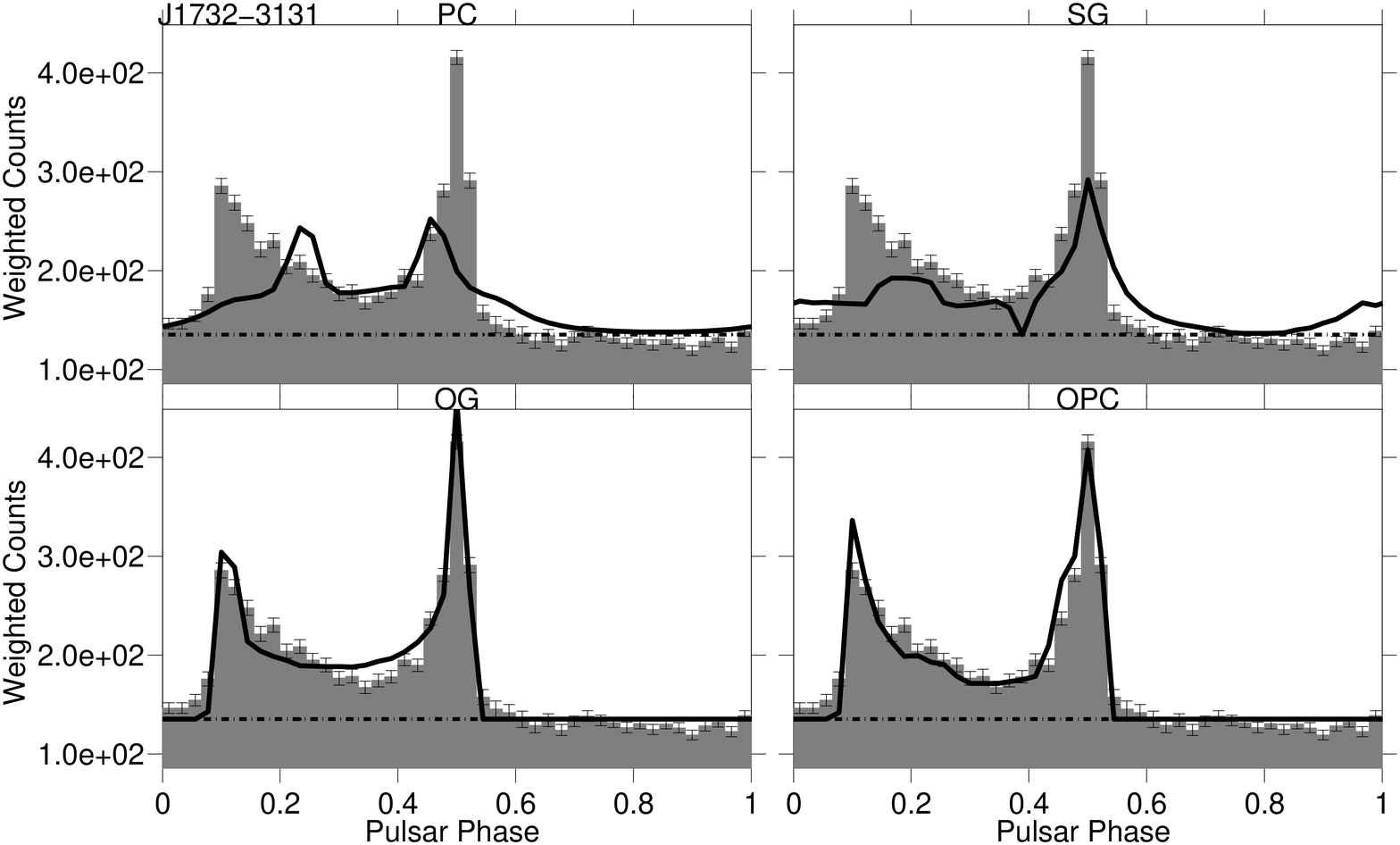}
\caption{\emph{Top:} PSR J1620-4927; \emph{bottom:} PSR J1732-3131. For each model the best $\gamma$-ray light-curve (thick black line) is superimposed on the LAT  pulsar light-curve (shaded histogram). The estimated background is indicated by the dash-dot line.}
\label{fitGm15}
\end{figure}
  
\clearpage
\begin{figure}[htbp!]
\centering
\includegraphics[width=0.9\textwidth]{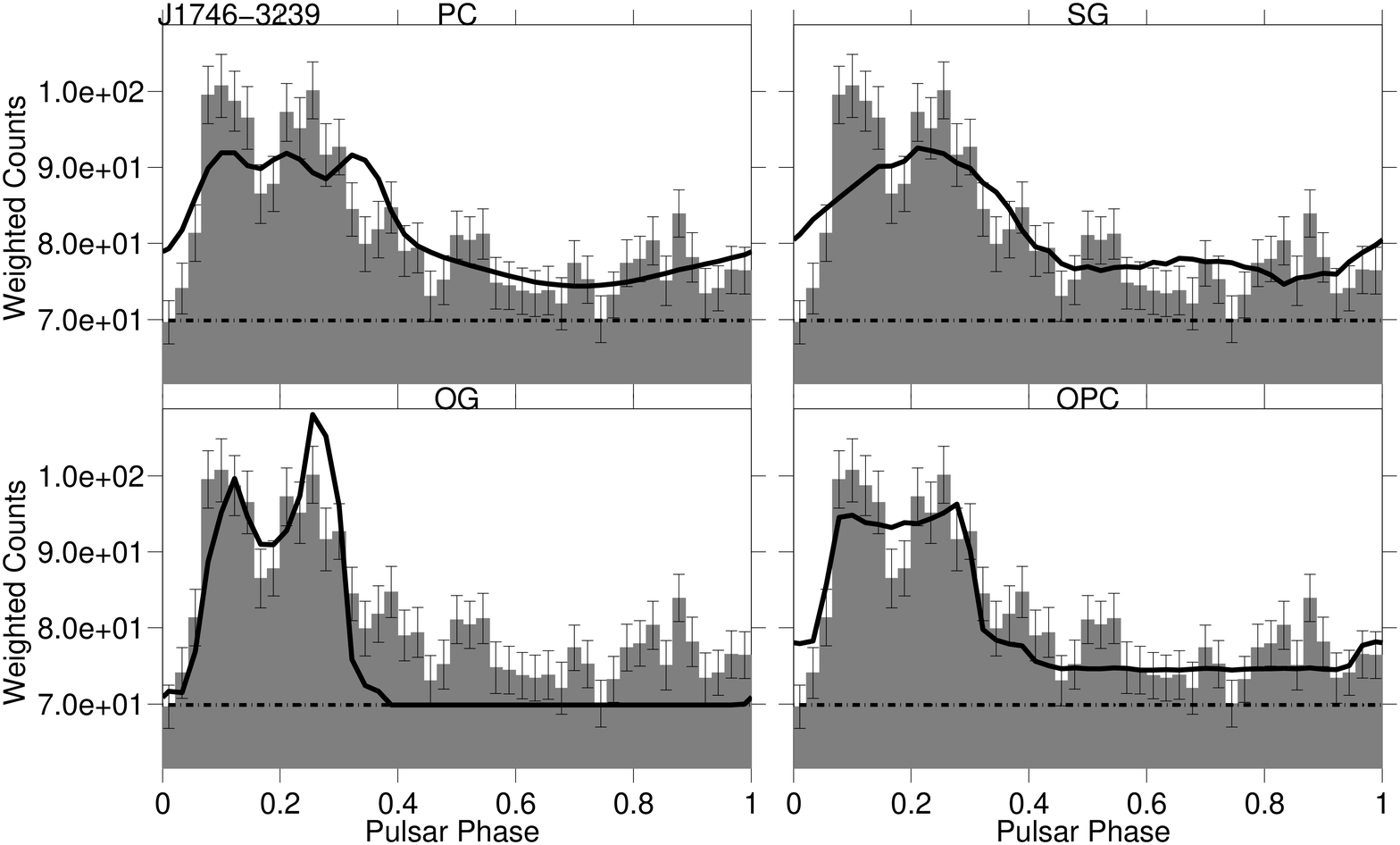}
\includegraphics[width=0.9\textwidth]{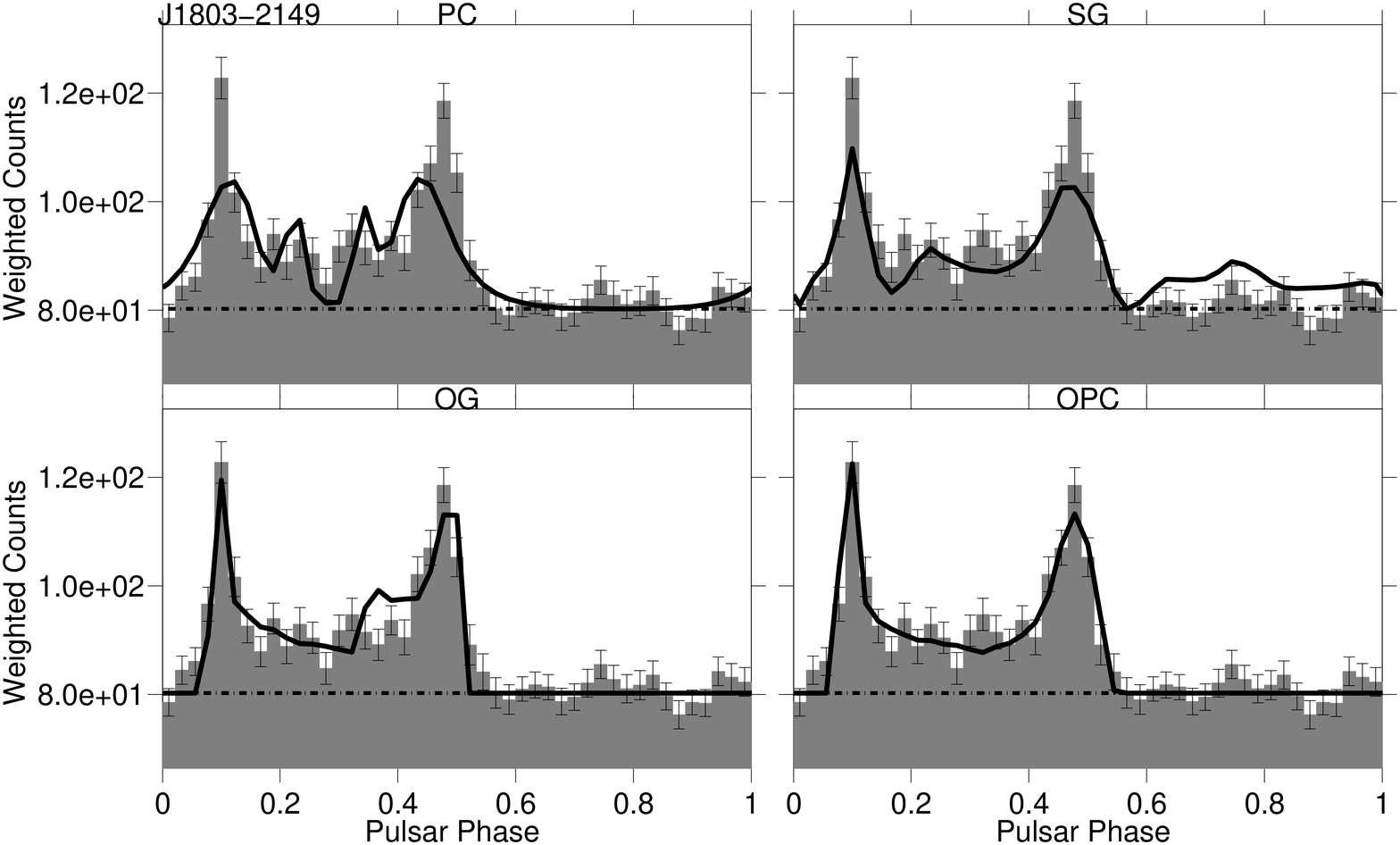}
\caption{\emph{Top:} PSR J1746-3239; \emph{bottom:} PSR J1803-2149. For each model the best $\gamma$-ray light-curve (thick black line) is superimposed on the LAT  pulsar light-curve (shaded histogram). The estimated background is indicated by the dash-dot line.}
\label{fitGm17}
\end{figure}
  
\clearpage
\begin{figure}[htbp!]
\centering
\includegraphics[width=0.9\textwidth]{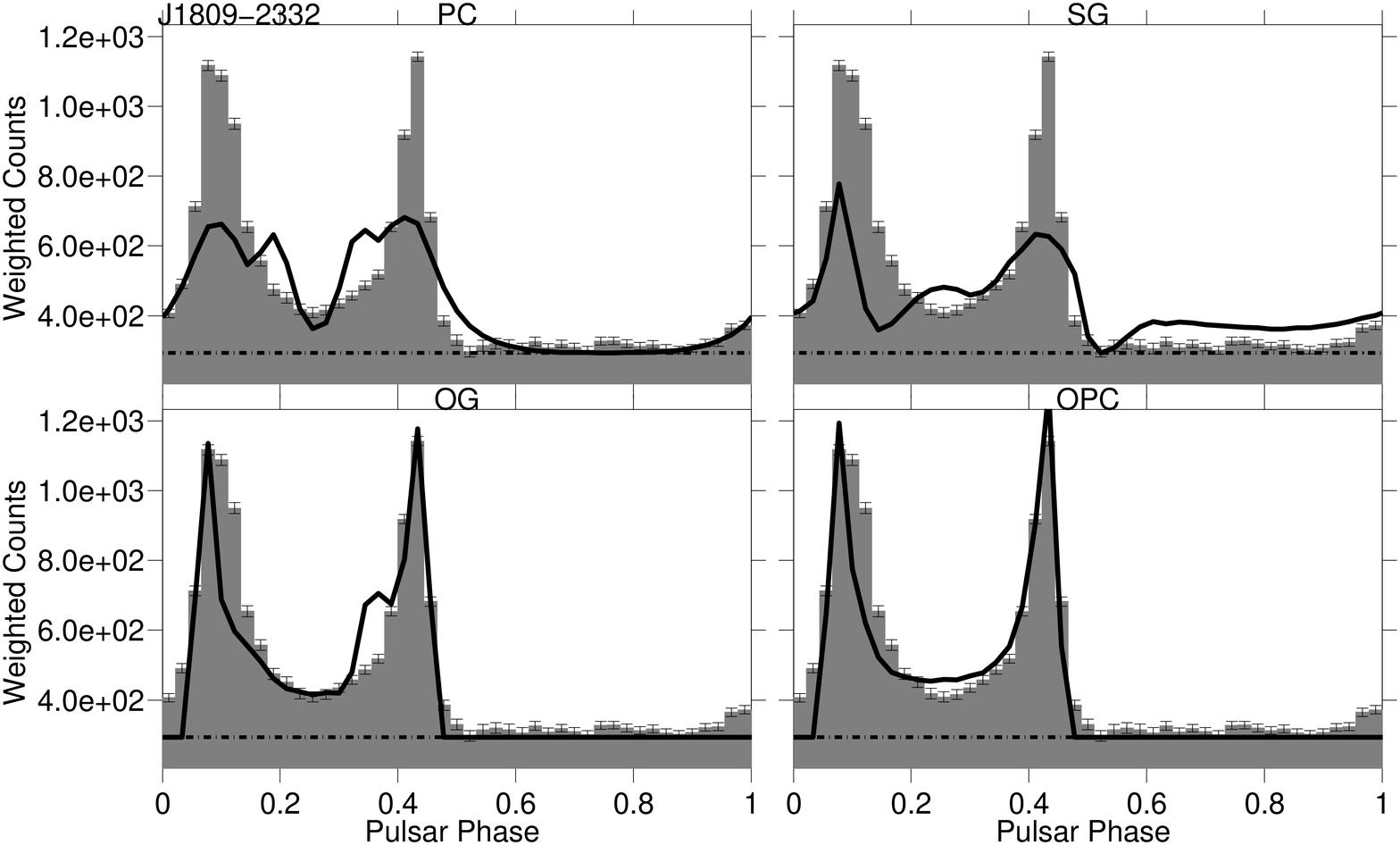}
\includegraphics[width=0.9\textwidth]{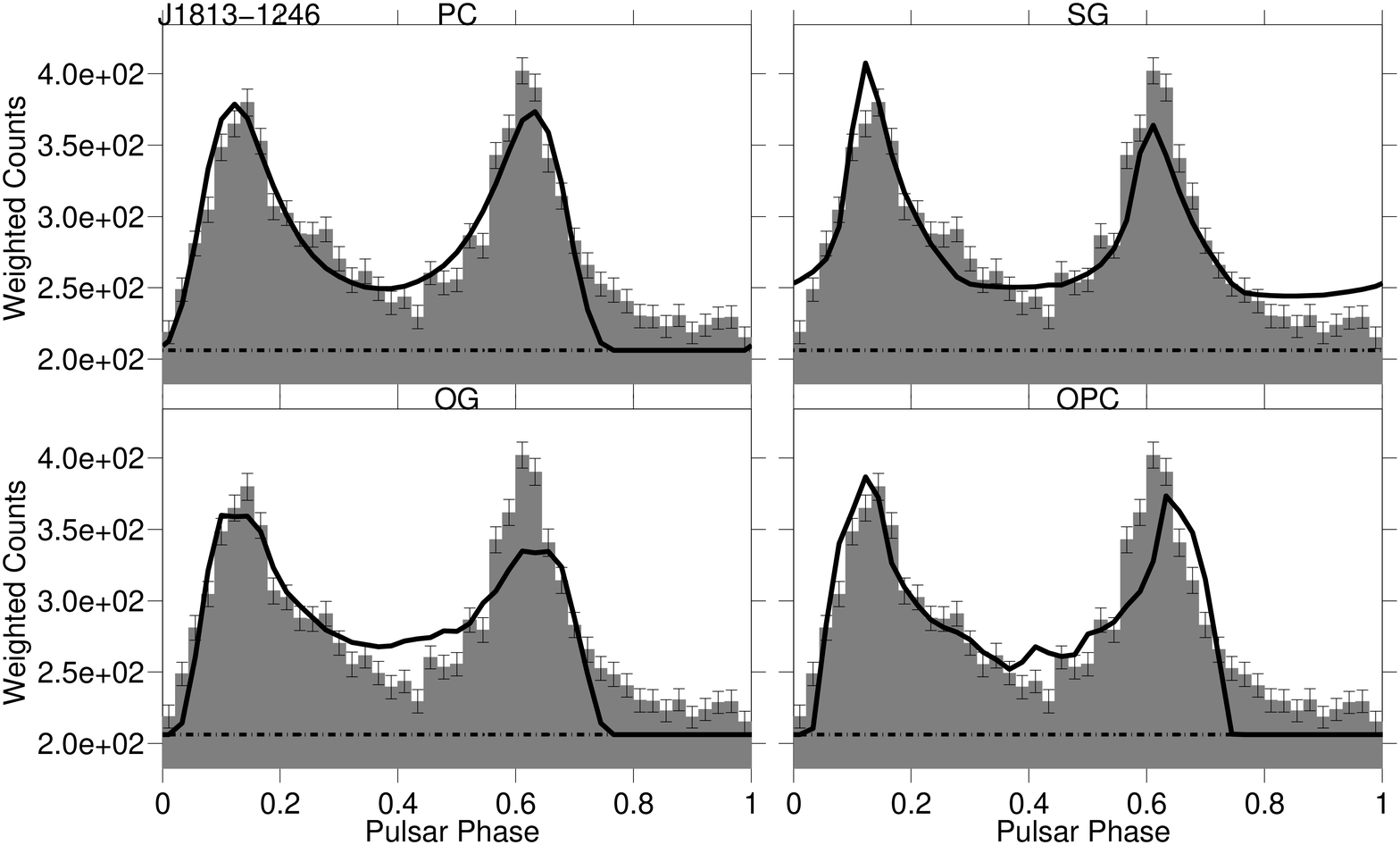}
\caption{\emph{Top:} PSR J1809-2332; \emph{bottom:} PSR J1813-1246. For each model the best $\gamma$-ray light-curve (thick black line) is superimposed on the LAT  pulsar light-curve (shaded histogram). The estimated background is indicated by the dash-dot line.}
\label{fitGm19}
\end{figure}
  
\clearpage
\begin{figure}[htbp!]
\centering
\includegraphics[width=0.9\textwidth]{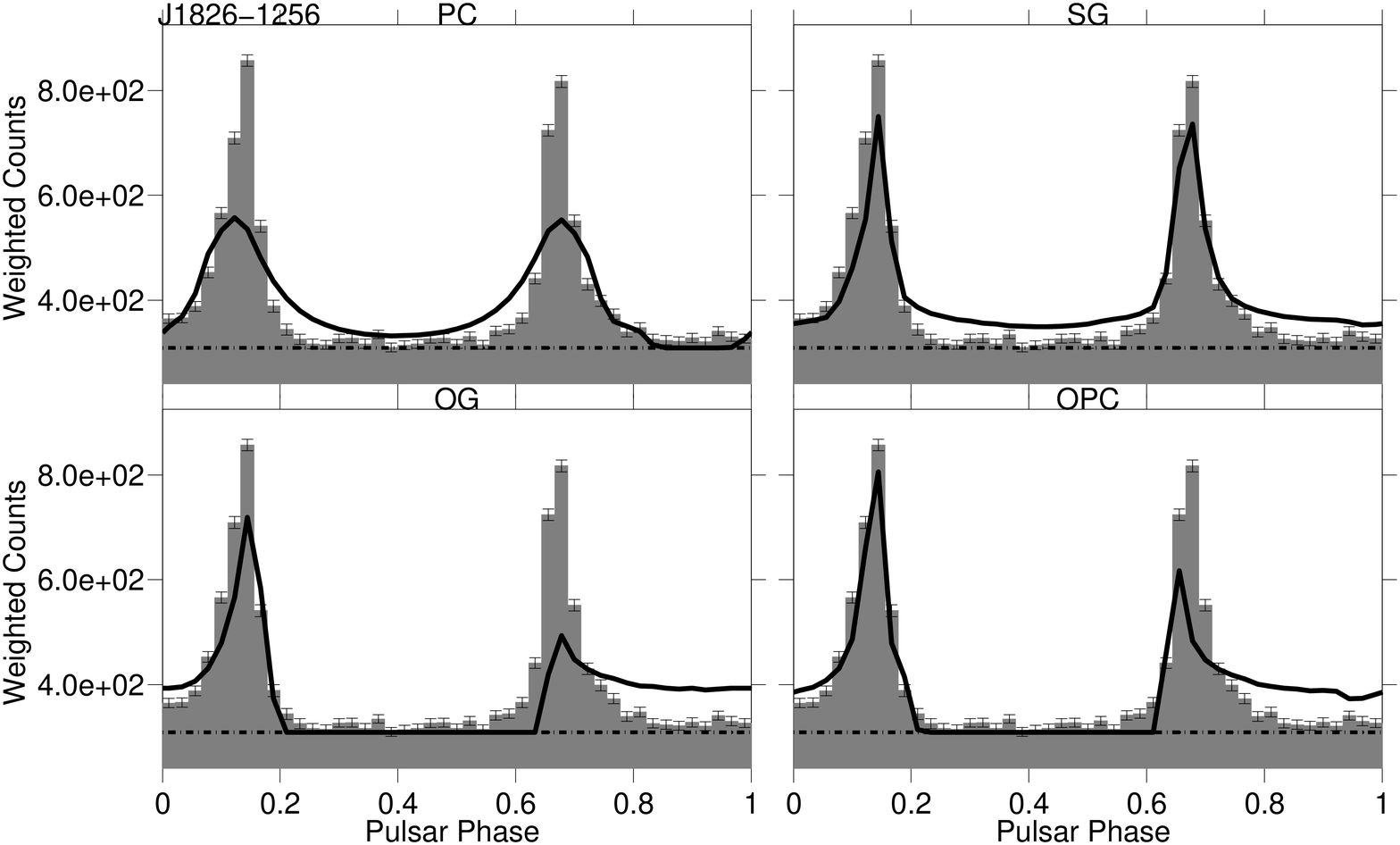}
\includegraphics[width=0.9\textwidth]{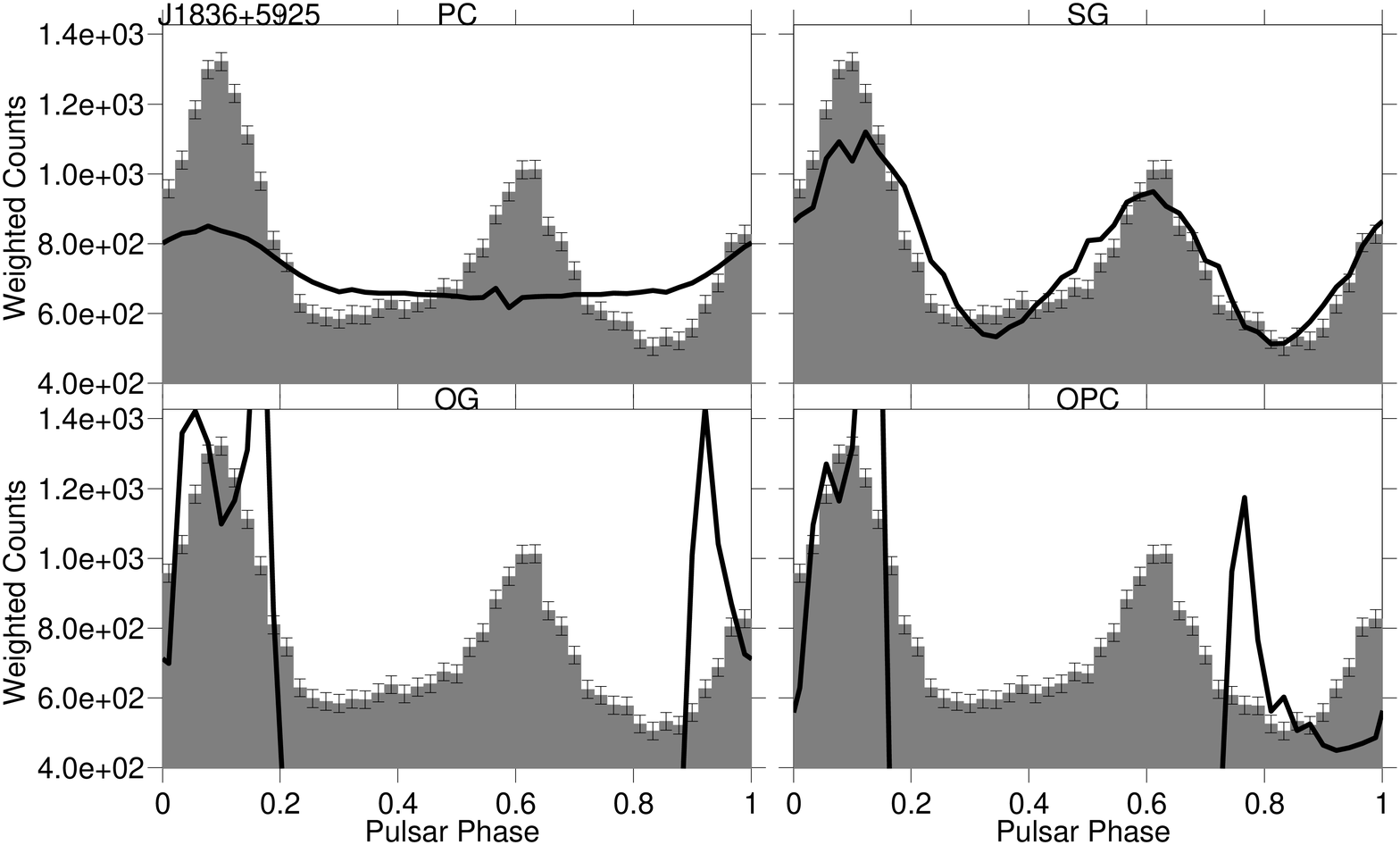}
\caption{\emph{Top:} PSR J1826-1256; \emph{bottom:} PSR J1836+5925. For each model the best $\gamma$-ray light-curve (thick black line) is superimposed on the LAT  pulsar light-curve (shaded histogram). The estimated background is indicated by the dash-dot line. For PSR J1836+5925 the SG is the only model that predicts enough off-pulse emission while OG and OPC models completely fail in explaining the observation probably because they do not predict enough off-pulse emission.}
\label{fitGm21}
\end{figure}
  
\clearpage
\begin{figure}[htbp!]
\centering
\includegraphics[width=0.9\textwidth]{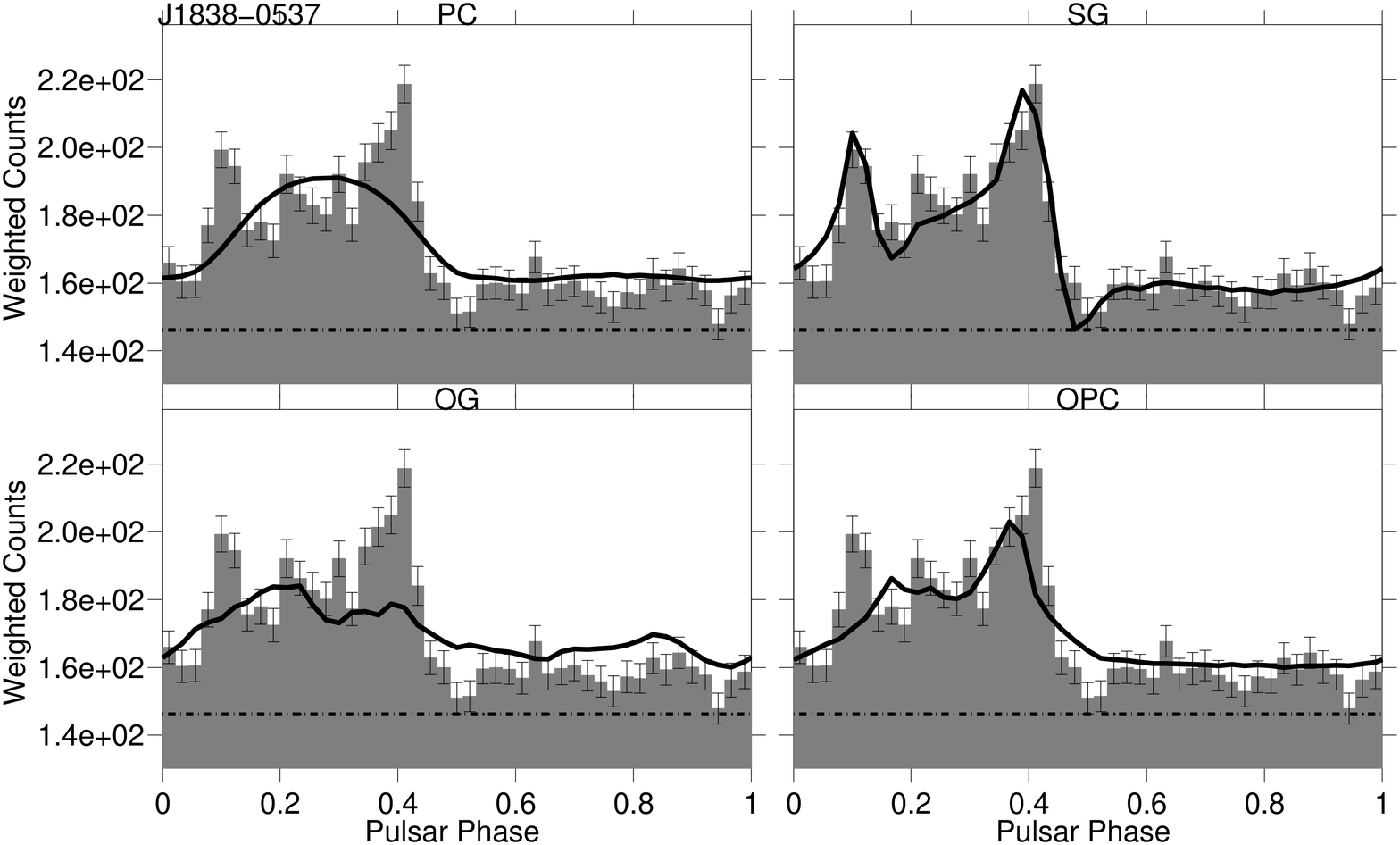}
\includegraphics[width=0.9\textwidth]{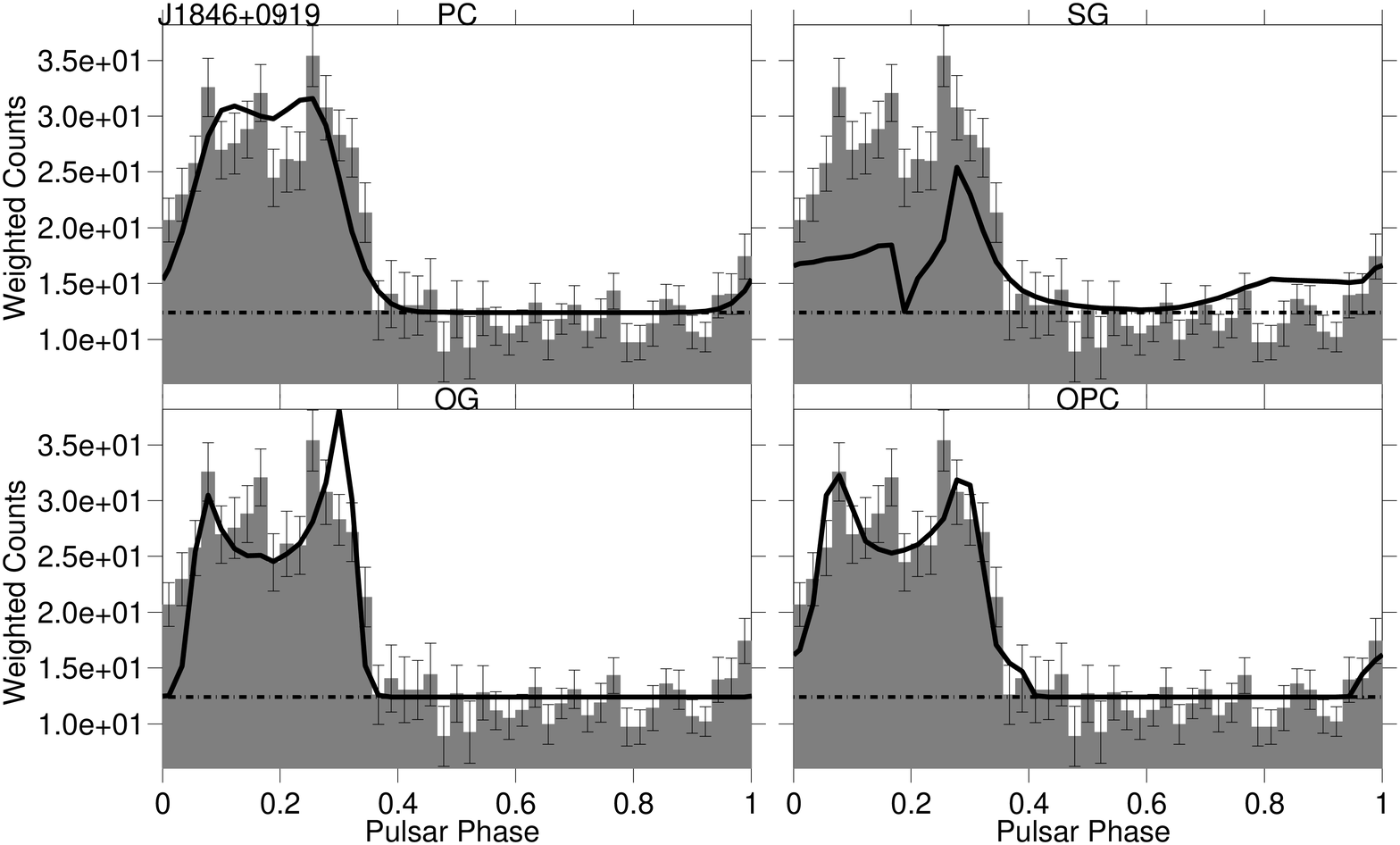}
\caption{\emph{Top:} PSR J1838-0537; \emph{bottom:} PSR J1846+0919. For each model the best $\gamma$-ray light-curve (thick black line) is superimposed on the LAT  pulsar light-curve (shaded histogram). The estimated background is indicated by the dash-dot line.}
\label{fitGm23}
\end{figure}
  
\clearpage
\begin{figure}[htbp!]
\centering
\includegraphics[width=0.9\textwidth]{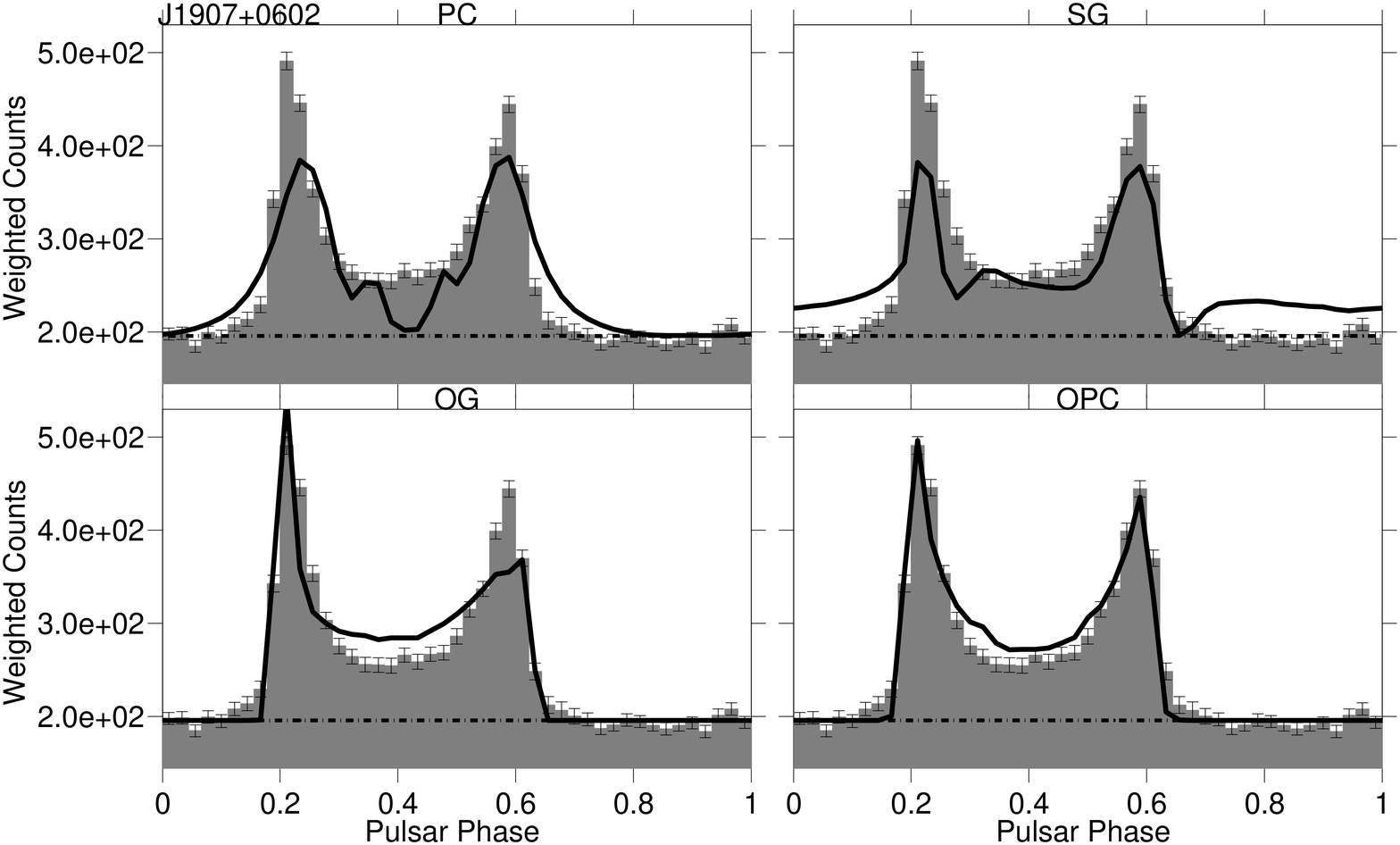}
\includegraphics[width=0.9\textwidth]{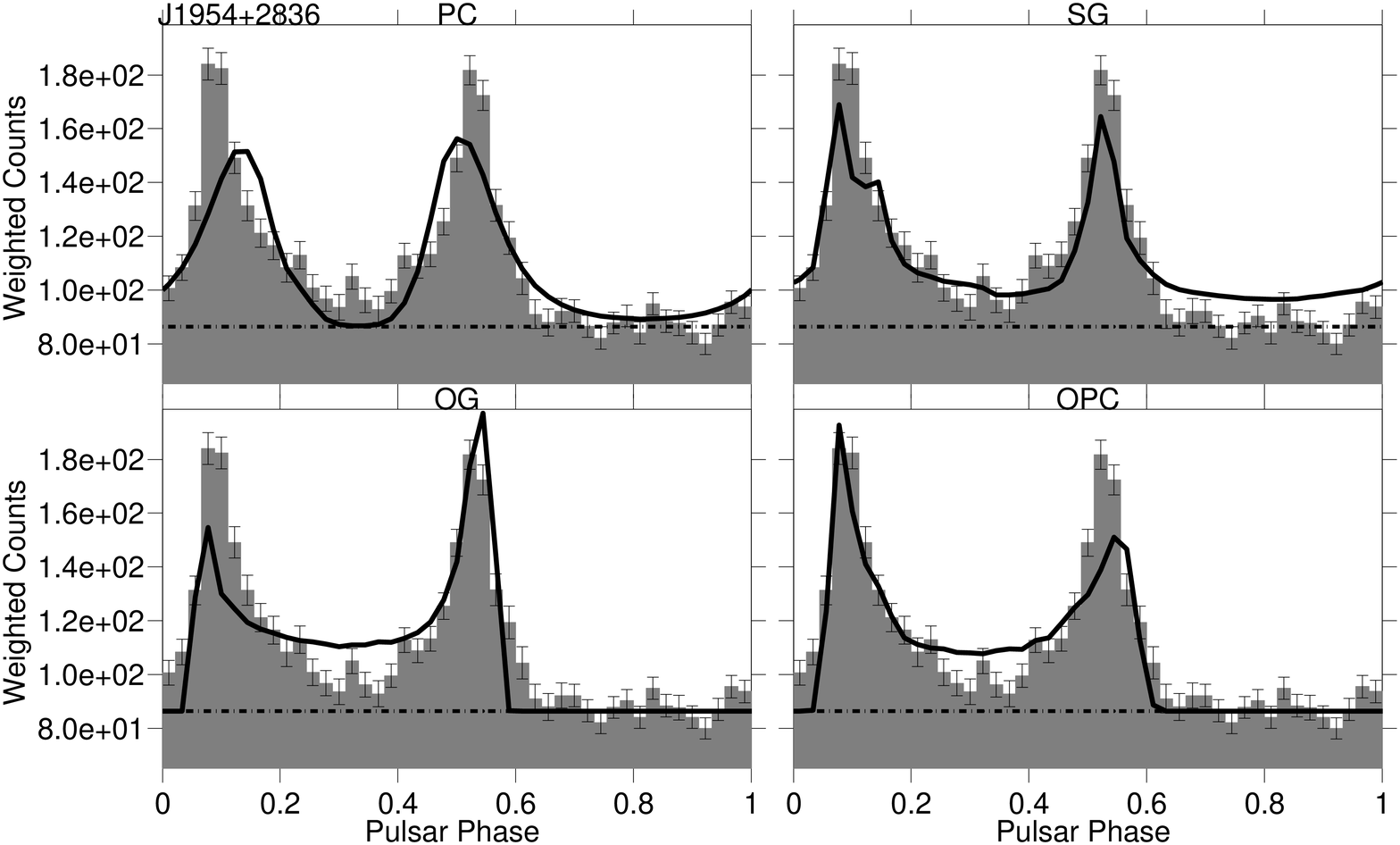}
\caption{\emph{Top:} PSR J1907+0602; \emph{bottom:} PSR J1954+2836. For each model the best $\gamma$-ray light-curve (thick black line) is superimposed on the LAT  pulsar light-curve (shaded histogram). The estimated background is indicated by the dash-dot line.}
\label{fitGm25}
\end{figure}
  
\clearpage
\begin{figure}[htbp!]
\centering
\includegraphics[width=0.9\textwidth]{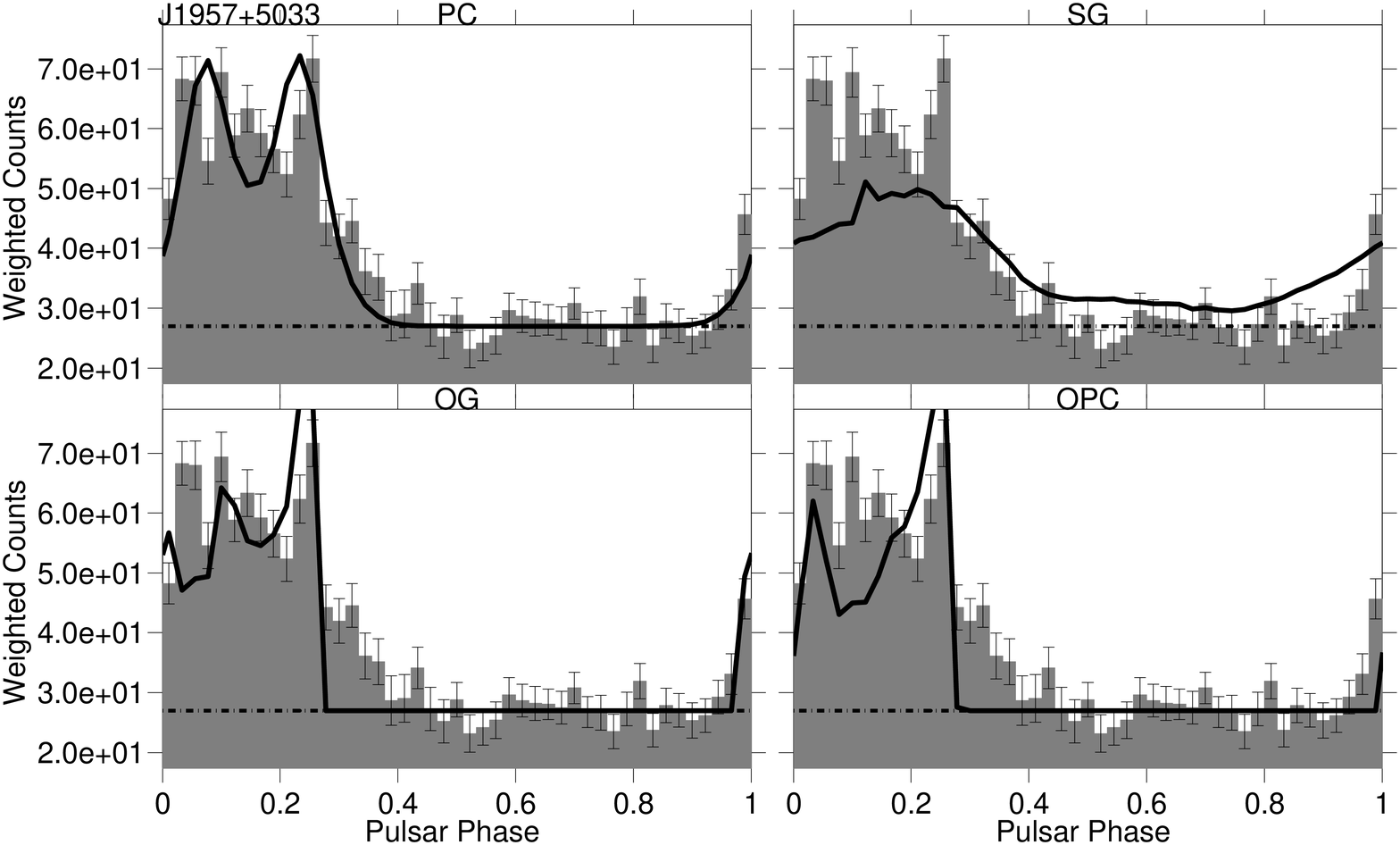}
\includegraphics[width=0.9\textwidth]{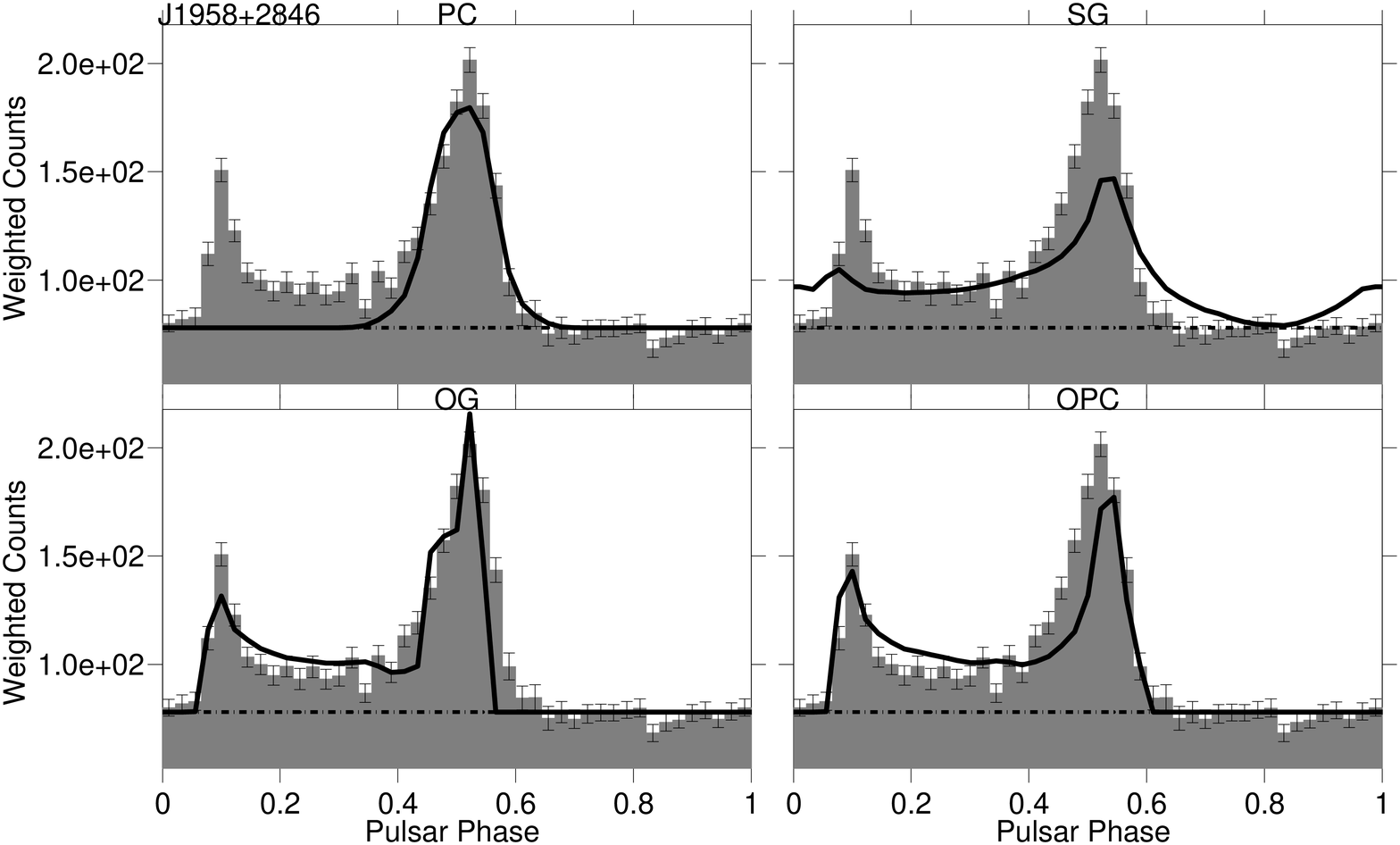}
\caption{\emph{Top:} PSR J1957+5033; \emph{bottom:} PSR J1958+2846. For each model the best $\gamma$-ray light-curve (thick black line) is superimposed on the LAT  pulsar light-curve (shaded histogram). The estimated background is indicated by the dash-dot line.}
\label{fitGm27}
\end{figure}
  
\clearpage
\begin{figure}[htbp!]
\centering
\includegraphics[width=0.9\textwidth]{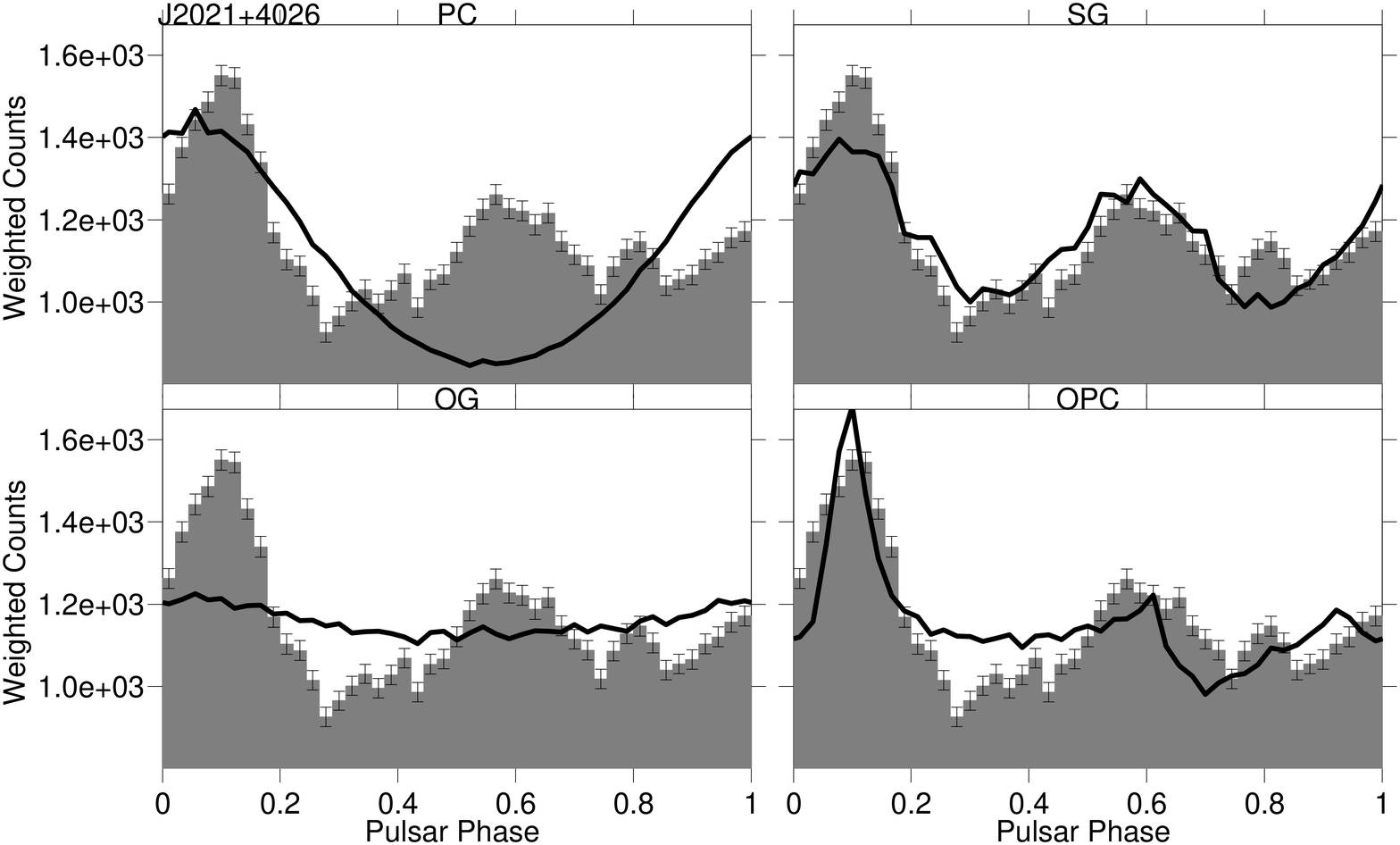}
\includegraphics[width=0.9\textwidth]{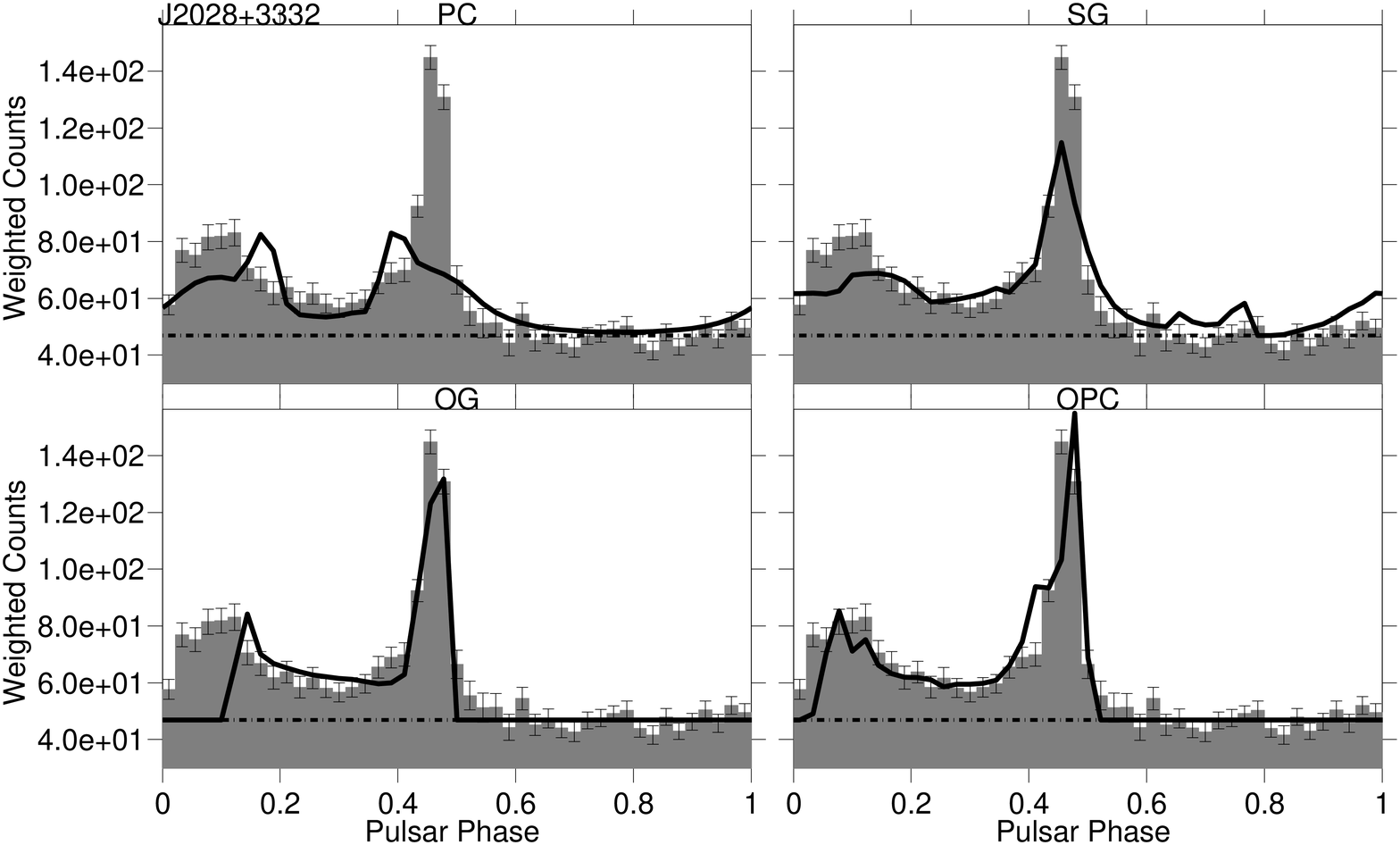}
\caption{\emph{Top:} PSR J2021+4026; \emph{bottom:} PSR J2028+3332. For each model the best $\gamma$-ray light-curve (thick black line) is superimposed on the LAT  pulsar light-curve (shaded histogram). The estimated background is indicated by the dash-dot line.}
\label{fitGm29}
\end{figure}
  
\clearpage
\begin{figure}[htbp!]
\centering
\includegraphics[width=0.9\textwidth]{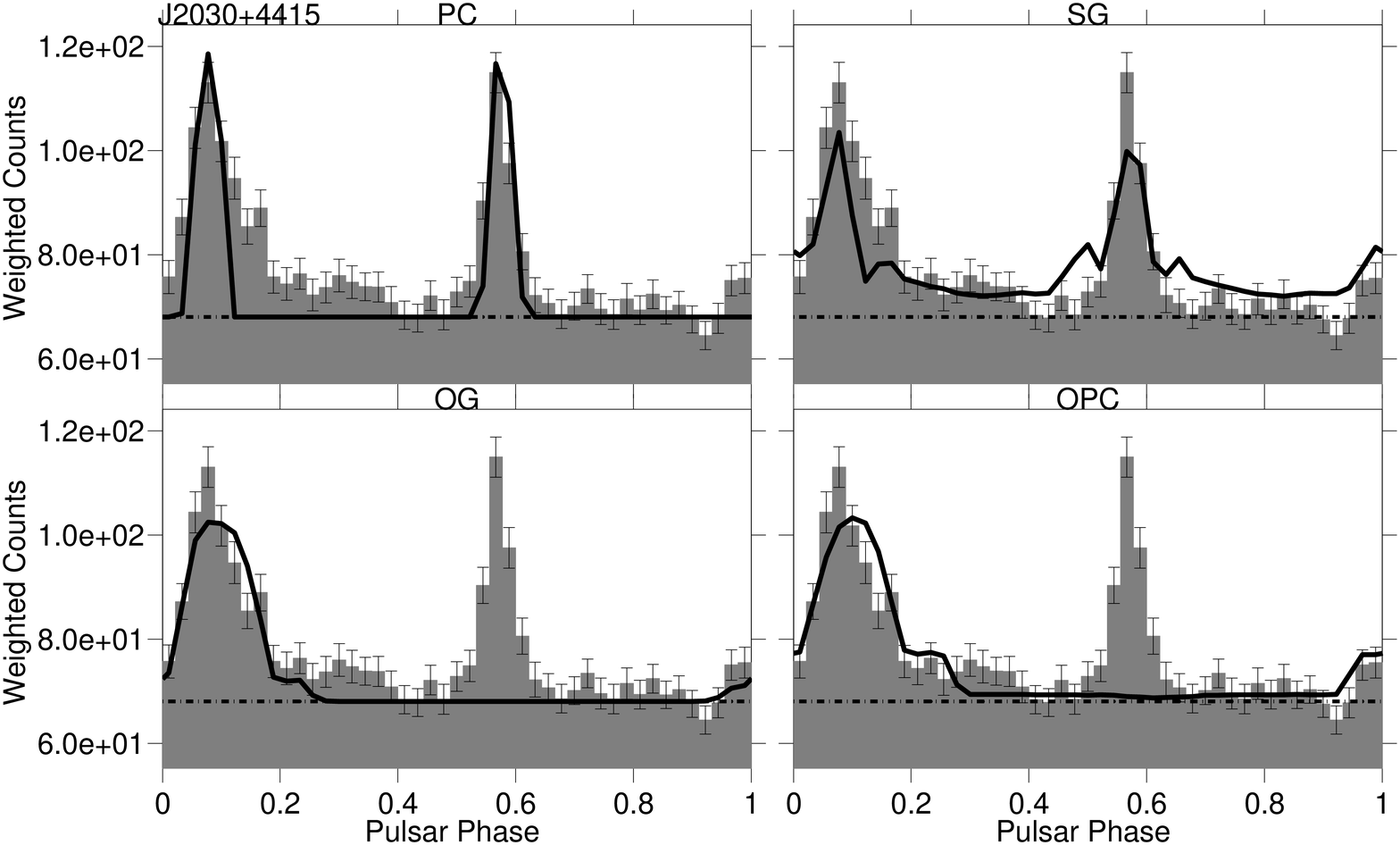}
\includegraphics[width=0.9\textwidth]{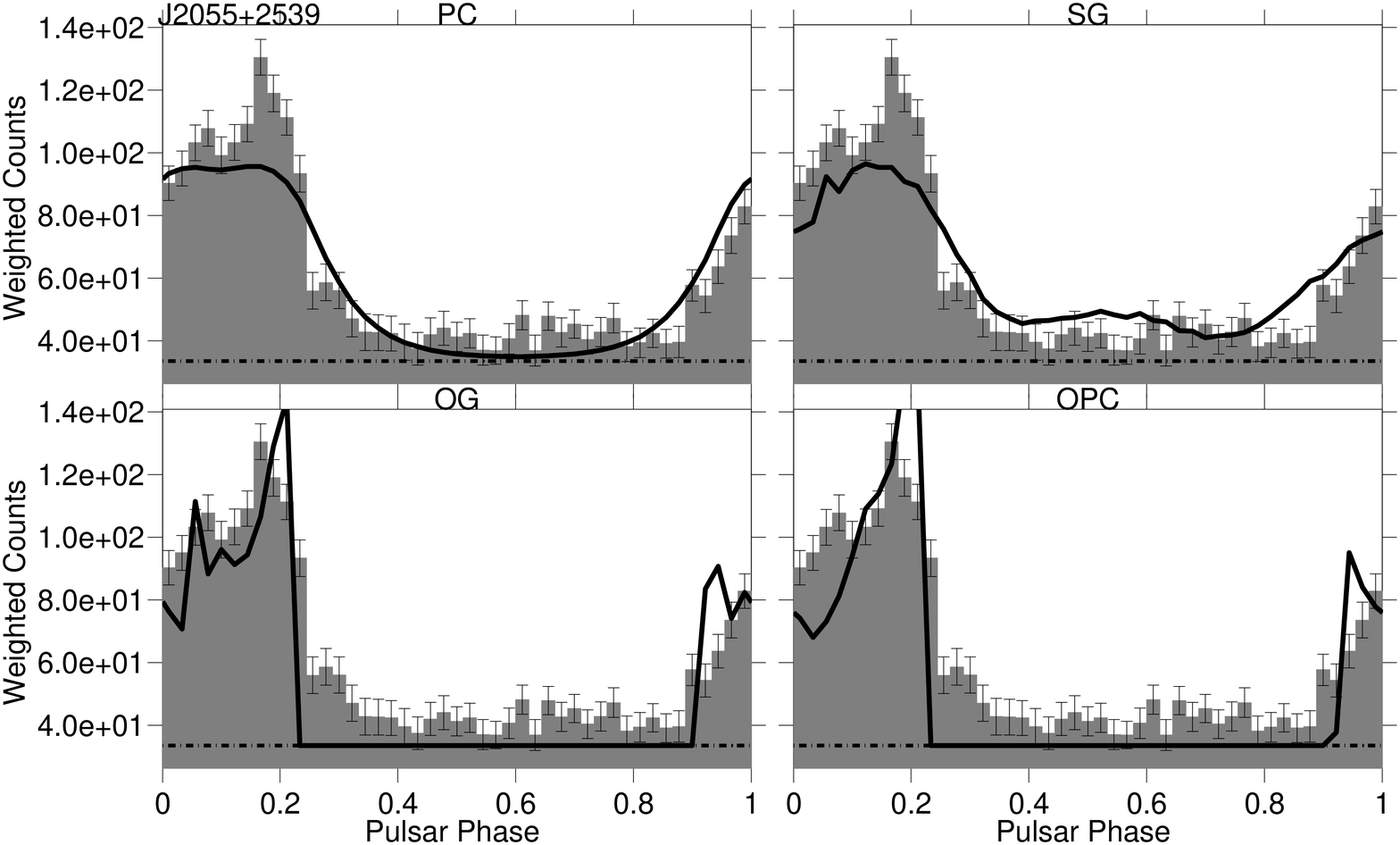}
\caption{\emph{Top:} PSR J2030+4415; \emph{bottom:} PSR J2055+2539. For each model the best $\gamma$-ray light-curve (thick black line) is superimposed on the LAT  pulsar light-curve (shaded histogram). The estimated background is indicated by the dash-dot line.}
\label{fitGm31}
\end{figure}
  
\clearpage
\begin{figure}[htbp!]
\centering
\includegraphics[width=0.9\textwidth]{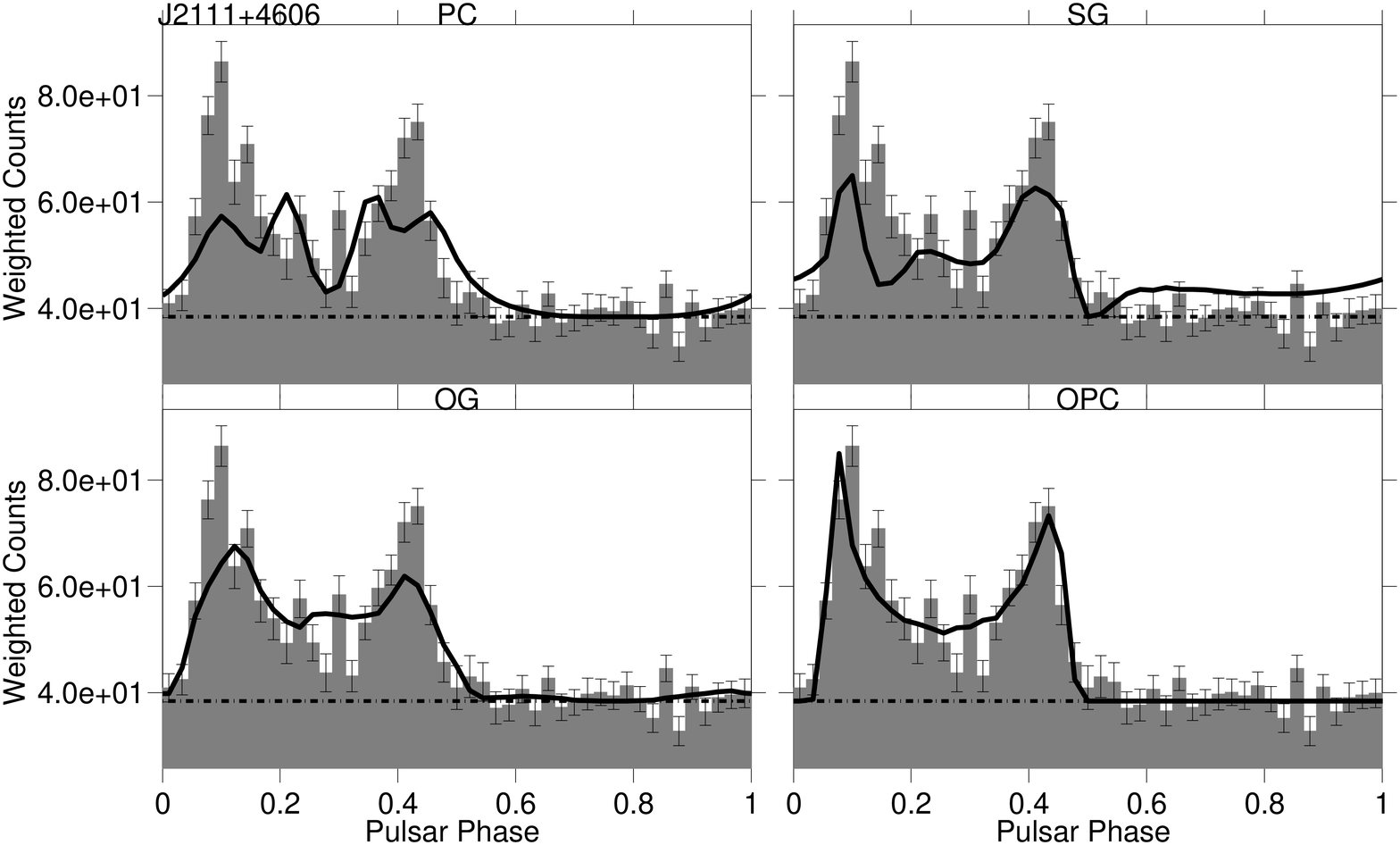}
\includegraphics[width=0.9\textwidth]{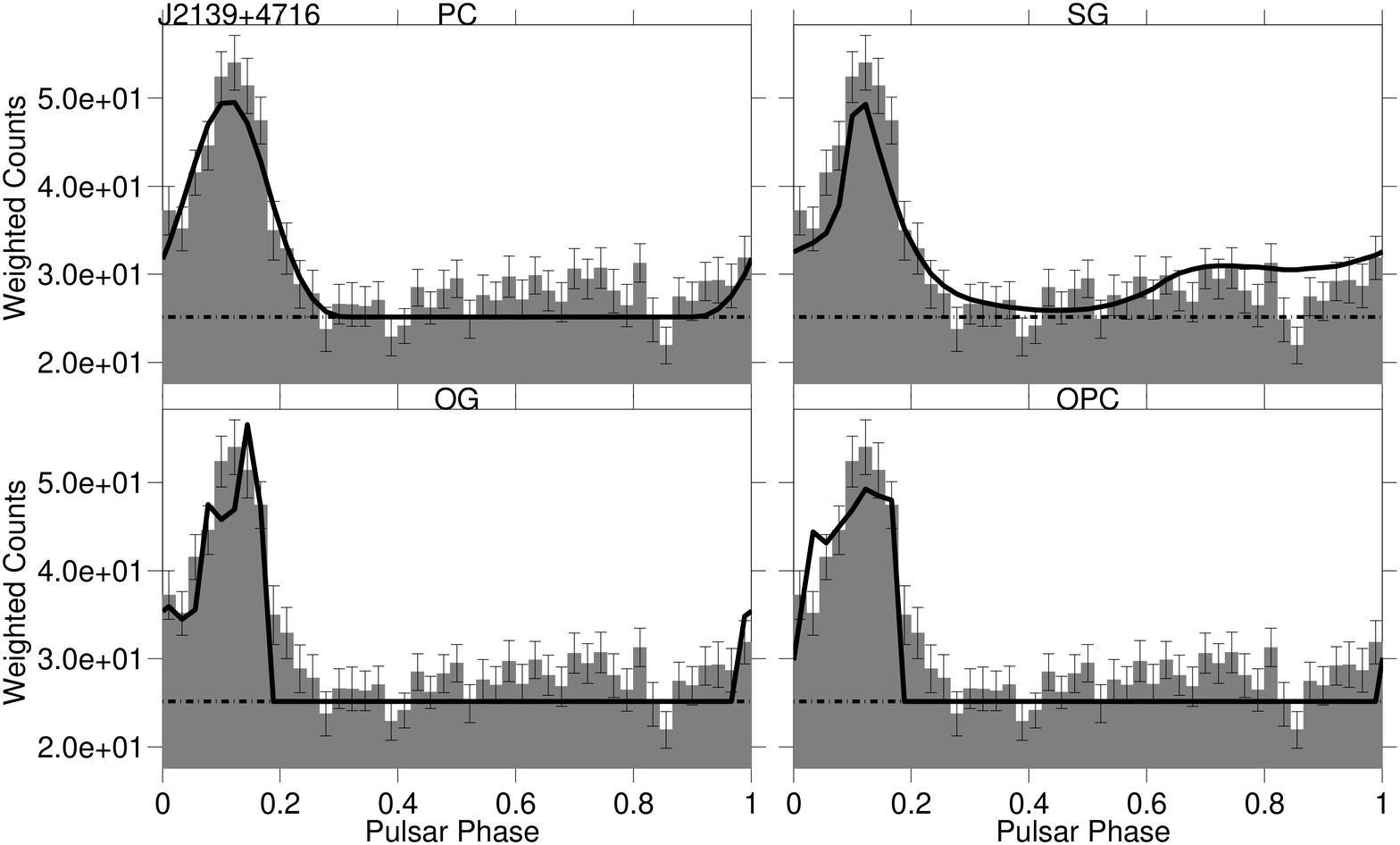}
\caption{\emph{Top:} PSR J2111+4606; \emph{bottom:} PSR J2139+4716. For each model the best $\gamma$-ray light-curve (thick black line) is superimposed on the LAT  pulsar light-curve (shaded histogram). The estimated background is indicated by the dash-dot line.}
\label{fitGm33}
\end{figure}

\clearpage
\begin{figure}[htbp!]
\centering
\includegraphics[width=0.9\textwidth]{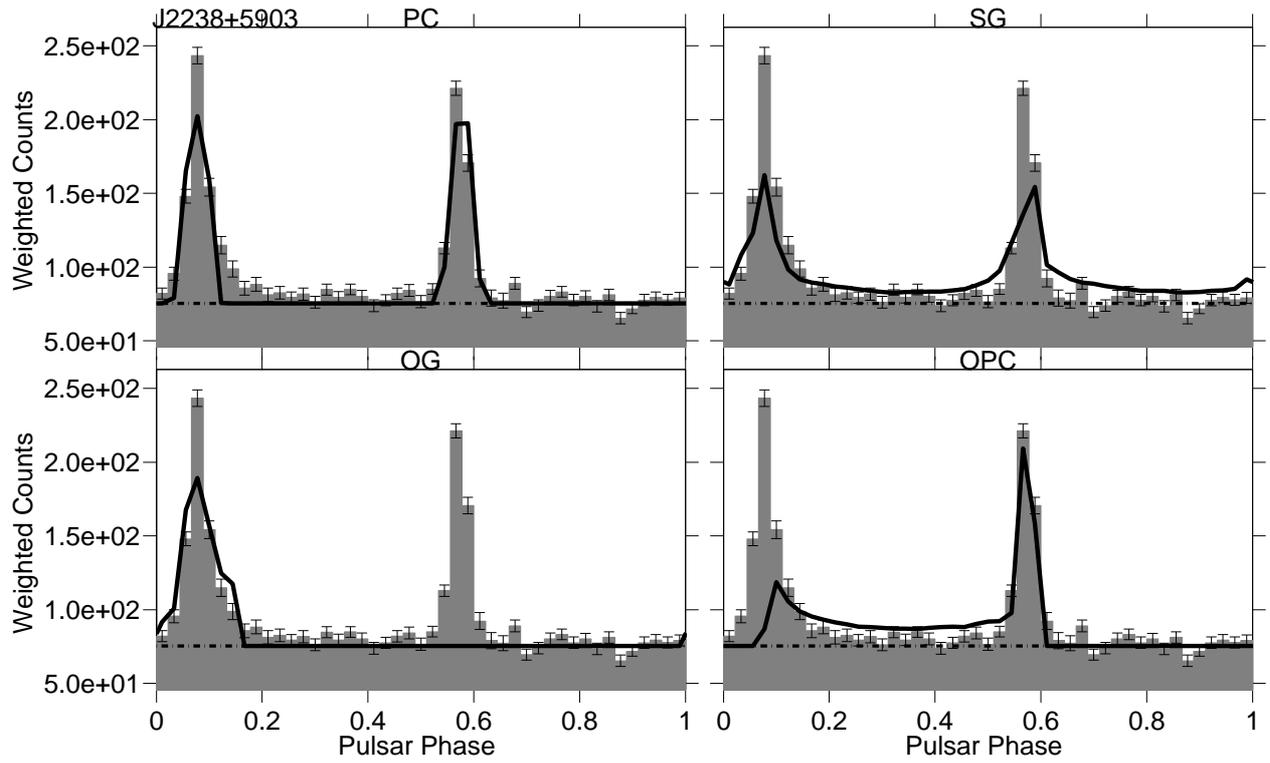}
\caption{PSR J2238+5903. For each model the best $\gamma$-ray light-curve (thick black line) is superimposed on the LAT  pulsar light-curve (shaded histogram). The estimated background is indicated by the dash-dot line.}
\label{fitGm35}
\end{figure}

%% file: JointFitsRes.tex
\begin{table*}[htbp!]
\centering
\begin{tabular}{| c || c | c | c | c | c || c | c | c | c | c |}
\hline
& $ \ln L_{PC}$ & $ \ln L_{SG}$  & $ \ln L_{OG}$  & $ \ln L_{OPC}$ \\
\hline
\hline
J0205$+$6449 & $ -237 $  & $ -301 $  & $ -280 $  & $ -206 $  \\
\hline
J0248$+$6021 & $ -275 $  & $ -149 $  & $ -137 $  & $ -190 $  \\
\hline
J0534$+$2200 & $ -8171 $  & $ -12294 $  & $ -8146 $  & $ -8367 $  \\
\hline
J0631$+$1036 & $ -220 $  & $ -95 $  & $ -83 $  & $ -107 $  \\
\hline
J0659$+$1414 & $ -247 $  & $ -250 $  & $ -300 $  & $ -350 $  \\
\hline
J0729$-$1448 & $ -104 $  & $ -25 $  & $ -37 $  & $ -27 $  \\
\hline
J0742$-$2822 & $ -137 $  & $ -62 $  & $ -72 $  & $ -53 $  \\
\hline
J0835$-$4510 & $ -85102 $  & $ -115612 $  & $ -26995 $  & $ -21028 $  \\
\hline
J0908$-$4913 & $ -492 $  & $ -80 $  & $ -185 $  & $ -138 $  \\
\hline
J0940$-$5428 & $ -50 $  & $ -58 $  & $ -34 $  & $ -34 $  \\
\hline
J1016$-$5857 & $ -232 $  & $ -86 $  & $ -83 $  & $ -93 $  \\
\hline
J1019$-$5749 & $ -51 $  & $ -47 $  & $ -113 $  & $ -120 $  \\
\hline
J1028$-$5819 & $ -941 $  & $ -887 $  & $ -1240 $  & $ -669 $  \\
\hline
J1048$-$5832 & $ -1255 $  & $ -1058 $  & $ -876 $  & $ -355 $  \\
\hline
J1057$-$5226 & $ -793 $  & $ -3160 $  & $ -785 $  & $ -1571 $  \\
\hline
J1105$-$6107 & $ -386 $  & $ -55 $  & $ -151 $  & $ -102 $  \\
\hline
J1112$-$6103 & $ -135 $  & $ -76 $  & $ -162 $  & $ -154 $  \\
\hline
J1119$-$6127 & $ -642 $  & $ -146 $  & $ -174 $  & $ -179 $  \\
\hline
J1124$-$5916 & $ -262 $  & $ -189 $  & $ -431 $  & $ -303 $  \\
\hline
J1357$-$6429 & $ -475 $  & $ -143 $  & $ -144 $  & $ -135 $  \\
\hline
J1410$-$6132 & $ -70 $  & $ -43 $  & $ -438 $  & $ -444 $  \\
\hline
J1420$-$6048 & $ -319 $  & $ -114 $  & $ -336 $  & $ -373 $  \\
\hline
J1509$-$5850 & $ -233 $  & $ -360 $  & $ -202 $  & $ -242 $  \\
\hline
J1513$-$5908 & $ -287 $  & $ -154 $  & $ -173 $  & $ -166 $  \\
\hline
J1648$-$4611 & $ -413 $  & $ -139 $  & $ -120 $  & $ -124 $  \\
\hline
J1702$-$4128 & $ -318 $  & $ -75 $  & $ -97 $  & $ -118 $  \\
\hline
J1709$-$4429 & $ -6164 $  & $ -10884 $  & $ -5132 $  & $ -6006 $  \\
\hline
J1718$-$3825 & $ -282 $  & $ -260 $  & $ -155 $  & $ -197 $  \\
\hline
J1730$-$3350 & $ -447 $  & $ -60 $  & $ -138 $  & $ -117 $  \\
\hline
J1741$-$2054 & $ -176 $  & $ -939 $  & $ -672 $  & $ -1075 $  \\
\hline
J1747$-$2958 & $ -432 $  & $ -265 $  & $ -280 $  & $ -241 $  \\
\hline
J1801$-$2451 & $ -180 $  & $ -134 $  & $ -188 $  & $ -177 $  \\
\hline
J1833$-$1034 & $ -609 $  & $ -257 $  & $ -146 $  & $ -141 $  \\
\hline
J1835$-$1106 & $ -129 $  & $ -38 $  & $ -24 $  & $ -26 $  \\
\hline
J1952$+$3252 & $ -1433 $  & $ -886 $  & $ -1113 $  & $ -871 $  \\
\hline
J2021$+$3651 & $ -2469 $  & $ -1809 $  & $ -2982 $  & $ -1699 $  \\
\hline
J2030$+$3641 & $ -239 $  & $ -227 $  & $ -143 $  & $ -167 $  \\
\hline
J2032$+$4127 & $ -663 $  & $ -569 $  & $ -1163 $  & $ -783 $  \\
\hline
J2043$+$2740 & $ -112 $  & $ -83 $  & $ -68 $  & $ -54 $  \\
\hline
J2229$+$6114 & $ -925 $  & $ -1602 $  & $ -1062 $  & $ -1319 $  \\
\hline
J2240$+$5832 & $ -140 $  & $ -41 $  & $ -53 $  & $ -29 $  \\
\hline
\end{tabular}
\caption{Best fit log-likelihood values resulting from the $\gamma$-ray fit of the 41 RL pulsars of the analysed sample.}
\label{JointLike}
\end{table*}
 \clearpage

\begin{figure}[htbp!]
\centering
\includegraphics[width=0.9\textwidth]{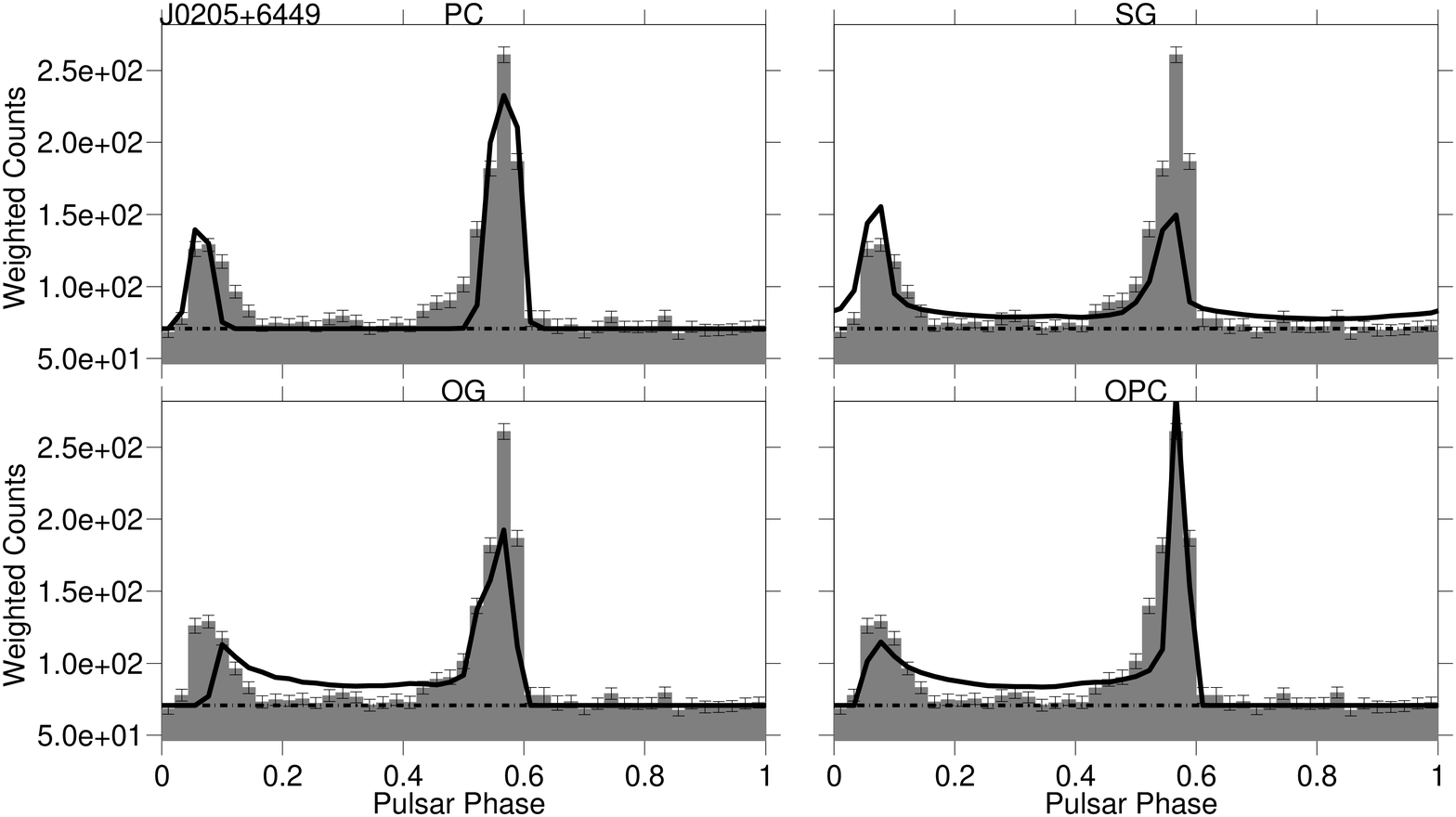}
\includegraphics[width=0.9\textwidth]{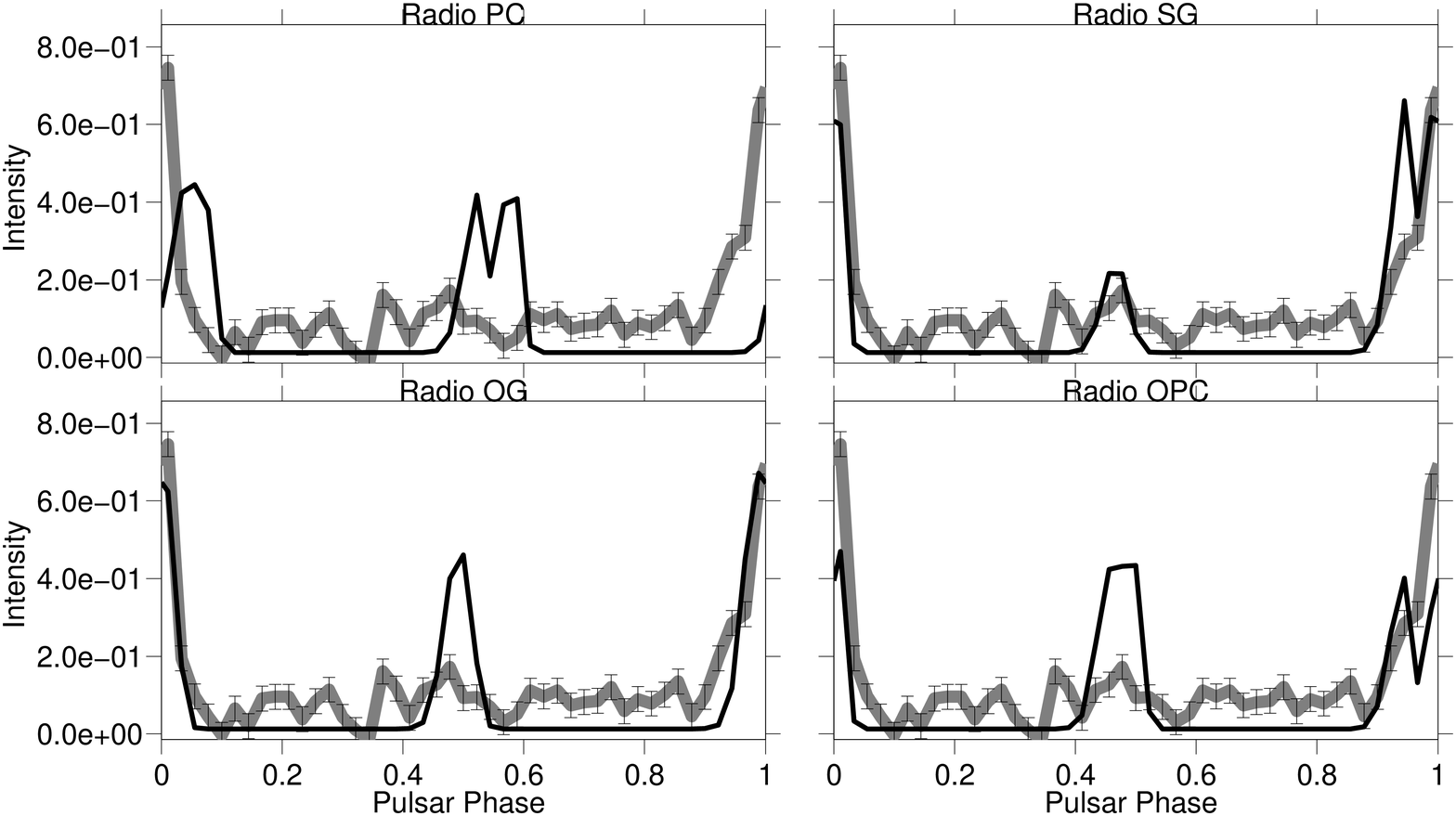}
\caption{PSR J0205+6449. \emph{Top}: for each model the best joint fit solution $\gamma$-ray light-curve (thick black line) is superimposed on the LAT pulsar $\gamma$-ray light-curve (shaded histogram). The estimated background is indicated by the dash-dot line. \emph{Bottom}: for each model the best joint fit solution radio light-curve (black line) is  is superimposed on the LAT pulsar radio light-curve (grey thick line).  The radio model is unique, but the $(\alpha,\zeta)$ solutions vary for each $\gamma$-ray model.}
\label{fitJoint_GmR1}
\end{figure}
  
\clearpage
\begin{figure}[htbp!]
\centering
\includegraphics[width=0.9\textwidth]{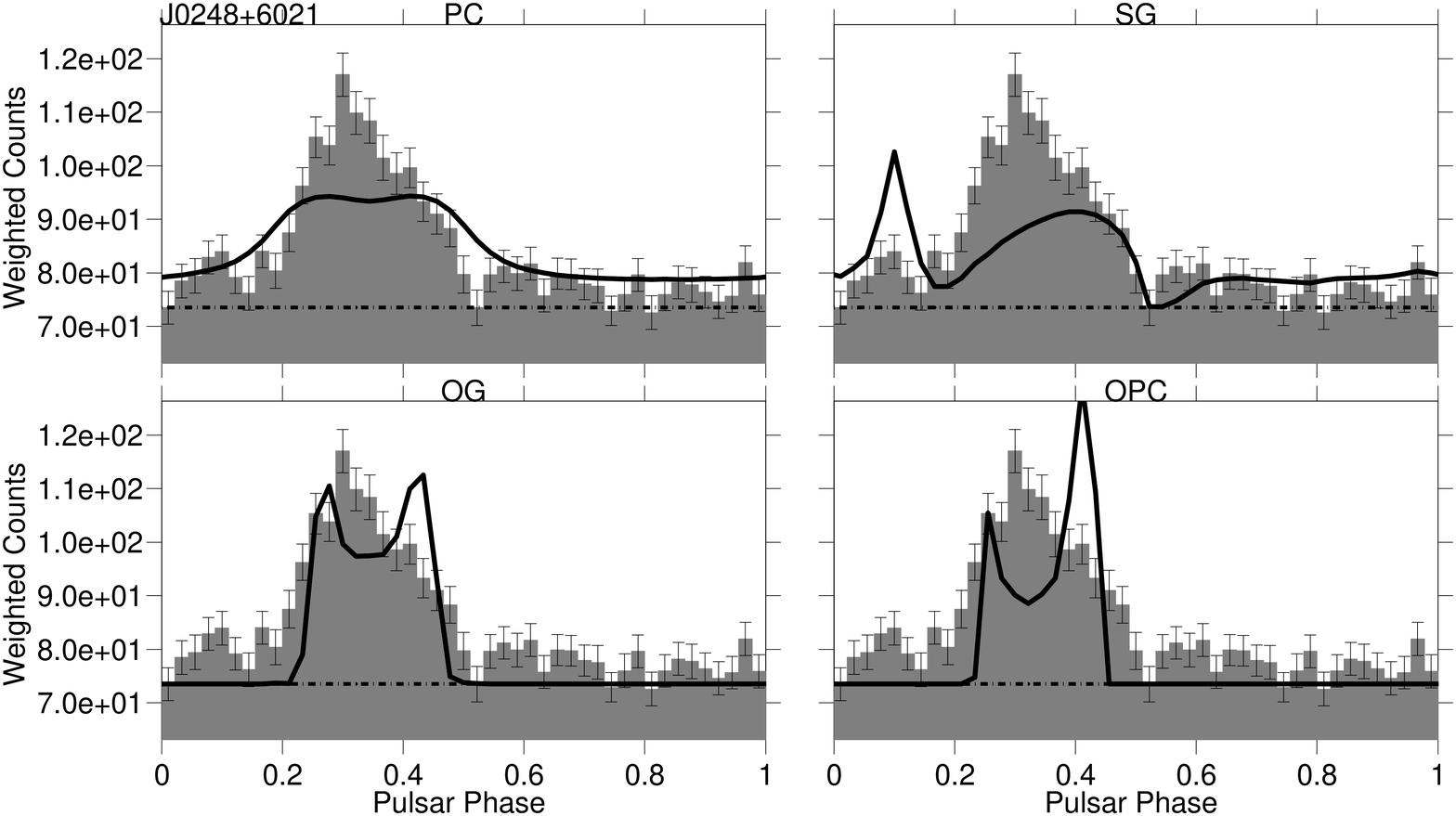}
\includegraphics[width=0.9\textwidth]{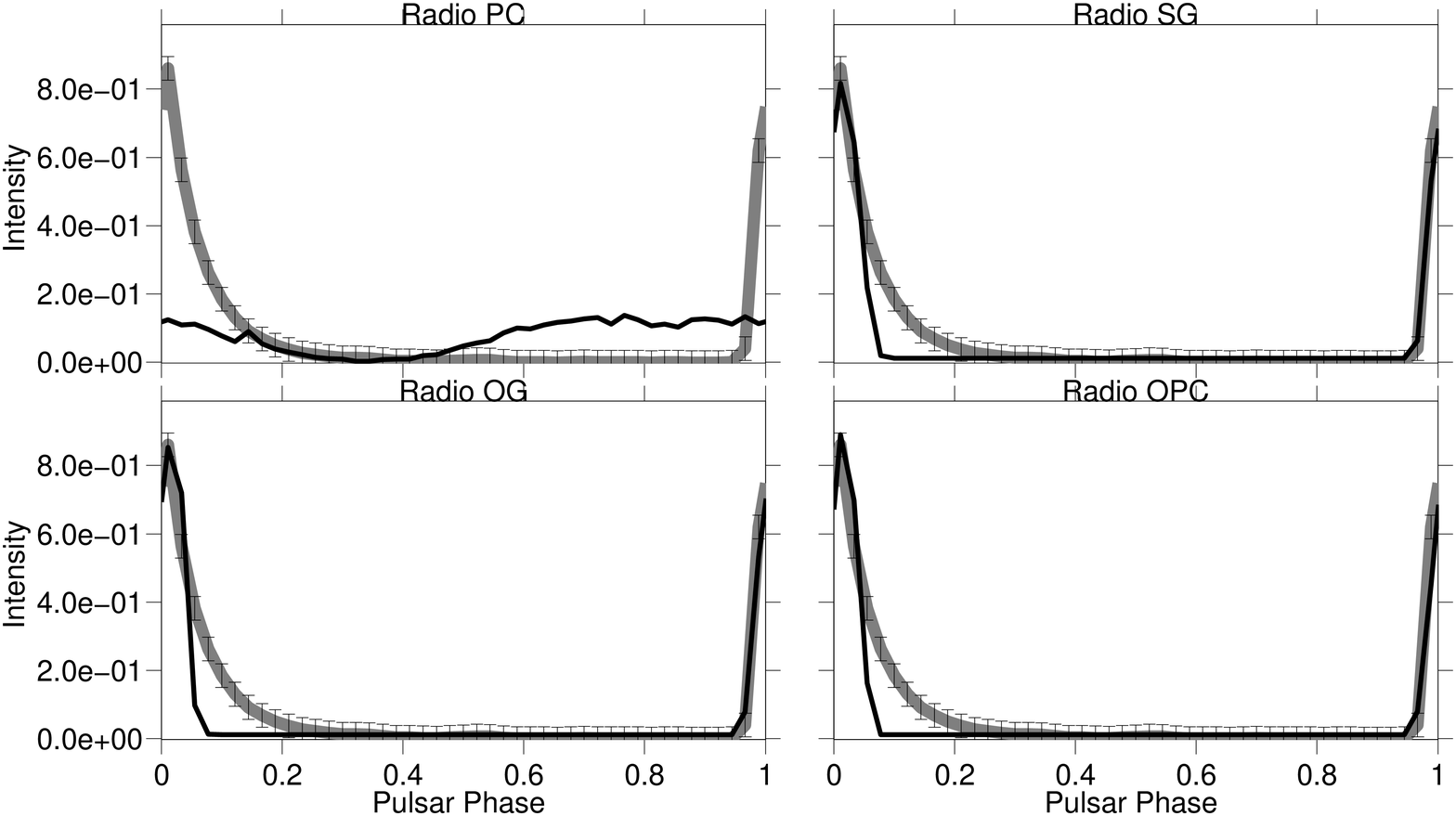}
\caption{PSR J0248+6021. \emph{Top}: for each model the best joint fit solution $\gamma$-ray light-curve (thick black line) is superimposed on the LAT pulsar $\gamma$-ray light-curve (shaded histogram). The estimated background is indicated by the dash-dot line. \emph{Bottom}: for each model the best joint fit solution radio light-curve (black line) is  is superimposed on the LAT pulsar radio light-curve (grey thick line).  The radio model is unique, but the $(\alpha,\zeta)$ solutions vary for each $\gamma$-ray model.}
\label{fitJoint_GmR2}
\end{figure}
  
\clearpage
\begin{figure}[htbp!]
\centering
\includegraphics[width=0.9\textwidth]{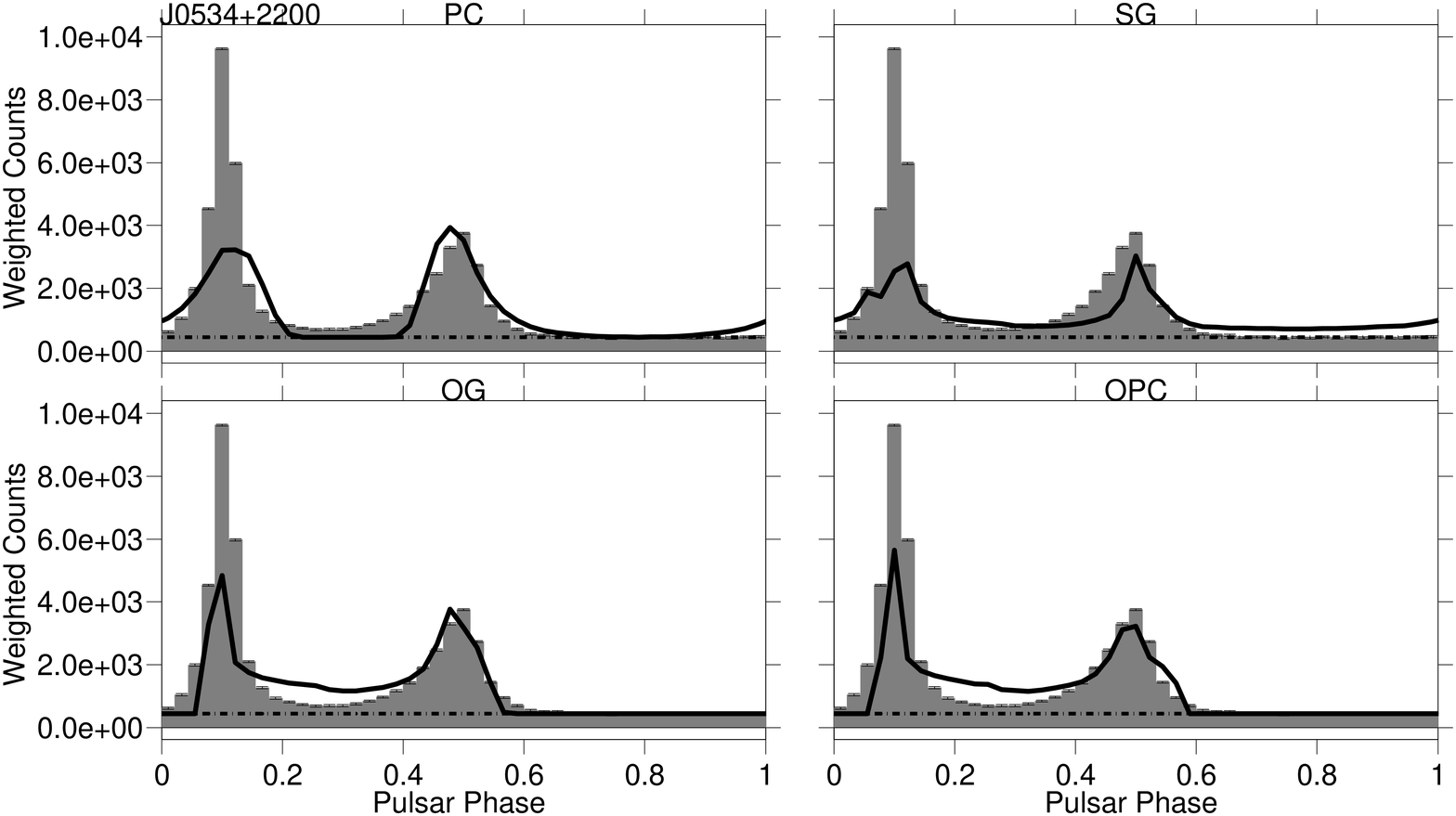}
\includegraphics[width=0.9\textwidth]{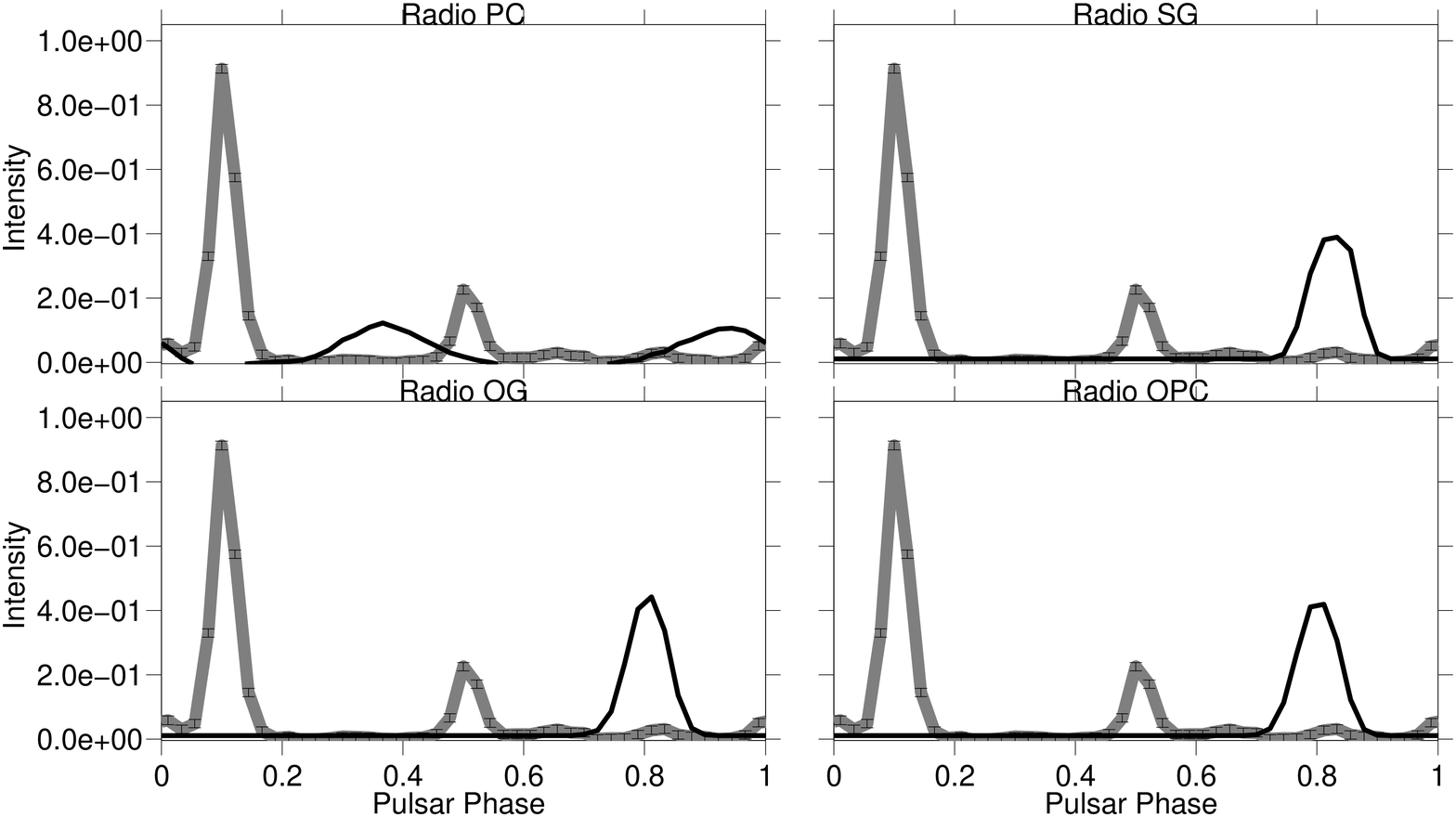}
\caption{PSR J0534+2200. \emph{Top}: for each model the best joint fit solution $\gamma$-ray light-curve (thick black line) is superimposed on the LAT pulsar $\gamma$-ray light-curve (shaded histogram). The estimated background is indicated by the dash-dot line. \emph{Bottom}: for each model the best joint fit solution radio light-curve (black line) is  is superimposed on the LAT pulsar radio light-curve (grey thick line).  The radio model is unique, but the $(\alpha,\zeta)$ solutions vary for each $\gamma$-ray model. See Section \ref{Data} for a discussion on why we decided to show the joint $\gamma$-ray plus Radio fit result for the 
Crab pulsar.}
\label{fitJoint_GmR3}
\end{figure}
  
\clearpage
\begin{figure}[htbp!]
\centering
\includegraphics[width=0.9\textwidth]{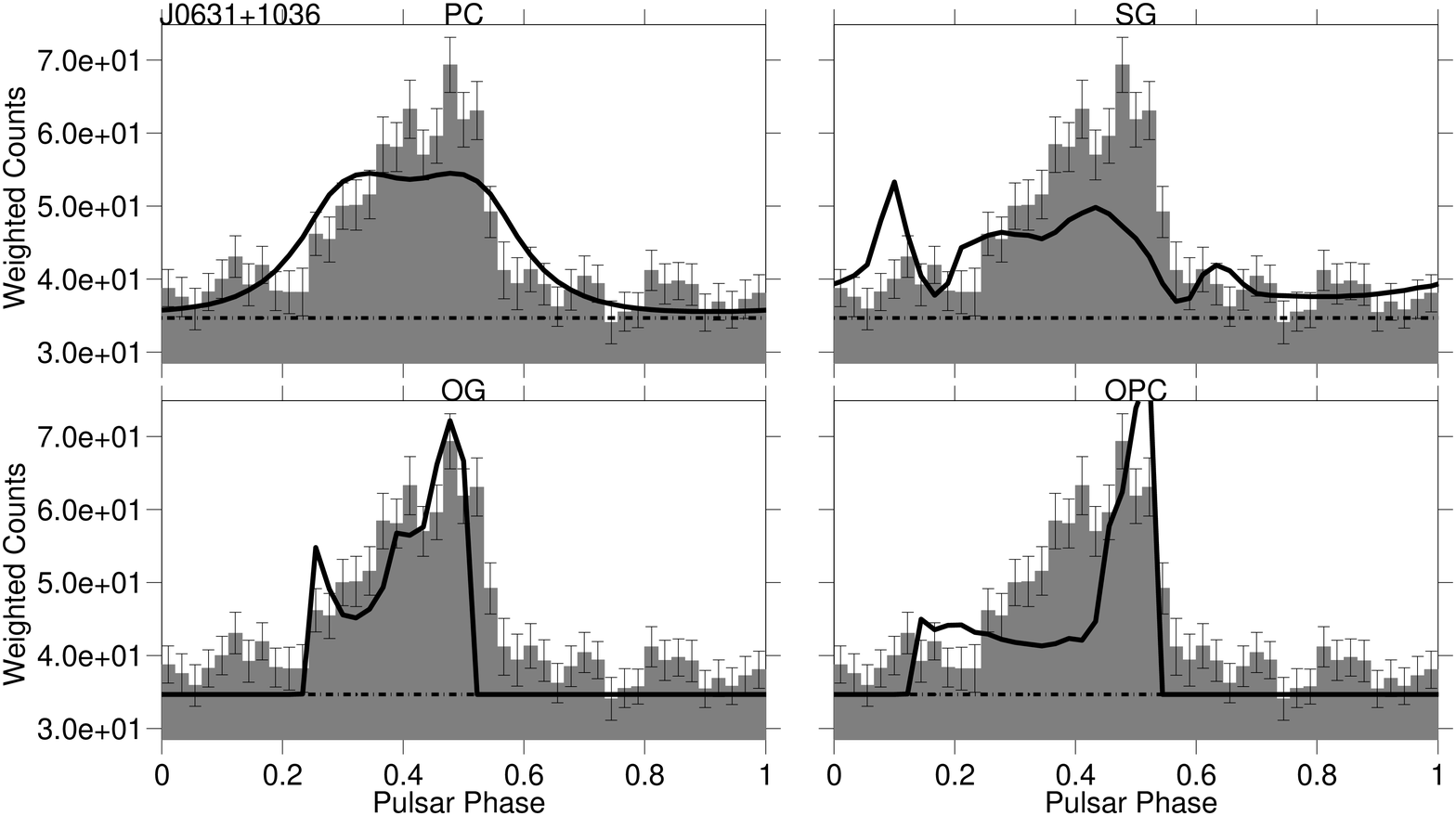}
\includegraphics[width=0.9\textwidth]{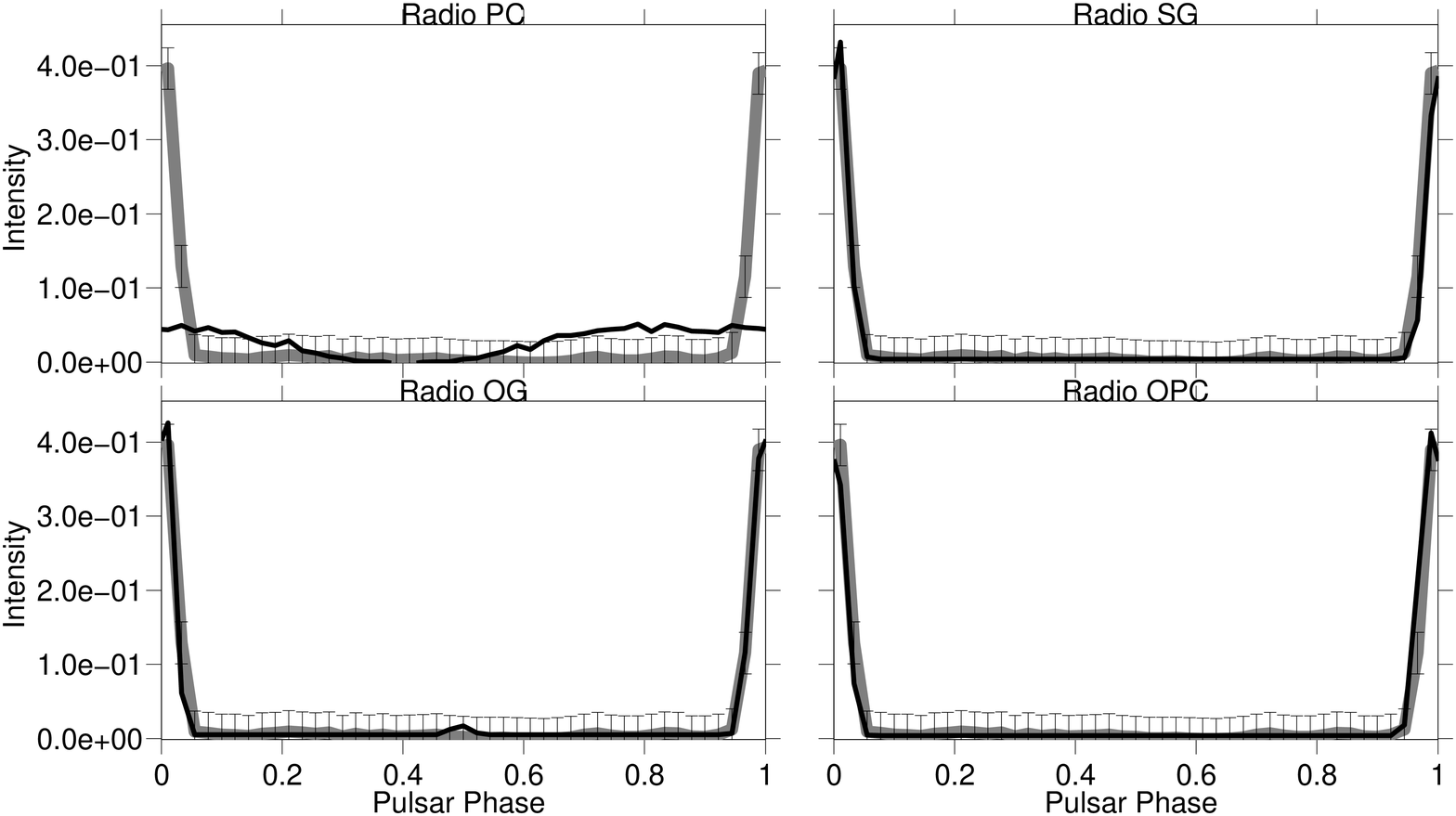}
\caption{PSR J0631+1036. \emph{Top}: for each model the best joint fit solution $\gamma$-ray light-curve (thick black line) is superimposed on the LAT pulsar $\gamma$-ray light-curve (shaded histogram). The estimated background is indicated by the dash-dot line. \emph{Bottom}: for each model the best joint fit solution radio light-curve (black line) is  is superimposed on the LAT pulsar radio light-curve (grey thick line).  The radio model is unique, but the $(\alpha,\zeta)$ solutions vary for each $\gamma$-ray model.}
\label{fitJoint_GmR4}
\end{figure}
  
\clearpage
\begin{figure}[htbp!]
\centering
\includegraphics[width=0.9\textwidth]{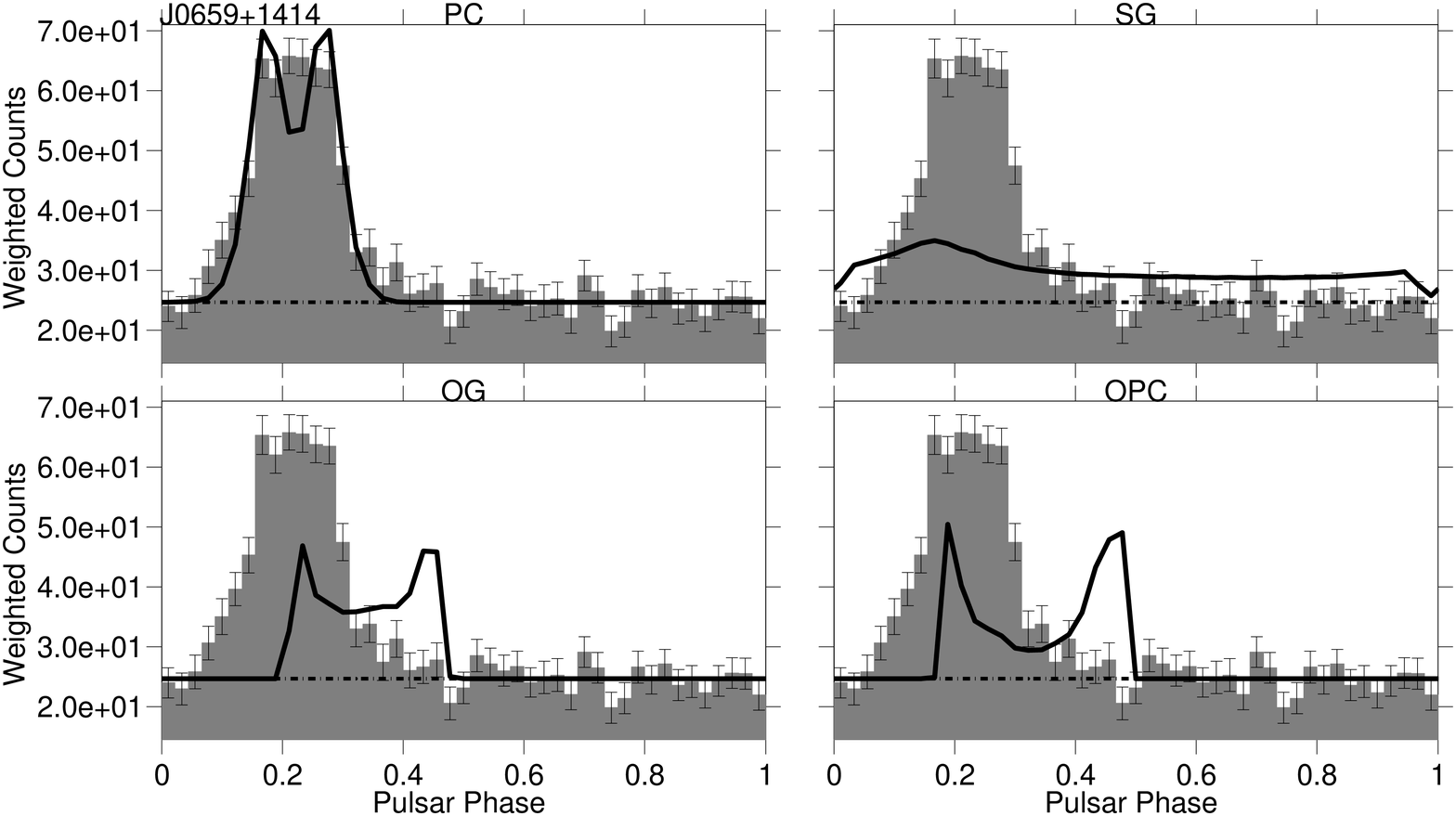}
\includegraphics[width=0.9\textwidth]{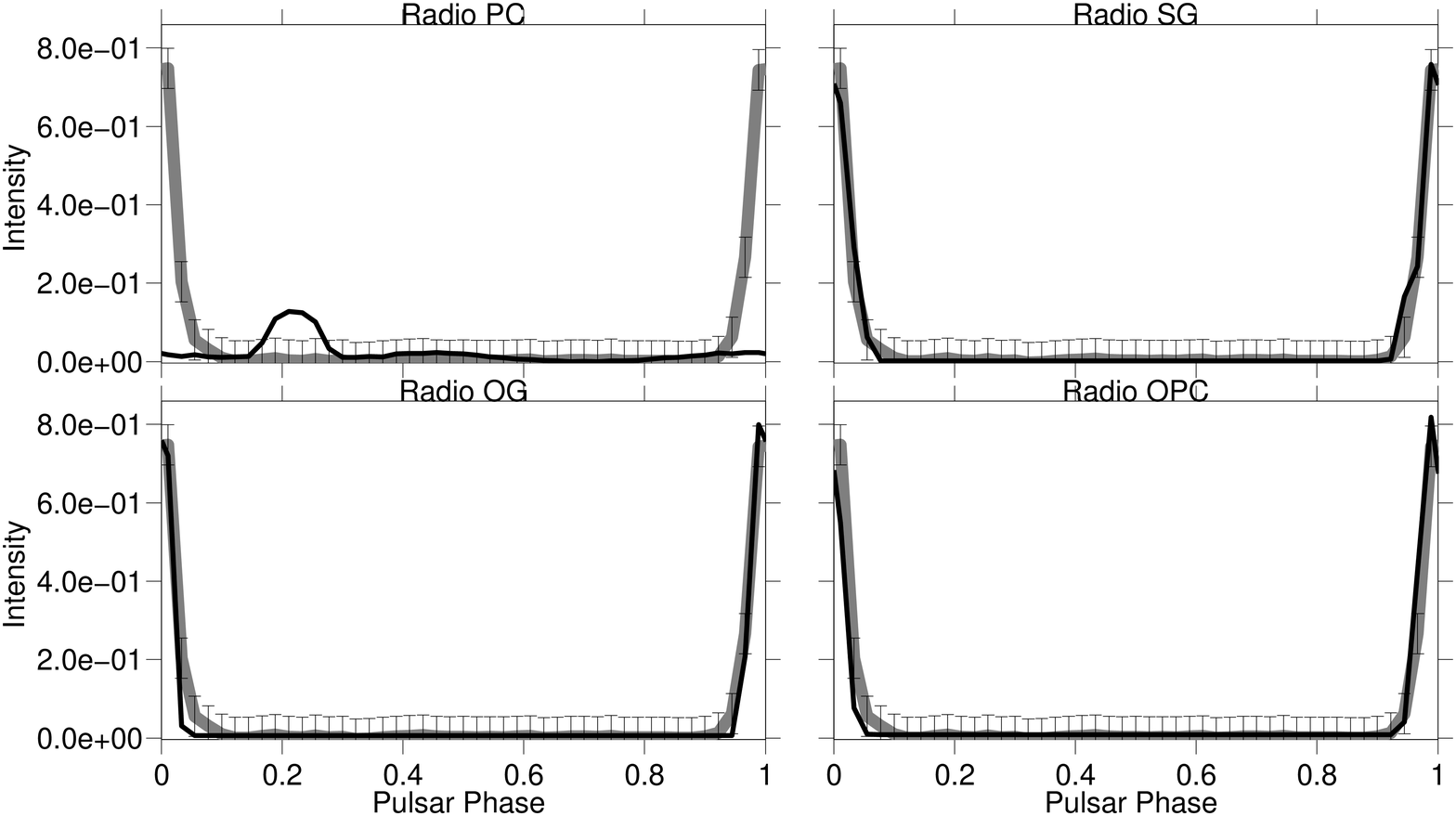}
\caption{PSR J0659+1414. \emph{Top}: for each model the best joint fit solution $\gamma$-ray light-curve (thick black line) is superimposed on the LAT pulsar $\gamma$-ray light-curve (shaded histogram). The estimated background is indicated by the dash-dot line. \emph{Bottom}: for each model the best joint fit solution radio light-curve (black line) is  is superimposed on the LAT pulsar radio light-curve (grey thick line).  The radio model is unique, but the $(\alpha,\zeta)$ solutions vary for each $\gamma$-ray model.}
\label{fitJoint_GmR5}
\end{figure}
  
\clearpage
\begin{figure}[htbp!]
\centering
\includegraphics[width=0.9\textwidth]{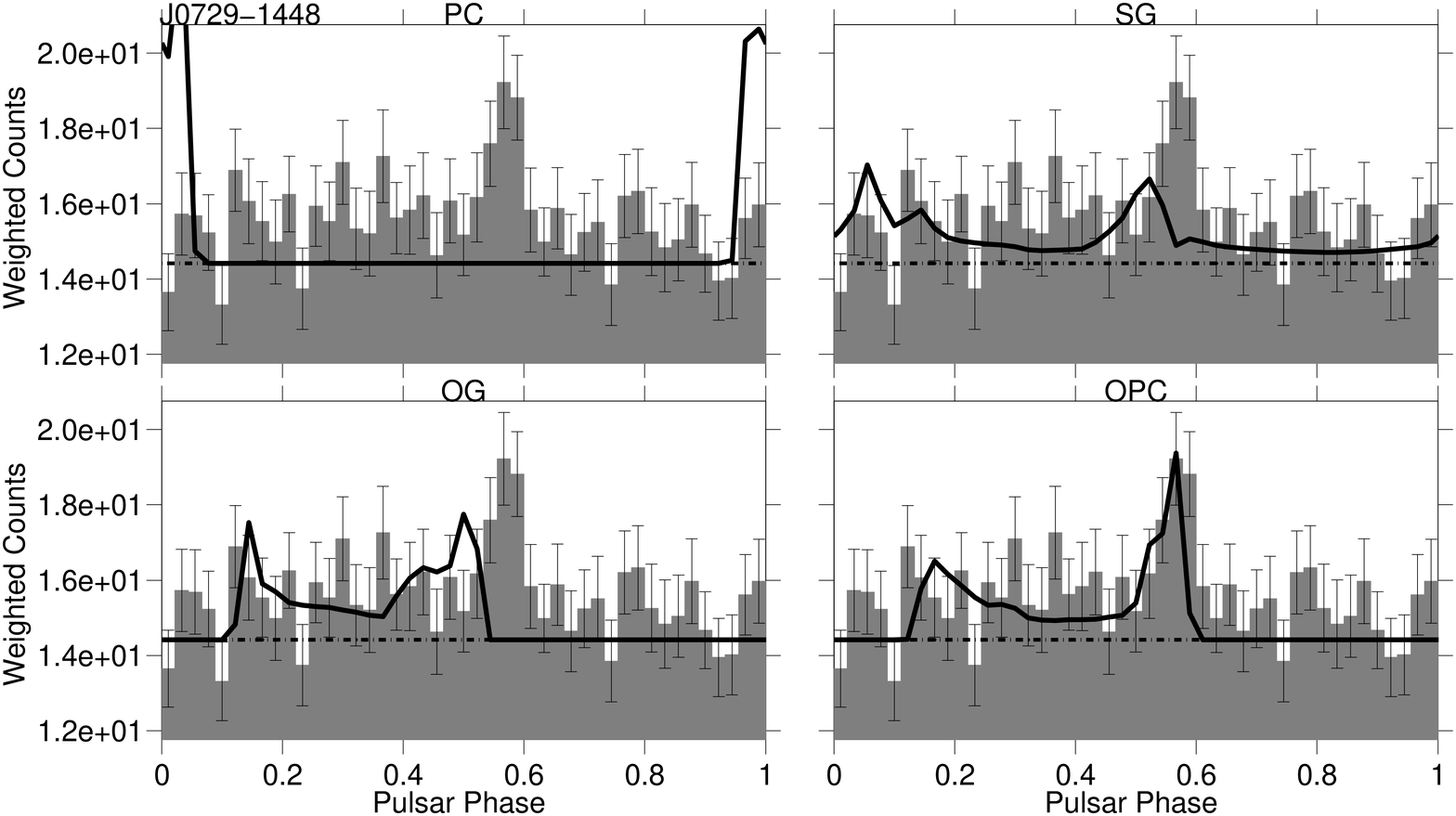}
\includegraphics[width=0.9\textwidth]{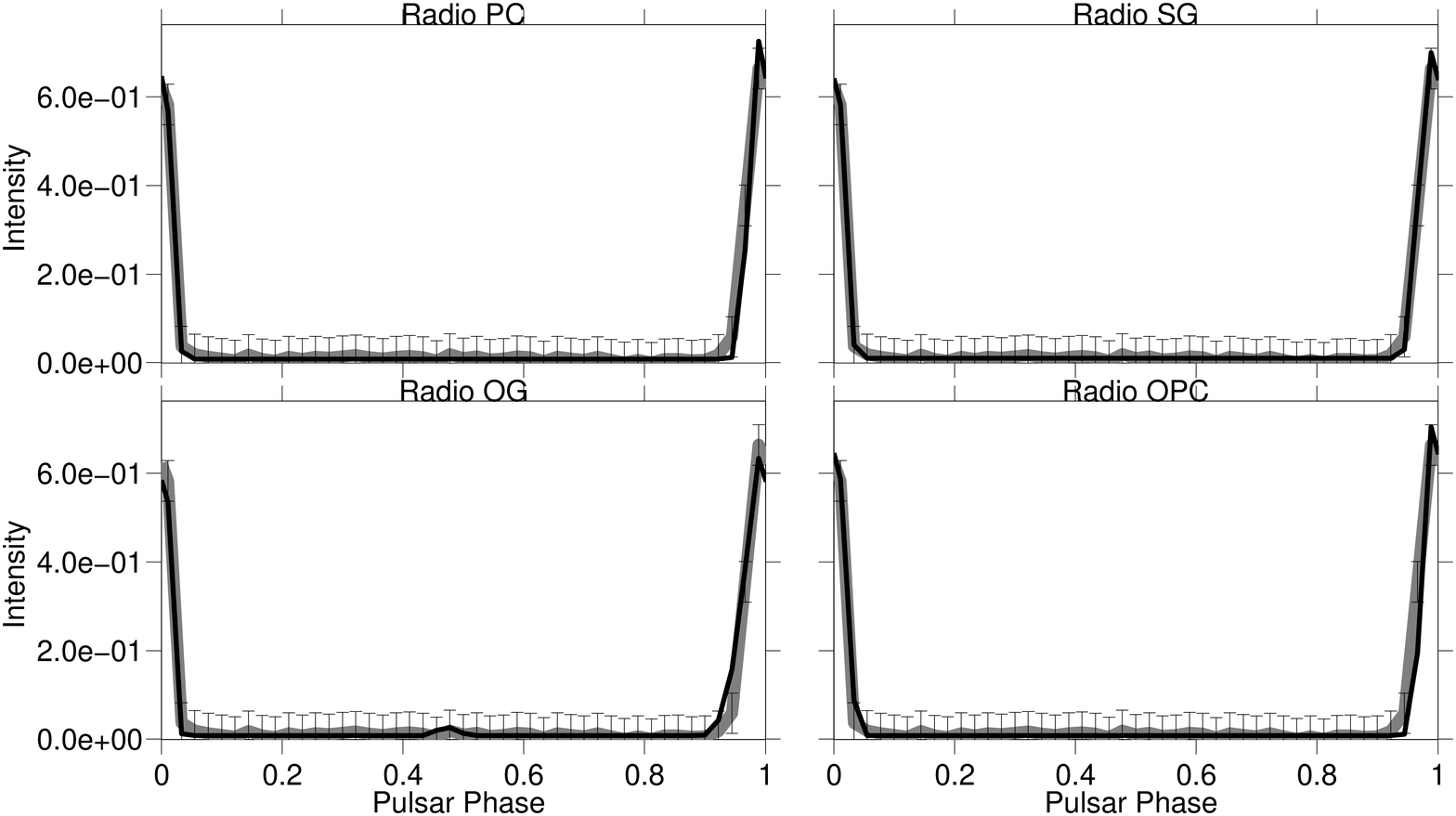}
\caption{PSR J0729-1448. \emph{Top}: for each model the best joint fit solution $\gamma$-ray light-curve (thick black line) is superimposed on the LAT pulsar $\gamma$-ray light-curve (shaded histogram). The estimated background is indicated by the dash-dot line. \emph{Bottom}: for each model the best joint fit solution radio light-curve (black line) is  is superimposed on the LAT pulsar radio light-curve (grey thick line).  The radio model is unique, but the $(\alpha,\zeta)$ solutions vary for each $\gamma$-ray model.}
\label{fitJoint_GmR6}
\end{figure}
  
\clearpage
\begin{figure}[htbp!]
\centering
\includegraphics[width=0.9\textwidth]{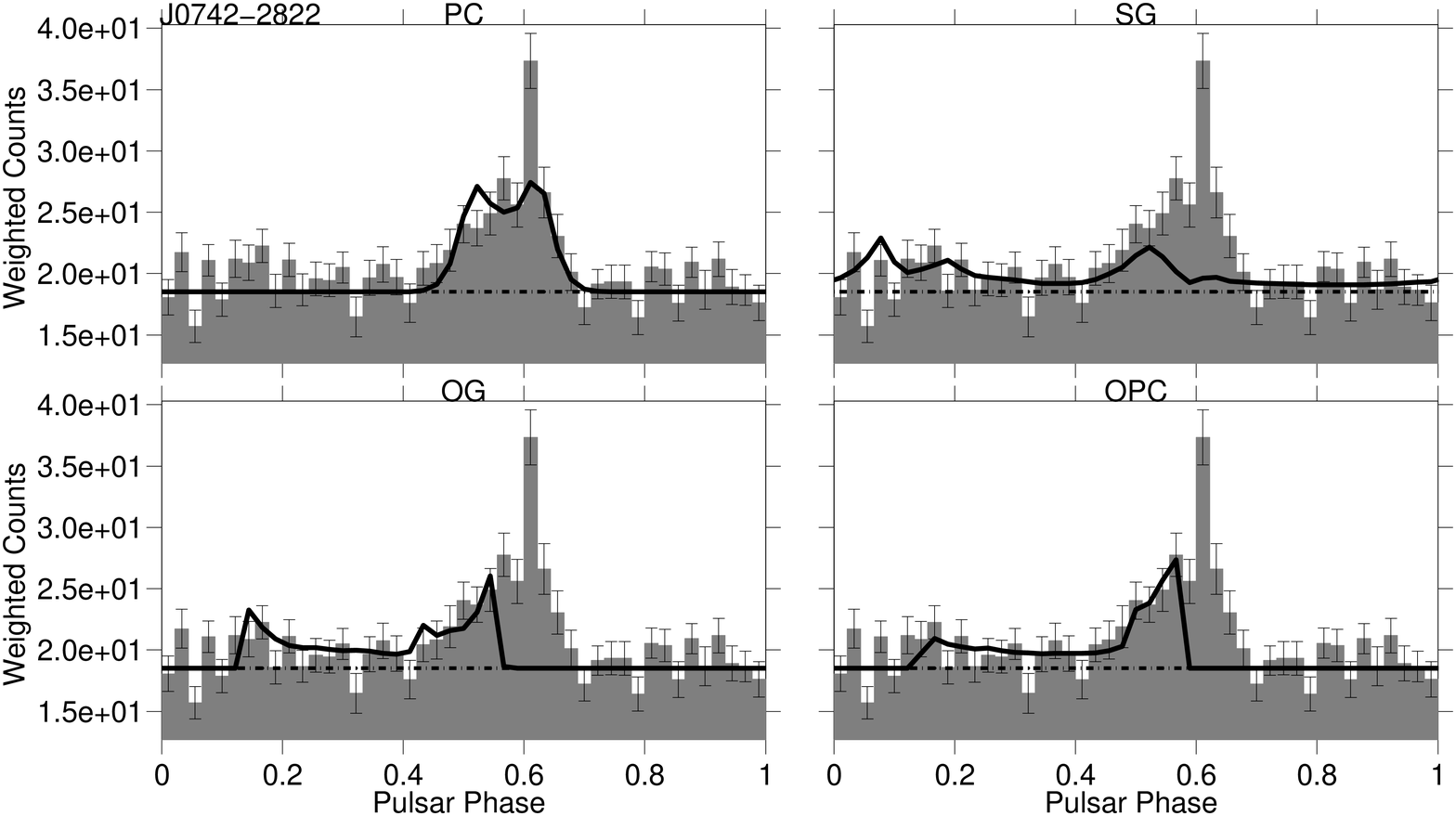}
\includegraphics[width=0.9\textwidth]{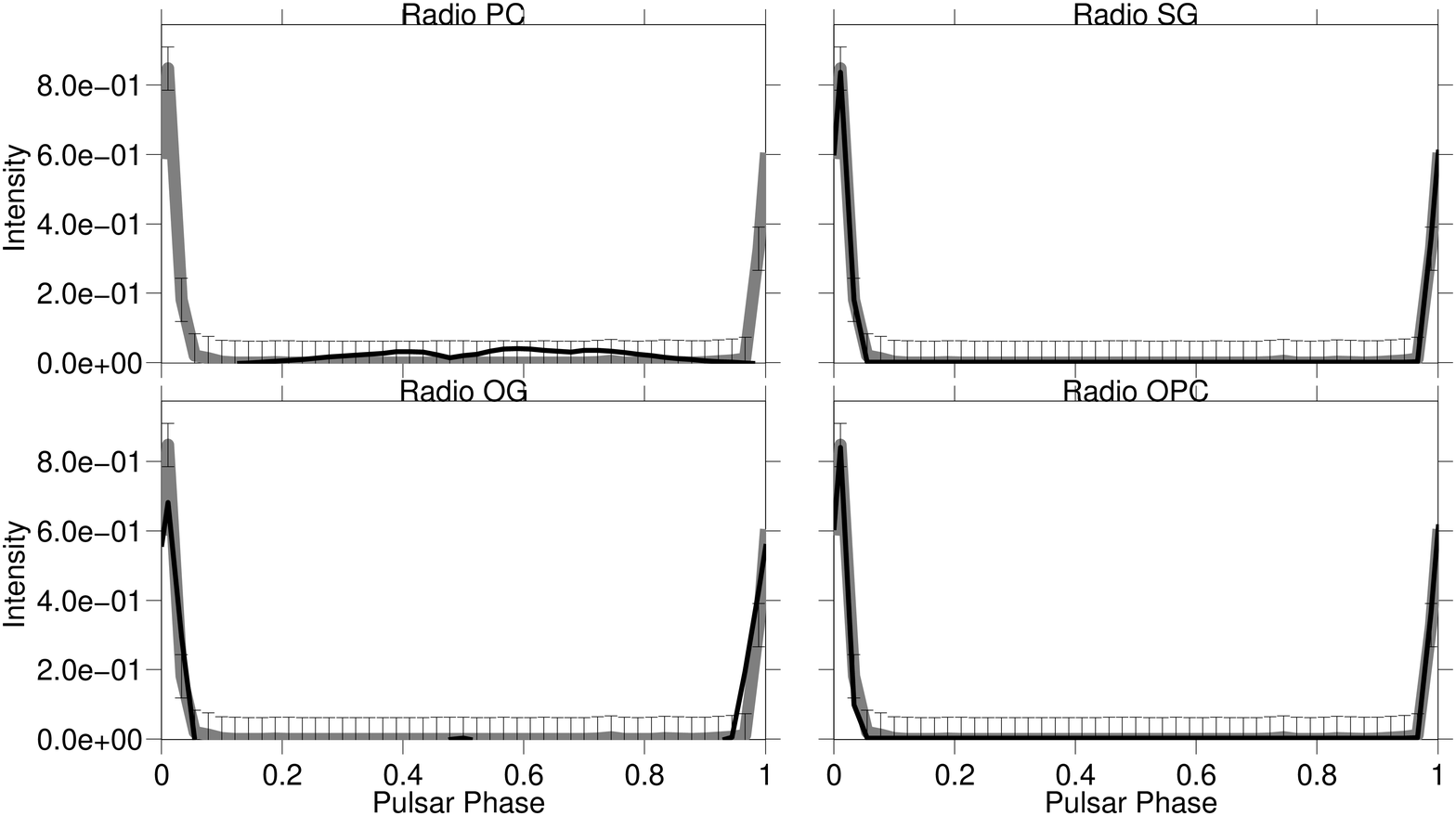}
\caption{PSR J0742-2822. \emph{Top}: for each model the best joint fit solution $\gamma$-ray light-curve (thick black line) is superimposed on the LAT pulsar $\gamma$-ray light-curve (shaded histogram). The estimated background is indicated by the dash-dot line. \emph{Bottom}: for each model the best joint fit solution radio light-curve (black line) is  is superimposed on the LAT pulsar radio light-curve (grey thick line).  The radio model is unique, but the $(\alpha,\zeta)$ solutions vary for each $\gamma$-ray model.}
\label{fitJoint_GmR7}
\end{figure}
  
\clearpage
\begin{figure}[htbp!]
\centering
\includegraphics[width=0.9\textwidth]{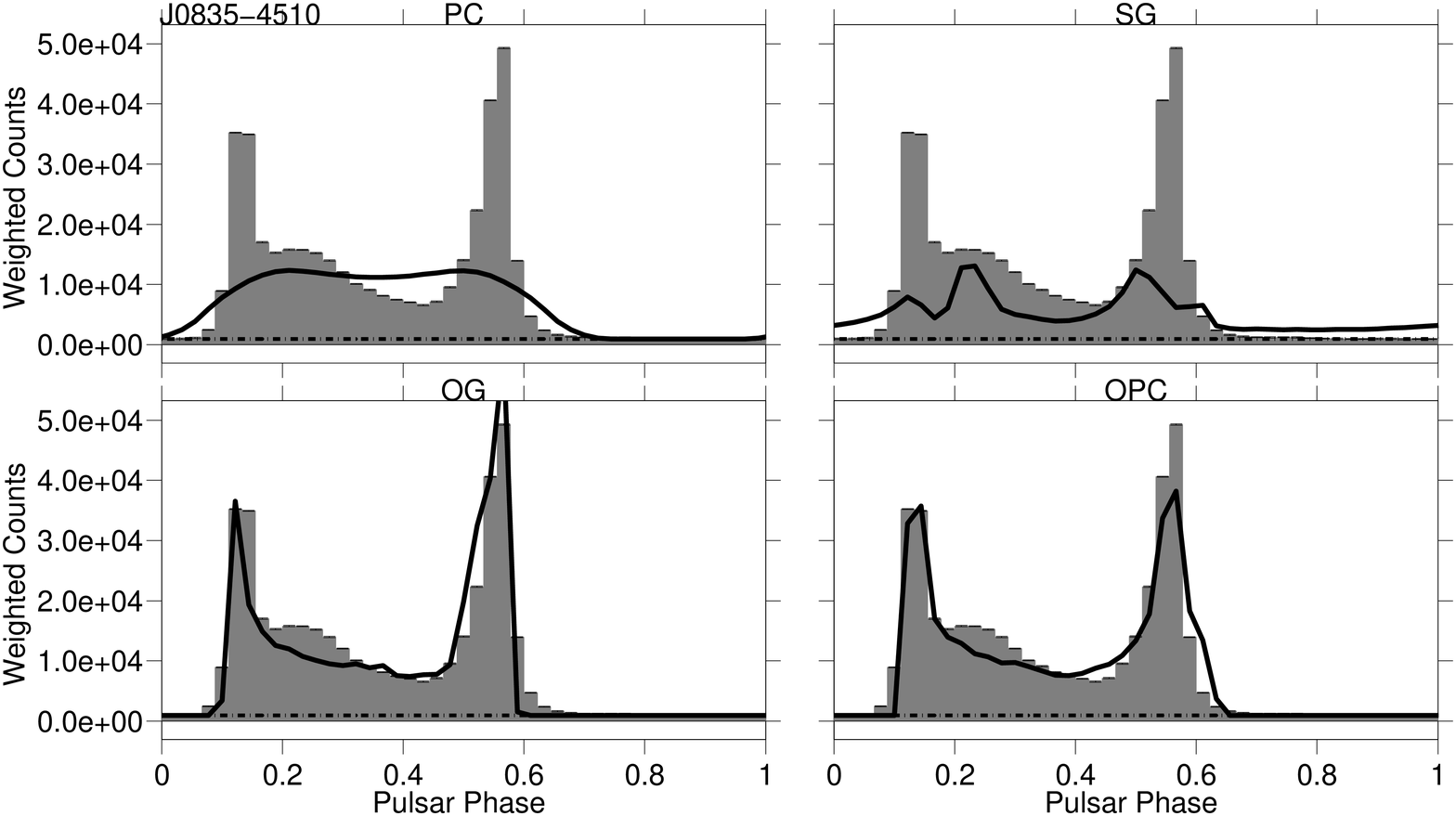}
\includegraphics[width=0.9\textwidth]{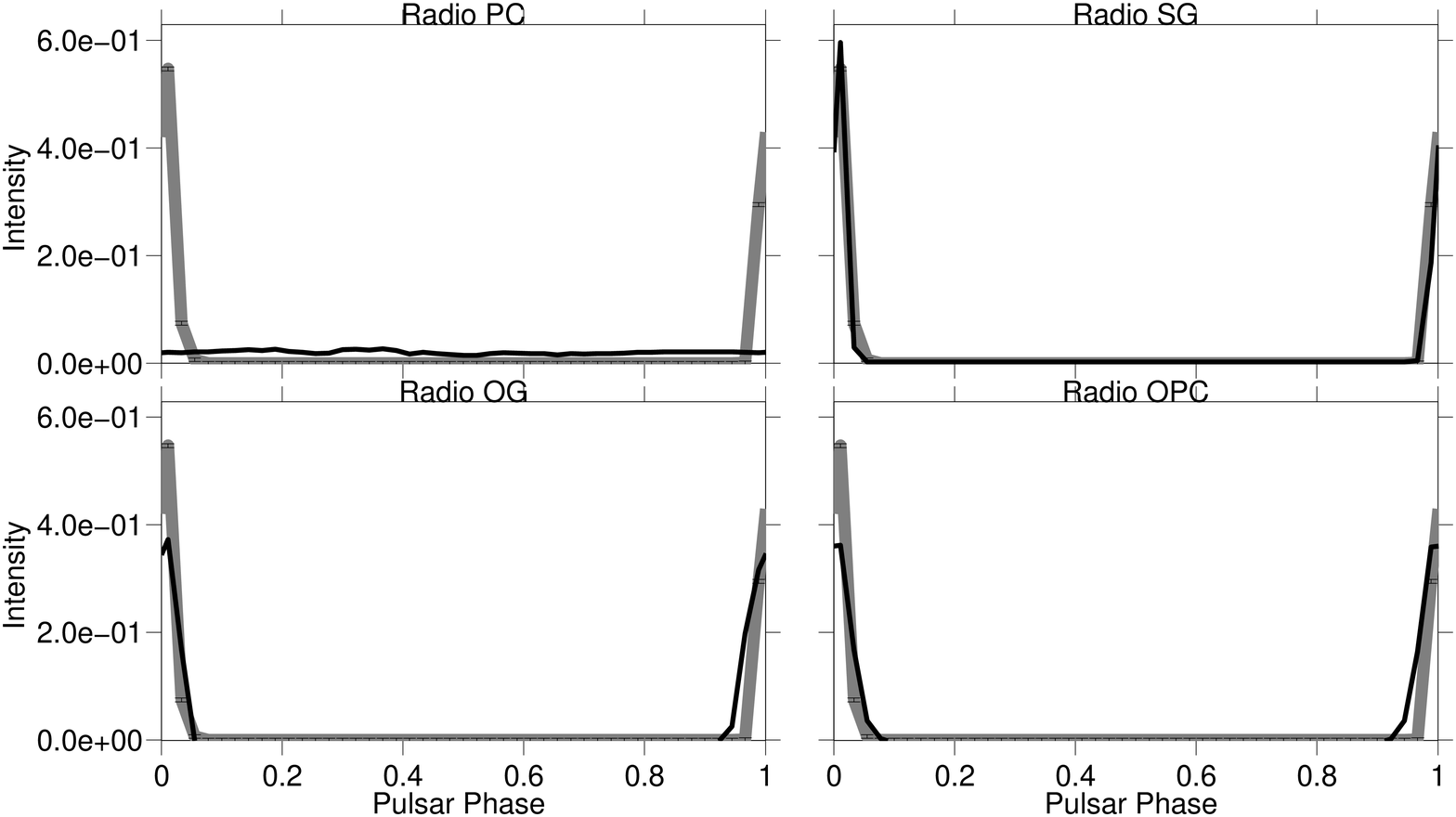}
\caption{PSR J0835-4510. \emph{Top}: for each model the best joint fit solution $\gamma$-ray light-curve (thick black line) is superimposed on the LAT pulsar $\gamma$-ray light-curve (shaded histogram). The estimated background is indicated by the dash-dot line. \emph{Bottom}: for each model the best joint fit solution radio light-curve (black line) is  is superimposed on the LAT pulsar radio light-curve (grey thick line).  The radio model is unique, but the $(\alpha,\zeta)$ solutions vary for each $\gamma$-ray model.}
\label{fitJoint_GmR8}
\end{figure}
  
\clearpage
\begin{figure}[htbp!]
\centering
\includegraphics[width=0.9\textwidth]{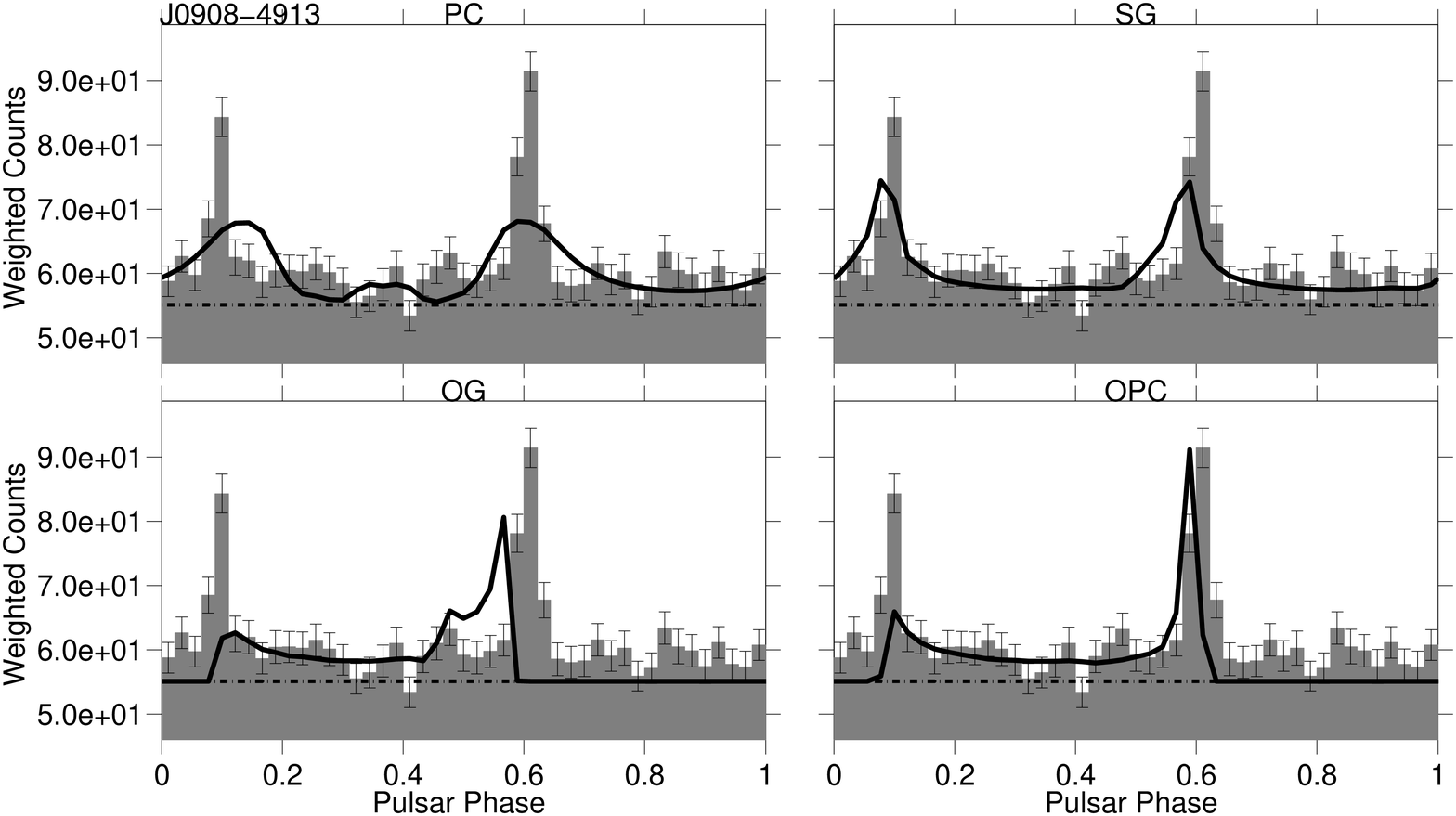}
\includegraphics[width=0.9\textwidth]{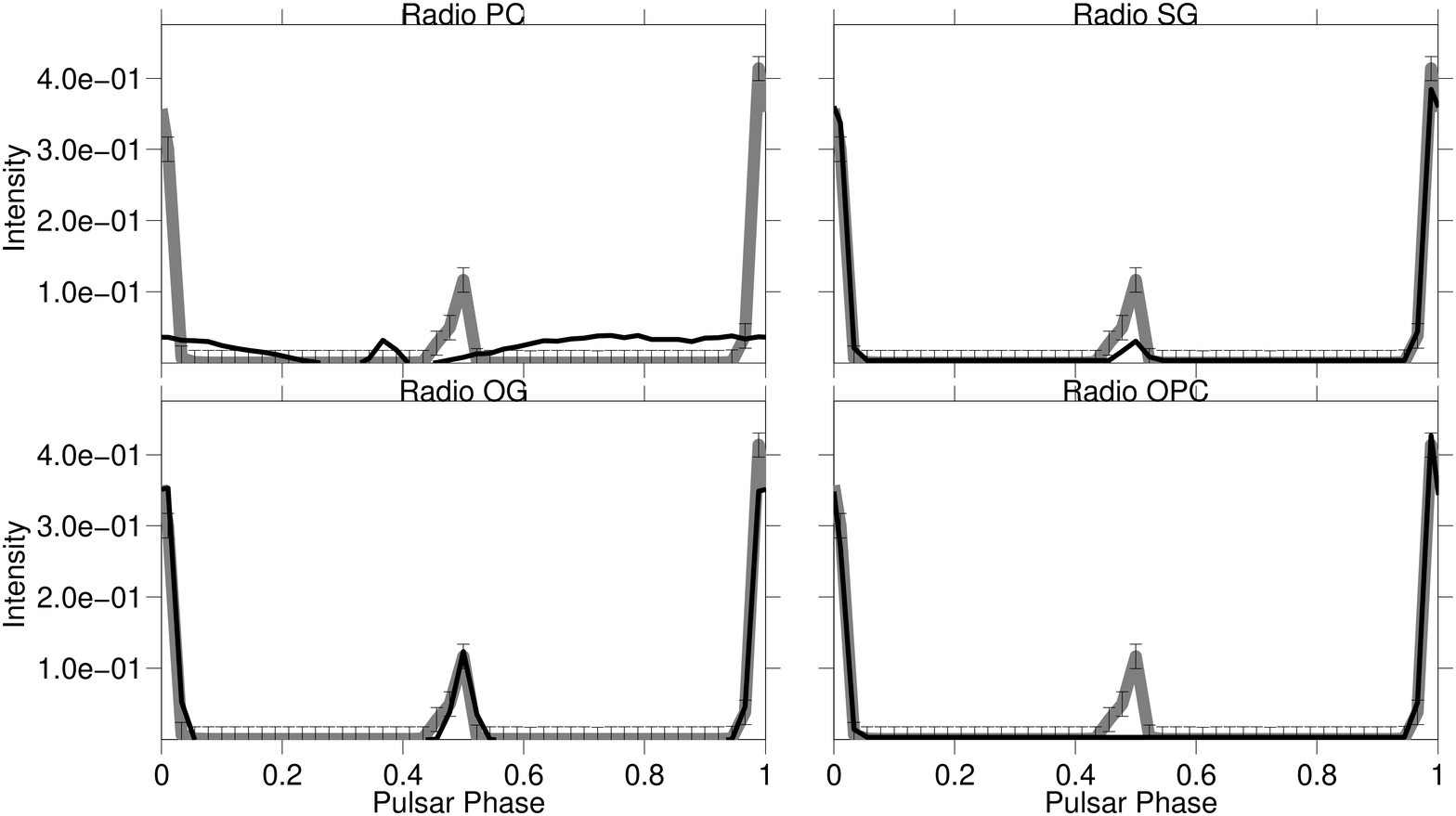}
\caption{PSR J0908-4913. \emph{Top}: for each model the best joint fit solution $\gamma$-ray light-curve (thick black line) is superimposed on the LAT pulsar $\gamma$-ray light-curve (shaded histogram). The estimated background is indicated by the dash-dot line. \emph{Bottom}: for each model the best joint fit solution radio light-curve (black line) is  is superimposed on the LAT pulsar radio light-curve (grey thick line).  The radio model is unique, but the $(\alpha,\zeta)$ solutions vary for each $\gamma$-ray model.}
\label{fitJoint_GmR9}
\end{figure}
  
\clearpage
\begin{figure}[htbp!]
\centering
\includegraphics[width=0.9\textwidth]{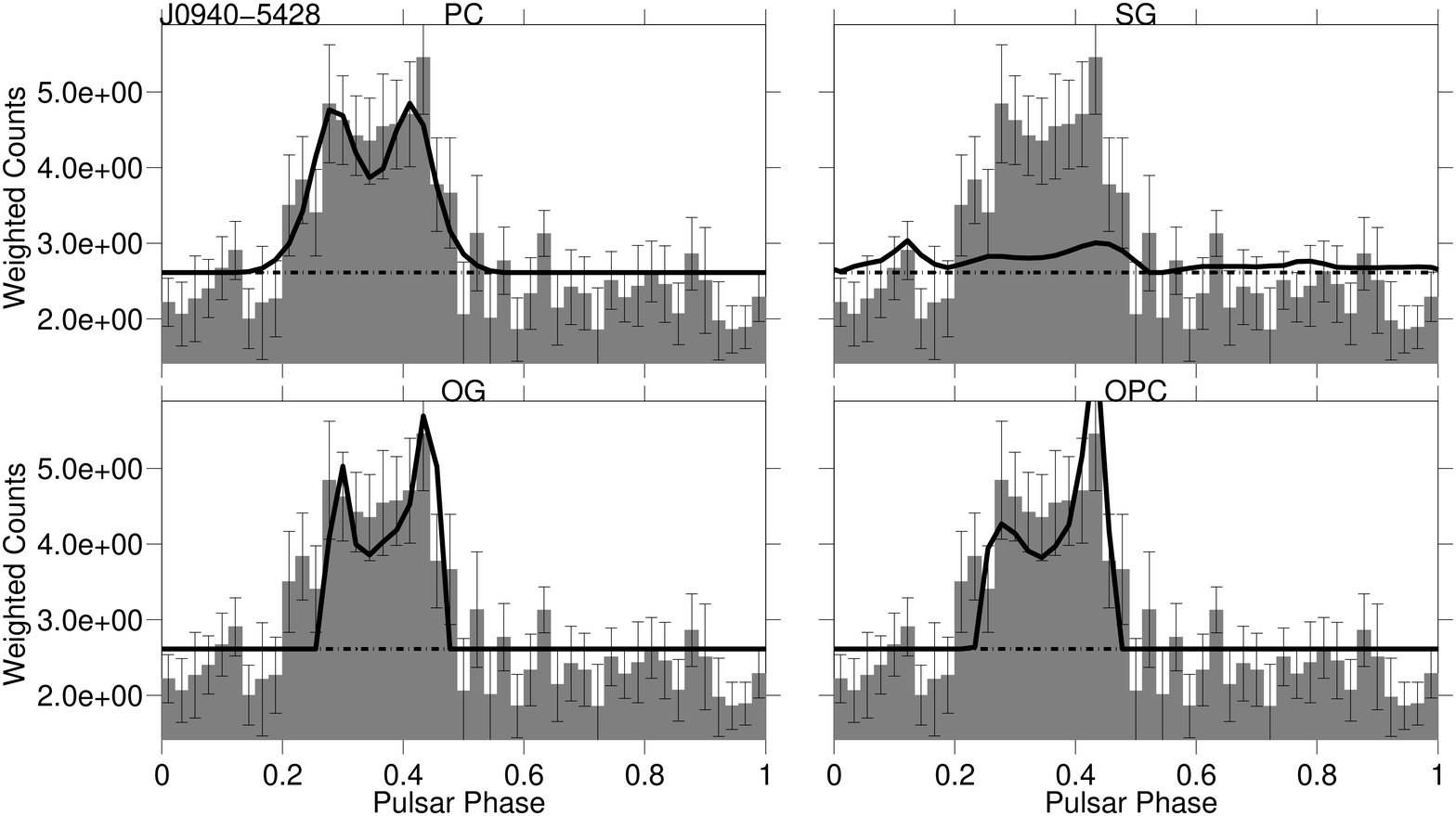}
\includegraphics[width=0.9\textwidth]{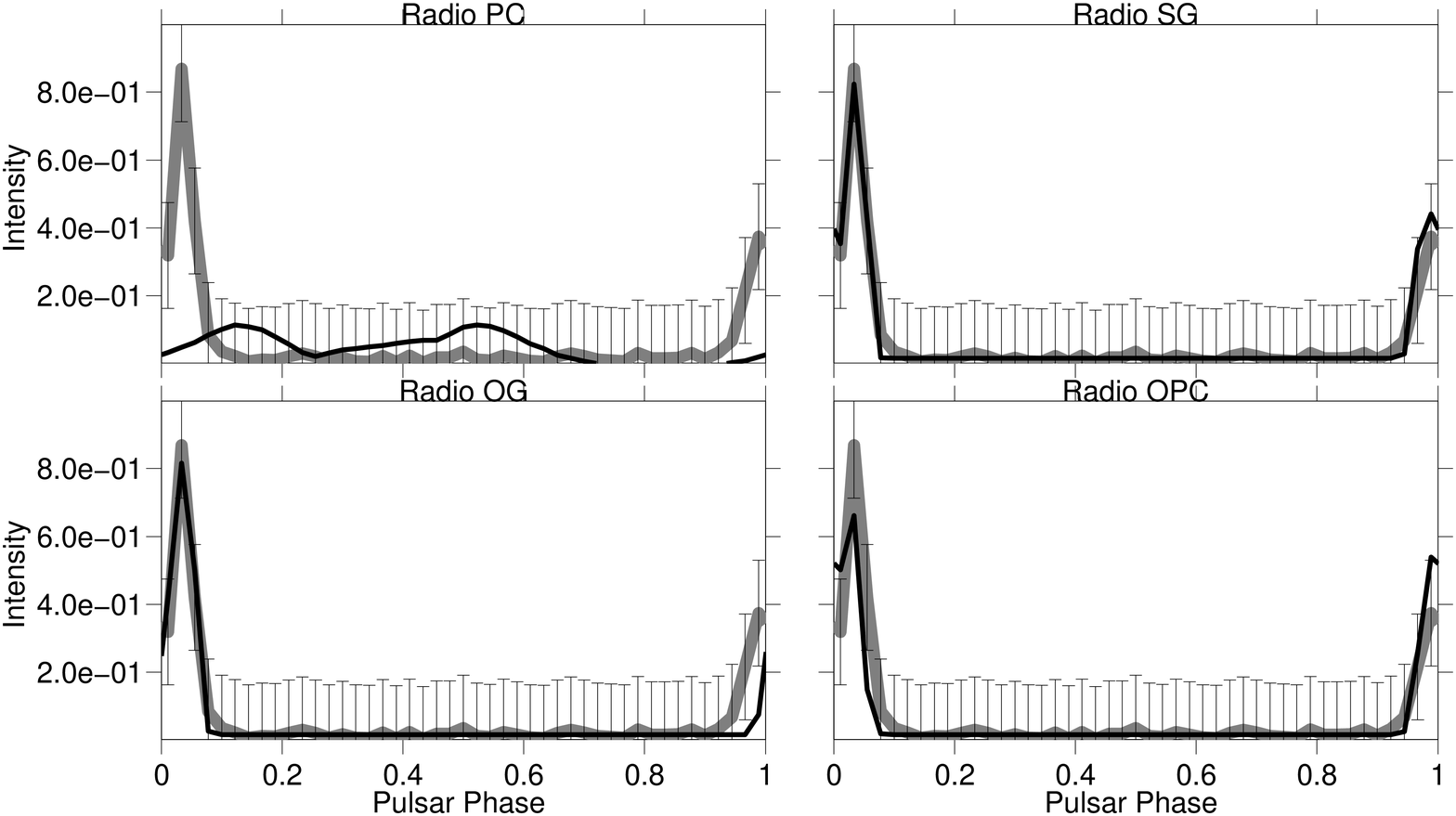}
\caption{PSR J0940-5428. \emph{Top}: for each model the best joint fit solution $\gamma$-ray light-curve (thick black line) is superimposed on the LAT pulsar $\gamma$-ray light-curve (shaded histogram). The estimated background is indicated by the dash-dot line. \emph{Bottom}: for each model the best joint fit solution radio light-curve (black line) is  is superimposed on the LAT pulsar radio light-curve (grey thick line).  The radio model is unique, but the $(\alpha,\zeta)$ solutions vary for each $\gamma$-ray model.}
\label{fitJoint_GmR10}
\end{figure}
  
\clearpage
\begin{figure}[htbp!]
\centering
\includegraphics[width=0.9\textwidth]{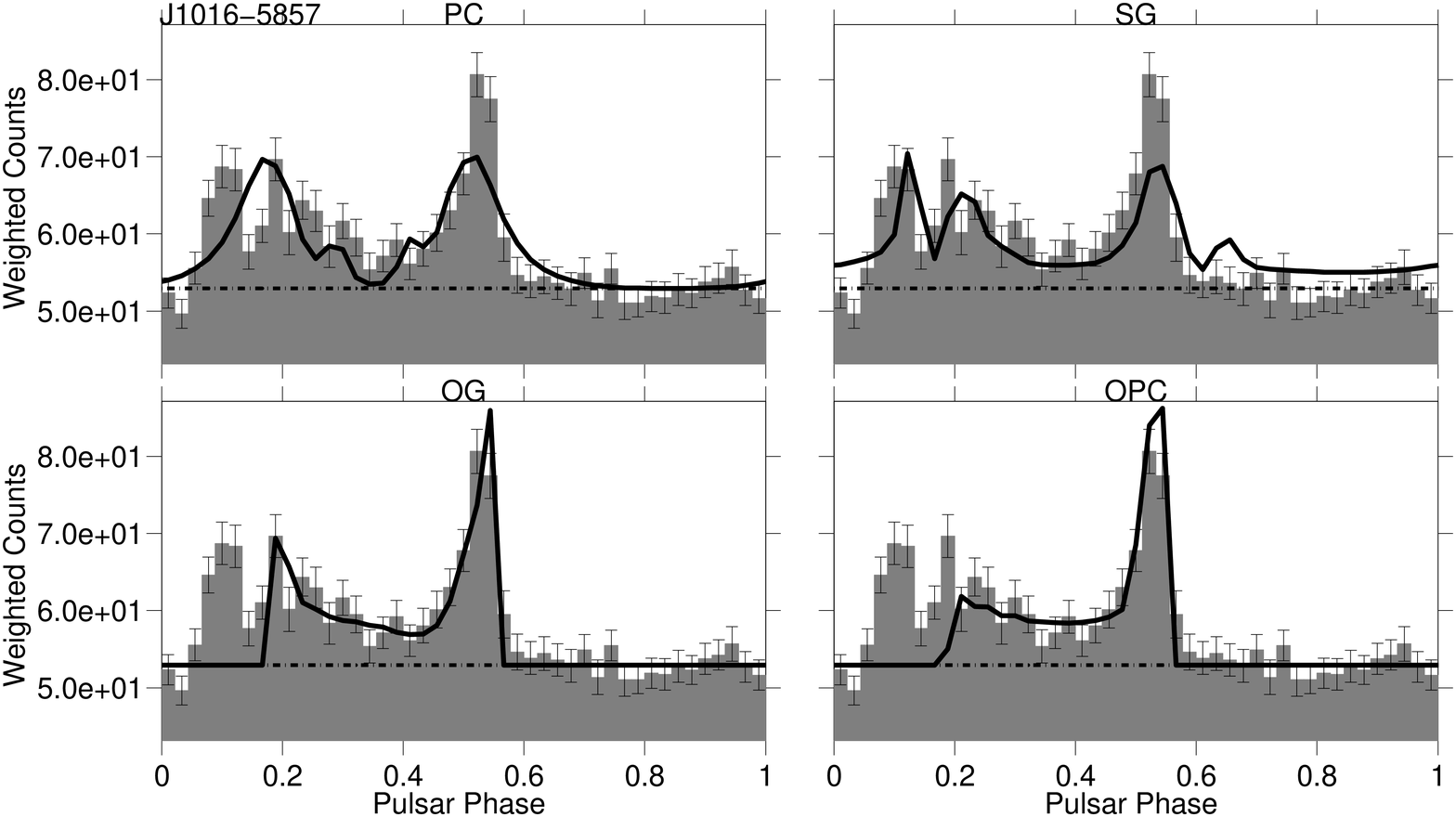}
\includegraphics[width=0.9\textwidth]{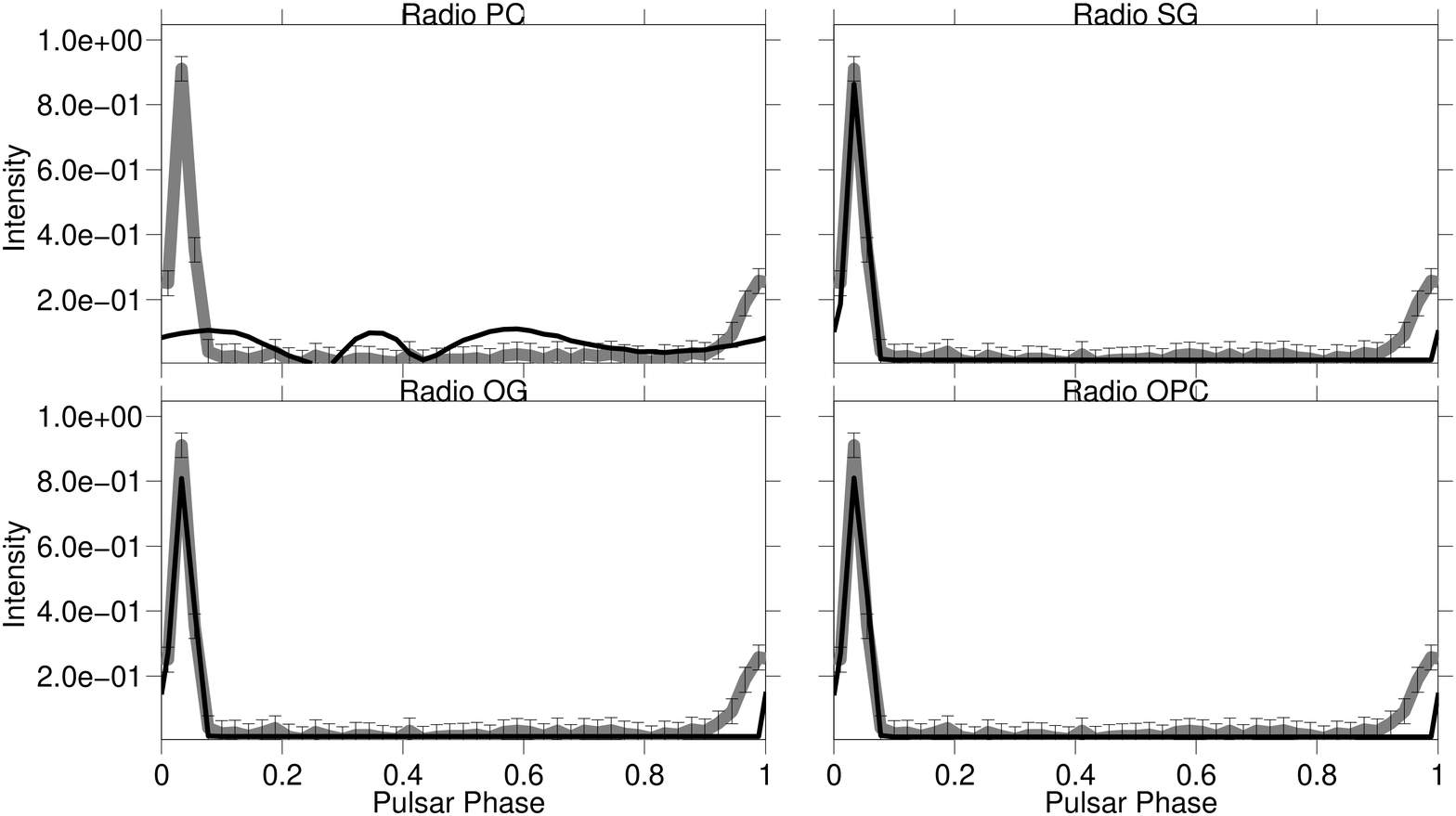}
\caption{PSR J1016-5857. \emph{Top}: for each model the best joint fit solution $\gamma$-ray light-curve (thick black line) is superimposed on the LAT pulsar $\gamma$-ray light-curve (shaded histogram). The estimated background is indicated by the dash-dot line. \emph{Bottom}: for each model the best joint fit solution radio light-curve (black line) is  is superimposed on the LAT pulsar radio light-curve (grey thick line).  The radio model is unique, but the $(\alpha,\zeta)$ solutions vary for each $\gamma$-ray model.}
\label{fitJoint_GmR11}
\end{figure}
  
\clearpage
\begin{figure}[htbp!]
\centering
\includegraphics[width=0.9\textwidth]{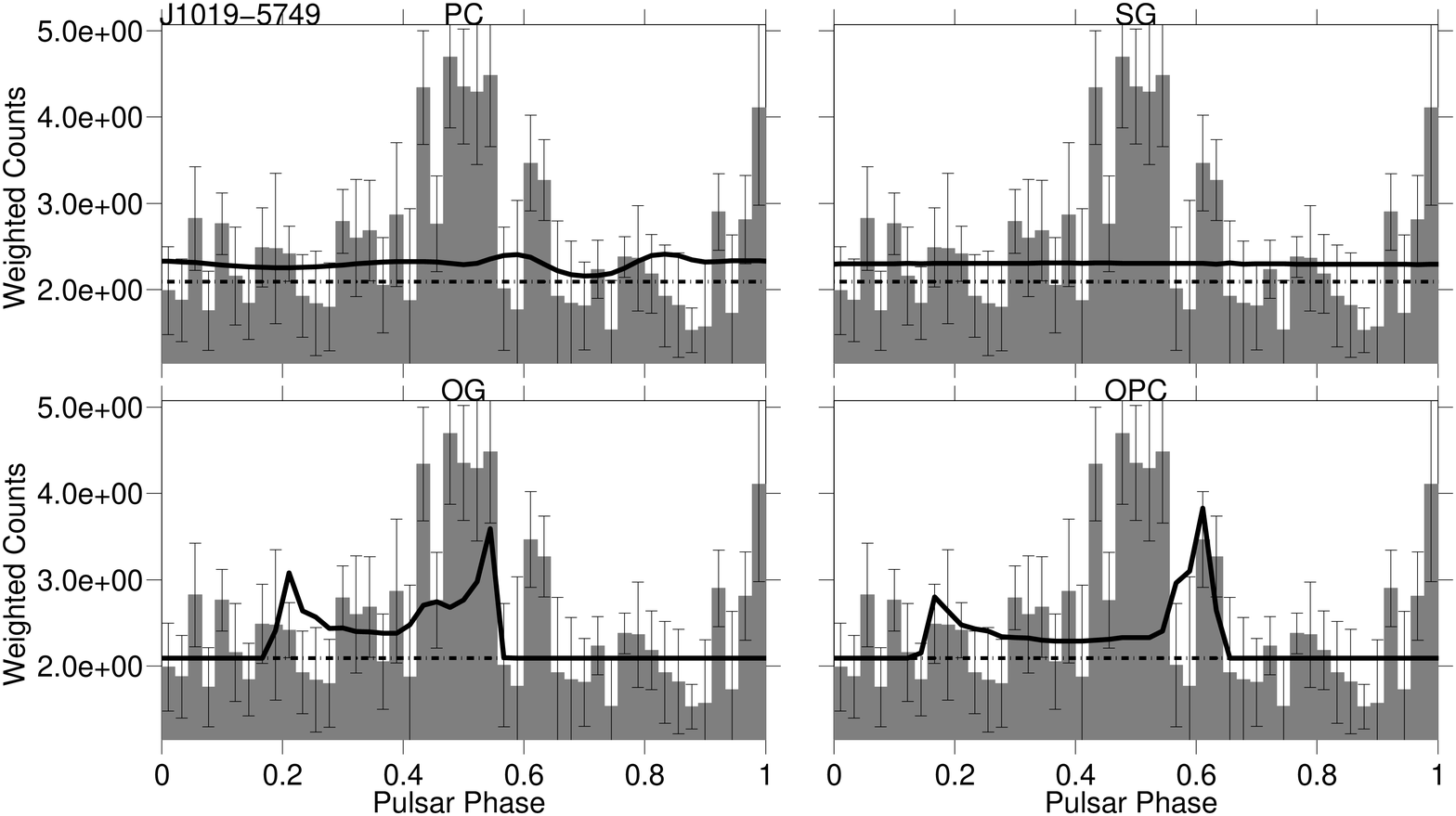}
\includegraphics[width=0.9\textwidth]{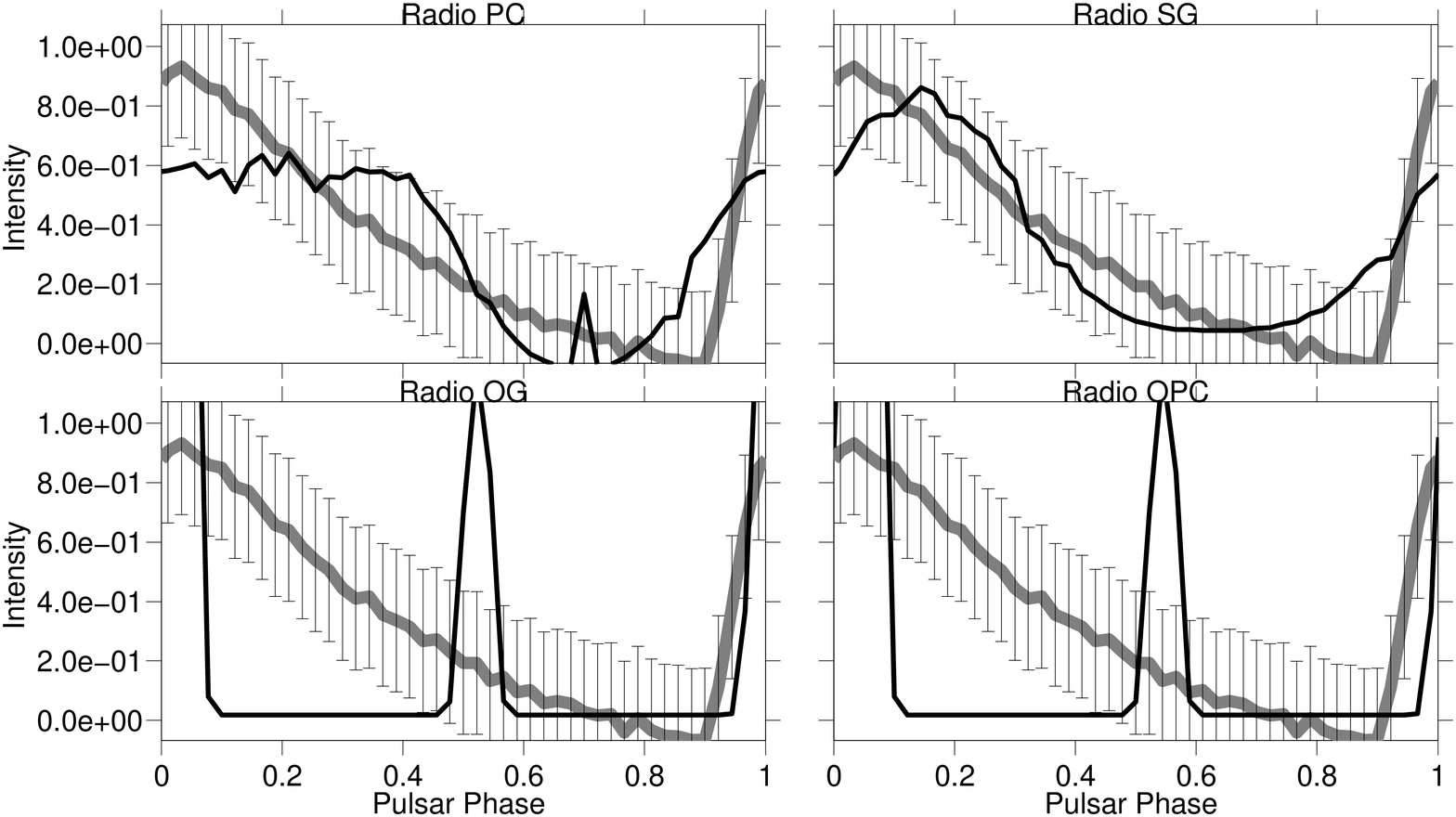}
\caption{PSR J1019-5749. \emph{Top}: for each model the best joint fit solution $\gamma$-ray light-curve (thick black line) is superimposed on the LAT pulsar $\gamma$-ray light-curve (shaded histogram). The estimated background is indicated by the dash-dot line. \emph{Bottom}: for each model the best joint fit solution radio light-curve (black line) is  is superimposed on the LAT pulsar radio light-curve (grey thick line).  The radio model is unique, but the $(\alpha,\zeta)$ solutions vary for each $\gamma$-ray model. 
Because of the low statistics of the $\gamma$-ray light curve, the best-fit solution of each model is dominated by the radio light curve. 
The optimum-solution is given by the SG model but it represents an unreliable result since the best fit $\gamma$-ray light curve corresponds to a flat profile.}
\label{fitJoint_GmR12}
\end{figure}
  
\clearpage
\begin{figure}[htbp!]
\centering
\includegraphics[width=0.9\textwidth]{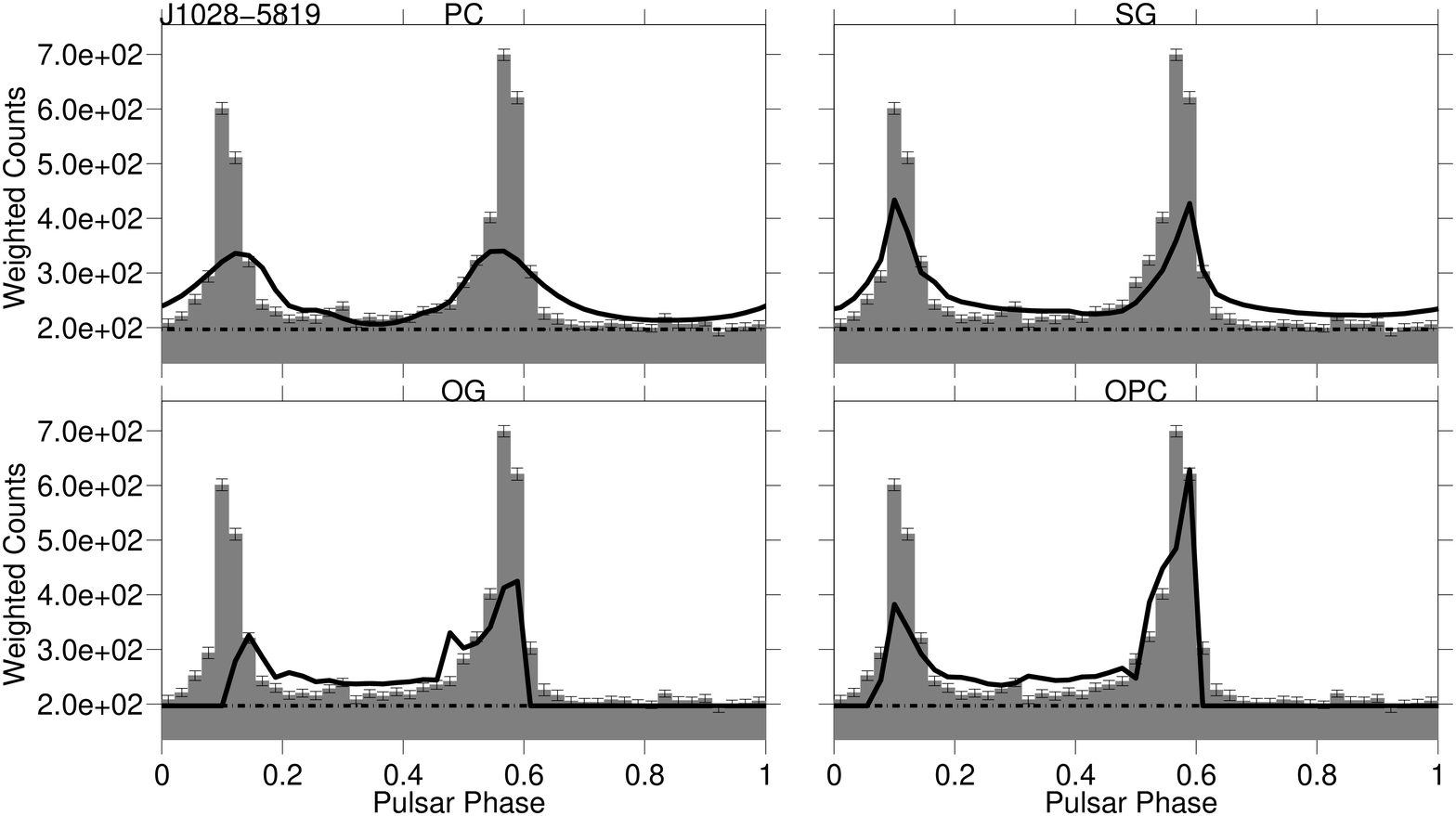}
\includegraphics[width=0.9\textwidth]{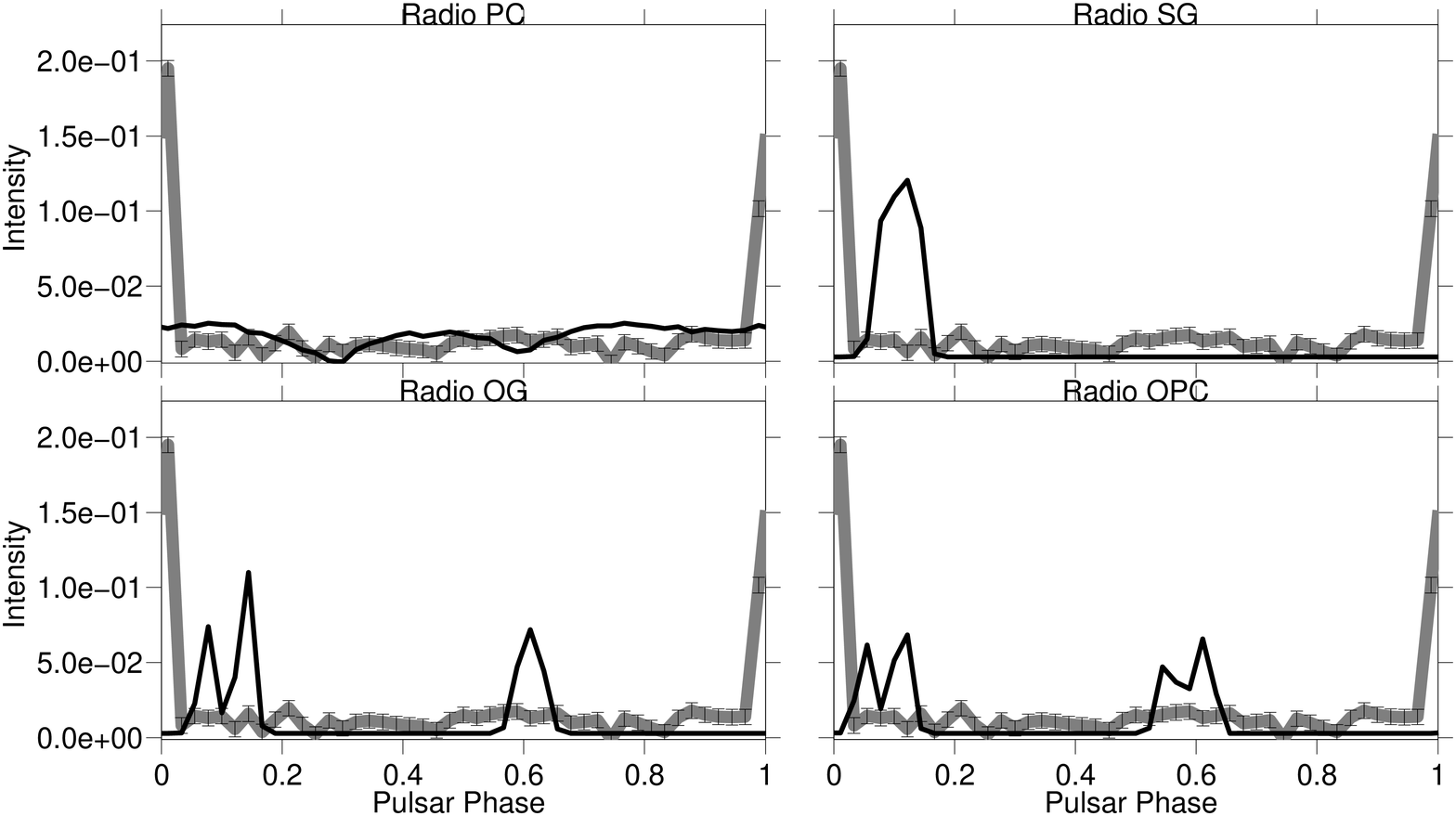}
\caption{PSR J1028-5819. \emph{Top}: for each model the best joint fit solution $\gamma$-ray light-curve (thick black line) is superimposed on the LAT pulsar $\gamma$-ray light-curve (shaded histogram). The estimated background is indicated by the dash-dot line. \emph{Bottom}: for each model the best joint fit solution radio light-curve (black line) is  is superimposed on the LAT pulsar radio light-curve (grey thick line).  The radio model is unique, but the $(\alpha,\zeta)$ solutions vary for each $\gamma$-ray model.}
\label{fitJoint_GmR13}
\end{figure}
  
\clearpage
\begin{figure}[htbp!]
\centering
\includegraphics[width=0.9\textwidth]{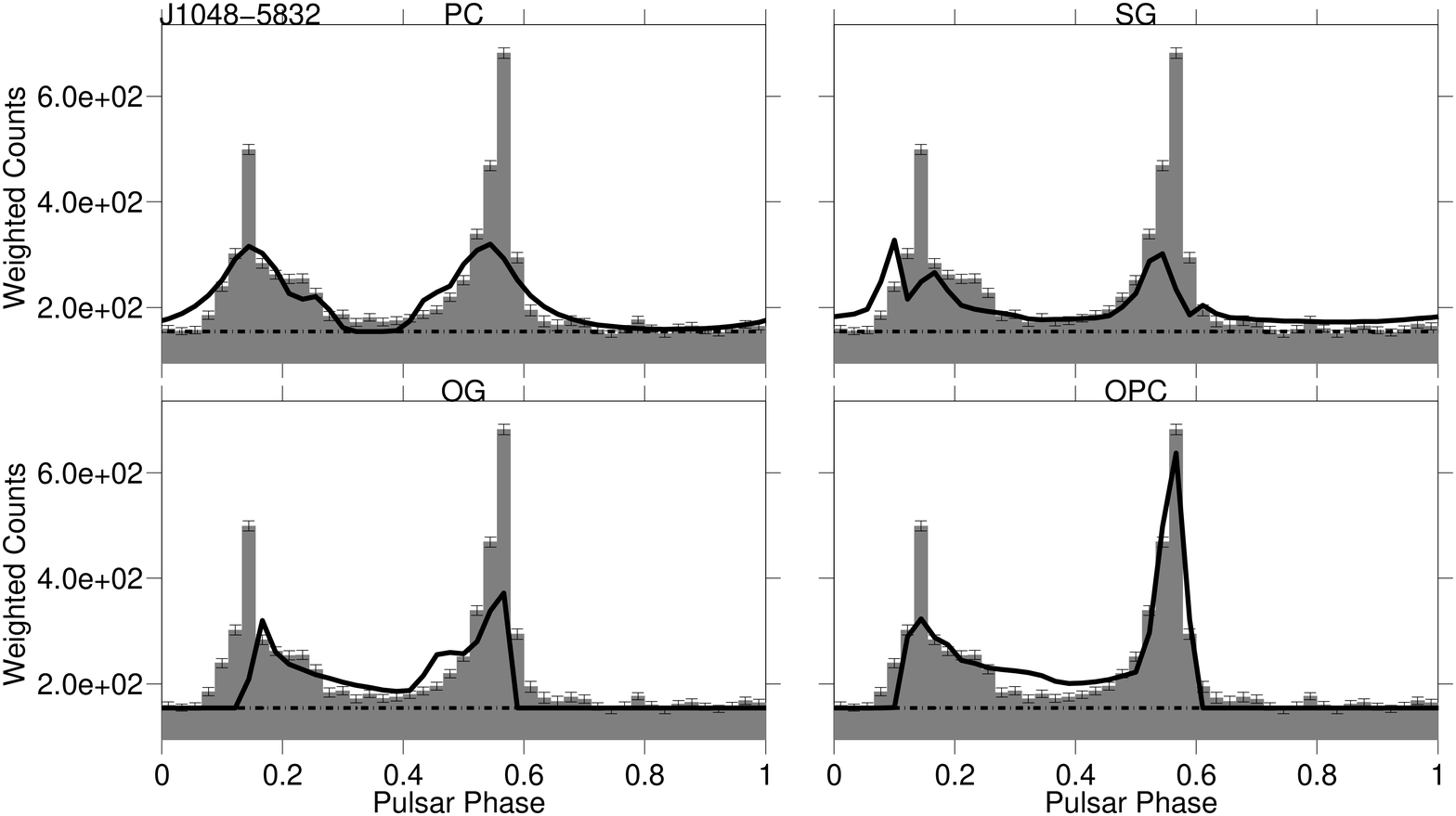}
\includegraphics[width=0.9\textwidth]{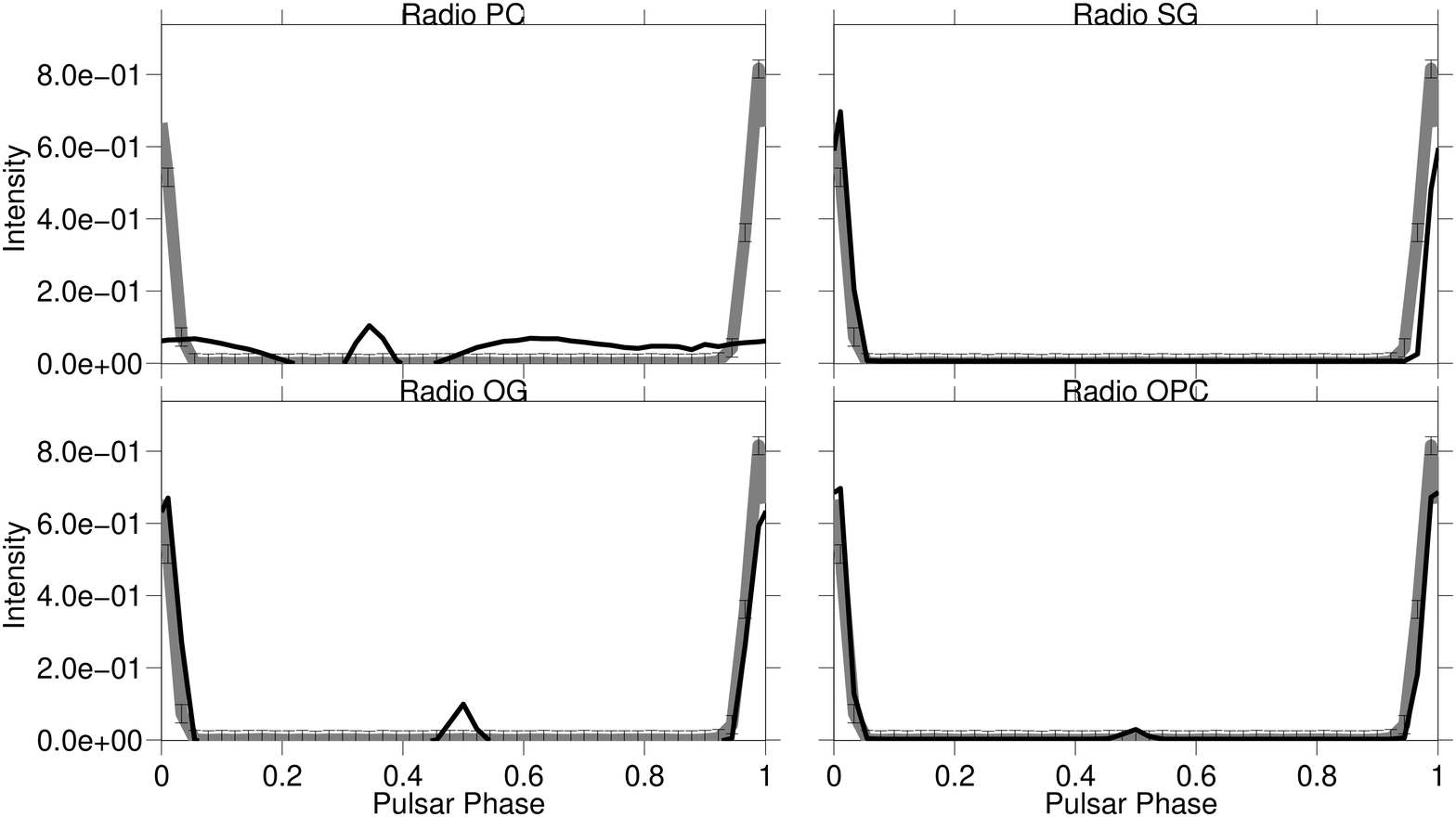}
\caption{PSR J1048-5832. \emph{Top}: for each model the best joint fit solution $\gamma$-ray light-curve (thick black line) is superimposed on the LAT pulsar $\gamma$-ray light-curve (shaded histogram). The estimated background is indicated by the dash-dot line. \emph{Bottom}: for each model the best joint fit solution radio light-curve (black line) is  is superimposed on the LAT pulsar radio light-curve (grey thick line).  The radio model is unique, but the $(\alpha,\zeta)$ solutions vary for each $\gamma$-ray model.}
\label{fitJoint_GmR14}
\end{figure}
  
\clearpage
\begin{figure}[htbp!]
\centering
\includegraphics[width=0.9\textwidth]{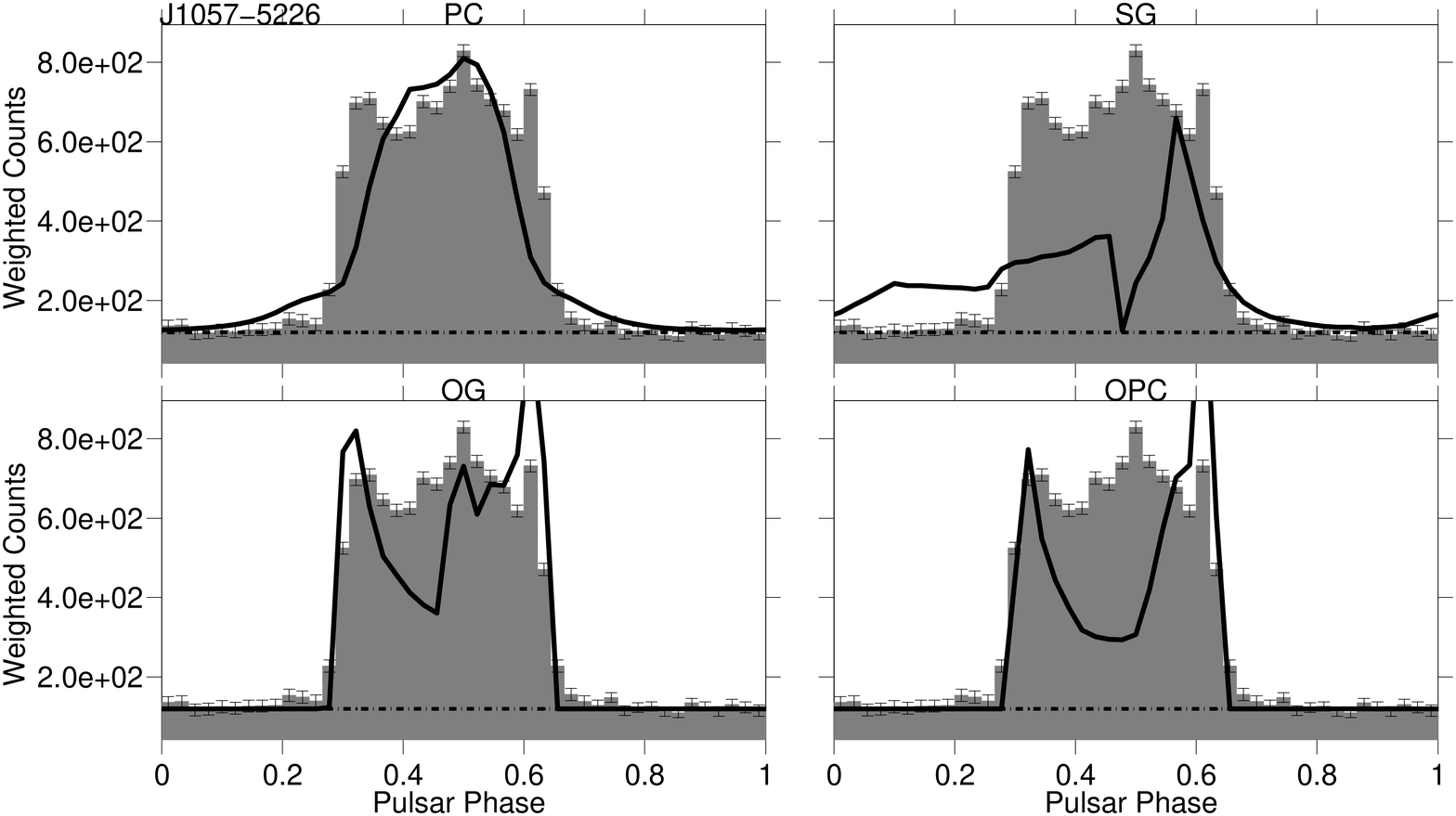}
\includegraphics[width=0.9\textwidth]{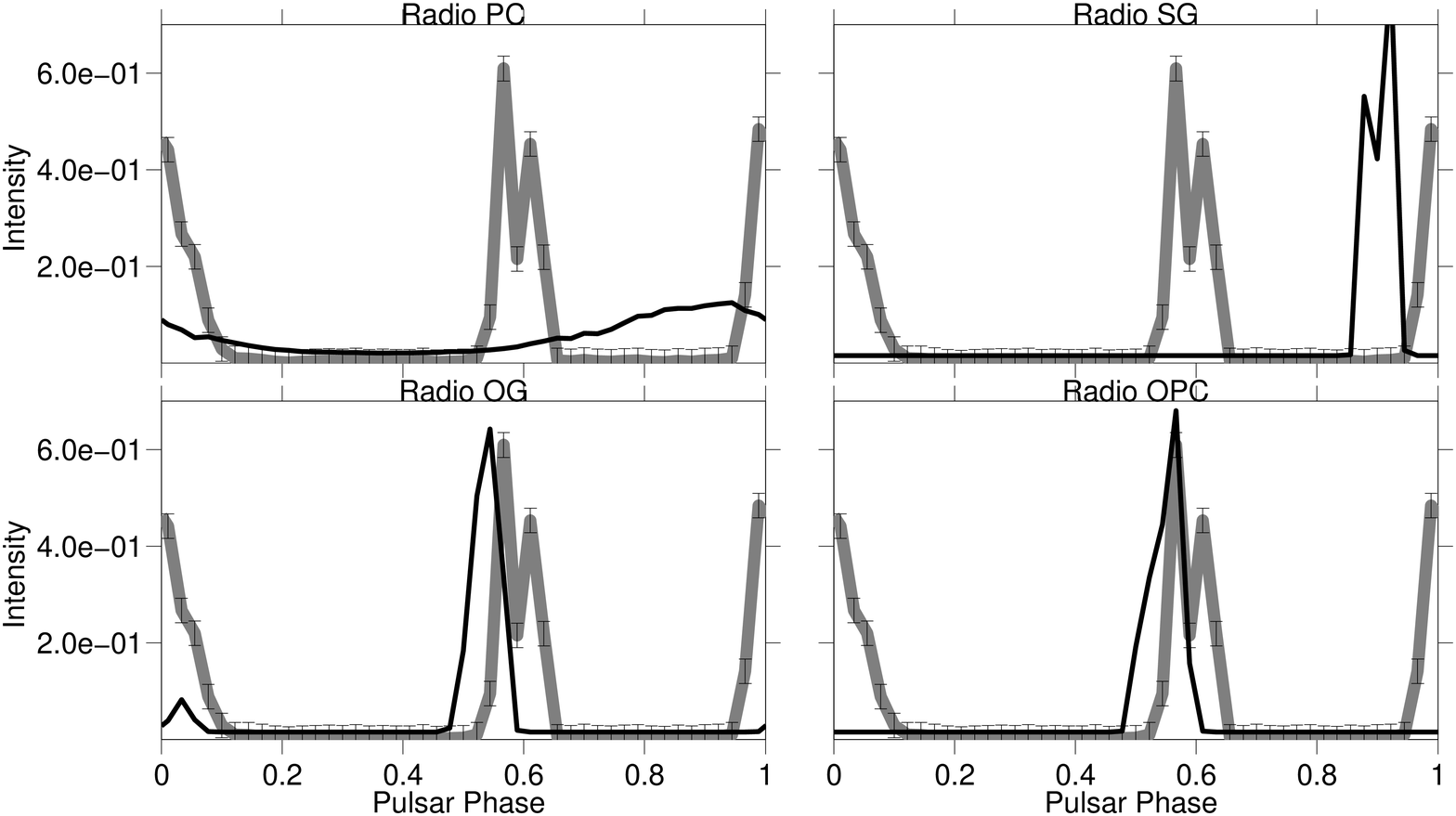}
\caption{PSR J1057-5226. \emph{Top}: for each model the best joint fit solution $\gamma$-ray light-curve (thick black line) is superimposed on the LAT pulsar $\gamma$-ray light-curve (shaded histogram). The estimated background is indicated by the dash-dot line. \emph{Bottom}: for each model the best joint fit solution radio light-curve (black line) is  is superimposed on the LAT pulsar radio light-curve (grey thick line).  The radio model is unique, but the $(\alpha,\zeta)$ solutions vary for each $\gamma$-ray model.}
\label{fitJoint_GmR15}
\end{figure}
  
\clearpage
\begin{figure}[htbp!]
\centering
\includegraphics[width=0.9\textwidth]{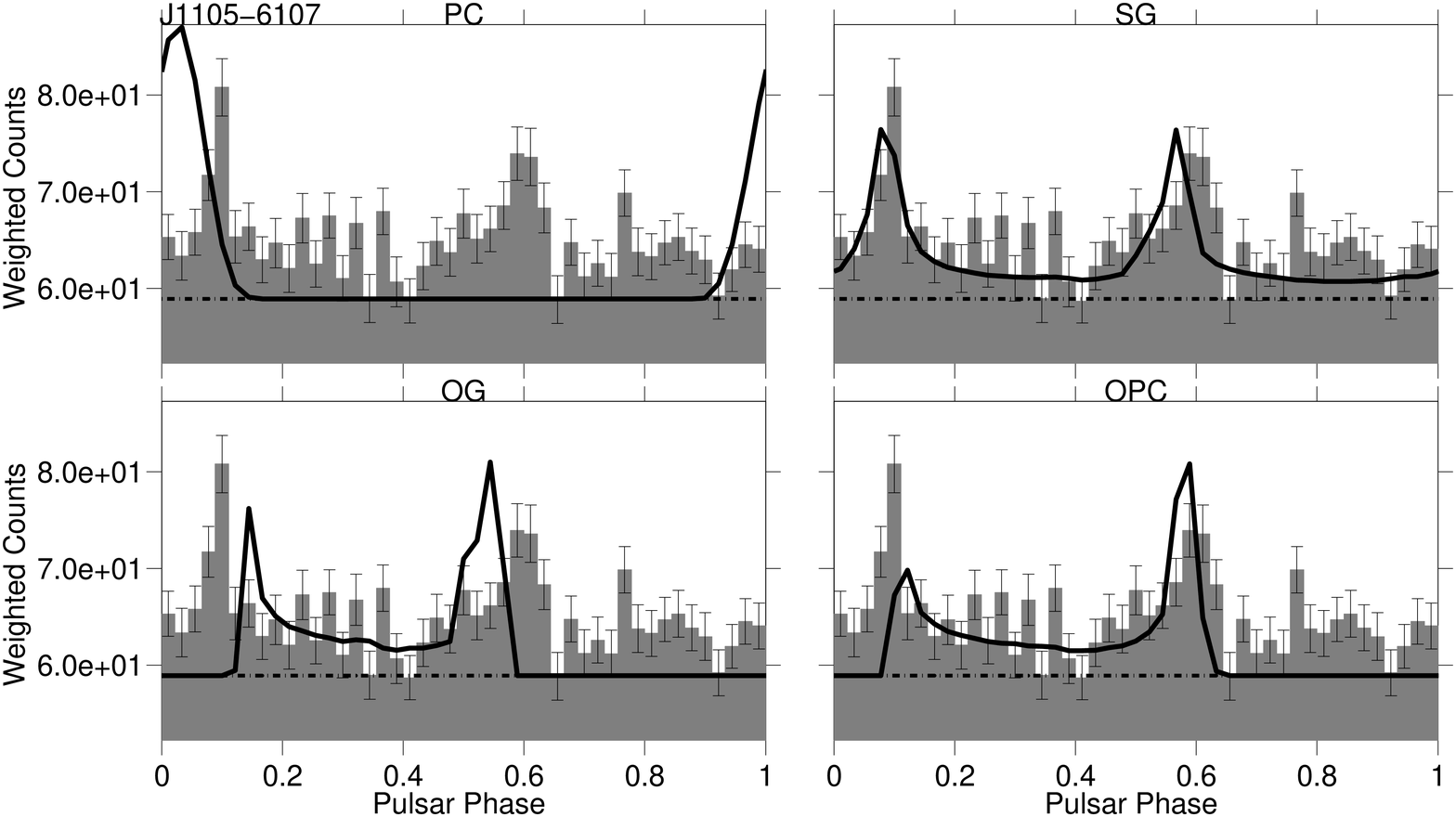}
\includegraphics[width=0.9\textwidth]{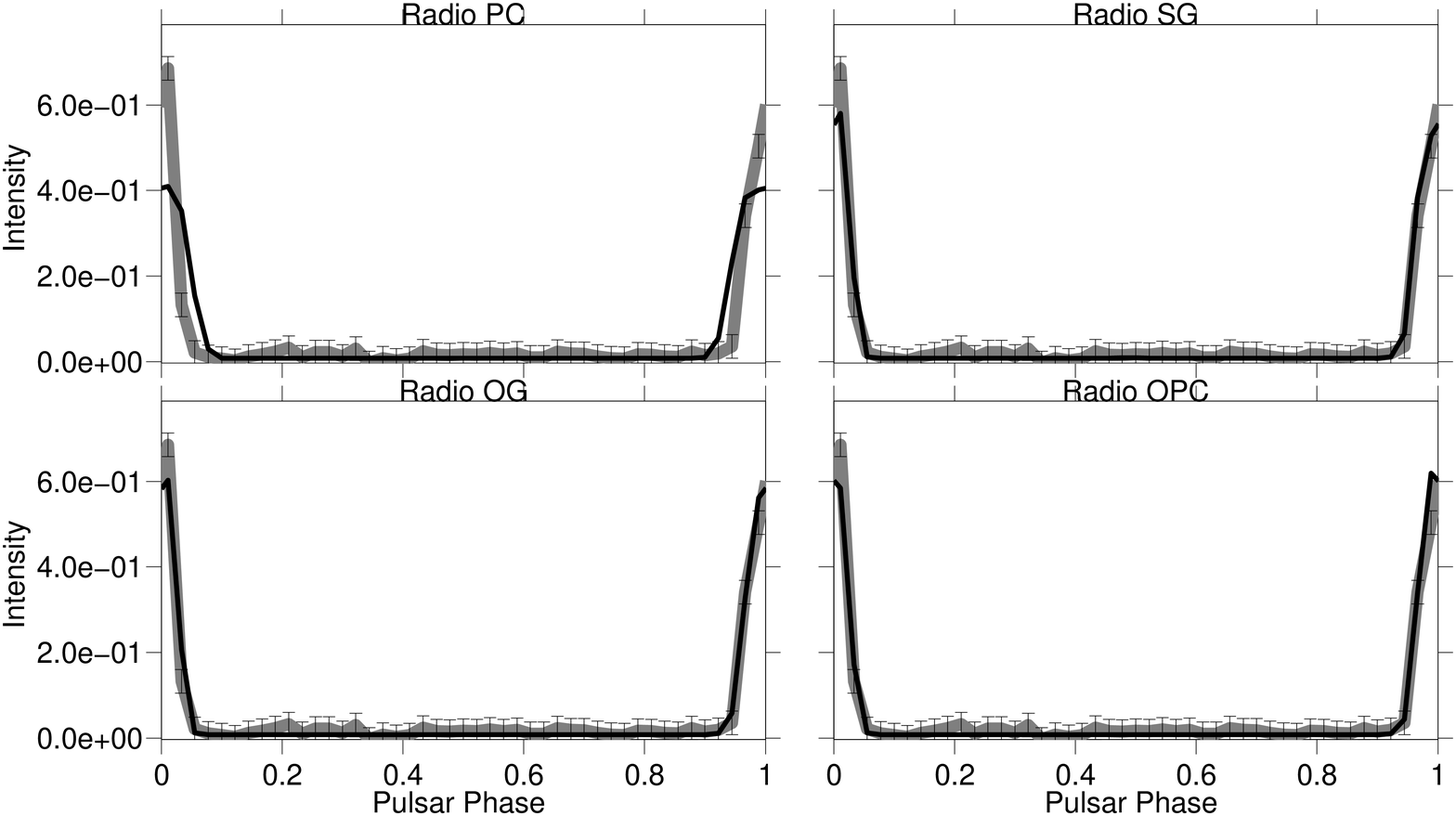}
\caption{PSR J1105-6107. \emph{Top}: for each model the best joint fit solution $\gamma$-ray light-curve (thick black line) is superimposed on the LAT pulsar $\gamma$-ray light-curve (shaded histogram). The estimated background is indicated by the dash-dot line. \emph{Bottom}: for each model the best joint fit solution radio light-curve (black line) is  is superimposed on the LAT pulsar radio light-curve (grey thick line).  The radio model is unique, but the $(\alpha,\zeta)$ solutions vary for each $\gamma$-ray model.}
\label{fitJoint_GmR16}
\end{figure}
  
\clearpage
\begin{figure}[htbp!]
\centering
\includegraphics[width=0.9\textwidth]{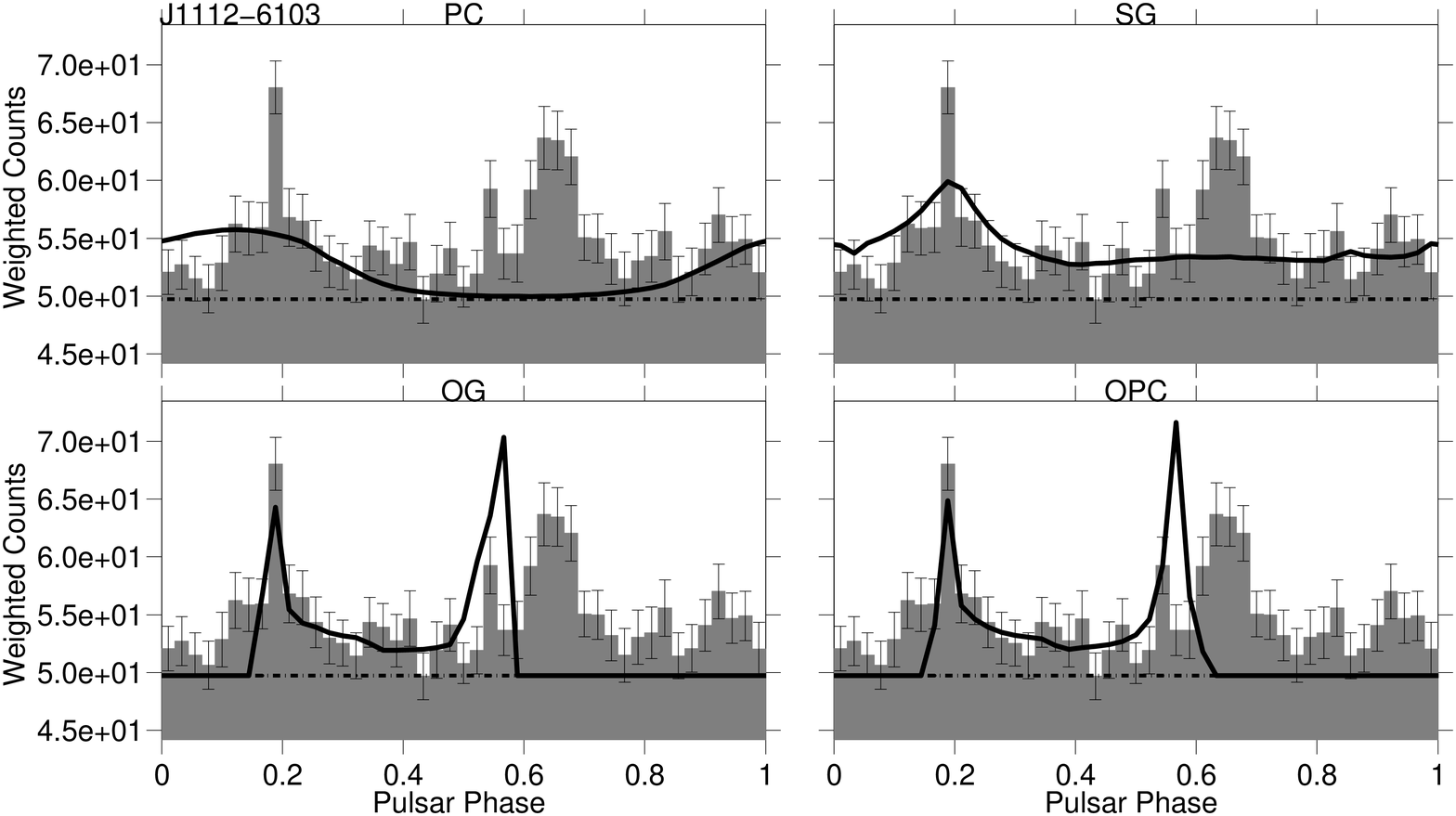}
\includegraphics[width=0.9\textwidth]{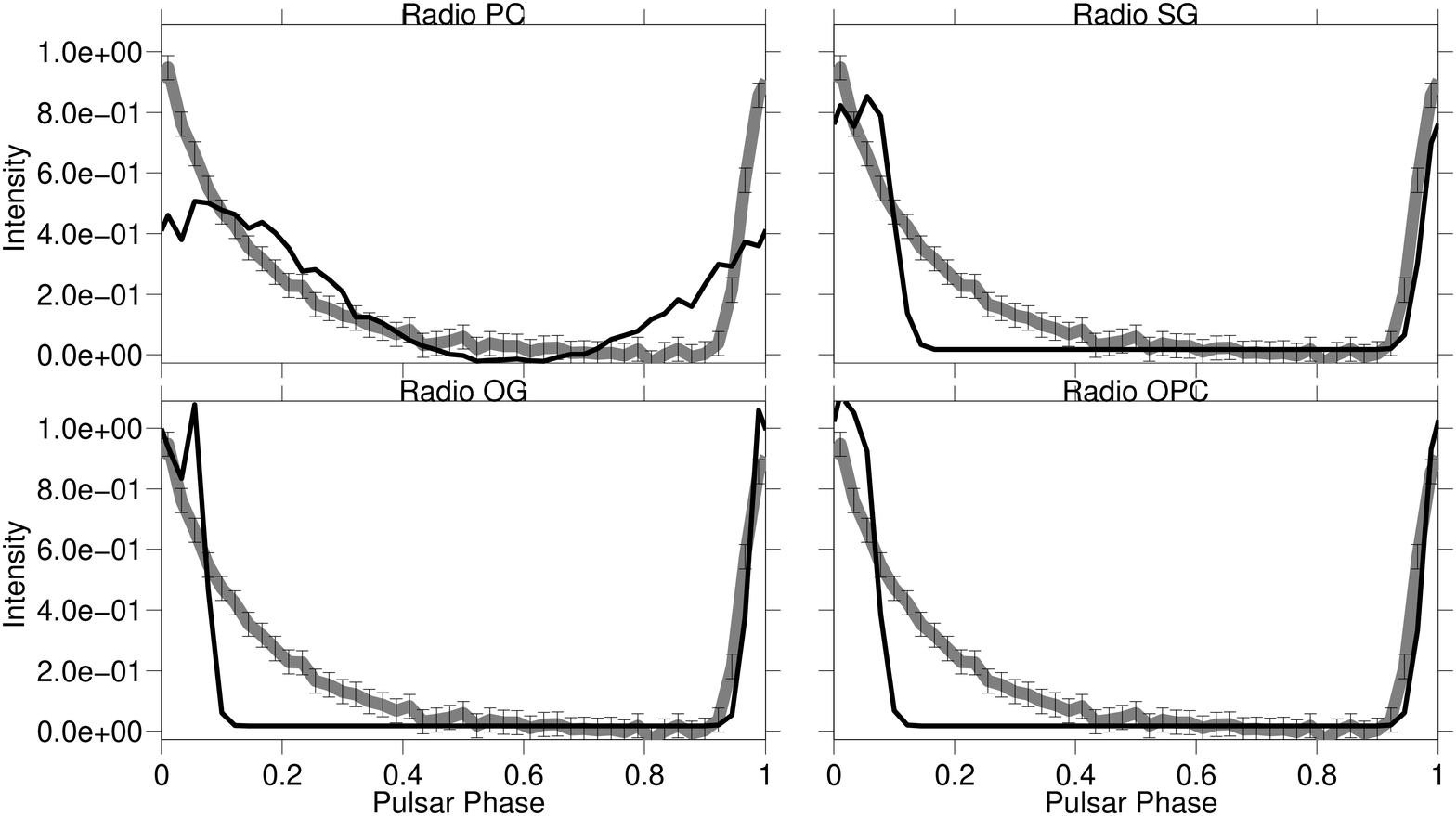}
\caption{PSR J1112-6103. \emph{Top}: for each model the best joint fit solution $\gamma$-ray light-curve (thick black line) is superimposed on the LAT pulsar $\gamma$-ray light-curve (shaded histogram). The estimated background is indicated by the dash-dot line. \emph{Bottom}: for each model the best joint fit solution radio light-curve (black line) is  is superimposed on the LAT pulsar radio light-curve (grey thick line).  The radio model is unique, but the $(\alpha,\zeta)$ solutions vary for each $\gamma$-ray model.}
\label{fitJoint_GmR17}
\end{figure}
  
\clearpage
\begin{figure}[htbp!]
\centering
\includegraphics[width=0.9\textwidth]{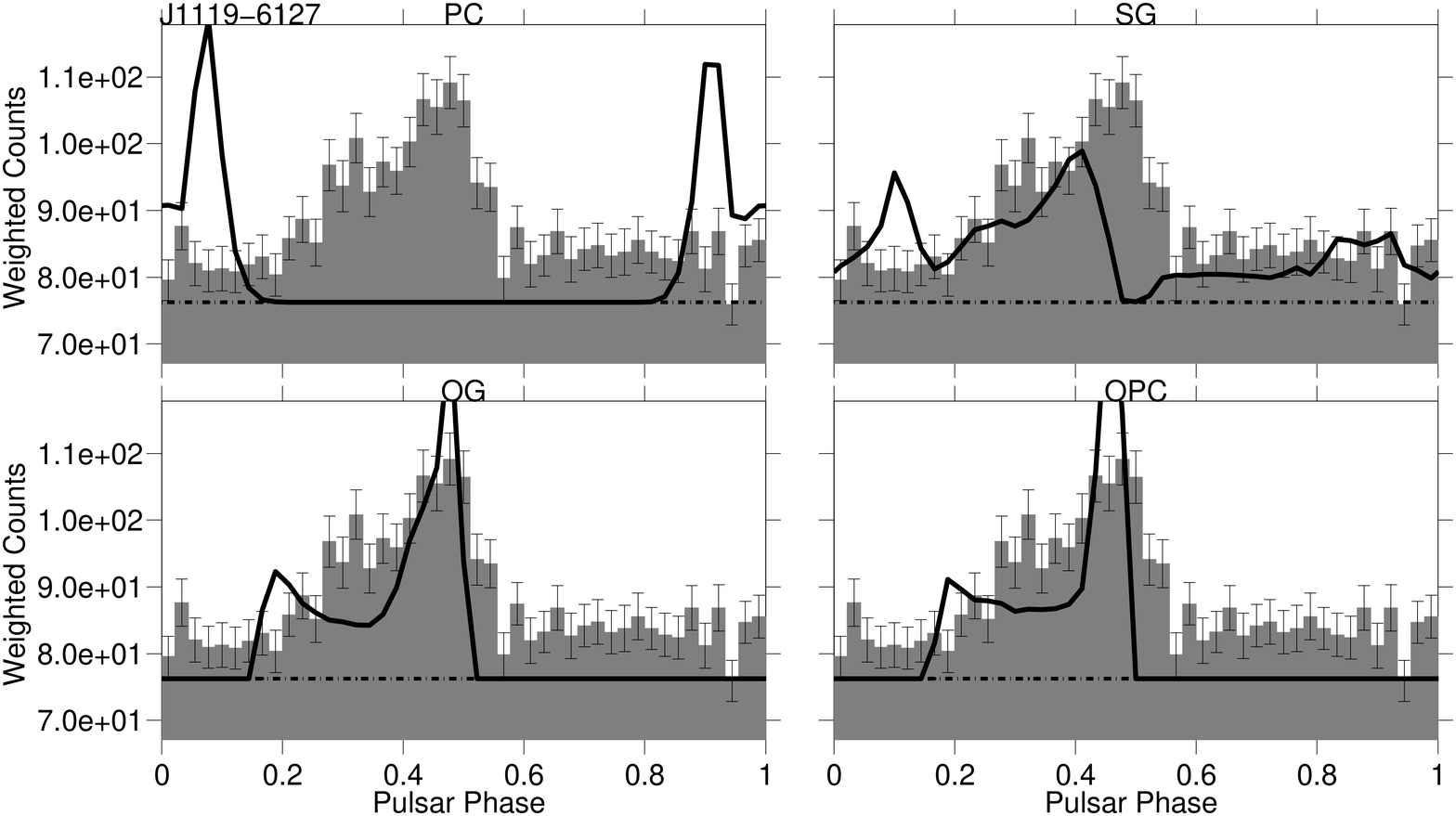}
\includegraphics[width=0.9\textwidth]{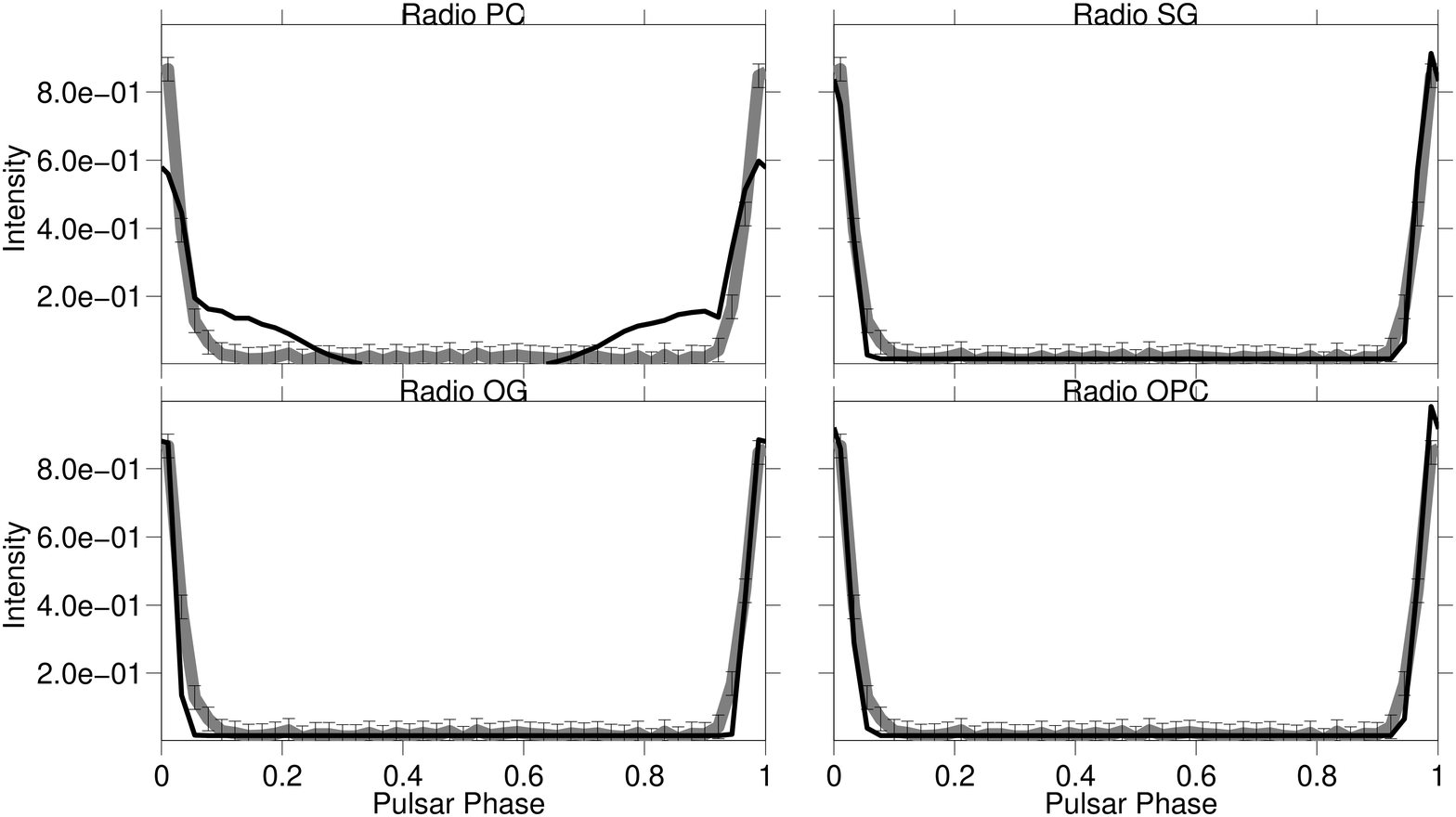}
\caption{PSR J1119-6127. \emph{Top}: for each model the best joint fit solution $\gamma$-ray light-curve (thick black line) is superimposed on the LAT pulsar $\gamma$-ray light-curve (shaded histogram). The estimated background is indicated by the dash-dot line. \emph{Bottom}: for each model the best joint fit solution radio light-curve (black line) is  is superimposed on the LAT pulsar radio light-curve (grey thick line).  The radio model is unique, but the $(\alpha,\zeta)$ solutions vary for each $\gamma$-ray model.}
\label{fitJoint_GmR18}
\end{figure}
  
\clearpage
\begin{figure}[htbp!]
\centering
\includegraphics[width=0.9\textwidth]{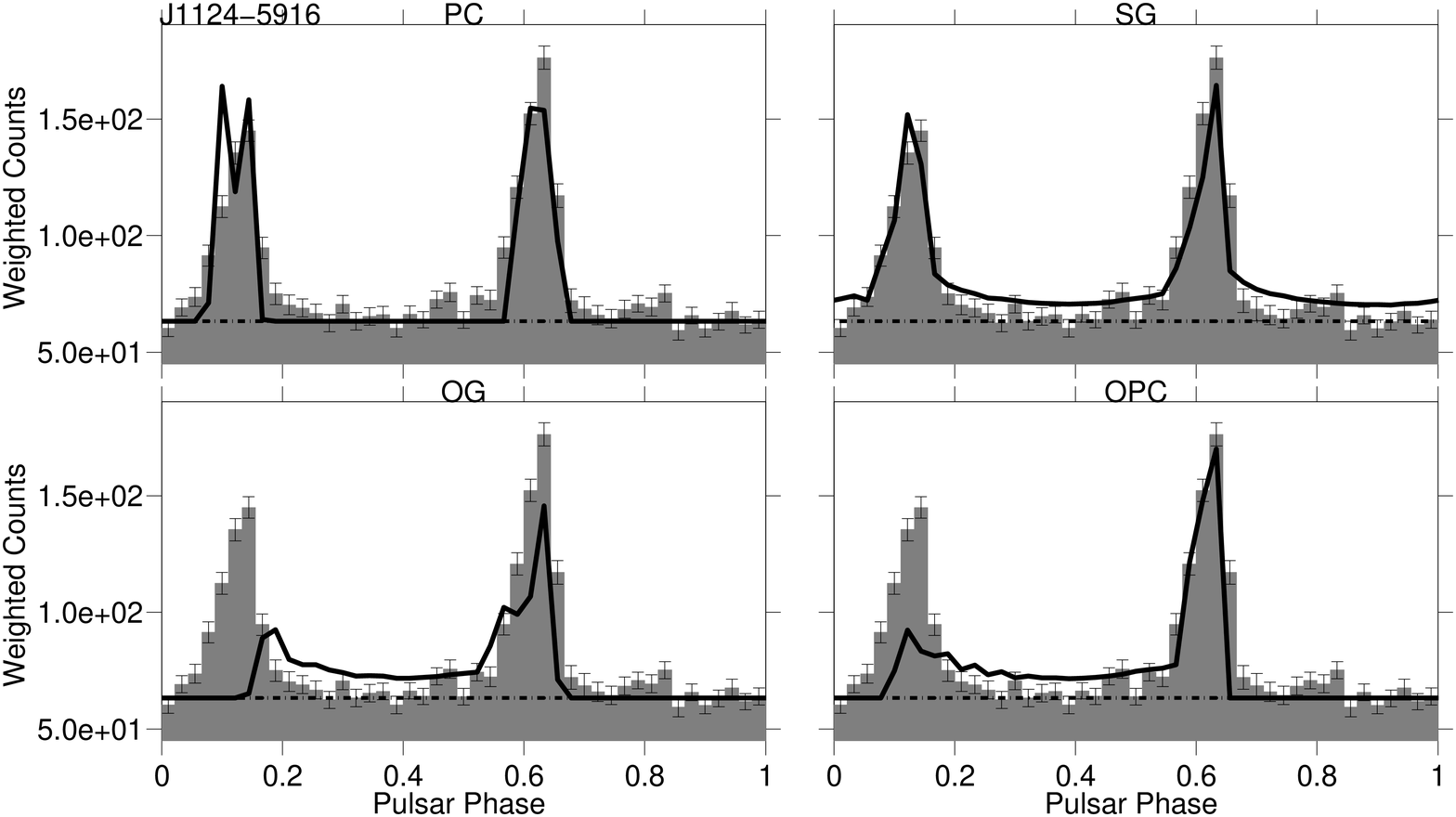}
\includegraphics[width=0.9\textwidth]{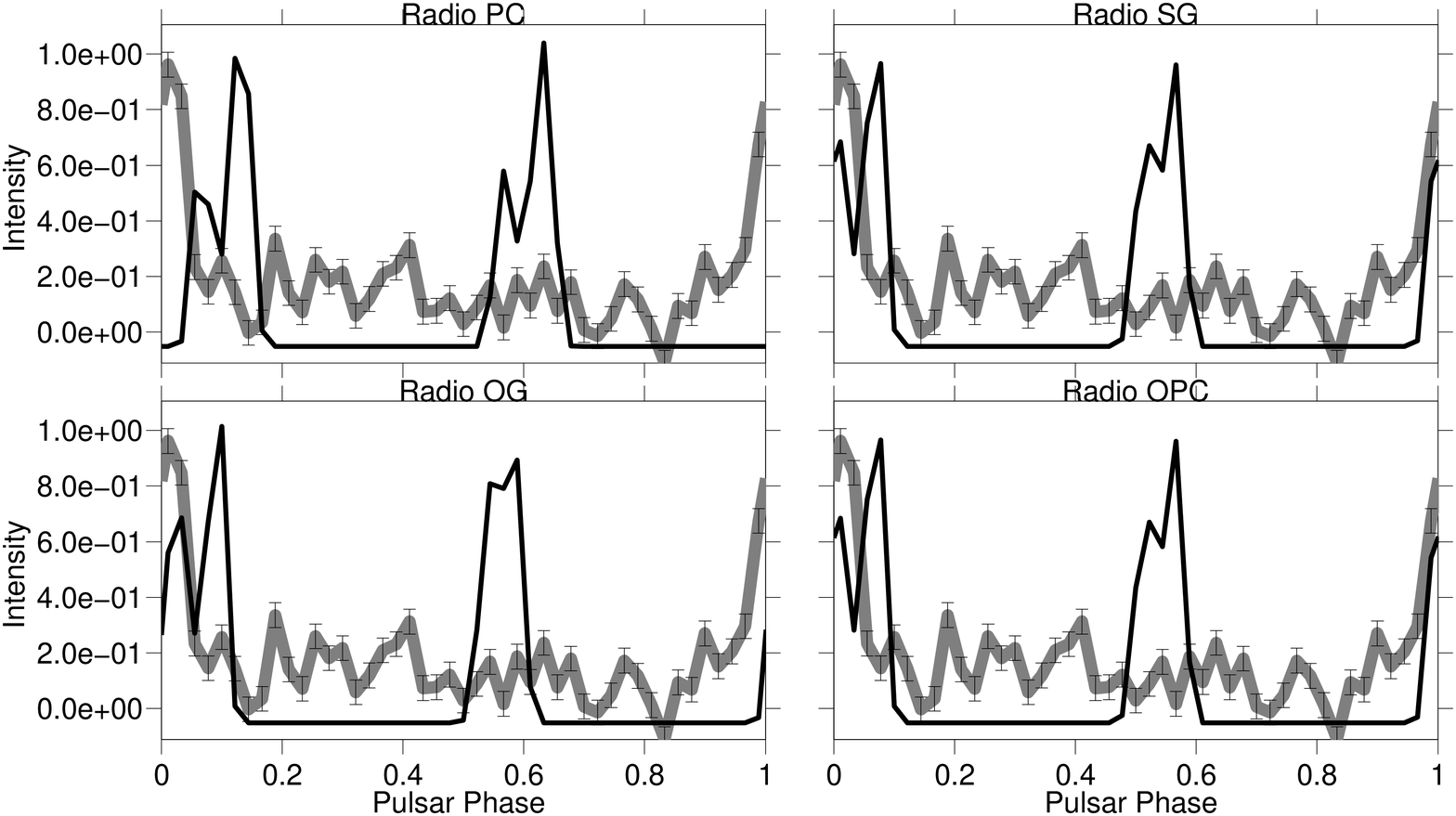}
\caption{PSR J1124-5916. \emph{Top}: for each model the best joint fit solution $\gamma$-ray light-curve (thick black line) is superimposed on the LAT pulsar $\gamma$-ray light-curve (shaded histogram). The estimated background is indicated by the dash-dot line. \emph{Bottom}: for each model the best joint fit solution radio light-curve (black line) is  is superimposed on the LAT pulsar radio light-curve (grey thick line).  The radio model is unique, but the $(\alpha,\zeta)$ solutions vary for each $\gamma$-ray model.}
\label{fitJoint_GmR19}
\end{figure}
  
\clearpage
\begin{figure}[htbp!]
\centering
\includegraphics[width=0.9\textwidth]{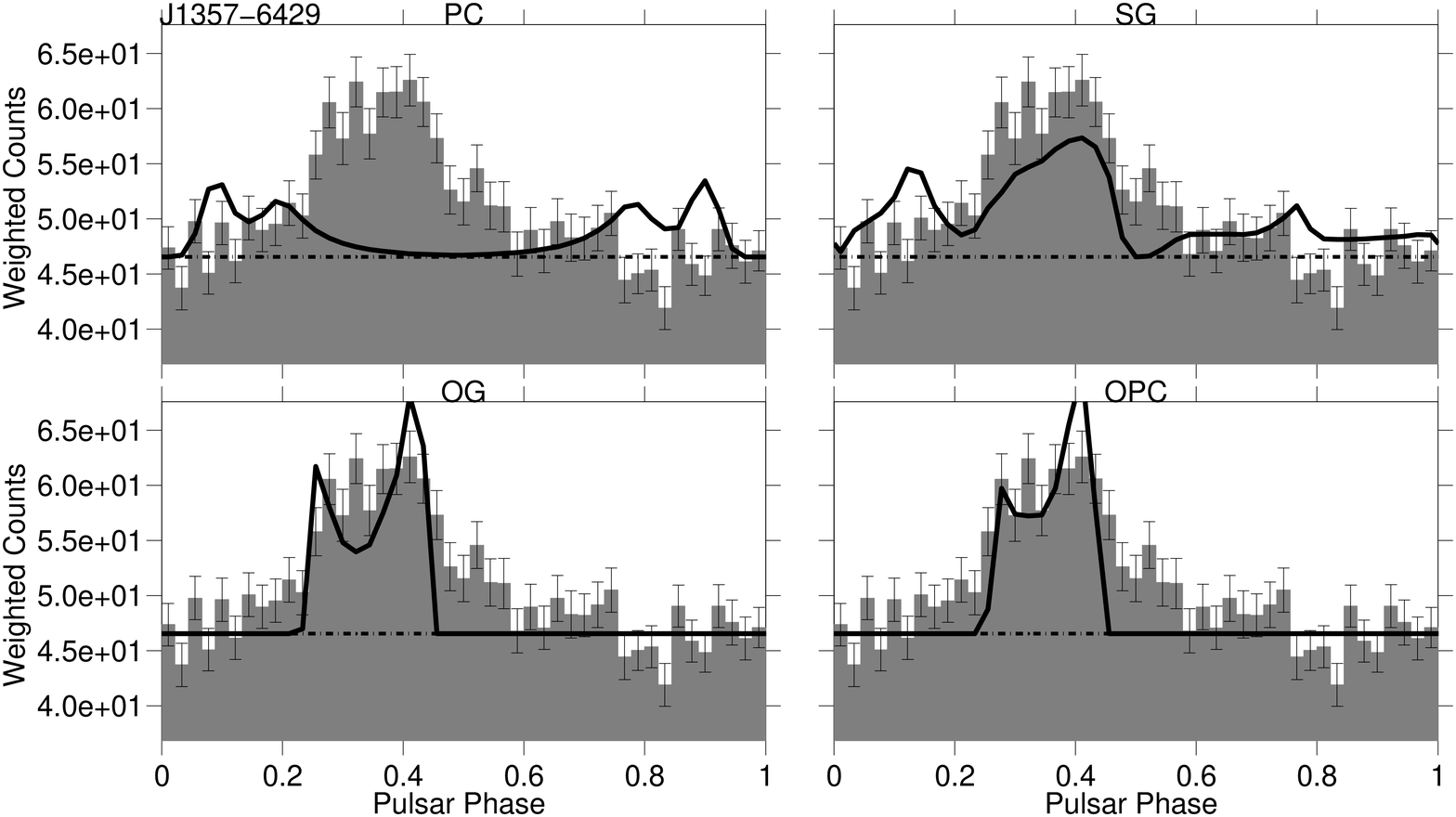}
\includegraphics[width=0.9\textwidth]{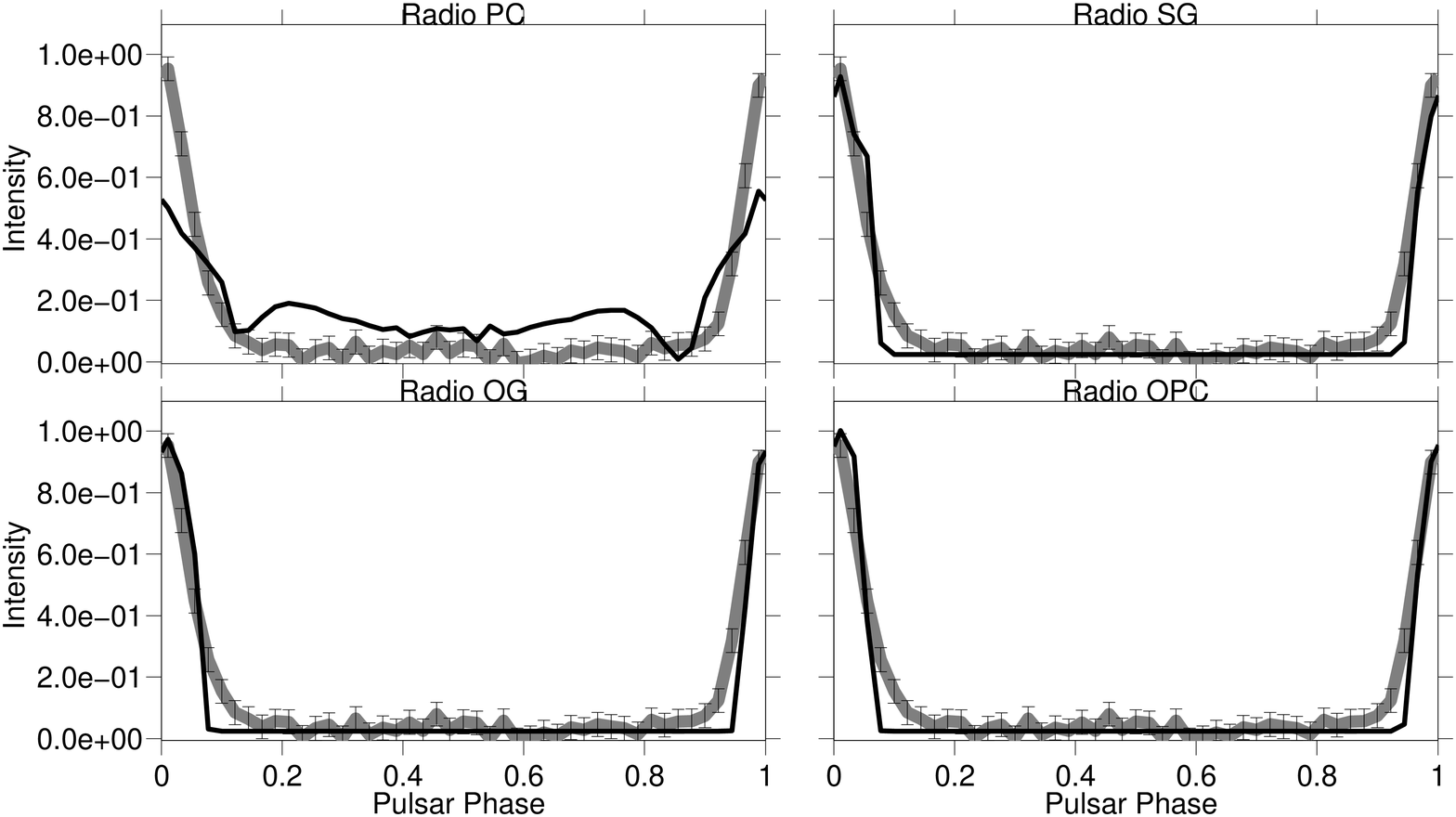}
\caption{PSR J1357-6429. \emph{Top}: for each model the best joint fit solution $\gamma$-ray light-curve (thick black line) is superimposed on the LAT pulsar $\gamma$-ray light-curve (shaded histogram). The estimated background is indicated by the dash-dot line. \emph{Bottom}: for each model the best joint fit solution radio light-curve (black line) is  is superimposed on the LAT pulsar radio light-curve (grey thick line).  The radio model is unique, but the $(\alpha,\zeta)$ solutions vary for each $\gamma$-ray model.}
\label{fitJoint_GmR20}
\end{figure}
  
\clearpage
\begin{figure}[htbp!]
\centering
\includegraphics[width=0.9\textwidth]{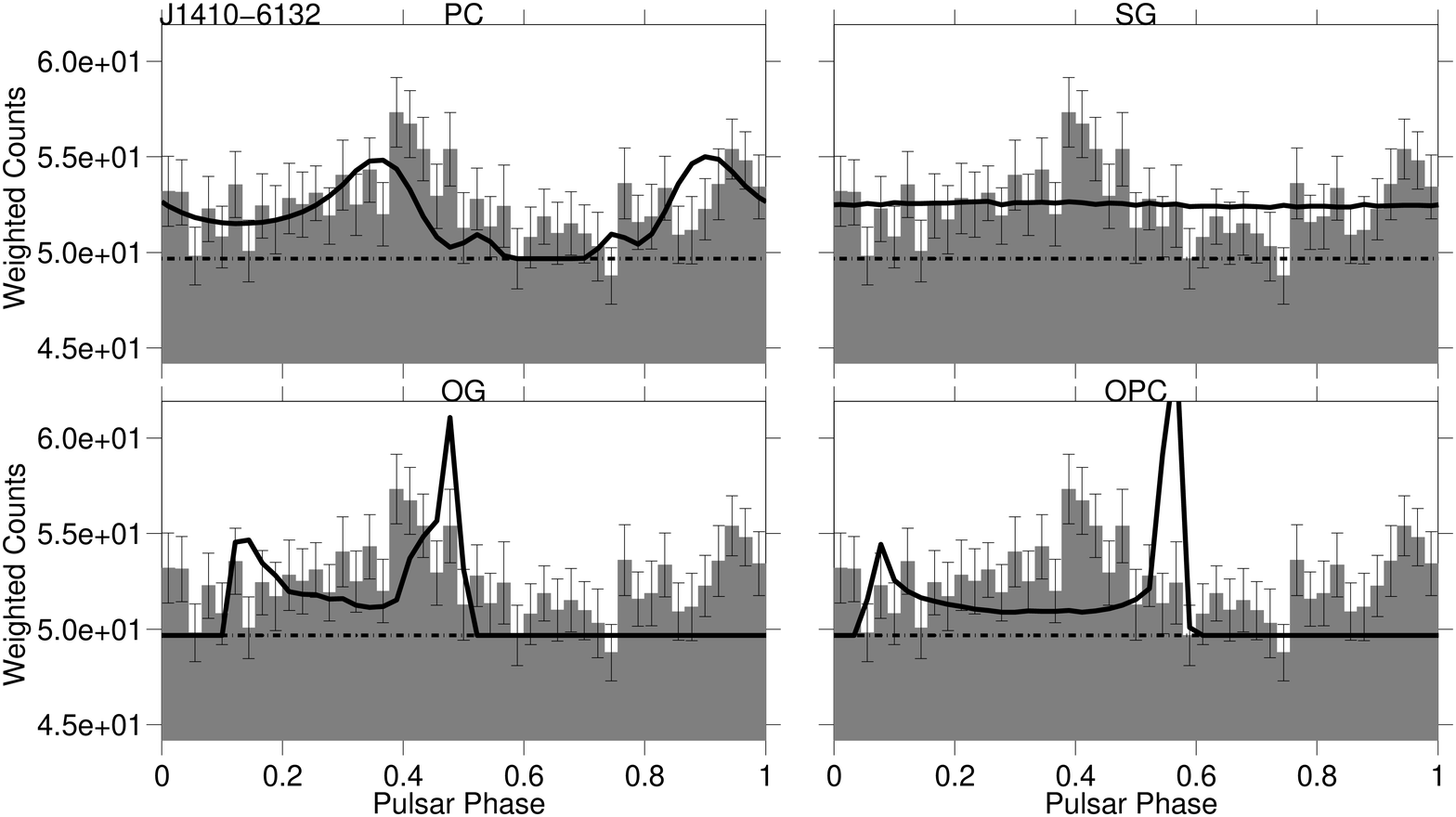}
\includegraphics[width=0.9\textwidth]{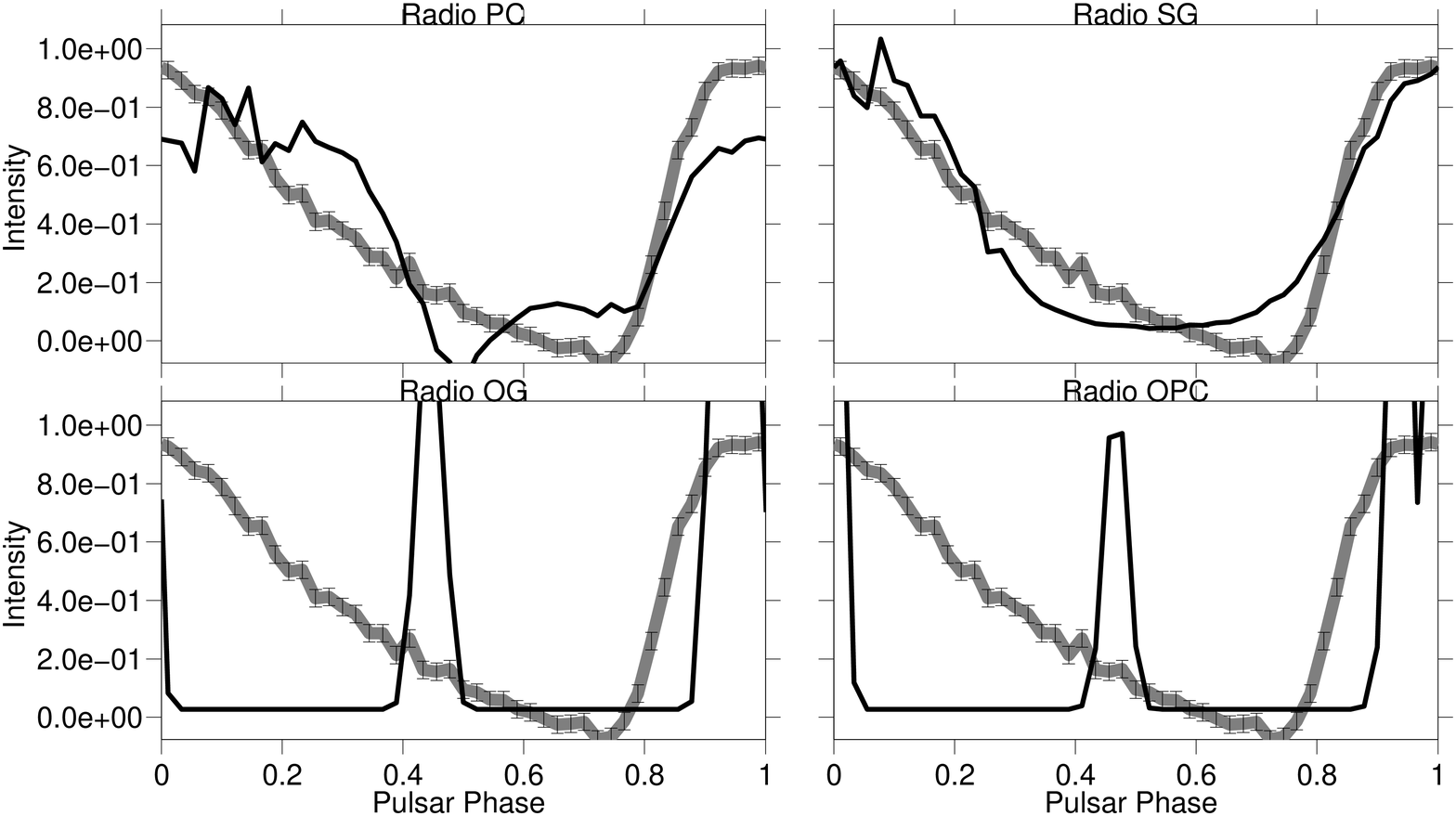}
\caption{PSR J1410-6132. \emph{Top}: for each model the best joint fit solution $\gamma$-ray light-curve (thick black line) is superimposed on the LAT pulsar $\gamma$-ray light-curve (shaded histogram). The estimated background is indicated by the dash-dot line. \emph{Bottom}: for each model the best joint fit solution radio light-curve (black line) is  is superimposed on the LAT pulsar radio light-curve (grey thick line).  The radio model is unique, but the $(\alpha,\zeta)$ solutions vary for each $\gamma$-ray model. For this pulsar the SG model gives the optimum-solution but it represents an unreliable result since the best fit $\gamma$-ray light curve correspond to a flat profile.}
\label{fitJoint_GmR21}
\end{figure}
  
\clearpage
\begin{figure}[htbp!]
\centering
\includegraphics[width=0.9\textwidth]{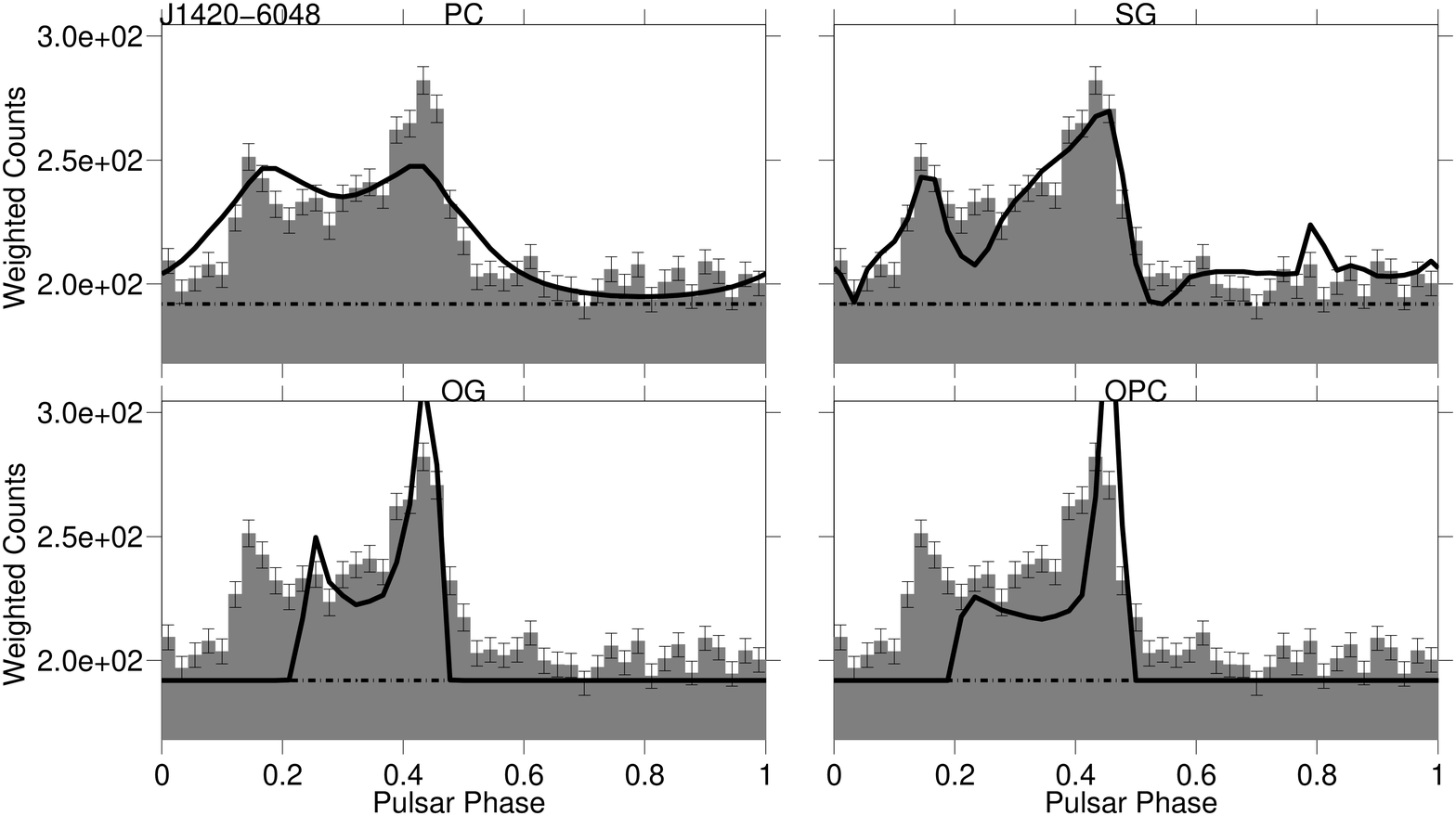}
\includegraphics[width=0.9\textwidth]{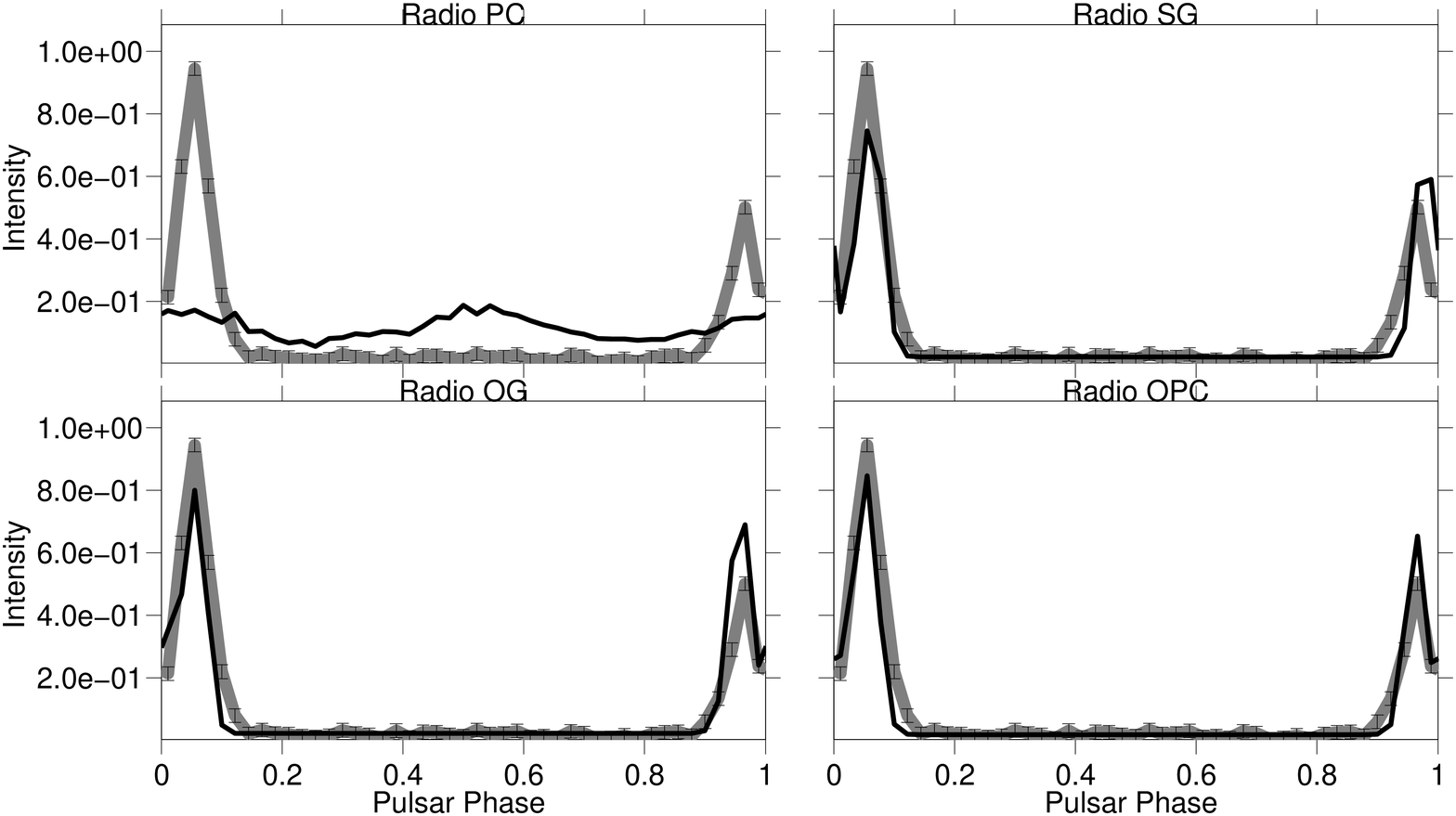}
\caption{PSR J1420-6048. \emph{Top}: for each model the best joint fit solution $\gamma$-ray light-curve (thick black line) is superimposed on the LAT pulsar $\gamma$-ray light-curve (shaded histogram). The estimated background is indicated by the dash-dot line. \emph{Bottom}: for each model the best joint fit solution radio light-curve (black line) is  is superimposed on the LAT pulsar radio light-curve (grey thick line).  The radio model is unique, but the $(\alpha,\zeta)$ solutions vary for each $\gamma$-ray model.}
\label{fitJoint_GmR22}
\end{figure}
  
\clearpage
\begin{figure}[htbp!]
\centering
\includegraphics[width=0.9\textwidth]{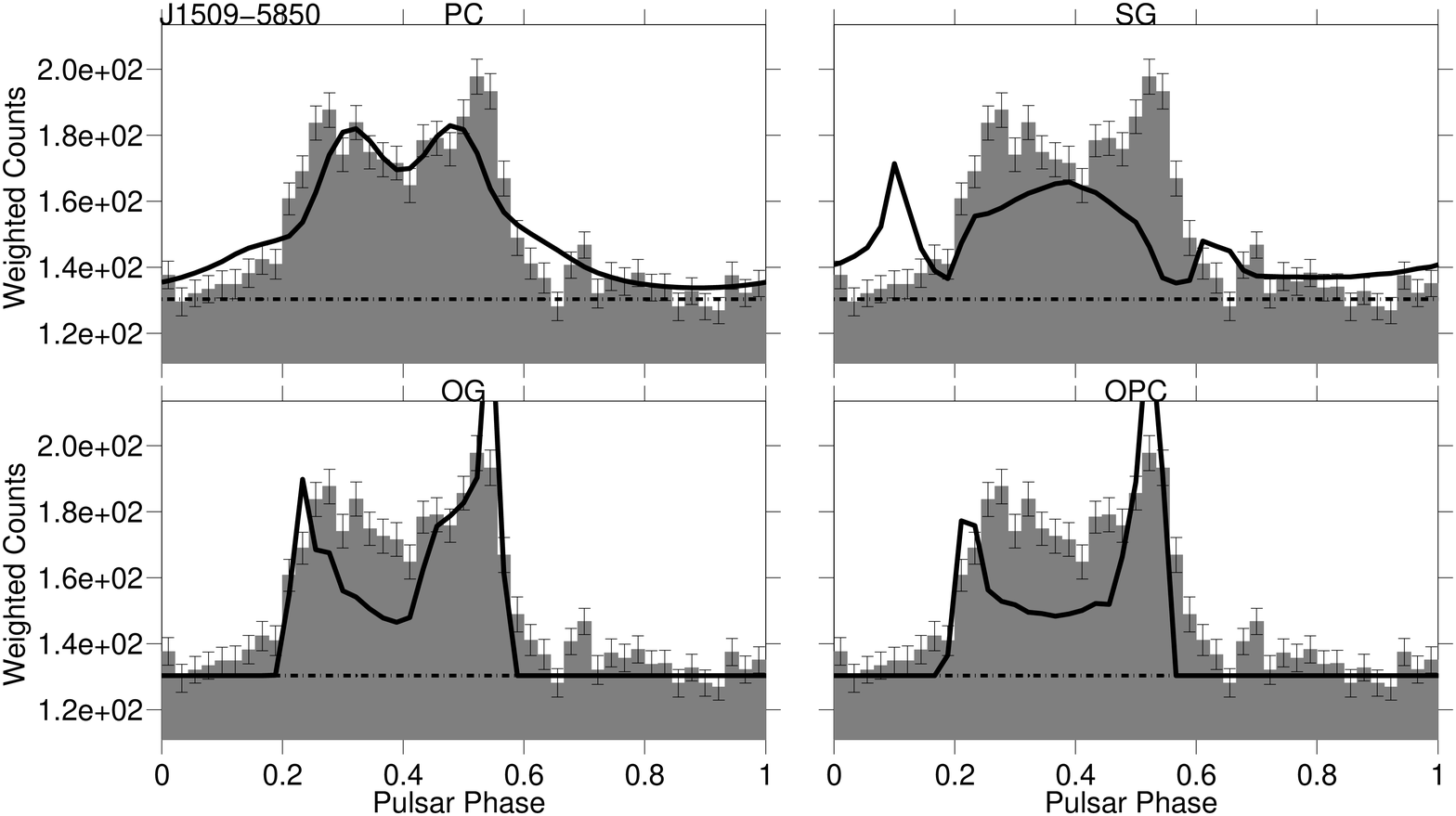}
\includegraphics[width=0.9\textwidth]{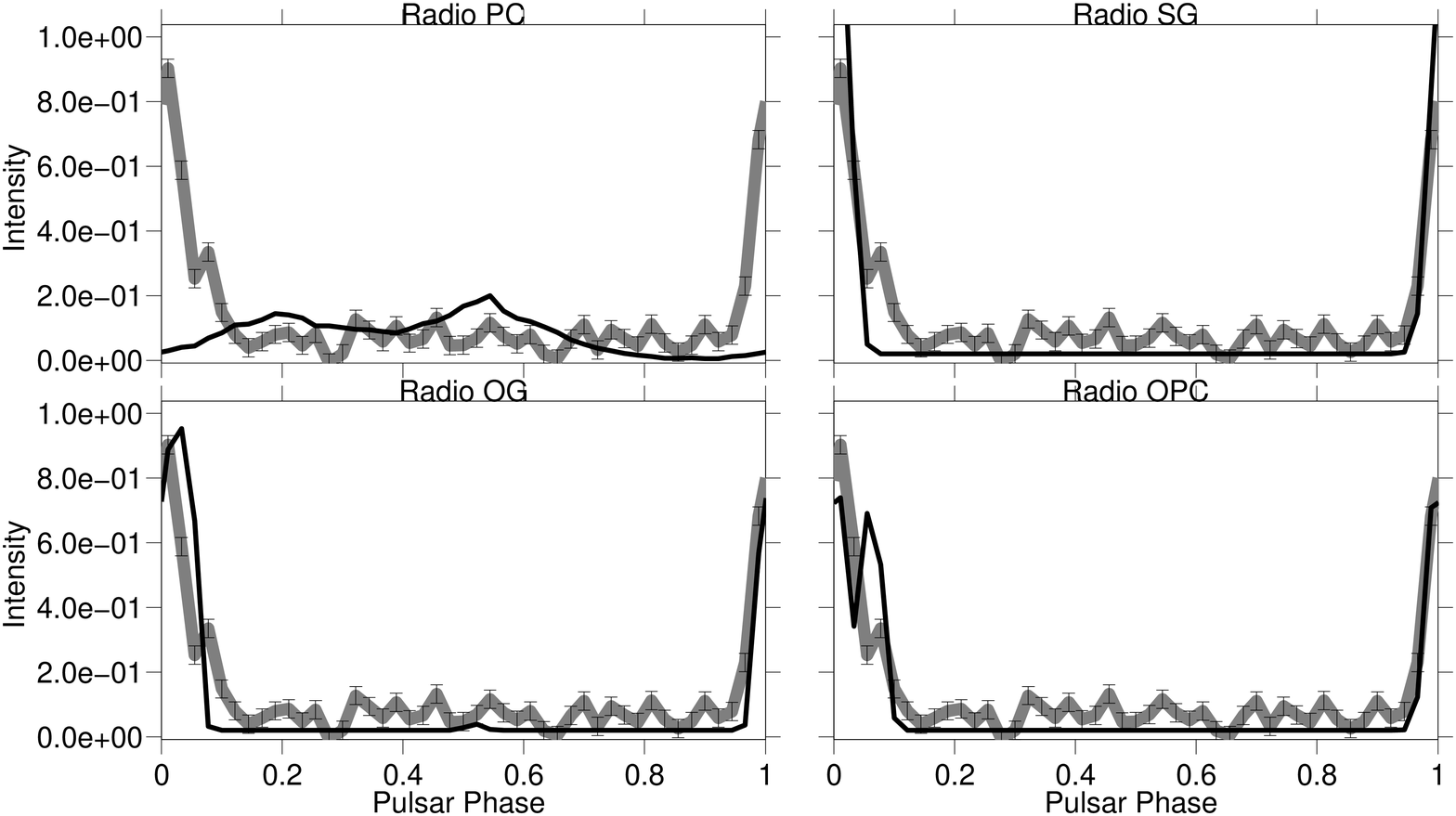}
\caption{PSR J1509-5850. \emph{Top}: for each model the best joint fit solution $\gamma$-ray light-curve (thick black line) is superimposed on the LAT pulsar $\gamma$-ray light-curve (shaded histogram). The estimated background is indicated by the dash-dot line. \emph{Bottom}: for each model the best joint fit solution radio light-curve (black line) is  is superimposed on the LAT pulsar radio light-curve (grey thick line).  The radio model is unique, but the $(\alpha,\zeta)$ solutions vary for each $\gamma$-ray model.}
\label{fitJoint_GmR23}
\end{figure}
  
\clearpage
\begin{figure}[htbp!]
\centering
\includegraphics[width=0.9\textwidth]{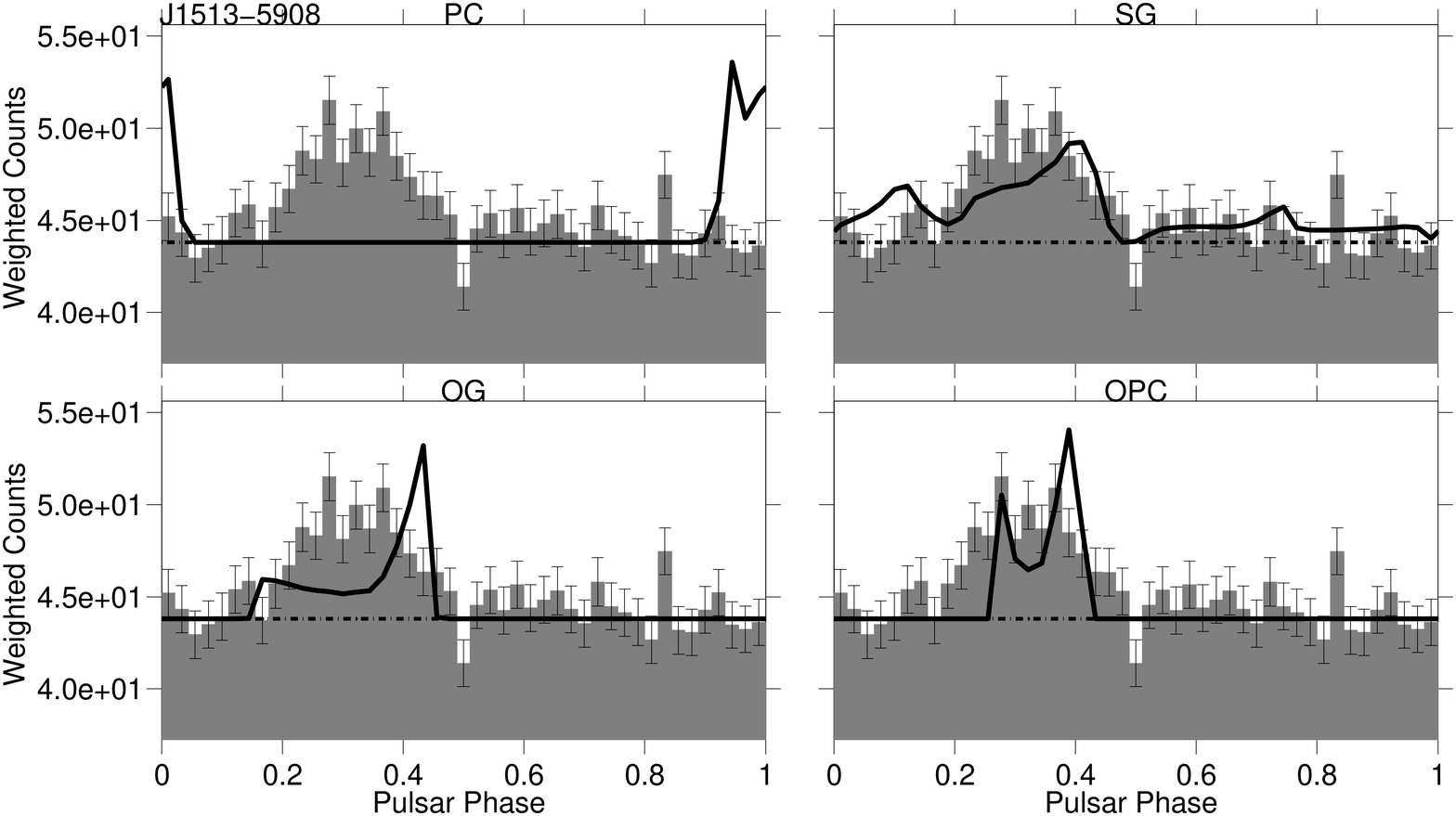}
\includegraphics[width=0.9\textwidth]{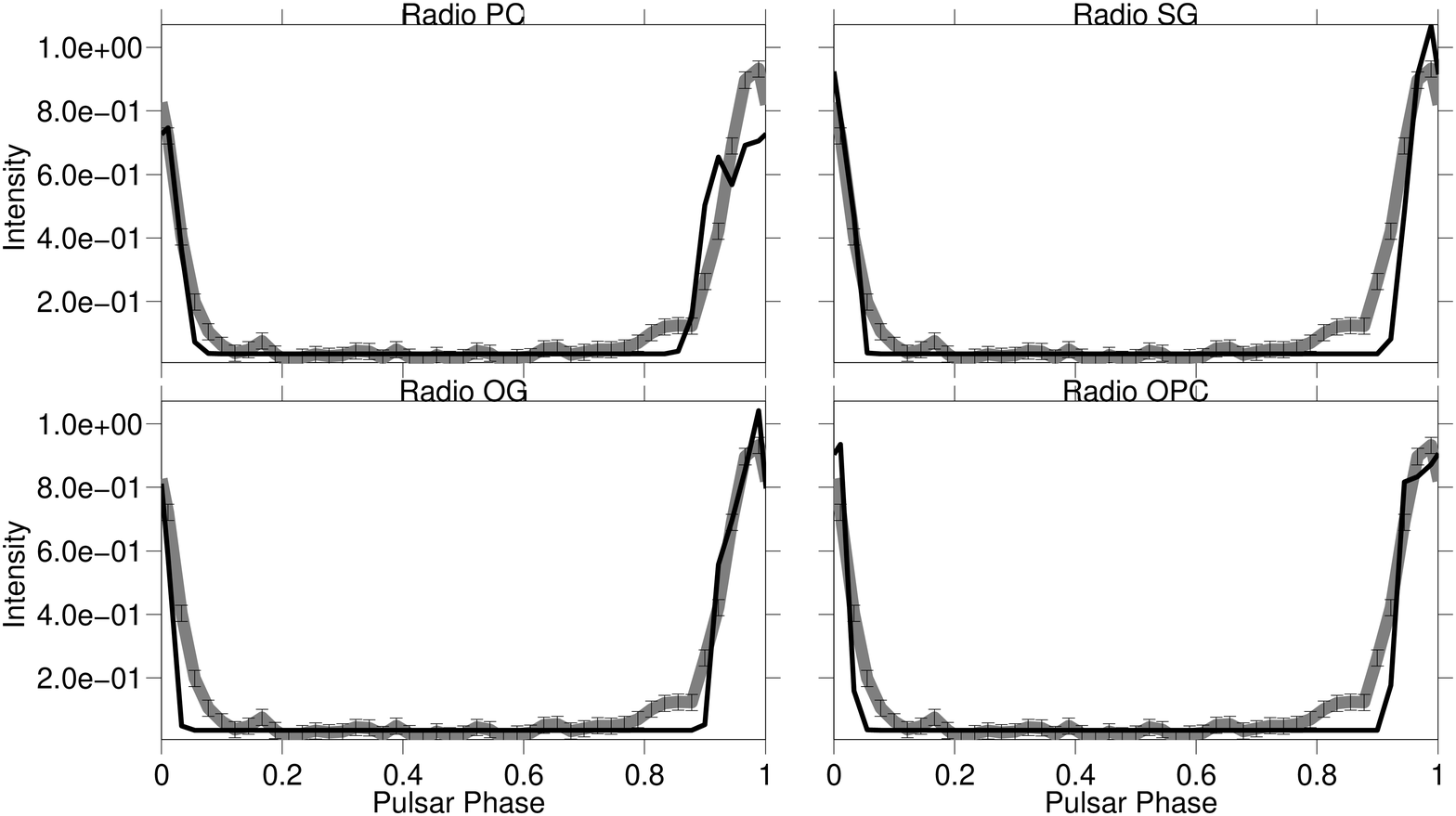}
\caption{PSR J1513-5908. \emph{Top}: for each model the best joint fit solution $\gamma$-ray light-curve (thick black line) is superimposed on the LAT pulsar $\gamma$-ray light-curve (shaded histogram). The estimated background is indicated by the dash-dot line. \emph{Bottom}: for each model the best joint fit solution radio light-curve (black line) is  is superimposed on the LAT pulsar radio light-curve (grey thick line).  The radio model is unique, but the $(\alpha,\zeta)$ solutions vary for each $\gamma$-ray model.}
\label{fitJoint_GmR24}
\end{figure}
  
\clearpage
\begin{figure}[htbp!]
\centering
\includegraphics[width=0.9\textwidth]{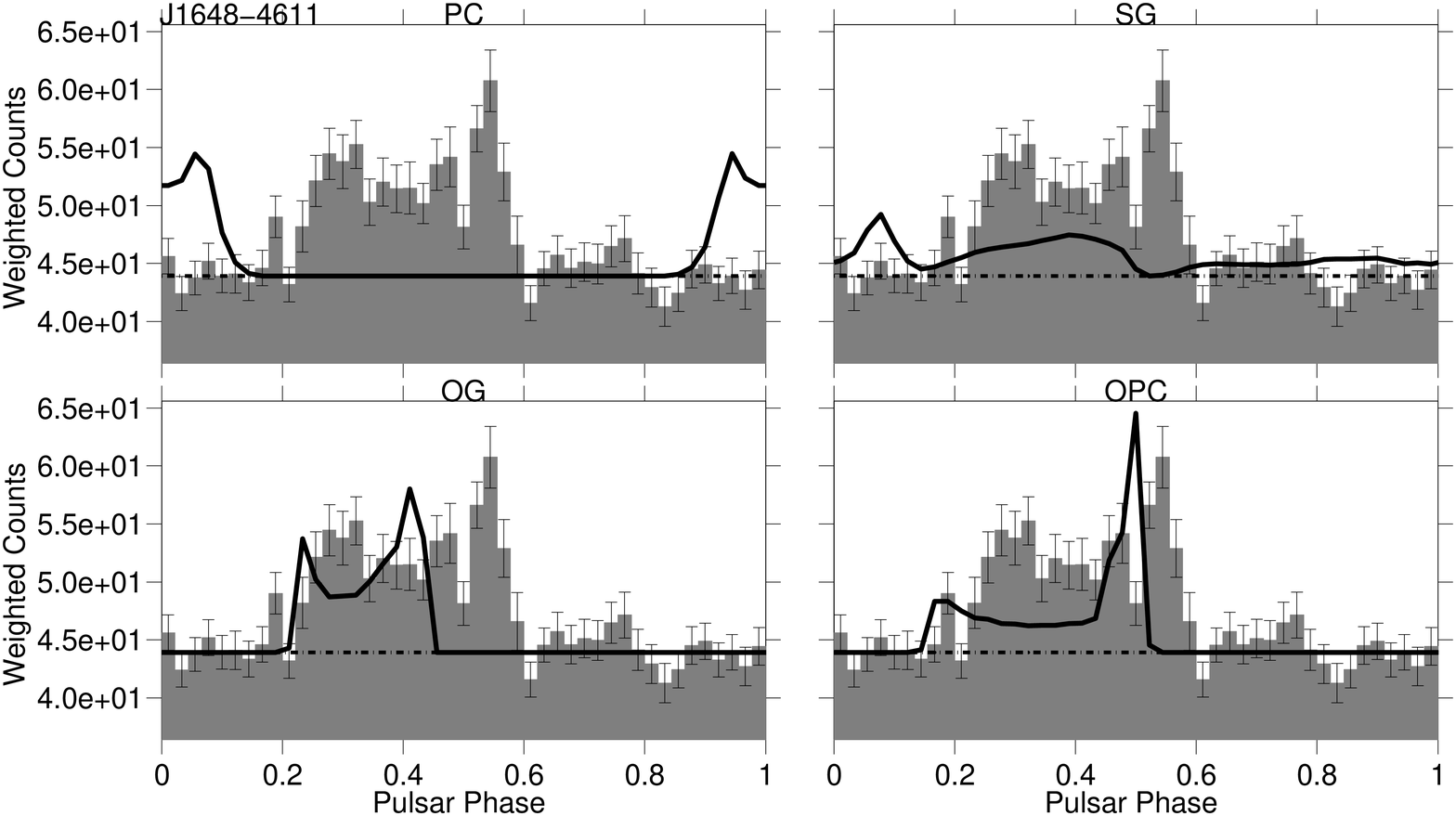}
\includegraphics[width=0.9\textwidth]{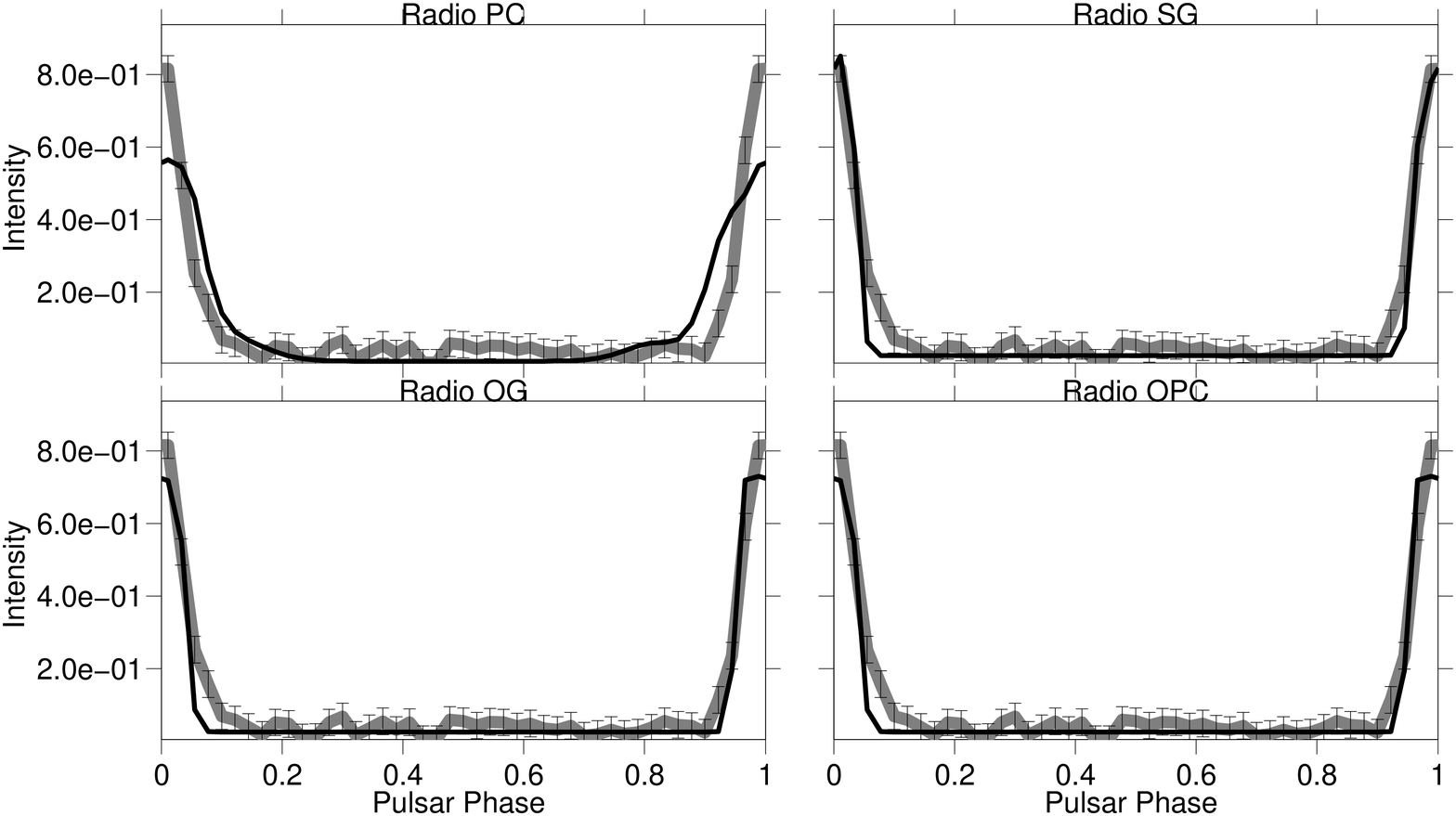}
\caption{PSR J1648-4611. \emph{Top}: for each model the best joint fit solution $\gamma$-ray light-curve (thick black line) is superimposed on the LAT pulsar $\gamma$-ray light-curve (shaded histogram). The estimated background is indicated by the dash-dot line. \emph{Bottom}: for each model the best joint fit solution radio light-curve (black line) is  is superimposed on the LAT pulsar radio light-curve (grey thick line).  The radio model is unique, but the $(\alpha,\zeta)$ solutions vary for each $\gamma$-ray model.}
\label{fitJoint_GmR25}
\end{figure}
  
\clearpage
\begin{figure}[htbp!]
\centering
\includegraphics[width=0.9\textwidth]{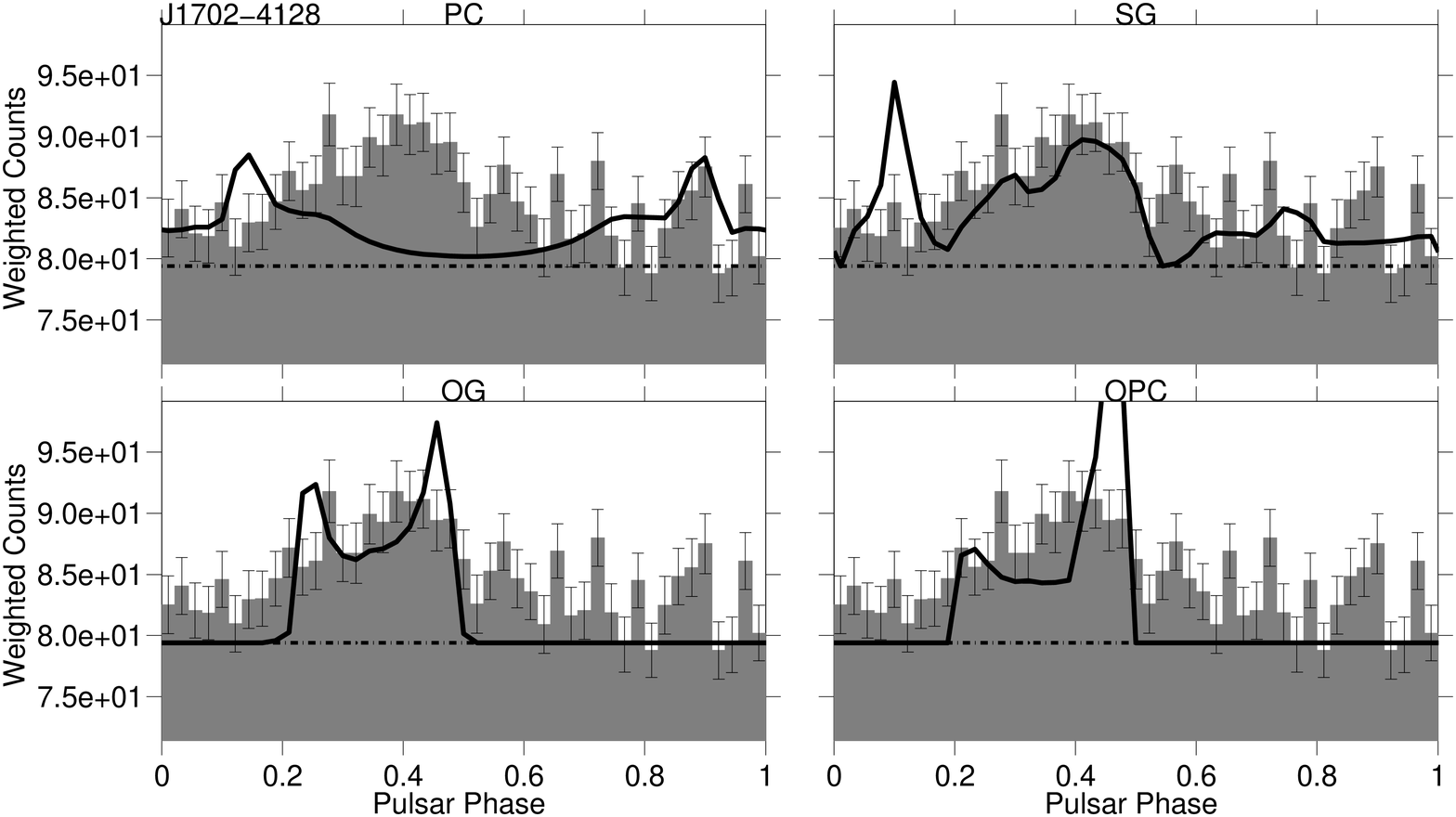}
\includegraphics[width=0.9\textwidth]{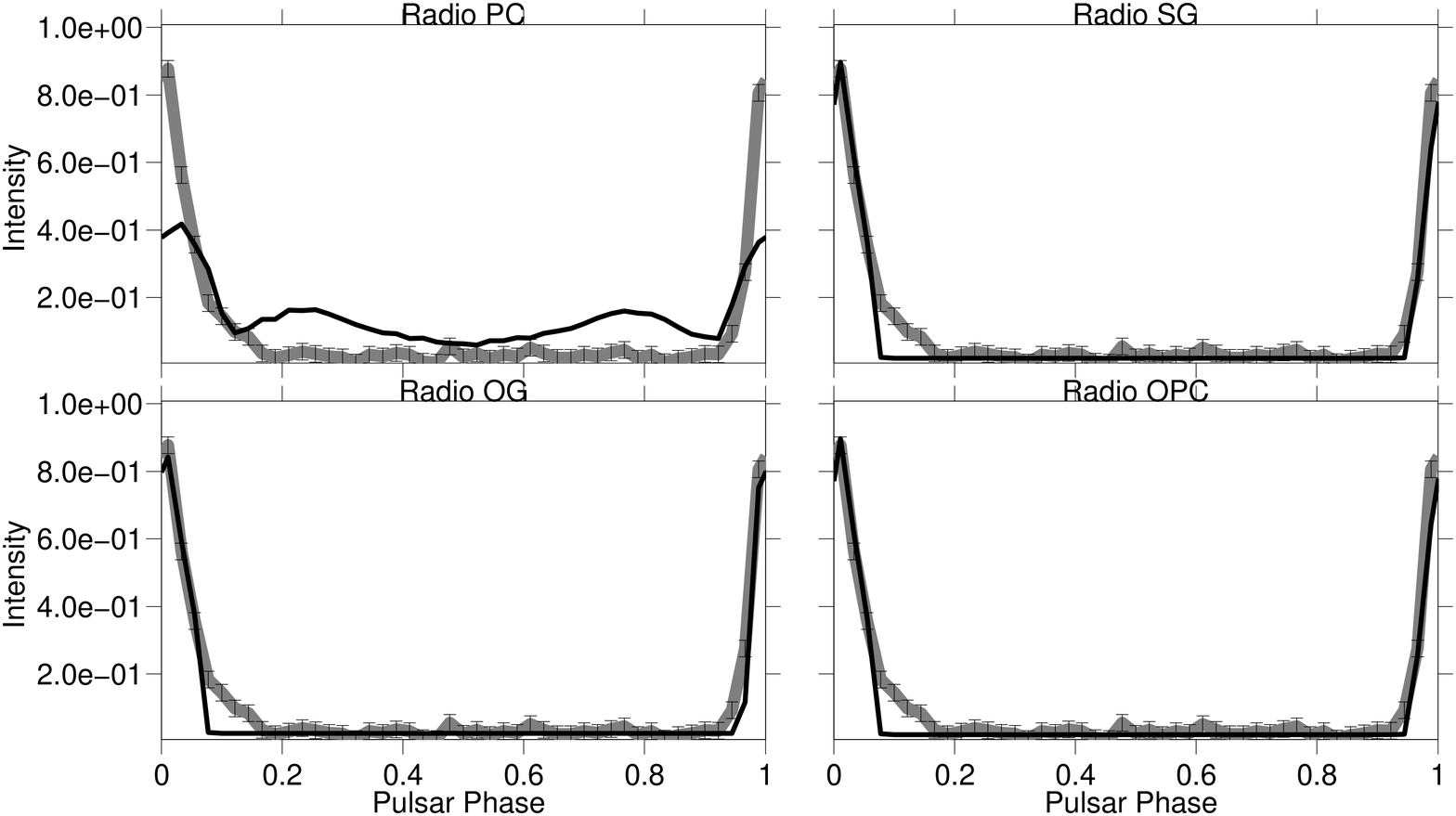}
\caption{PSR J1702-4128. \emph{Top}: for each model the best joint fit solution $\gamma$-ray light-curve (thick black line) is superimposed on the LAT pulsar $\gamma$-ray light-curve (shaded histogram). The estimated background is indicated by the dash-dot line. \emph{Bottom}: for each model the best joint fit solution radio light-curve (black line) is  is superimposed on the LAT pulsar radio light-curve (grey thick line).  The radio model is unique, but the $(\alpha,\zeta)$ solutions vary for each $\gamma$-ray model.}
\label{fitJoint_GmR26}
\end{figure}
  
\clearpage
\begin{figure}[htbp!]
\centering
\includegraphics[width=0.9\textwidth]{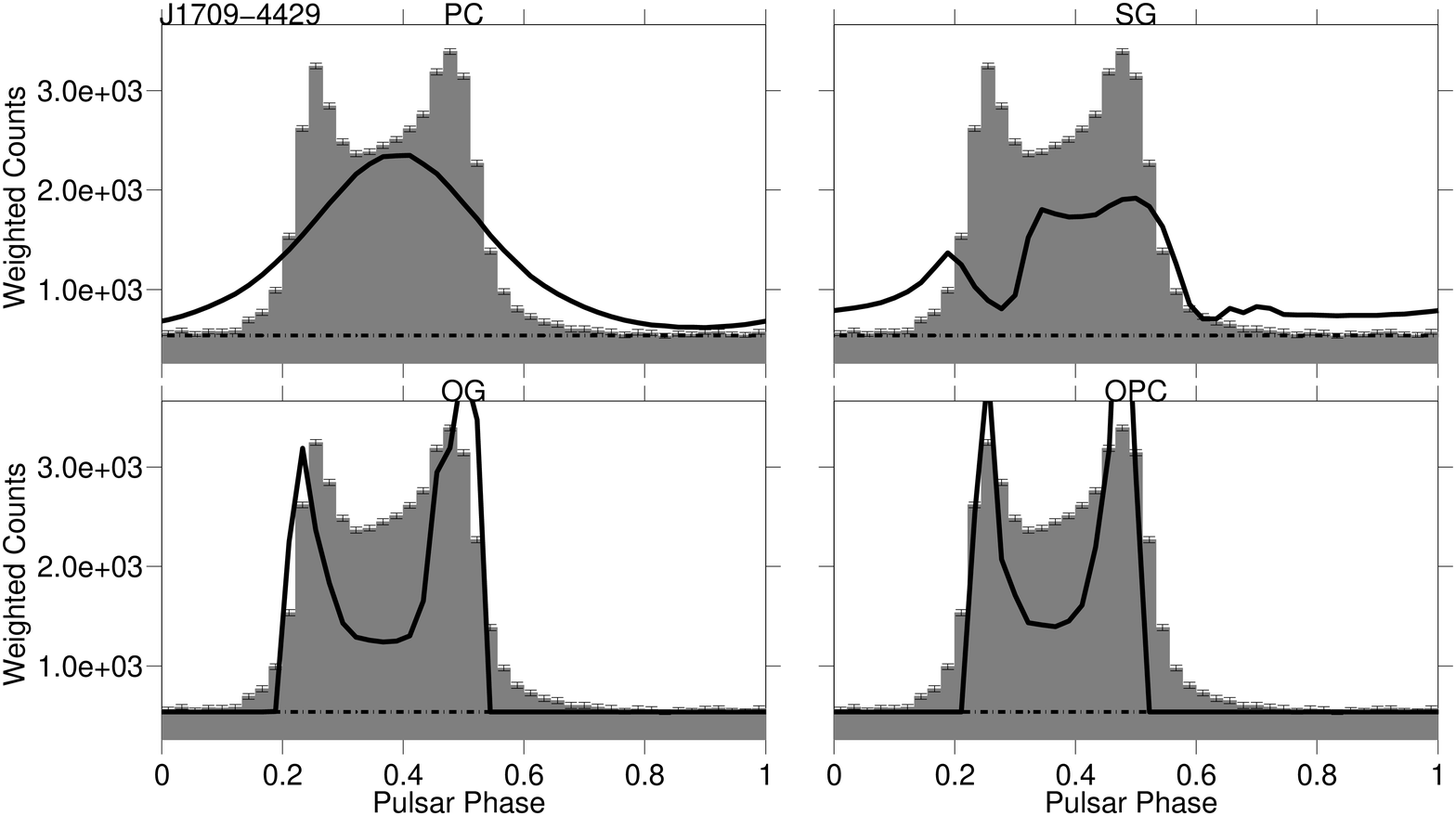}
\includegraphics[width=0.9\textwidth]{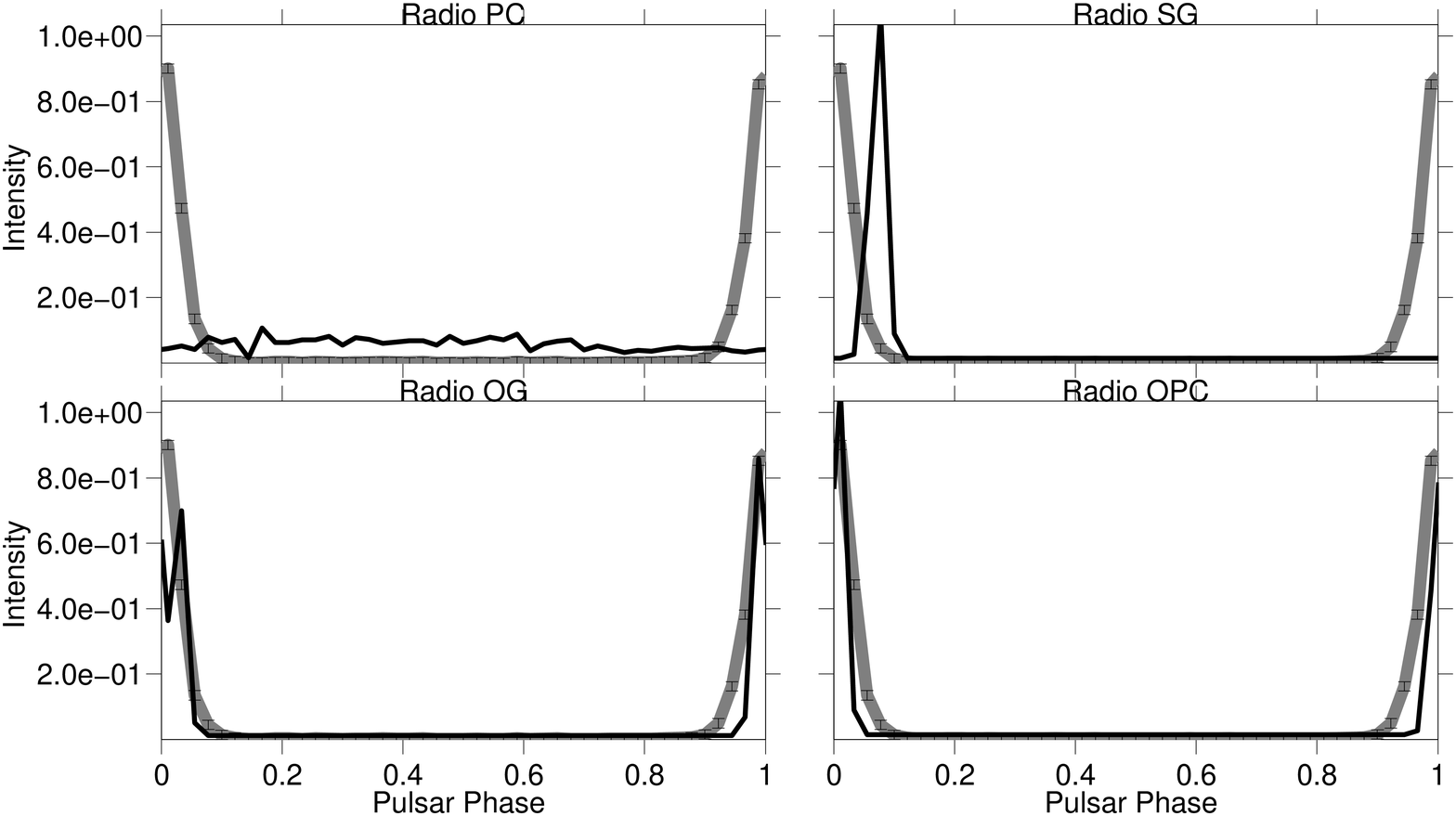}
\caption{PSR J1709-4429. \emph{Top}: for each model the best joint fit solution $\gamma$-ray light-curve (thick black line) is superimposed on the LAT pulsar $\gamma$-ray light-curve (shaded histogram). The estimated background is indicated by the dash-dot line. \emph{Bottom}: for each model the best joint fit solution radio light-curve (black line) is  is superimposed on the LAT pulsar radio light-curve (grey thick line).  The radio model is unique, but the $(\alpha,\zeta)$ solutions vary for each $\gamma$-ray model.}
\label{fitJoint_GmR27}
\end{figure}
  
\clearpage
\begin{figure}[htbp!]
\centering
\includegraphics[width=0.9\textwidth]{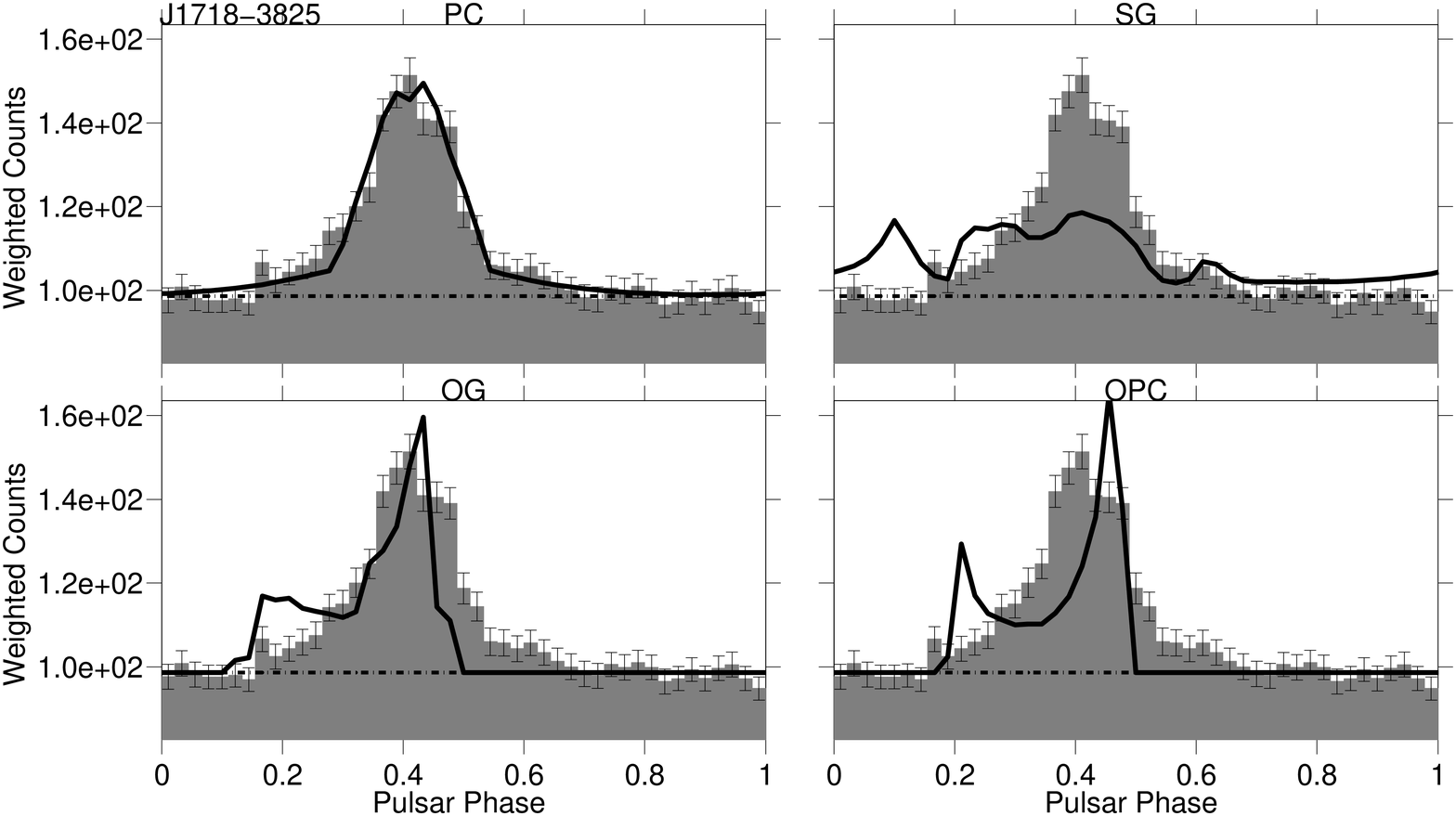}
\includegraphics[width=0.9\textwidth]{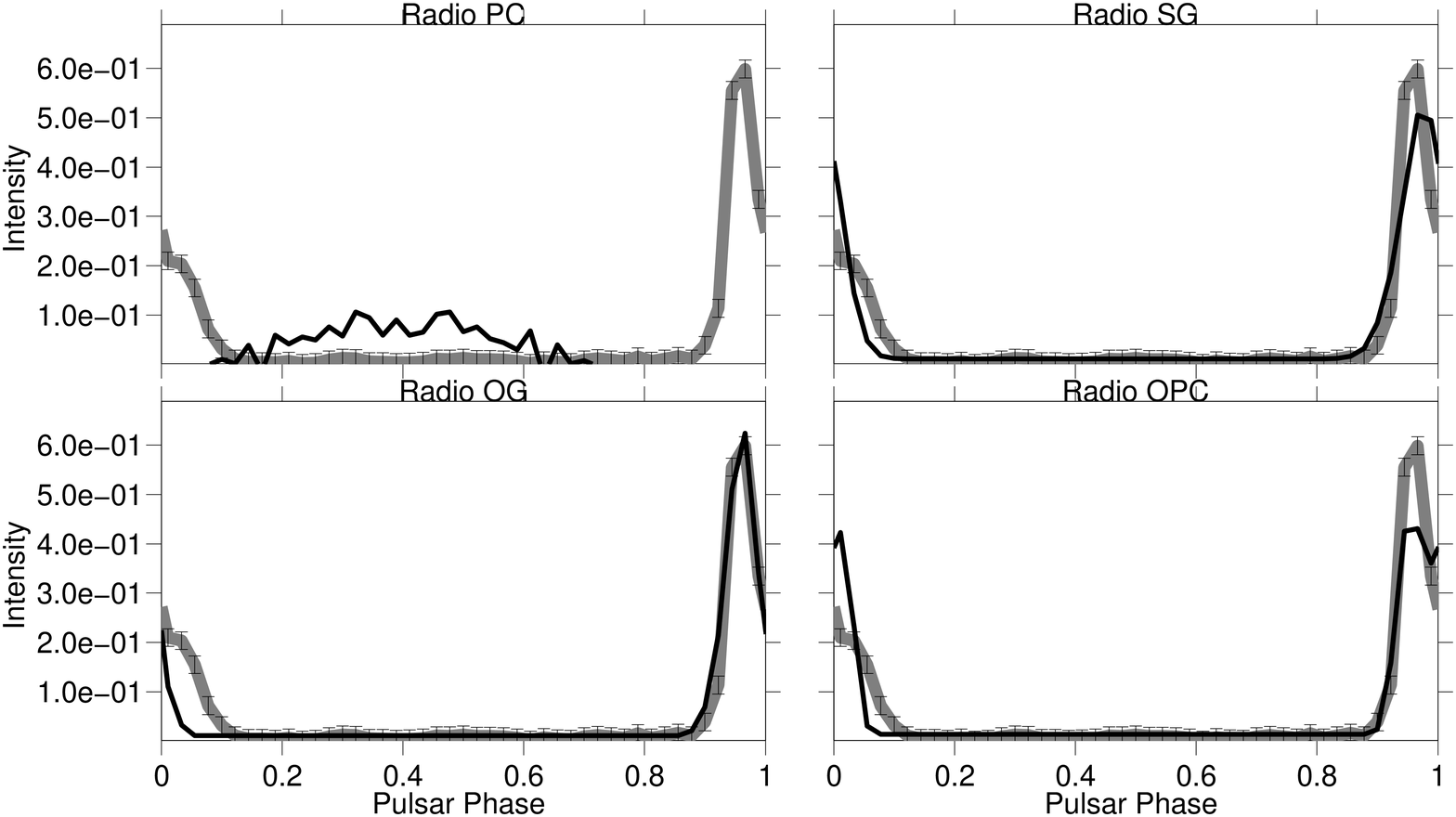}
\caption{PSR J1718-3825. \emph{Top}: for each model the best joint fit solution $\gamma$-ray light-curve (thick black line) is superimposed on the LAT pulsar $\gamma$-ray light-curve (shaded histogram). The estimated background is indicated by the dash-dot line. \emph{Bottom}: for each model the best joint fit solution radio light-curve (black line) is  is superimposed on the LAT pulsar radio light-curve (grey thick line).  The radio model is unique, but the $(\alpha,\zeta)$ solutions vary for each $\gamma$-ray model.}
\label{fitJoint_GmR28}
\end{figure}
  
\clearpage
\begin{figure}[htbp!]
\centering
\includegraphics[width=0.9\textwidth]{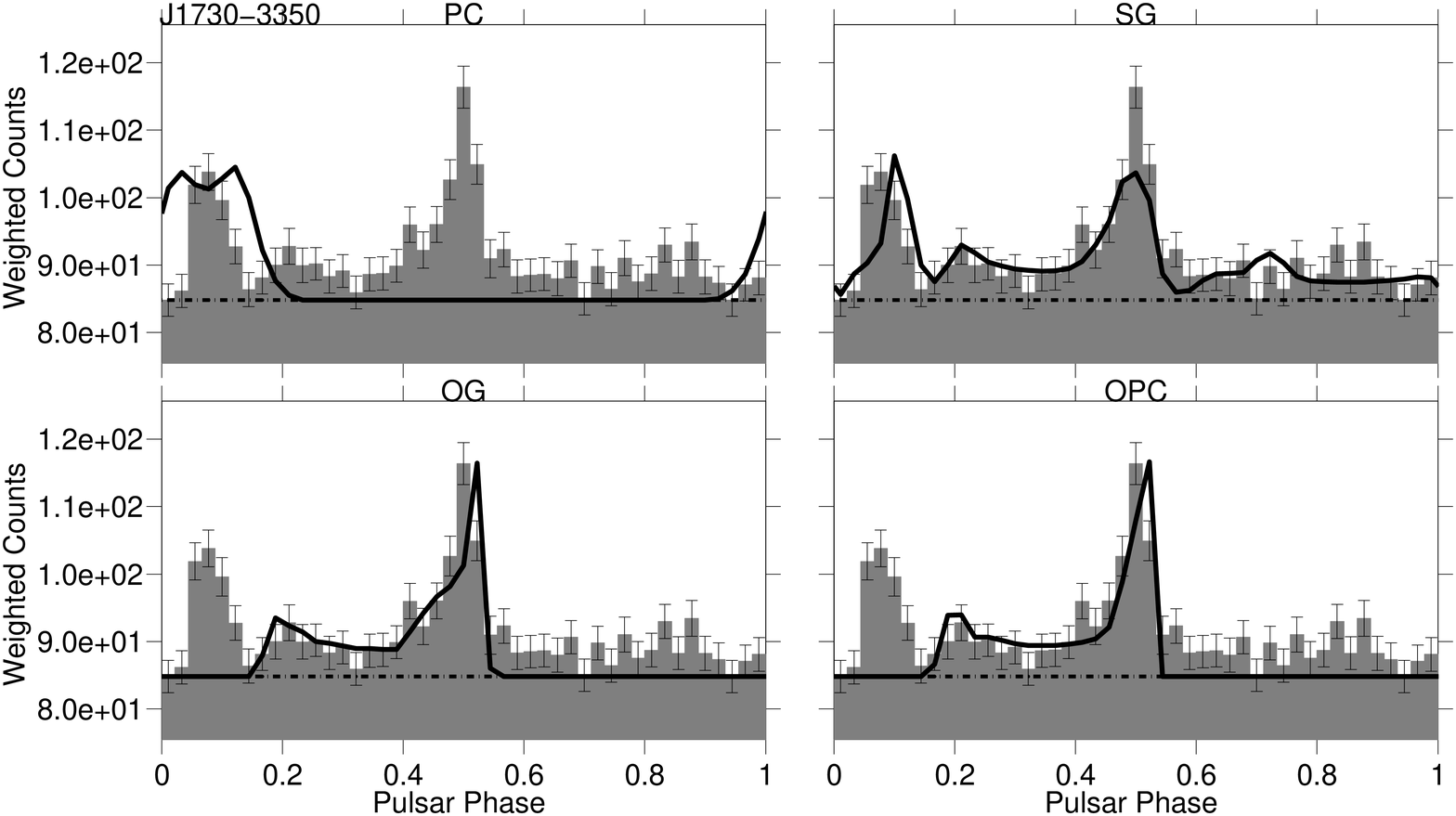}
\includegraphics[width=0.9\textwidth]{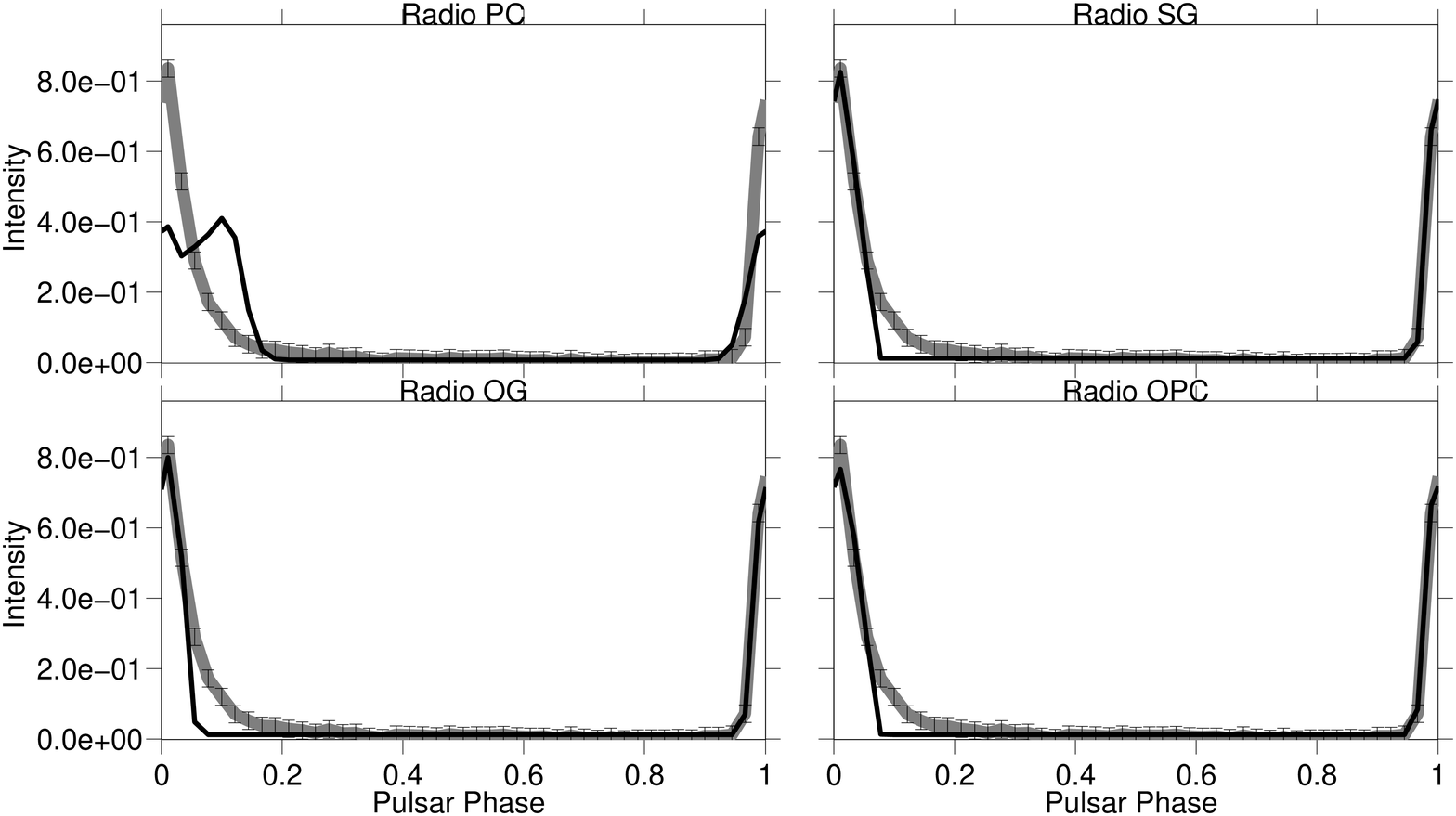}
\caption{PSR J1730-3350. \emph{Top}: for each model the best joint fit solution $\gamma$-ray light-curve (thick black line) is superimposed on the LAT pulsar $\gamma$-ray light-curve (shaded histogram). The estimated background is indicated by the dash-dot line. \emph{Bottom}: for each model the best joint fit solution radio light-curve (black line) is  is superimposed on the LAT pulsar radio light-curve (grey thick line).  The radio model is unique, but the $(\alpha,\zeta)$ solutions vary for each $\gamma$-ray model.}
\label{fitJoint_GmR29}
\end{figure}
  
\clearpage
\begin{figure}[htbp!]
\centering
\includegraphics[width=0.9\textwidth]{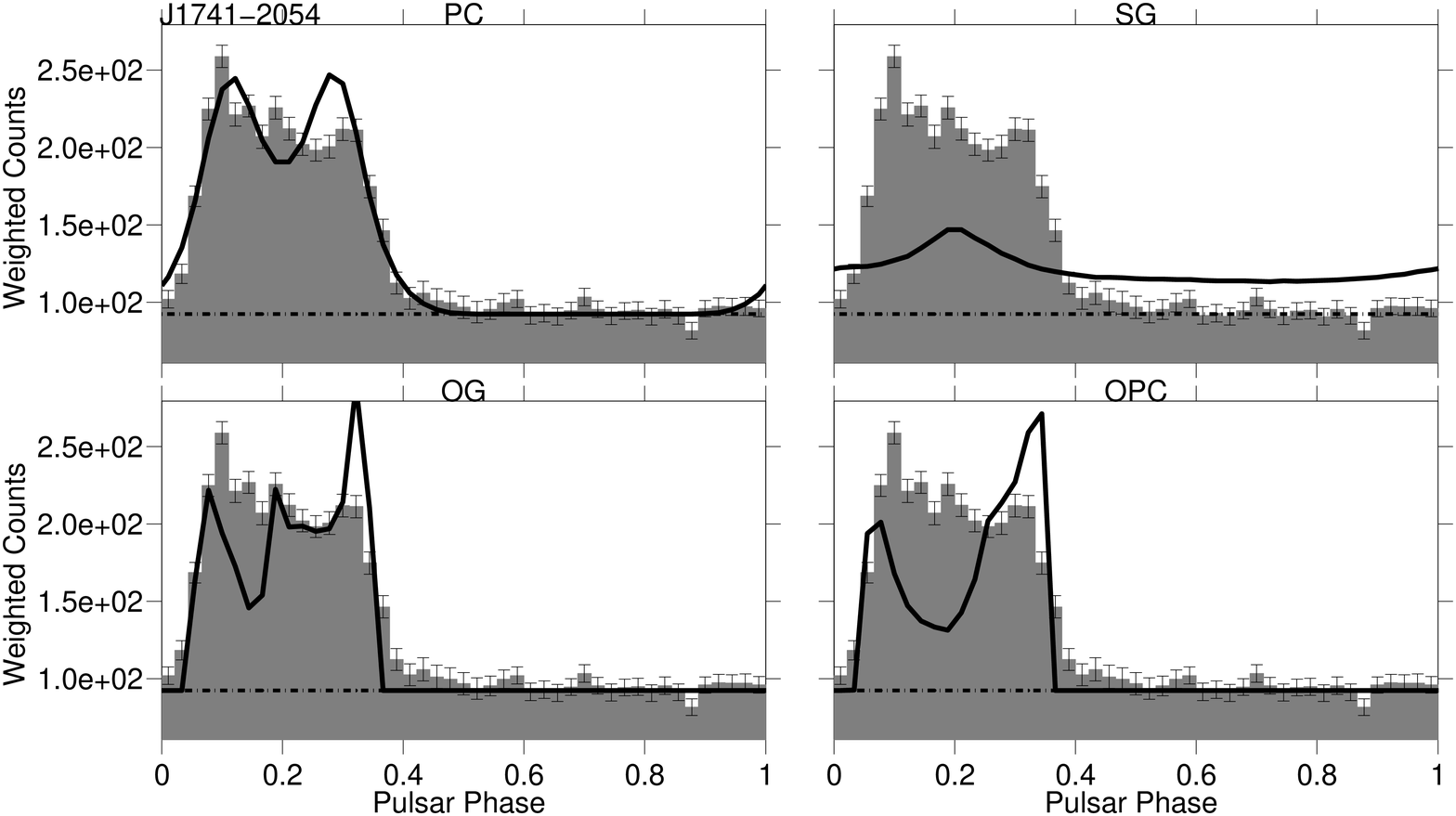}
\includegraphics[width=0.9\textwidth]{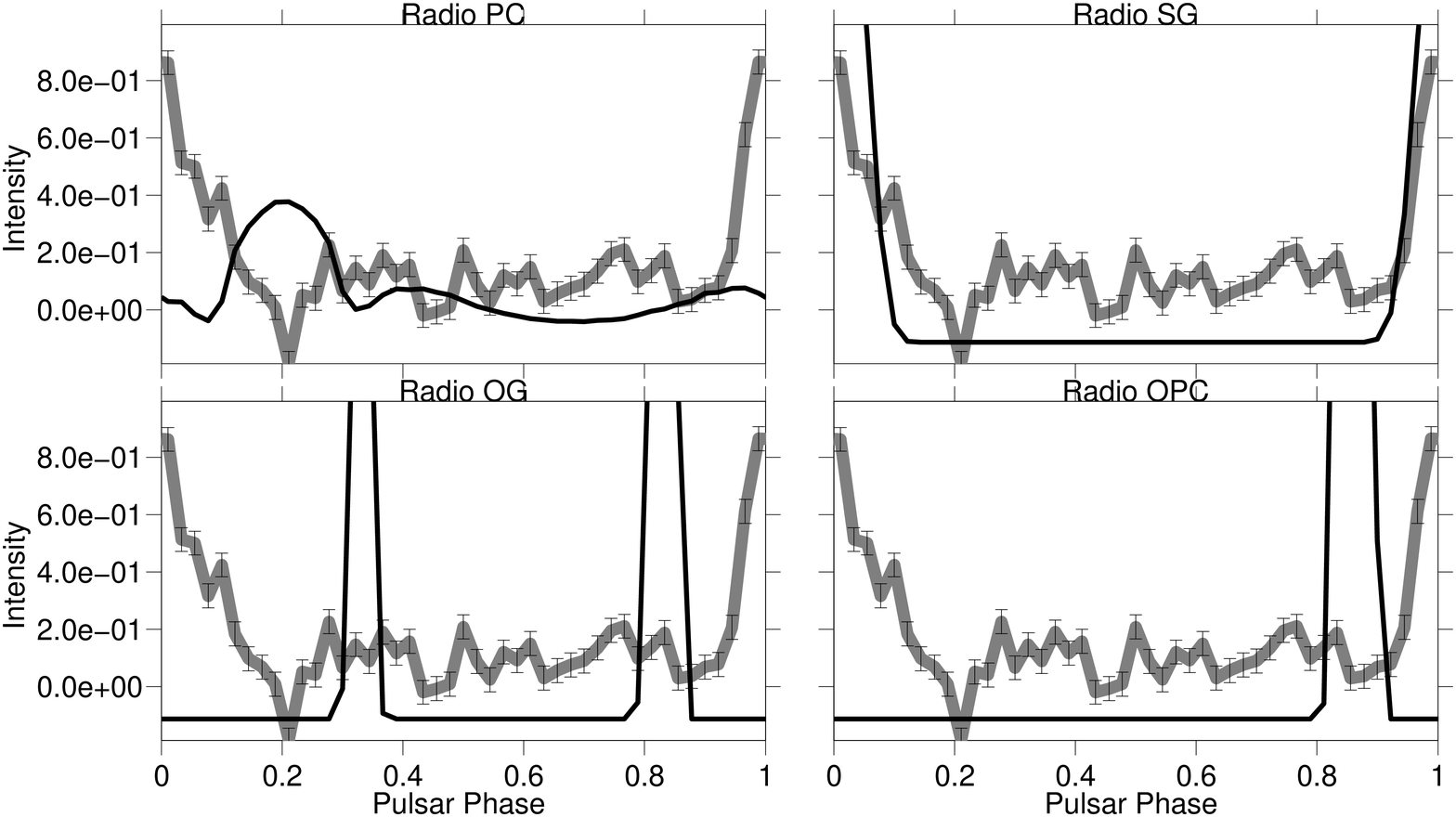}
\caption{PSR J1741-2054. \emph{Top}: for each model the best joint fit solution $\gamma$-ray light-curve (thick black line) is superimposed on the LAT pulsar $\gamma$-ray light-curve (shaded histogram). The estimated background is indicated by the dash-dot line. \emph{Bottom}: for each model the best joint fit solution radio light-curve (black line) is  is superimposed on the LAT pulsar radio light-curve (grey thick line).  The radio model is unique, but the $(\alpha,\zeta)$ solutions vary for each $\gamma$-ray model.}
\label{fitJoint_GmR30}
\end{figure}
  
\clearpage
\begin{figure}[htbp!]
\centering
\includegraphics[width=0.9\textwidth]{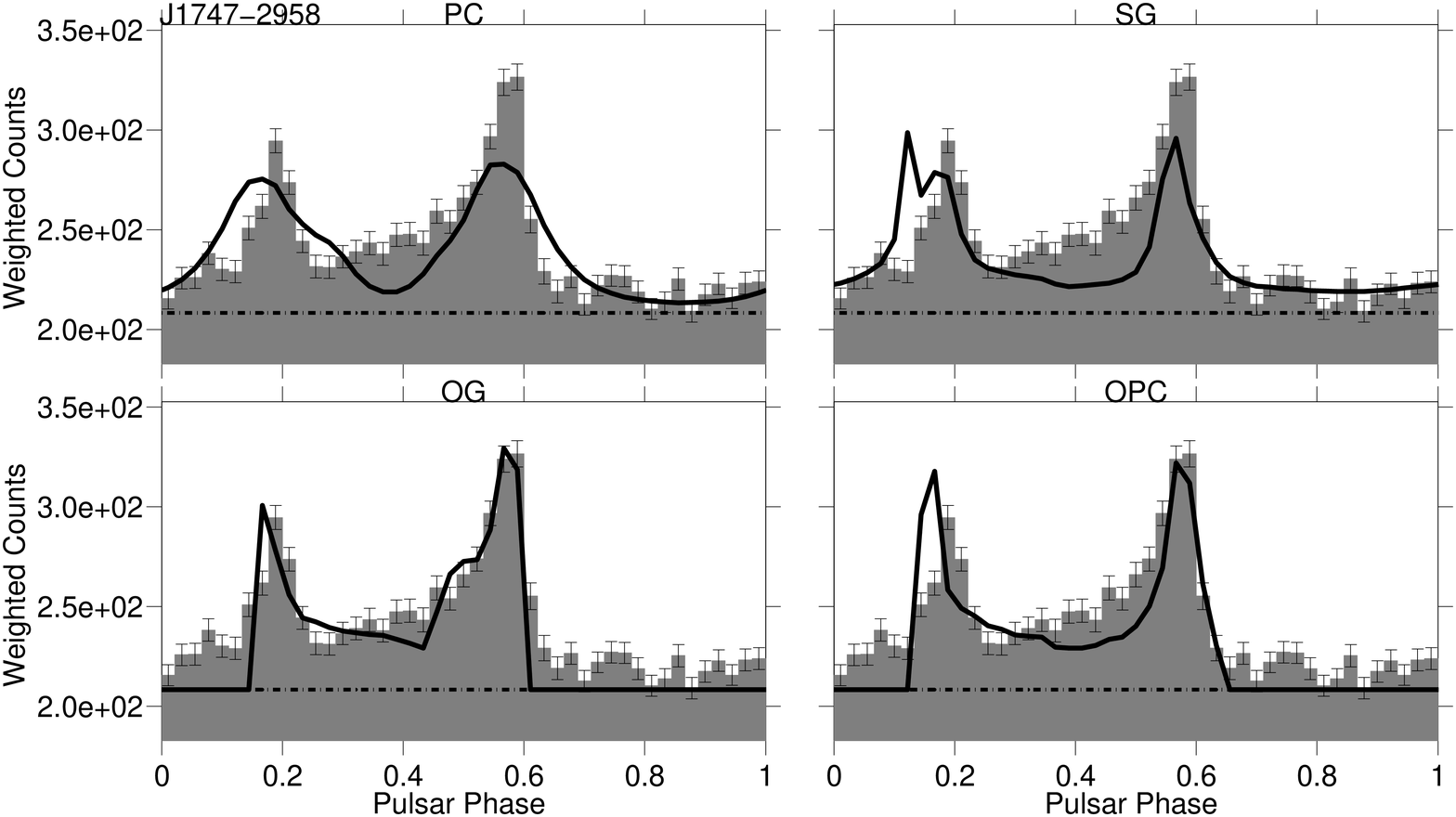}
\includegraphics[width=0.9\textwidth]{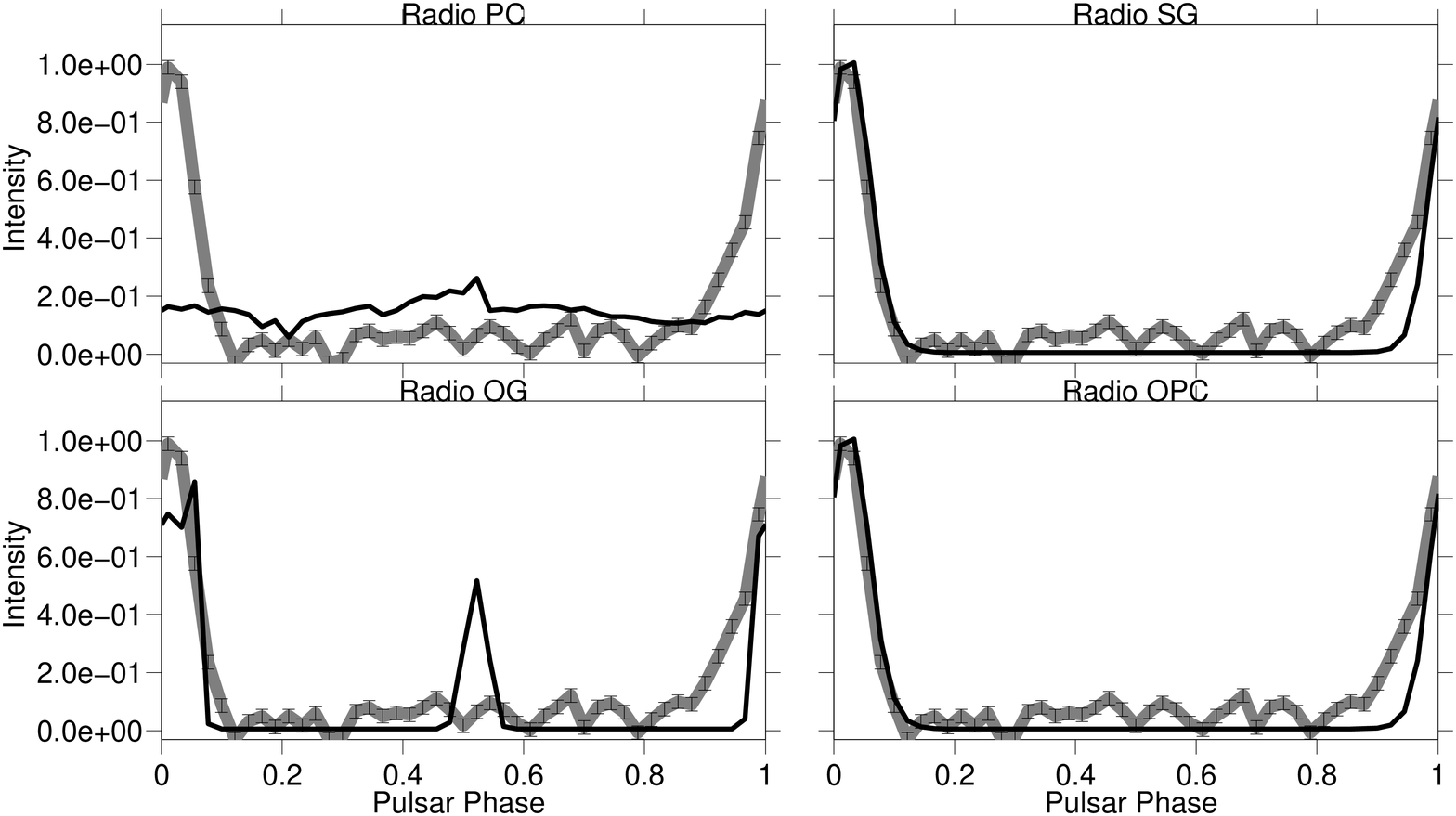}
\caption{PSR J1747-2958. \emph{Top}: for each model the best joint fit solution $\gamma$-ray light-curve (thick black line) is superimposed on the LAT pulsar $\gamma$-ray light-curve (shaded histogram). The estimated background is indicated by the dash-dot line. \emph{Bottom}: for each model the best joint fit solution radio light-curve (black line) is  is superimposed on the LAT pulsar radio light-curve (grey thick line).  The radio model is unique, but the $(\alpha,\zeta)$ solutions vary for each $\gamma$-ray model.}
\label{fitJoint_GmR31}
\end{figure}
  
\clearpage
\begin{figure}[htbp!]
\centering
\includegraphics[width=0.9\textwidth]{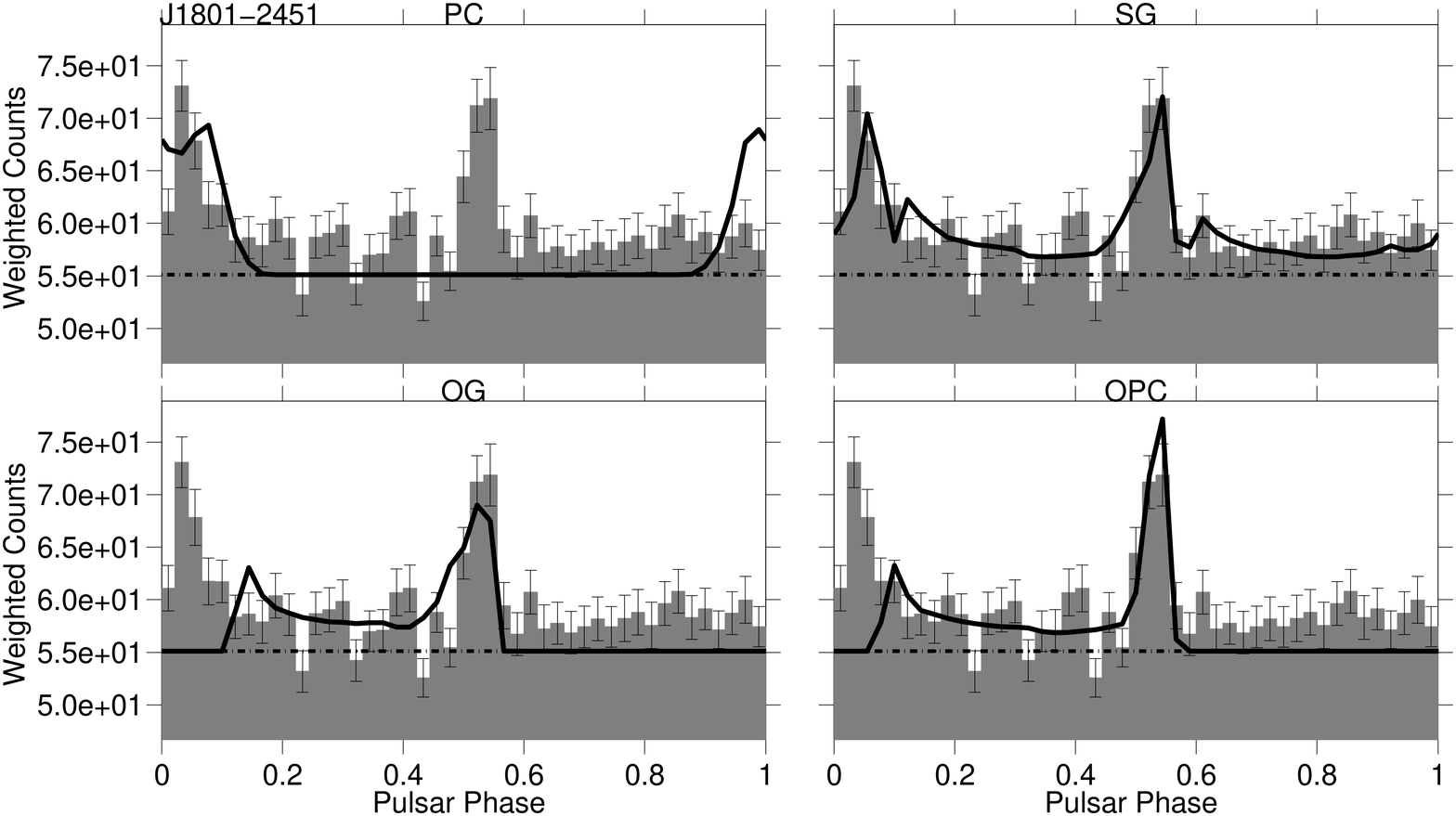}
\includegraphics[width=0.9\textwidth]{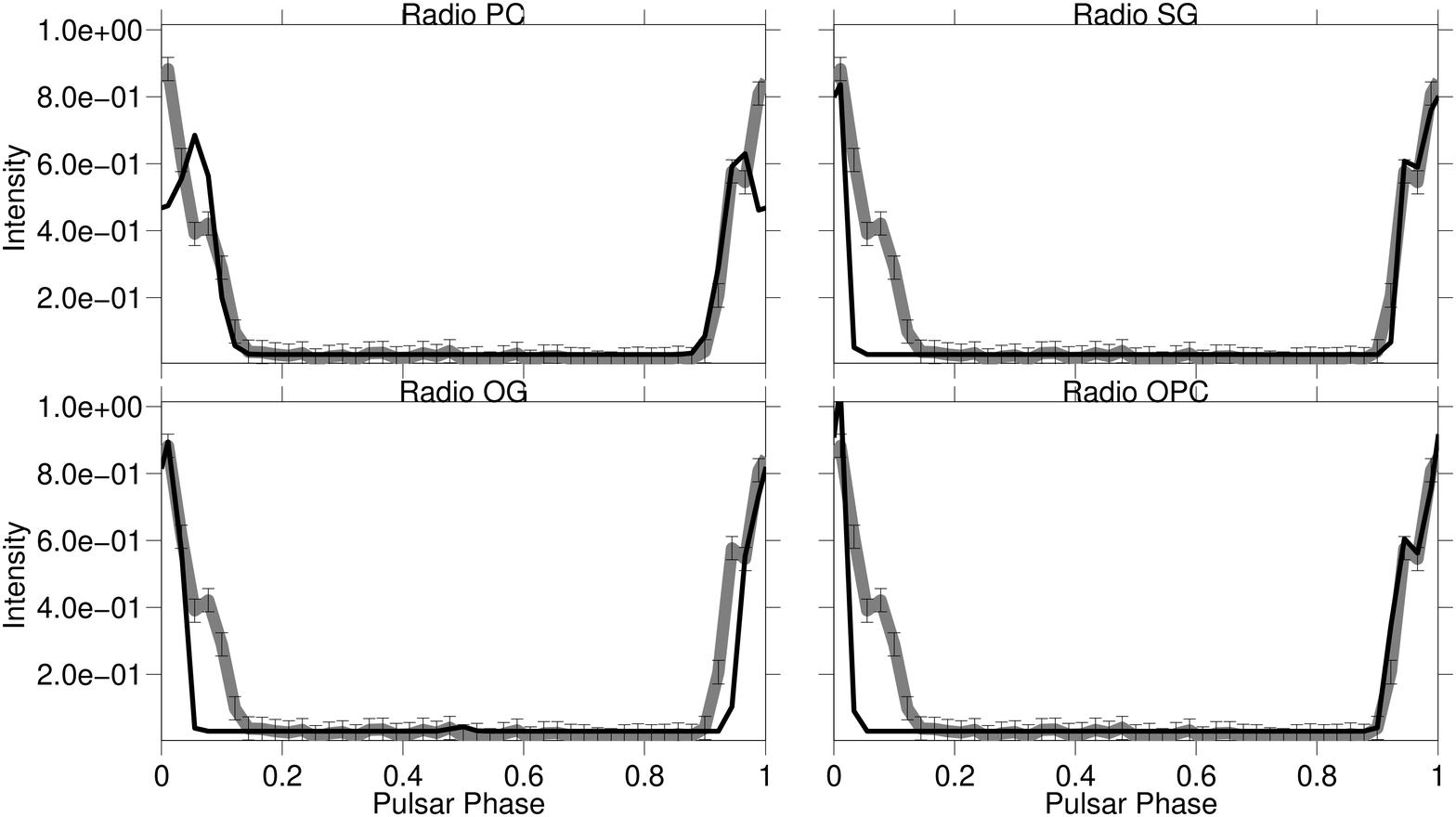}
\caption{PSR J1801-2451. \emph{Top}: for each model the best joint fit solution $\gamma$-ray light-curve (thick black line) is superimposed on the LAT pulsar $\gamma$-ray light-curve (shaded histogram). The estimated background is indicated by the dash-dot line. \emph{Bottom}: for each model the best joint fit solution radio light-curve (black line) is  is superimposed on the LAT pulsar radio light-curve (grey thick line).  The radio model is unique, but the $(\alpha,\zeta)$ solutions vary for each $\gamma$-ray model.}
\label{fitJoint_GmR32}
\end{figure}
  
\clearpage
\begin{figure}[htbp!]
\centering
\includegraphics[width=0.9\textwidth]{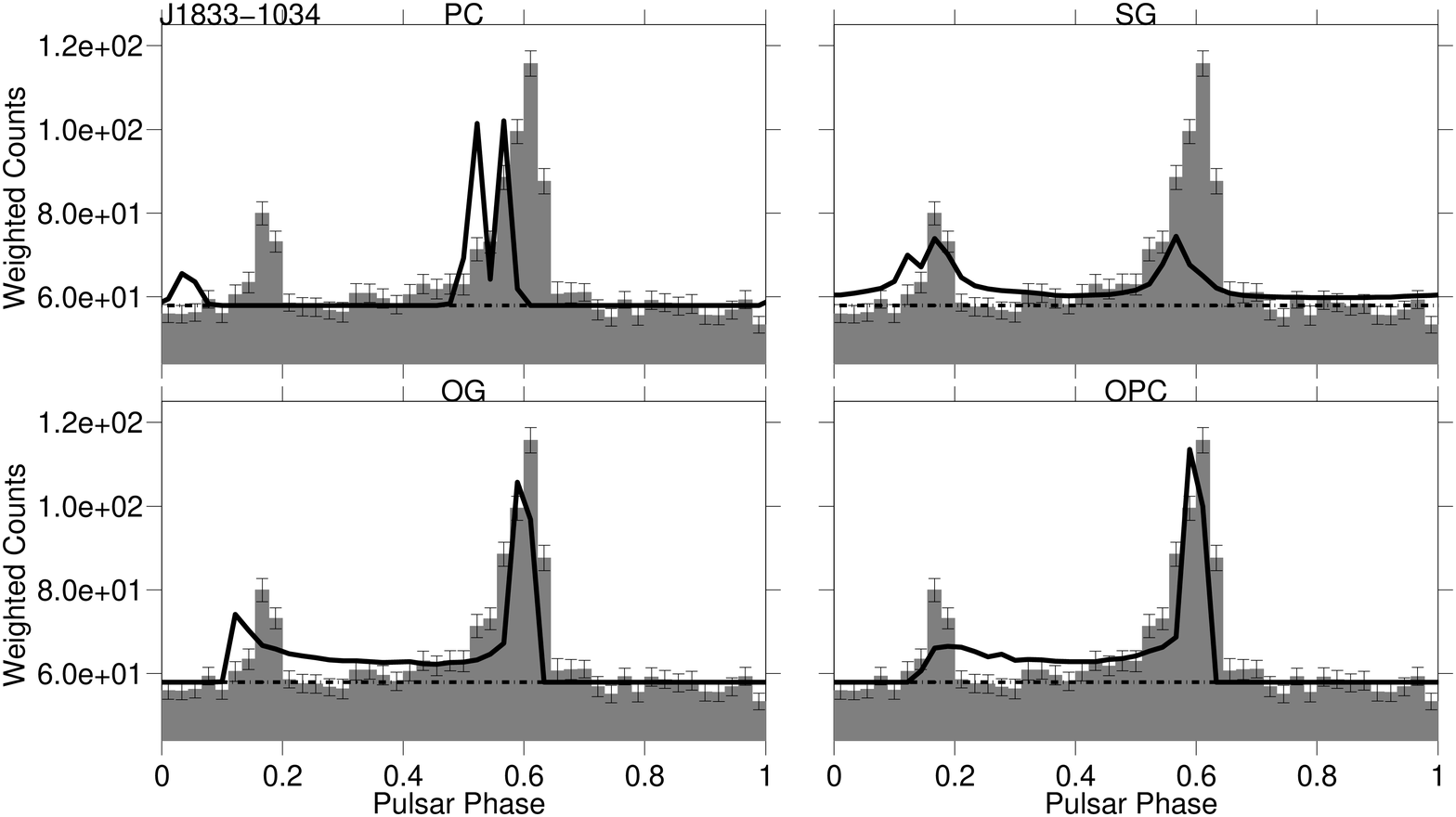}
\includegraphics[width=0.9\textwidth]{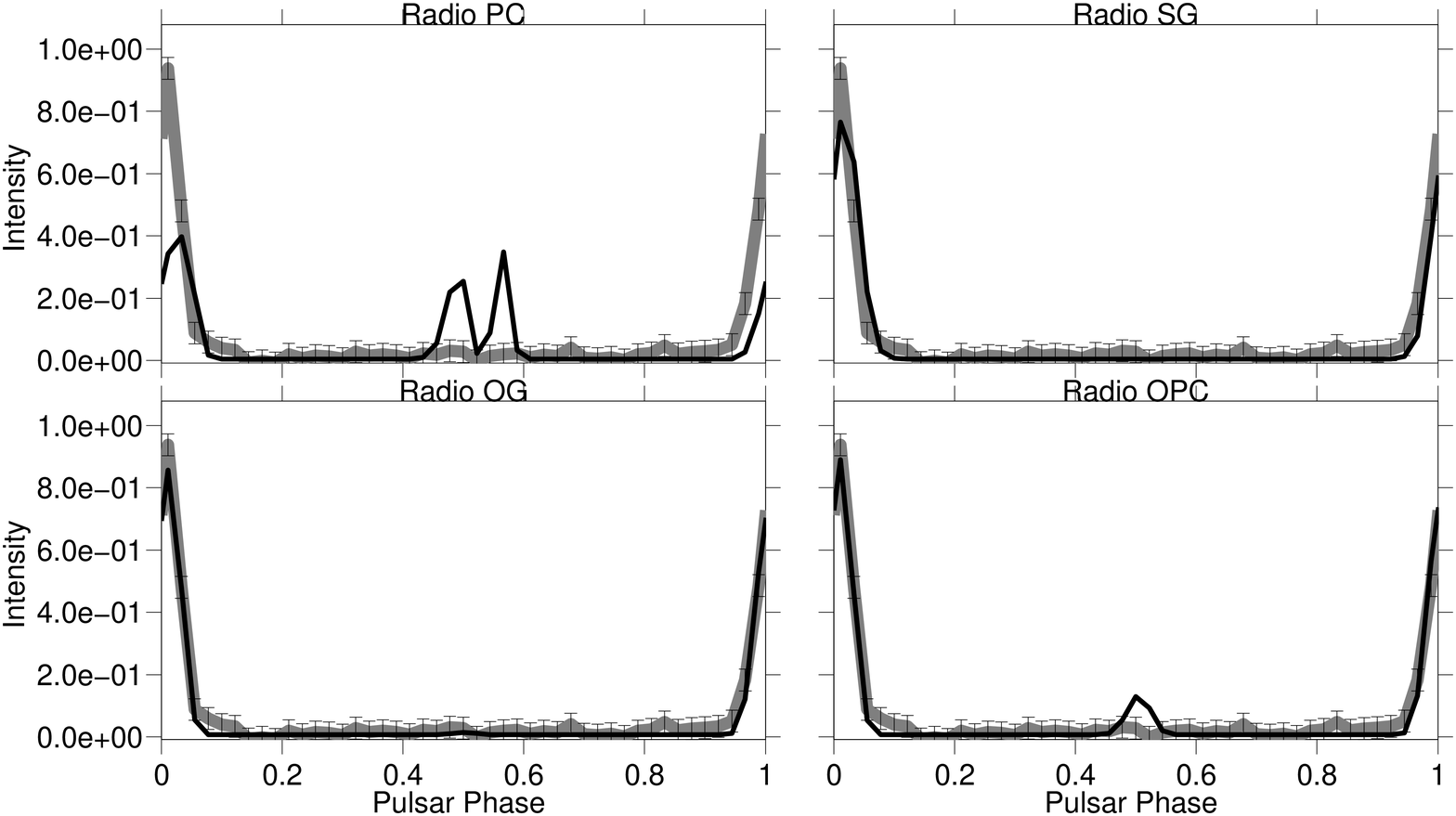}
\caption{PSR J1833-1034. \emph{Top}: for each model the best joint fit solution $\gamma$-ray light-curve (thick black line) is superimposed on the LAT pulsar $\gamma$-ray light-curve (shaded histogram). The estimated background is indicated by the dash-dot line. \emph{Bottom}: for each model the best joint fit solution radio light-curve (black line) is  is superimposed on the LAT pulsar radio light-curve (grey thick line).  The radio model is unique, but the $(\alpha,\zeta)$ solutions vary for each $\gamma$-ray model.}
\label{fitJoint_GmR33}
\end{figure}
  
\clearpage
\begin{figure}[htbp!]
\centering
\includegraphics[width=0.9\textwidth]{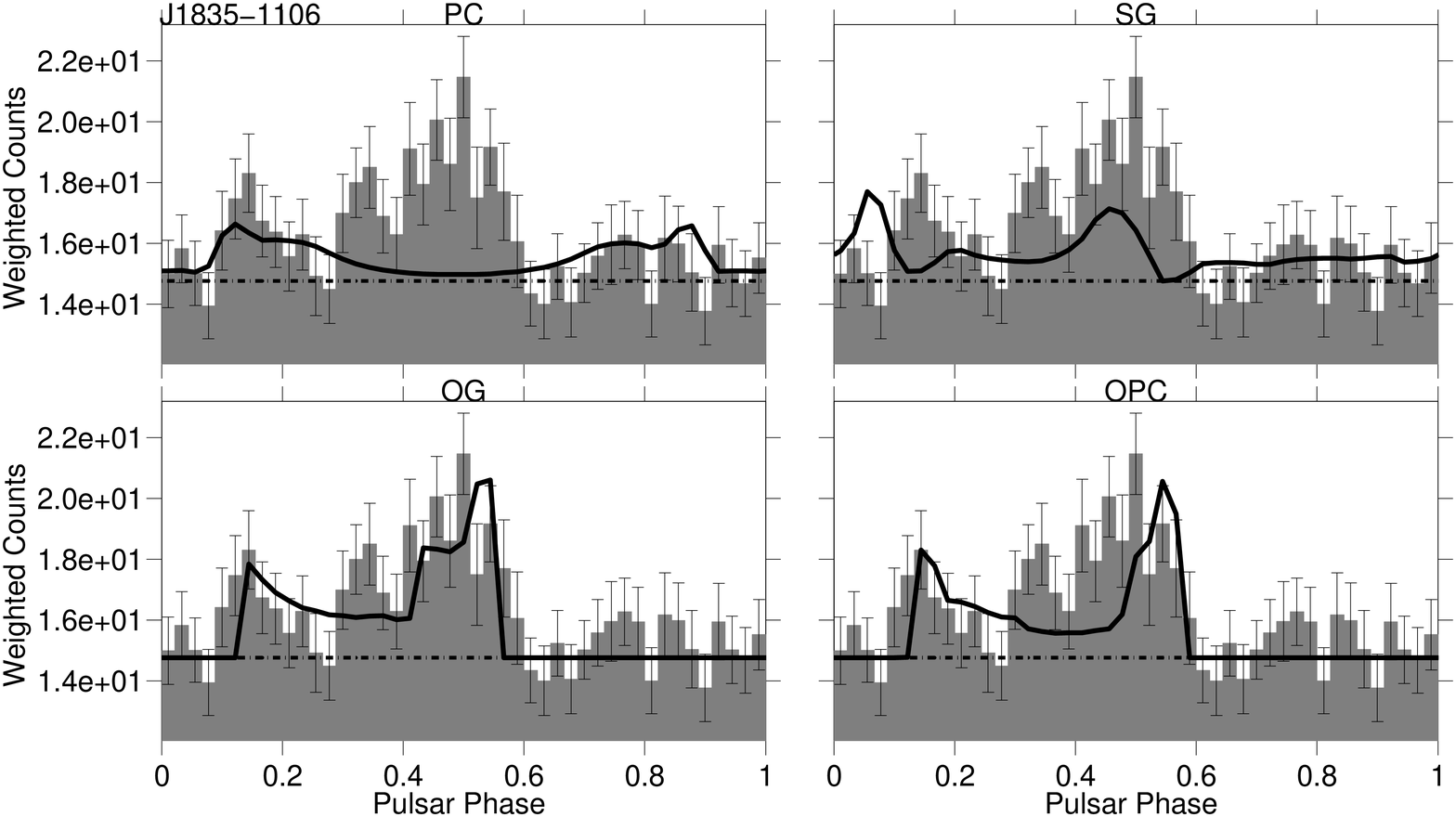}
\includegraphics[width=0.9\textwidth]{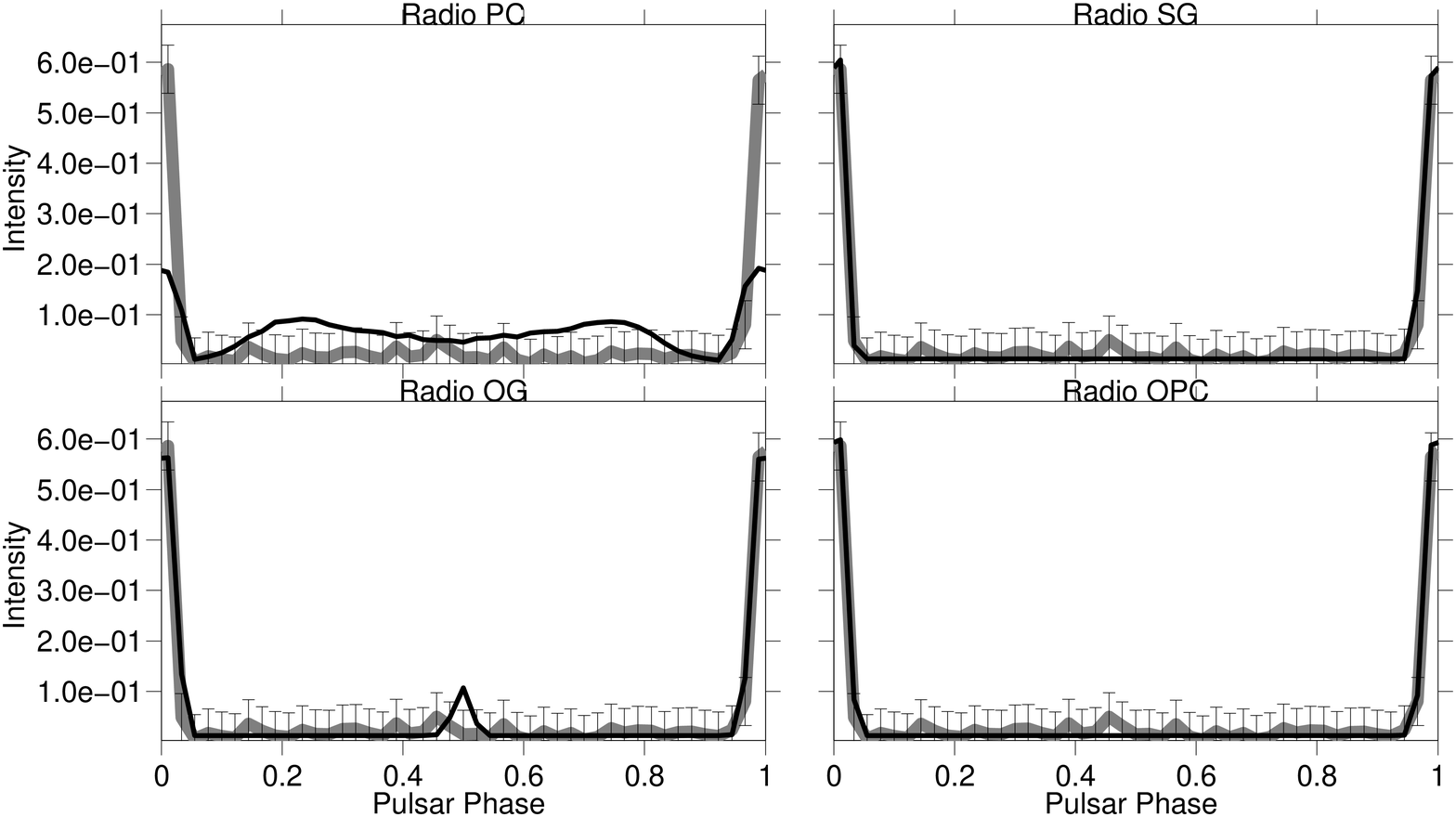}
\caption{PSR J1835-1106. \emph{Top}: for each model the best joint fit solution $\gamma$-ray light-curve (thick black line) is superimposed on the LAT pulsar $\gamma$-ray light-curve (shaded histogram). The estimated background is indicated by the dash-dot line. \emph{Bottom}: for each model the best joint fit solution radio light-curve (black line) is  is superimposed on the LAT pulsar radio light-curve (grey thick line).  The radio model is unique, but the $(\alpha,\zeta)$ solutions vary for each $\gamma$-ray model.}
\label{fitJoint_GmR34}
\end{figure}
  
\clearpage
\begin{figure}[htbp!]
\centering
\includegraphics[width=0.9\textwidth]{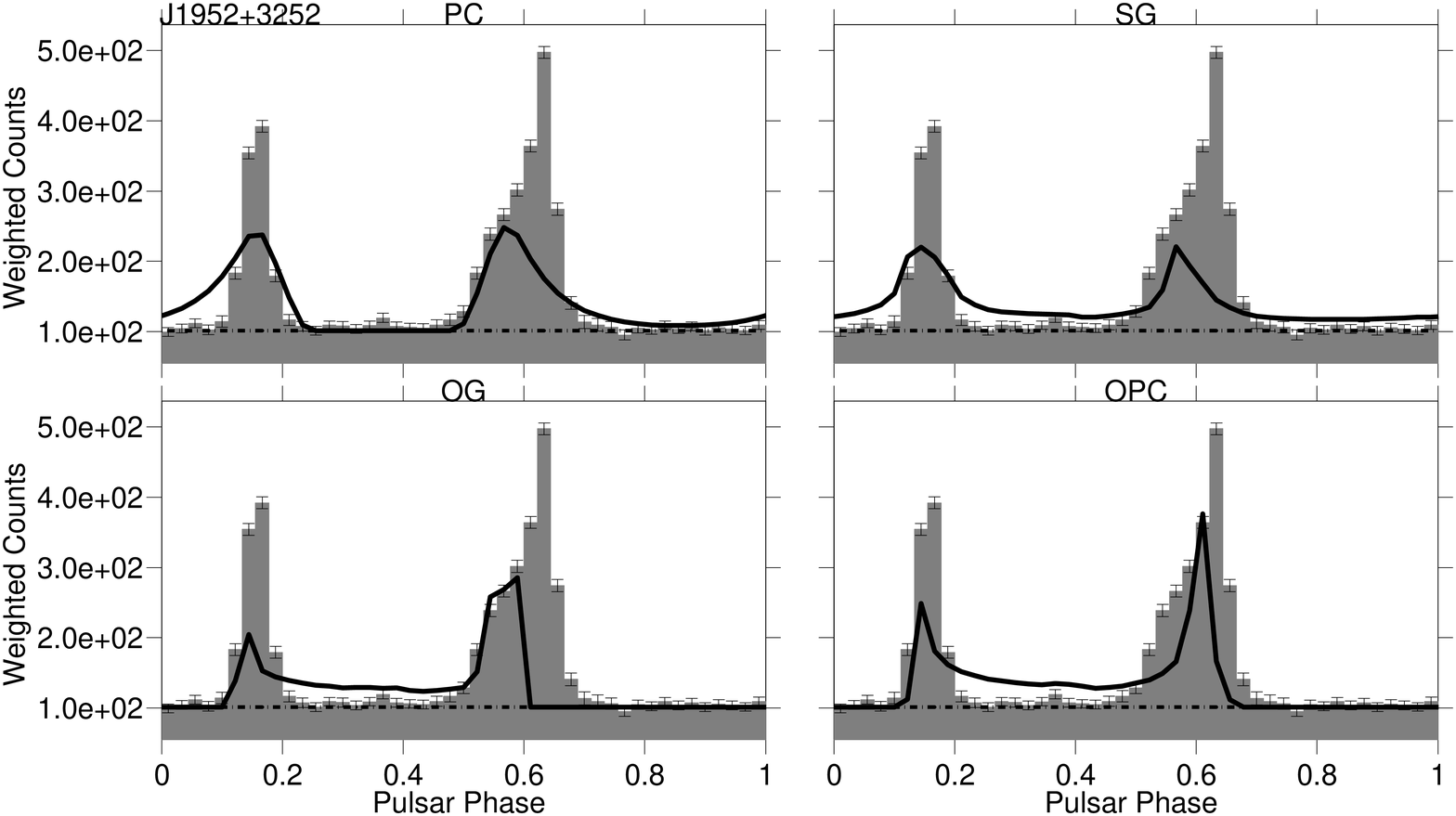}
\includegraphics[width=0.9\textwidth]{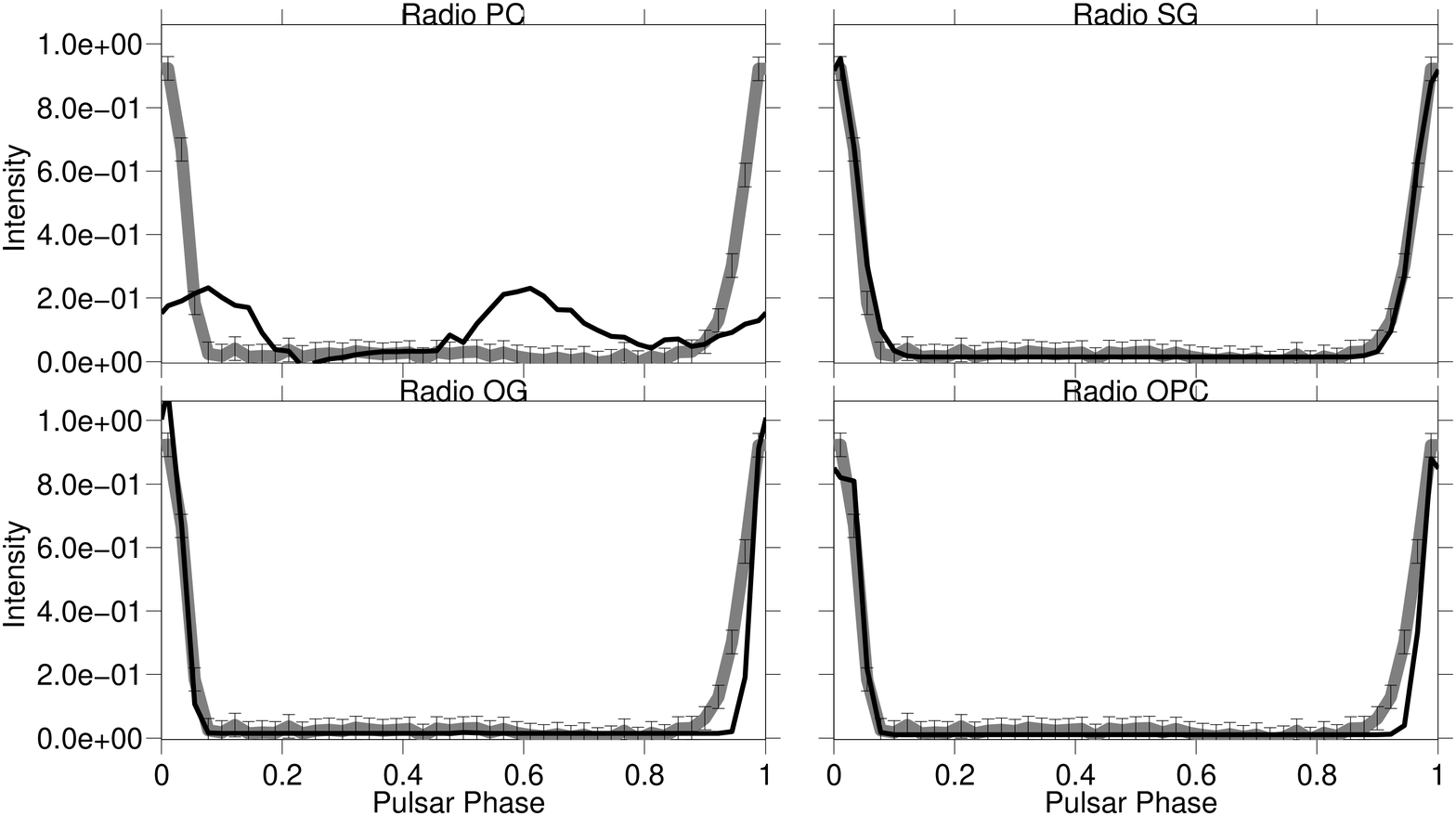}
\caption{PSR J1952+3252. \emph{Top}: for each model the best joint fit solution $\gamma$-ray light-curve (thick black line) is superimposed on the LAT pulsar $\gamma$-ray light-curve (shaded histogram). The estimated background is indicated by the dash-dot line. \emph{Bottom}: for each model the best joint fit solution radio light-curve (black line) is  is superimposed on the LAT pulsar radio light-curve (grey thick line).  The radio model is unique, but the $(\alpha,\zeta)$ solutions vary for each $\gamma$-ray model.}
\label{fitJoint_GmR35}
\end{figure}
  
\clearpage
\begin{figure}[htbp!]
\centering
\includegraphics[width=0.9\textwidth]{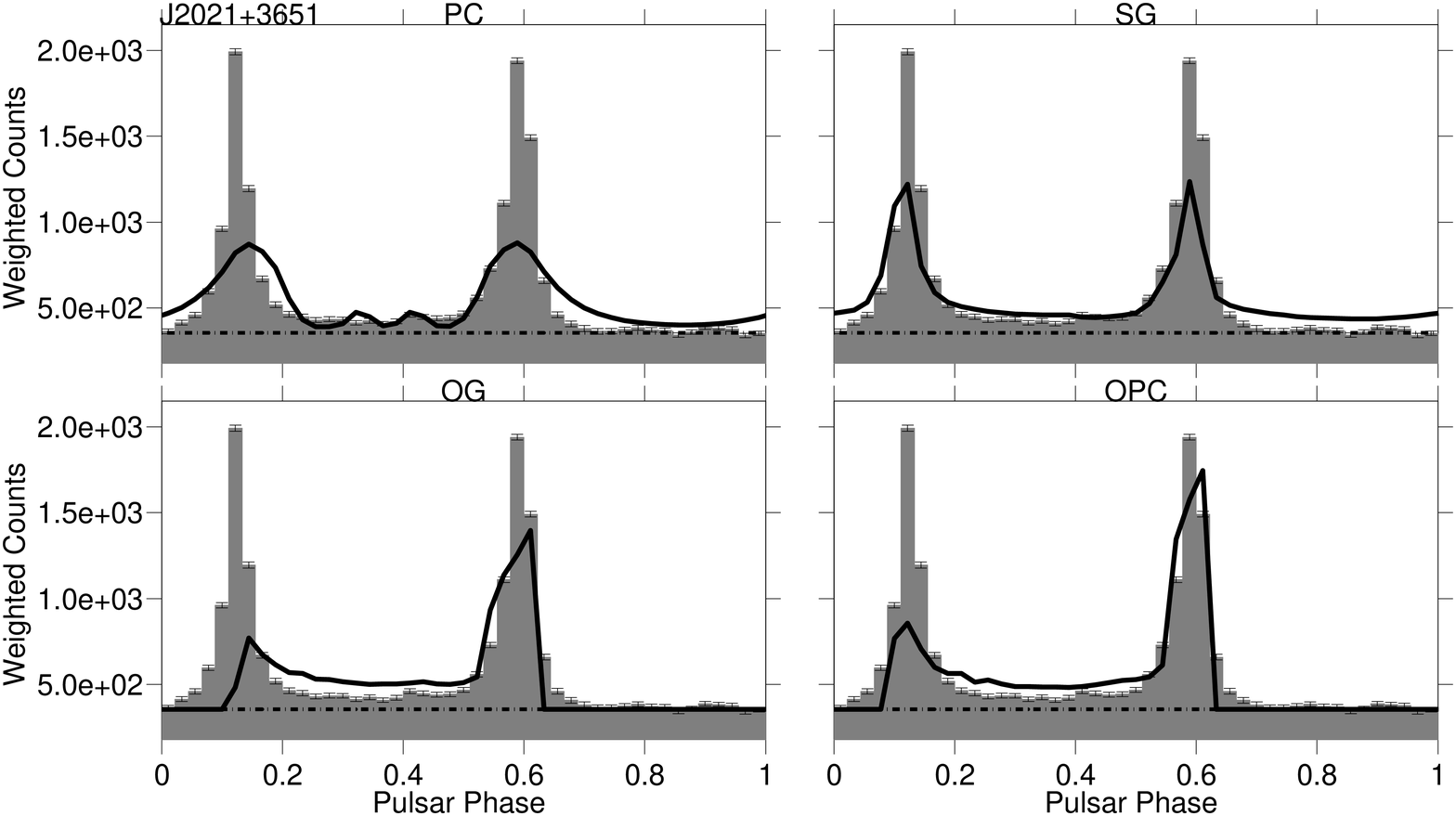}
\includegraphics[width=0.9\textwidth]{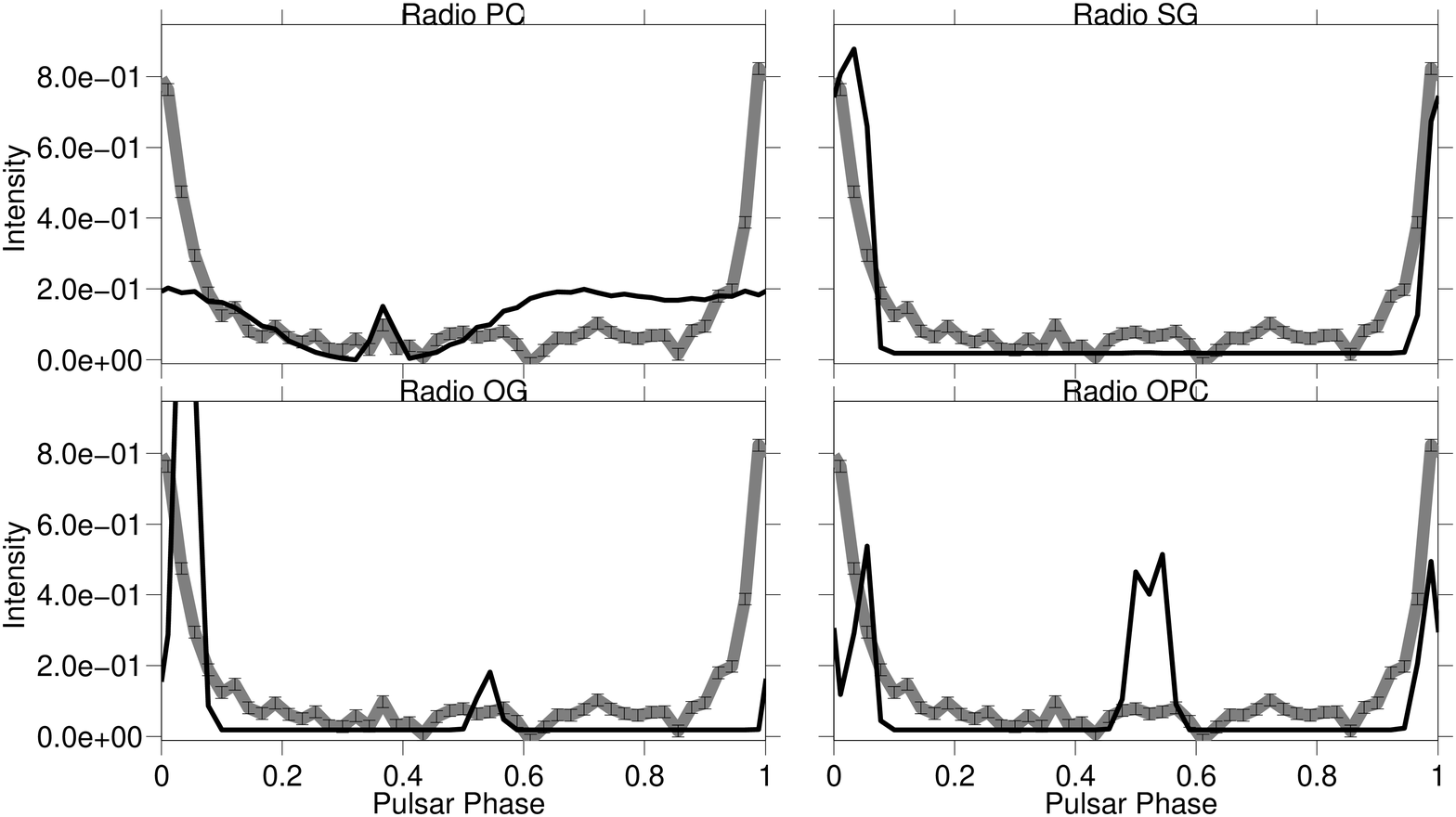}
\caption{PSR J2021+3651. \emph{Top}: for each model the best joint fit solution $\gamma$-ray light-curve (thick black line) is superimposed on the LAT pulsar $\gamma$-ray light-curve (shaded histogram). The estimated background is indicated by the dash-dot line. \emph{Bottom}: for each model the best joint fit solution radio light-curve (black line) is  is superimposed on the LAT pulsar radio light-curve (grey thick line).  The radio model is unique, but the $(\alpha,\zeta)$ solutions vary for each $\gamma$-ray model.}
\label{fitJoint_GmR36}
\end{figure}
  
\clearpage
\begin{figure}[htbp!]
\centering
\includegraphics[width=0.9\textwidth]{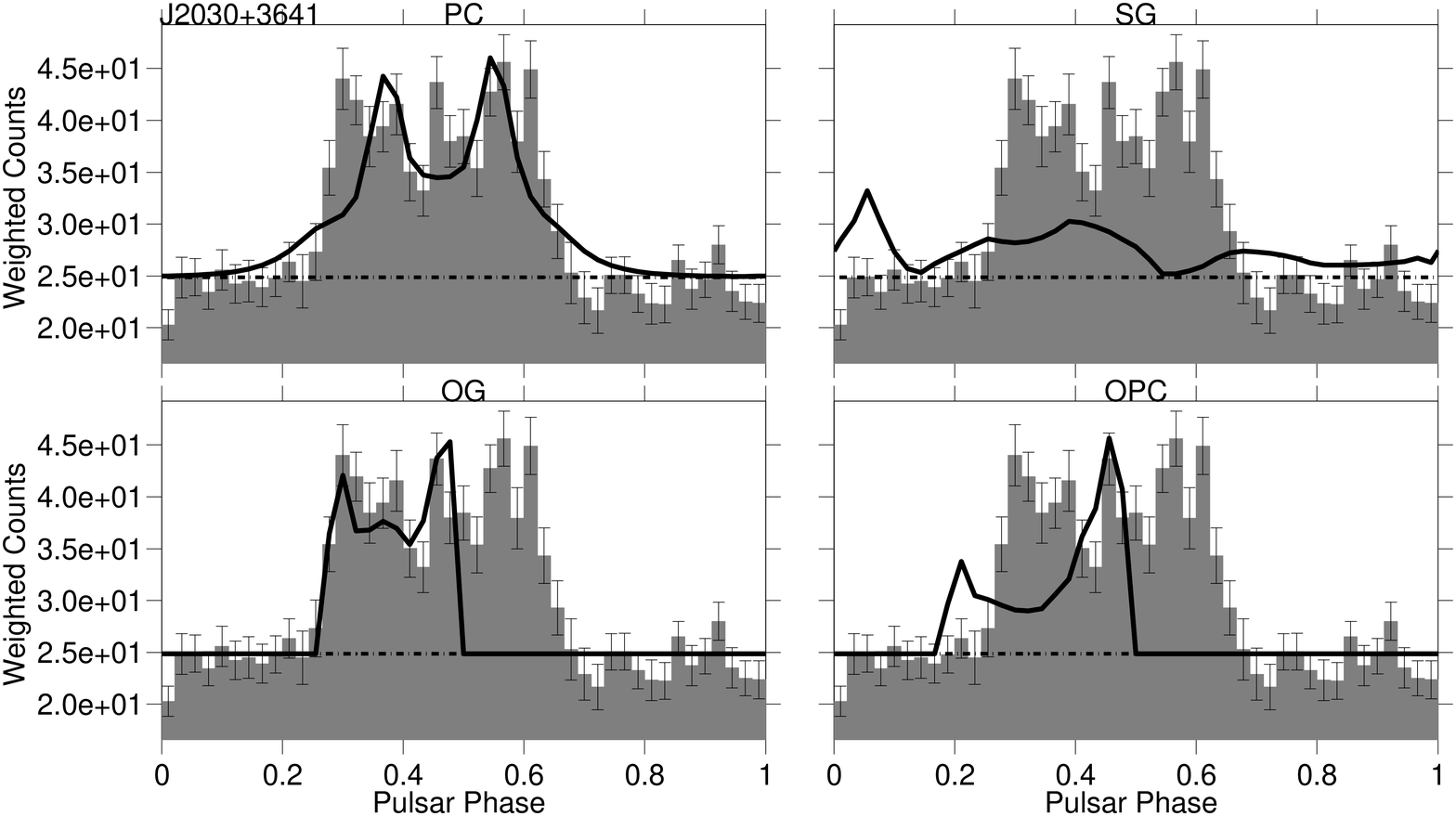}
\includegraphics[width=0.9\textwidth]{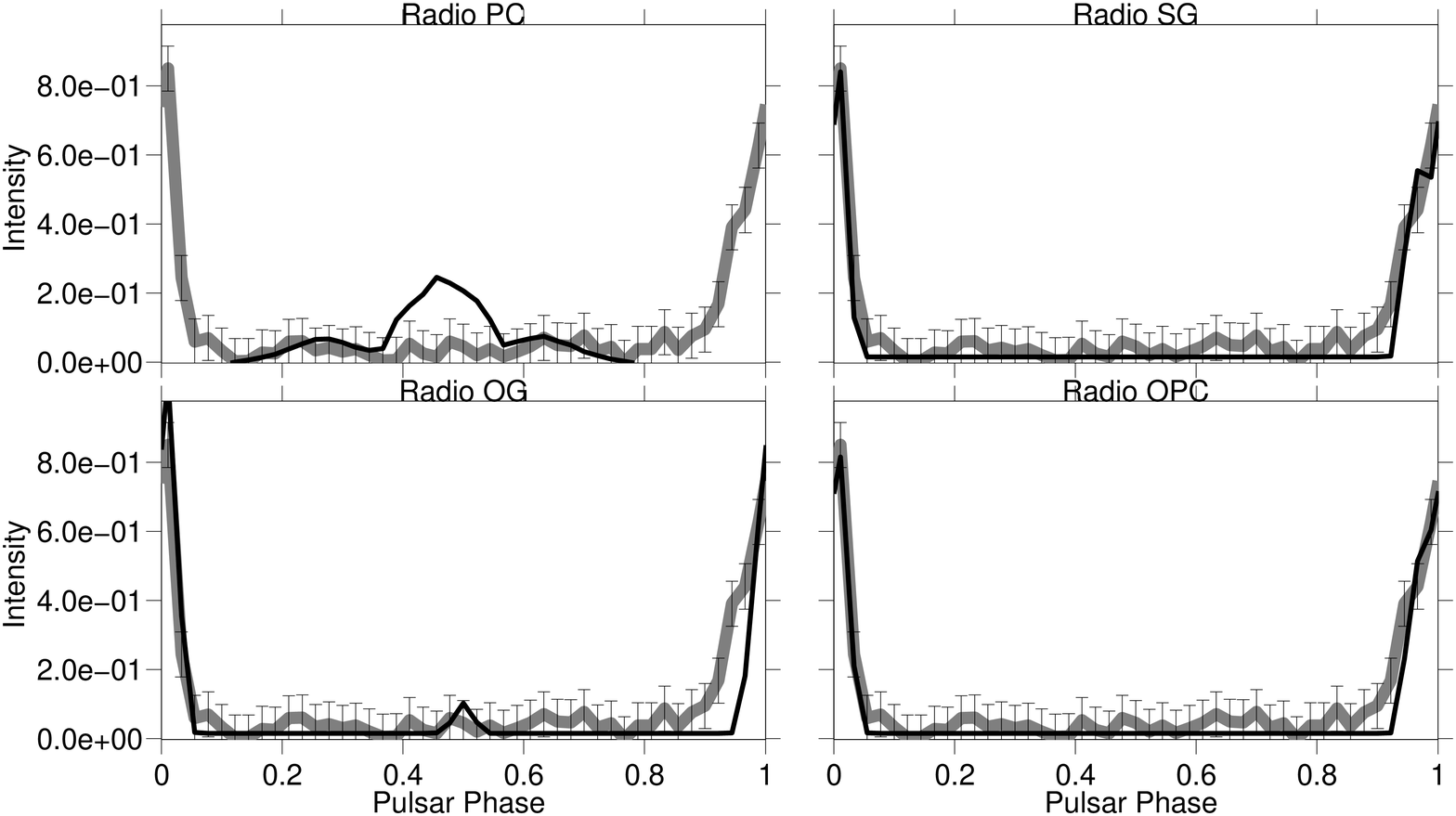}
\caption{PSR J2030+3641. \emph{Top}: for each model the best joint fit solution $\gamma$-ray light-curve (thick black line) is superimposed on the LAT pulsar $\gamma$-ray light-curve (shaded histogram). The estimated background is indicated by the dash-dot line. \emph{Bottom}: for each model the best joint fit solution radio light-curve (black line) is  is superimposed on the LAT pulsar radio light-curve (grey thick line).  The radio model is unique, but the $(\alpha,\zeta)$ solutions vary for each $\gamma$-ray model.}
\label{fitJoint_GmR37}
\end{figure}
  
\clearpage
\begin{figure}[htbp!]
\centering
\includegraphics[width=0.9\textwidth]{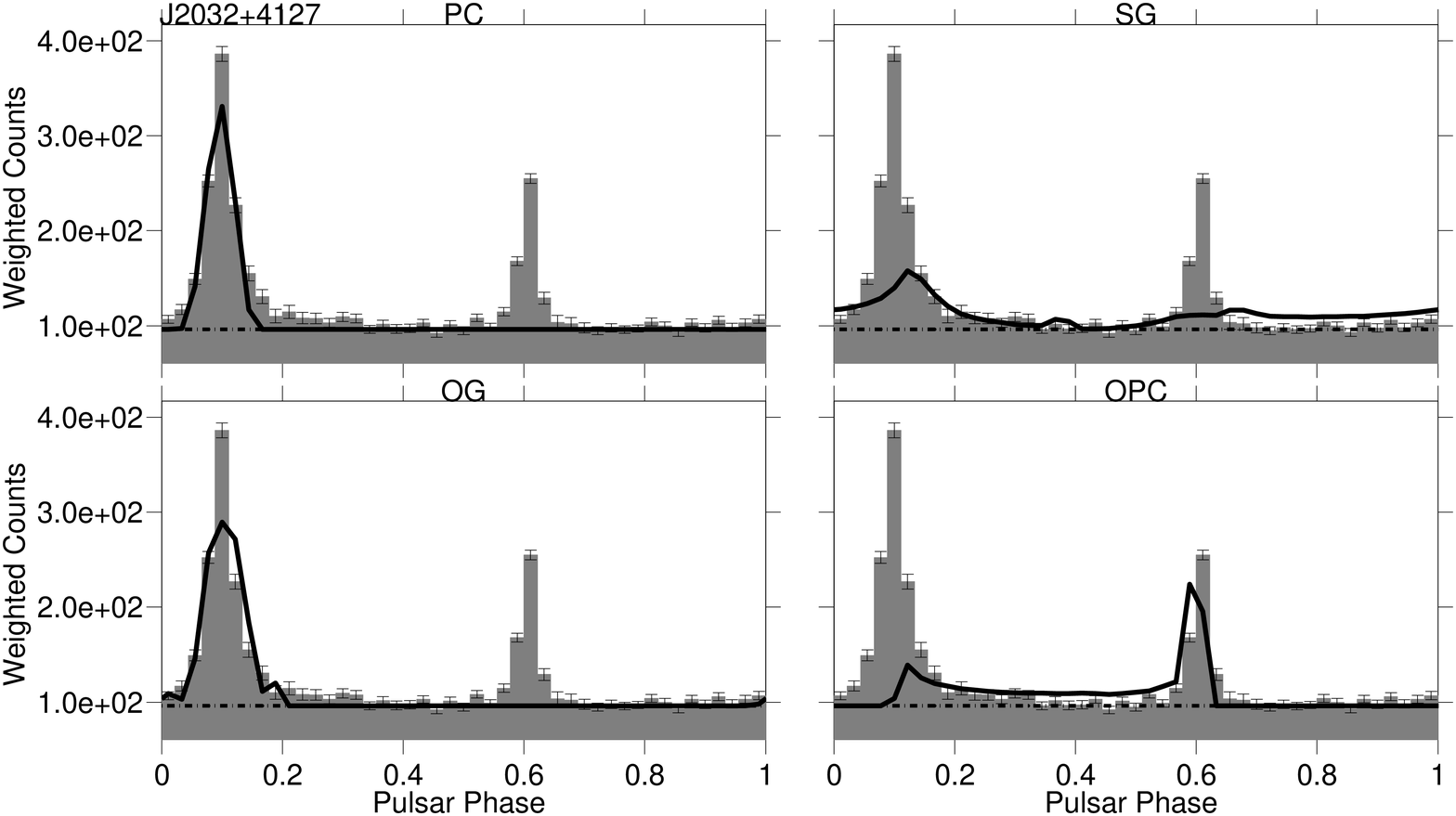}
\includegraphics[width=0.9\textwidth]{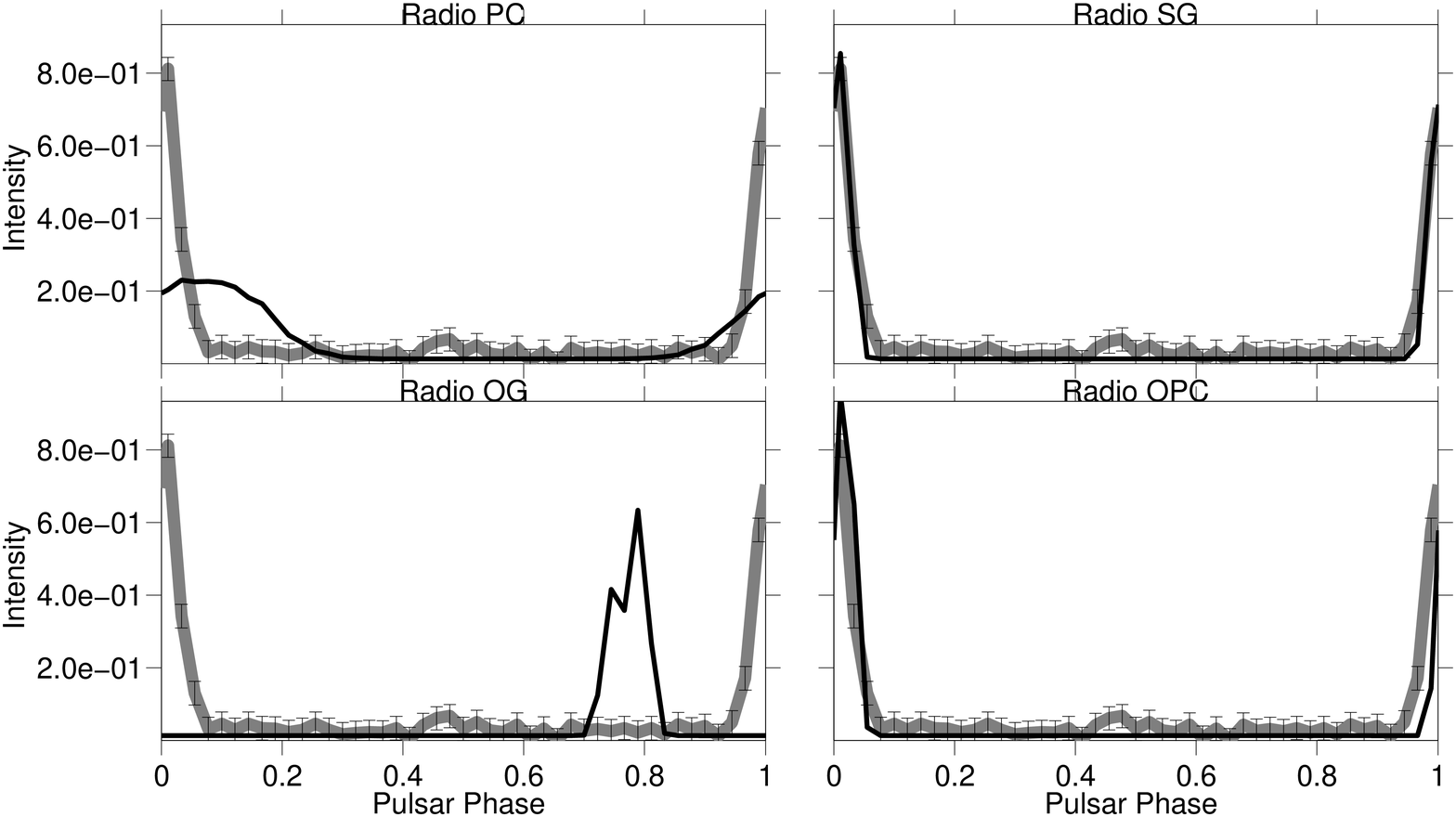}
\caption{PSR J2032+4127. \emph{Top}: for each model the best joint fit solution $\gamma$-ray light-curve (thick black line) is superimposed on the LAT pulsar $\gamma$-ray light-curve (shaded histogram). The estimated background is indicated by the dash-dot line. \emph{Bottom}: for each model the best joint fit solution radio light-curve (black line) is  is superimposed on the LAT pulsar radio light-curve (grey thick line).  The radio model is unique, but the $(\alpha,\zeta)$ solutions vary for each $\gamma$-ray model.}
\label{fitJoint_GmR38}
\end{figure}
  
\clearpage
\begin{figure}[htbp!]
\centering
\includegraphics[width=0.9\textwidth]{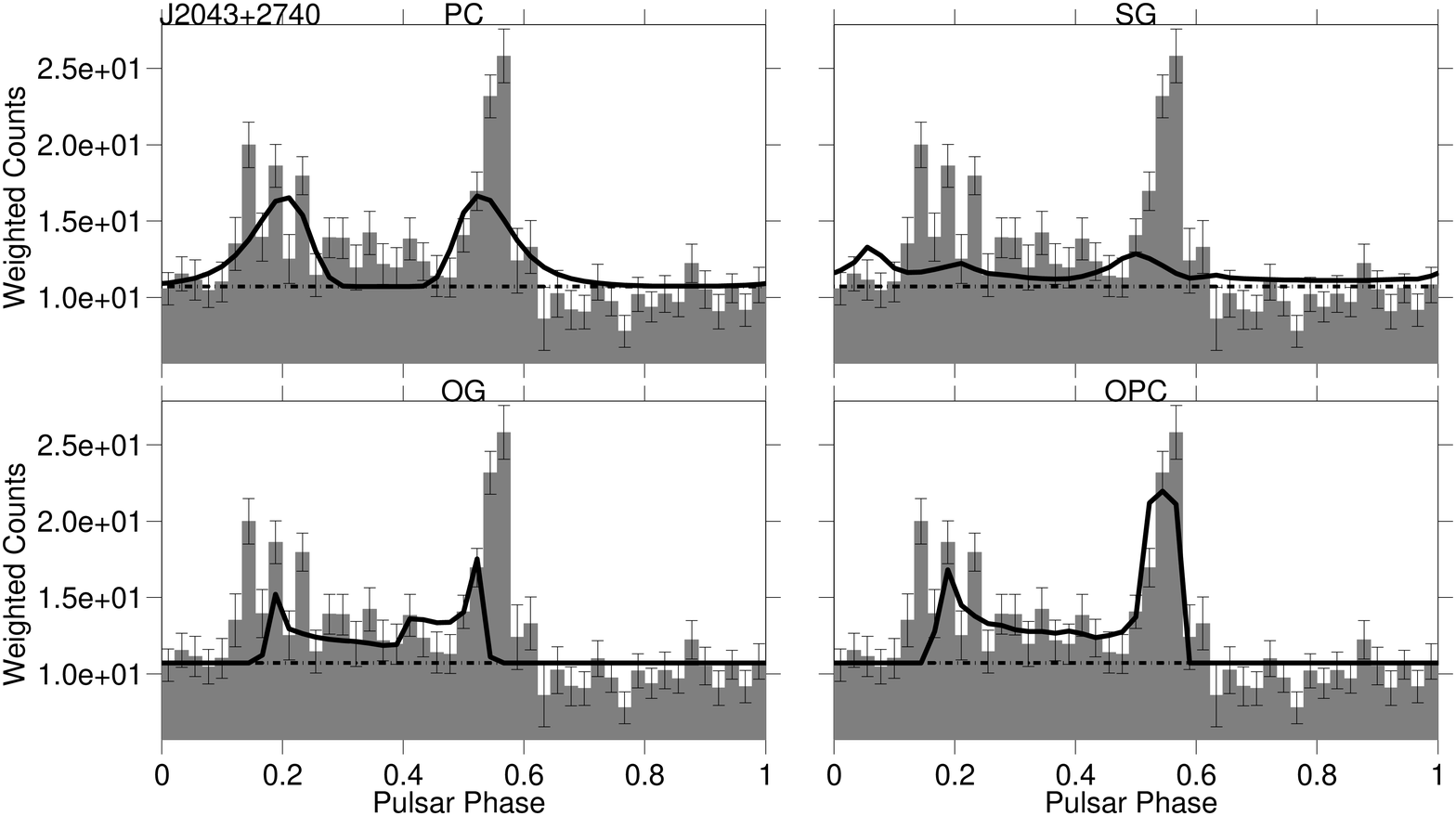}
\includegraphics[width=0.9\textwidth]{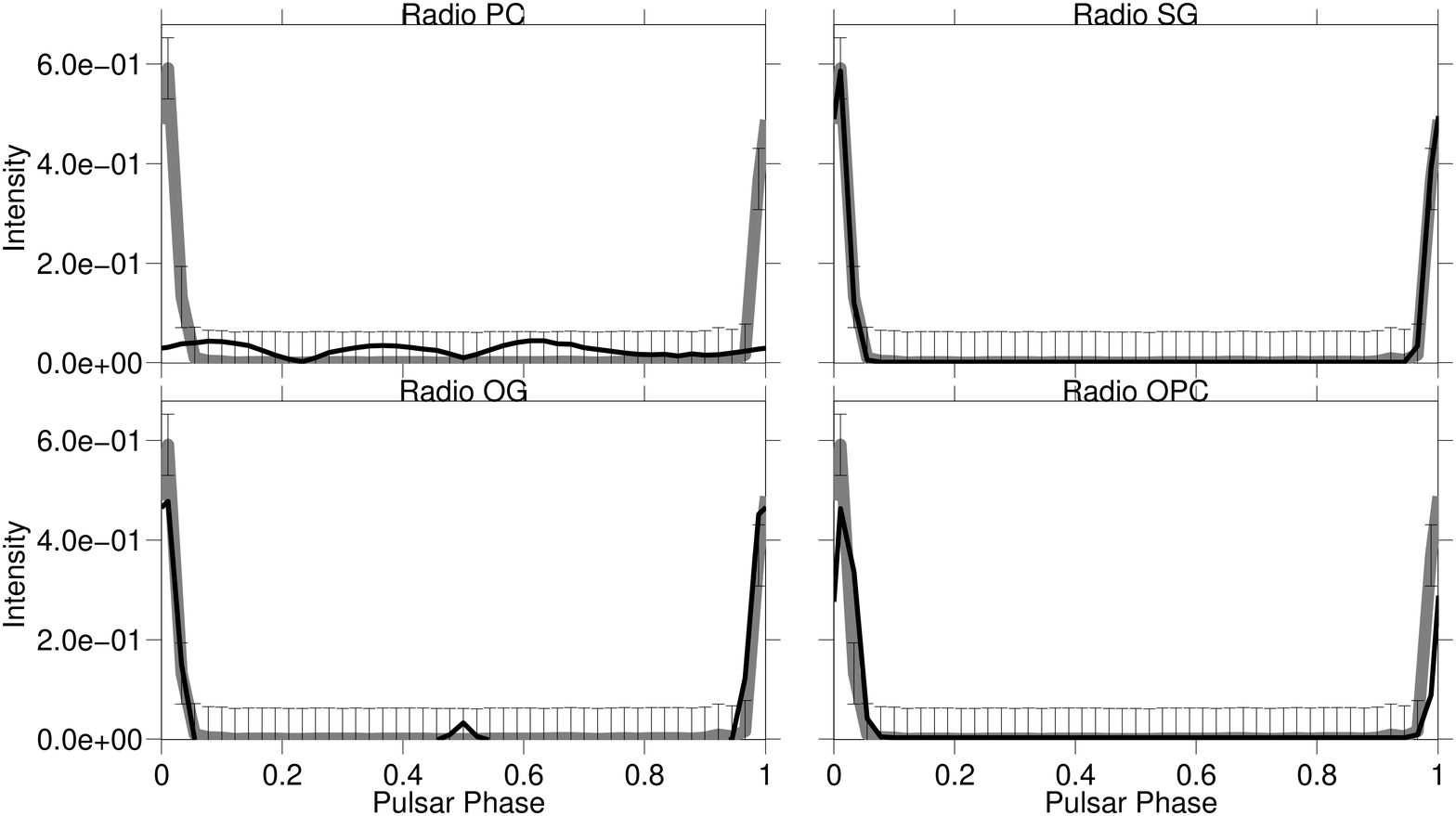}
\caption{PSR J2043+2740. \emph{Top}: for each model the best joint fit solution $\gamma$-ray light-curve (thick black line) is superimposed on the LAT pulsar $\gamma$-ray light-curve (shaded histogram). The estimated background is indicated by the dash-dot line. \emph{Bottom}: for each model the best joint fit solution radio light-curve (black line) is  is superimposed on the LAT pulsar radio light-curve (grey thick line).  The radio model is unique, but the $(\alpha,\zeta)$ solutions vary for each $\gamma$-ray model.}
\label{fitJoint_GmR39}
\end{figure}
  
\clearpage
\begin{figure}[htbp!]
\centering
\includegraphics[width=0.9\textwidth]{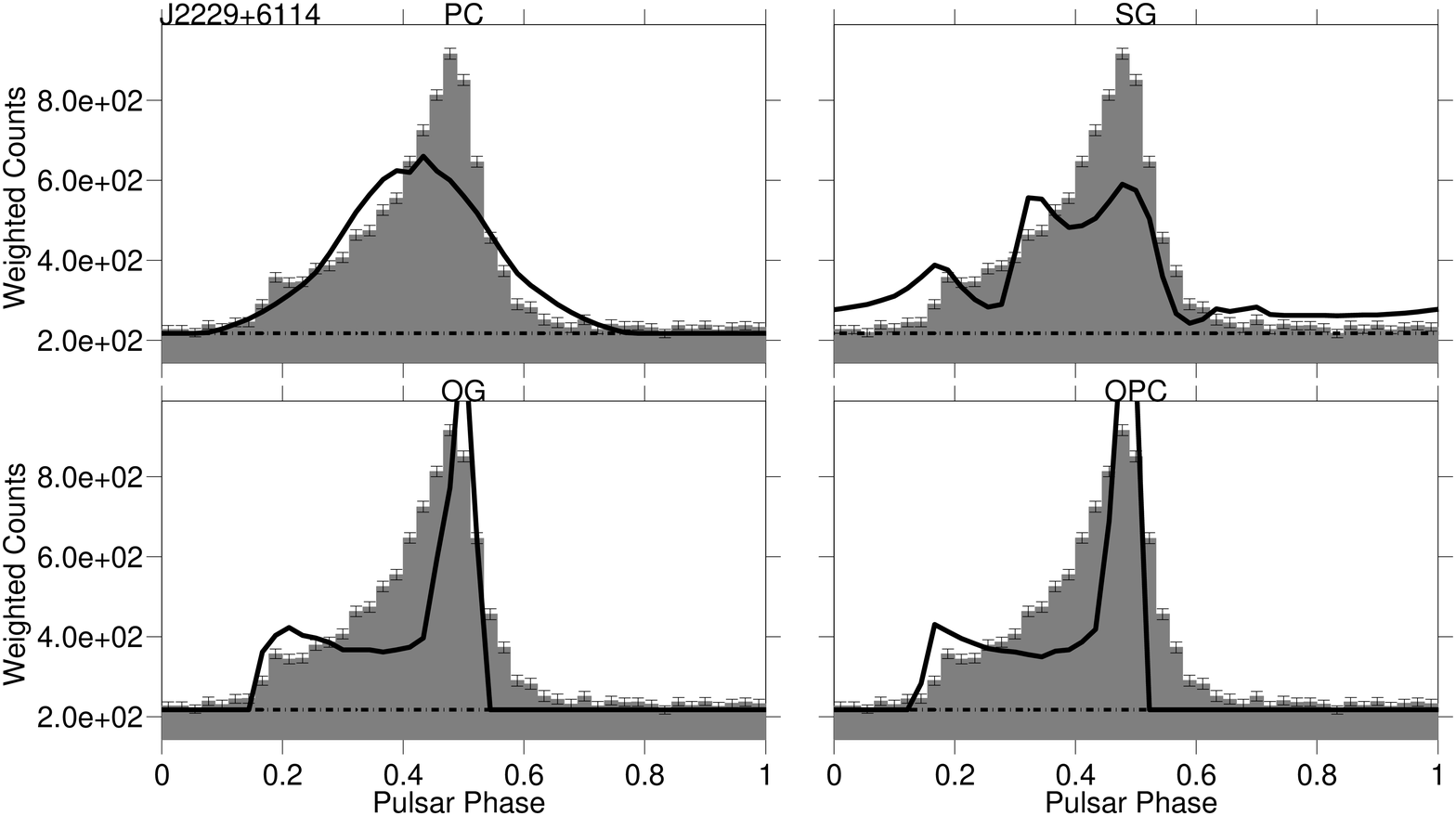}
\includegraphics[width=0.9\textwidth]{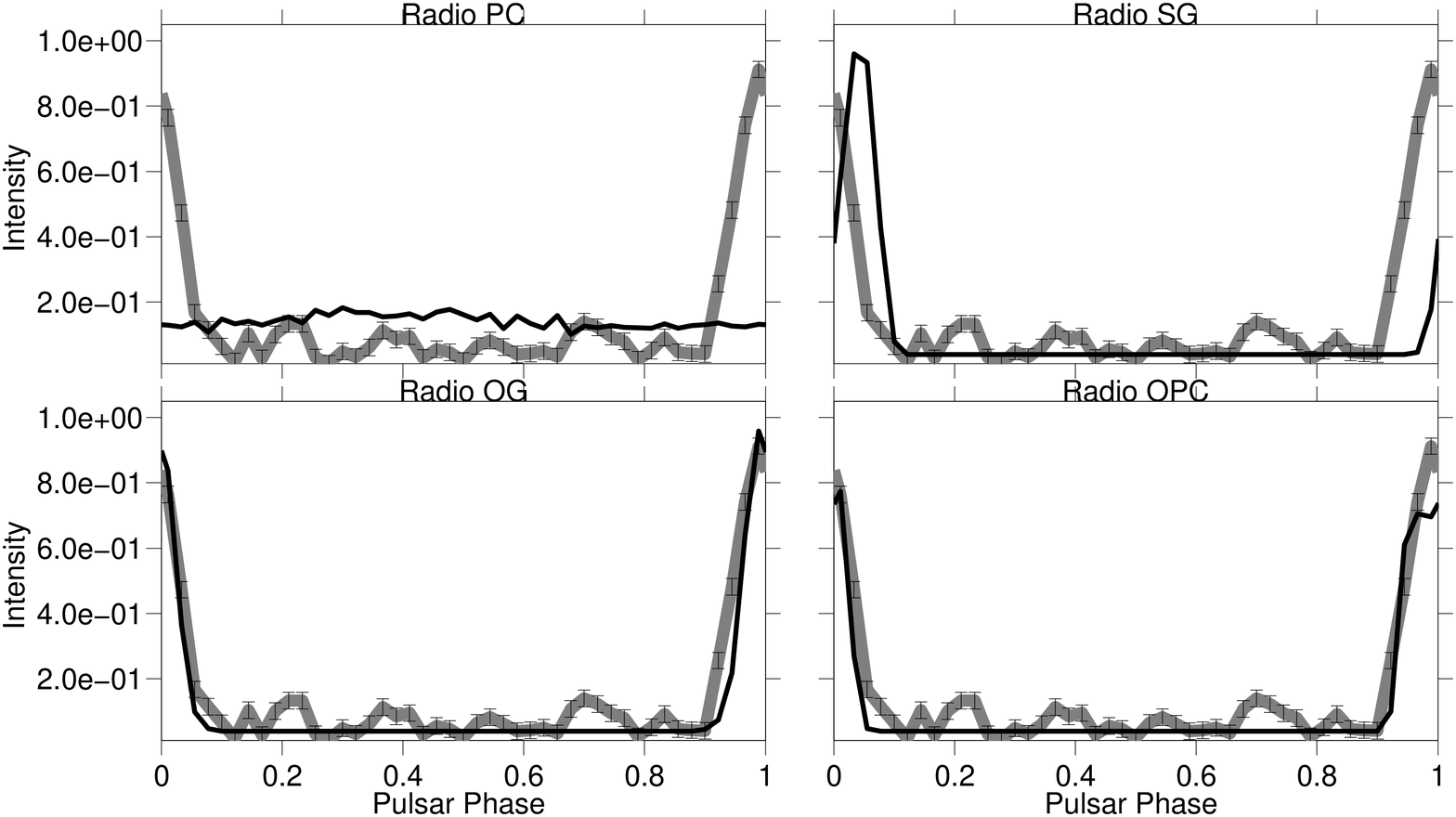}
\caption{PSR J2229+6114. \emph{Top}: for each model the best joint fit solution $\gamma$-ray light-curve (thick black line) is superimposed on the LAT pulsar $\gamma$-ray light-curve (shaded histogram). The estimated background is indicated by the dash-dot line. \emph{Bottom}: for each model the best joint fit solution radio light-curve (black line) is  is superimposed on the LAT pulsar radio light-curve (grey thick line).  The radio model is unique, but the $(\alpha,\zeta)$ solutions vary for each $\gamma$-ray model.}
\label{fitJoint_GmR40}
\end{figure}
  
\clearpage
\begin{figure}[htbp!]
\centering
\includegraphics[width=0.9\textwidth]{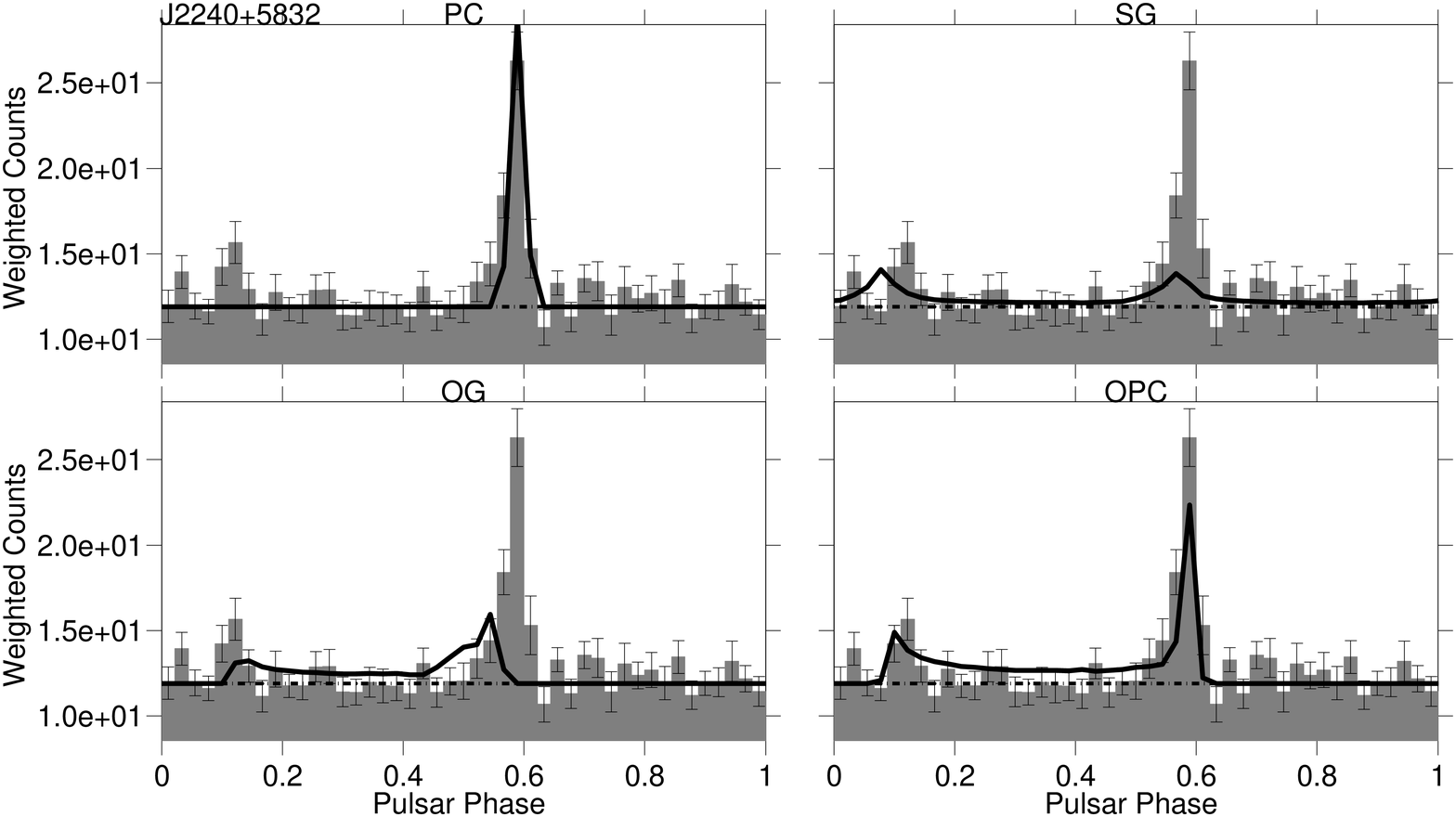}
\includegraphics[width=0.9\textwidth]{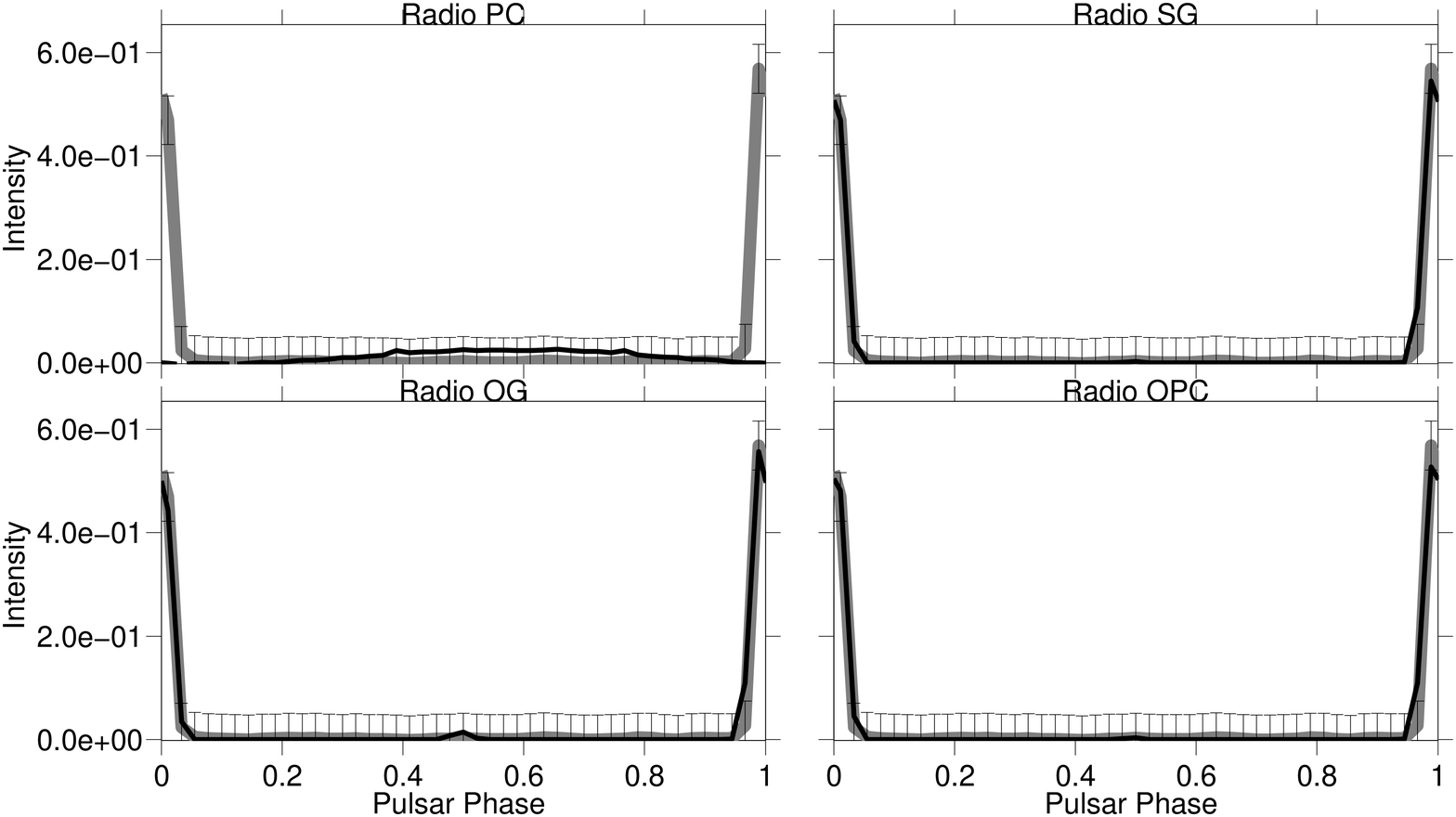}
\caption{PSR J2240+5832. \emph{Top}: for each model the best joint fit solution $\gamma$-ray light-curve (thick black line) is superimposed on the LAT pulsar $\gamma$-ray light-curve (shaded histogram). The estimated background is indicated by the dash-dot line. \emph{Bottom}: for each model the best joint fit solution radio light-curve (black line) is  is superimposed on the LAT pulsar radio light-curve (grey thick line).  The radio model is unique, but the $(\alpha,\zeta)$ solutions vary for each $\gamma$-ray model.}
\label{fitJoint_GmR41}
\end{figure}

%% file: JointFits_RQ2RL.tex
In this Appendix we give the results of the joint-fit of radio and $\gamma$-ray light curve for the 2 RF pulsars J0106$+$4855 and J1907$+$0602.

\subsection{J0106$+$4855}
\begin{figure}[htbp!]
\centering
\includegraphics[width=0.9\textwidth]{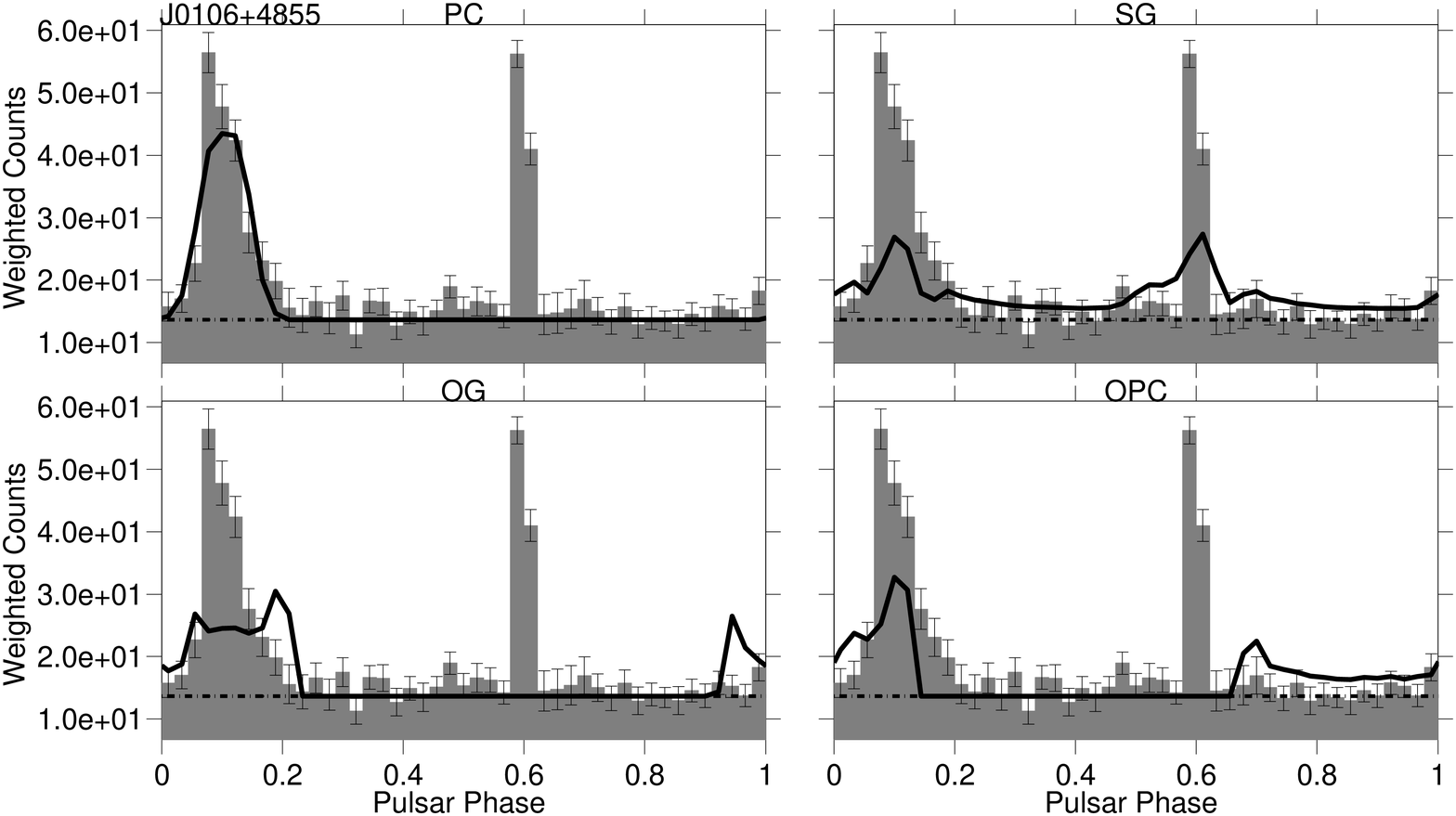}
\includegraphics[width=0.9\textwidth]{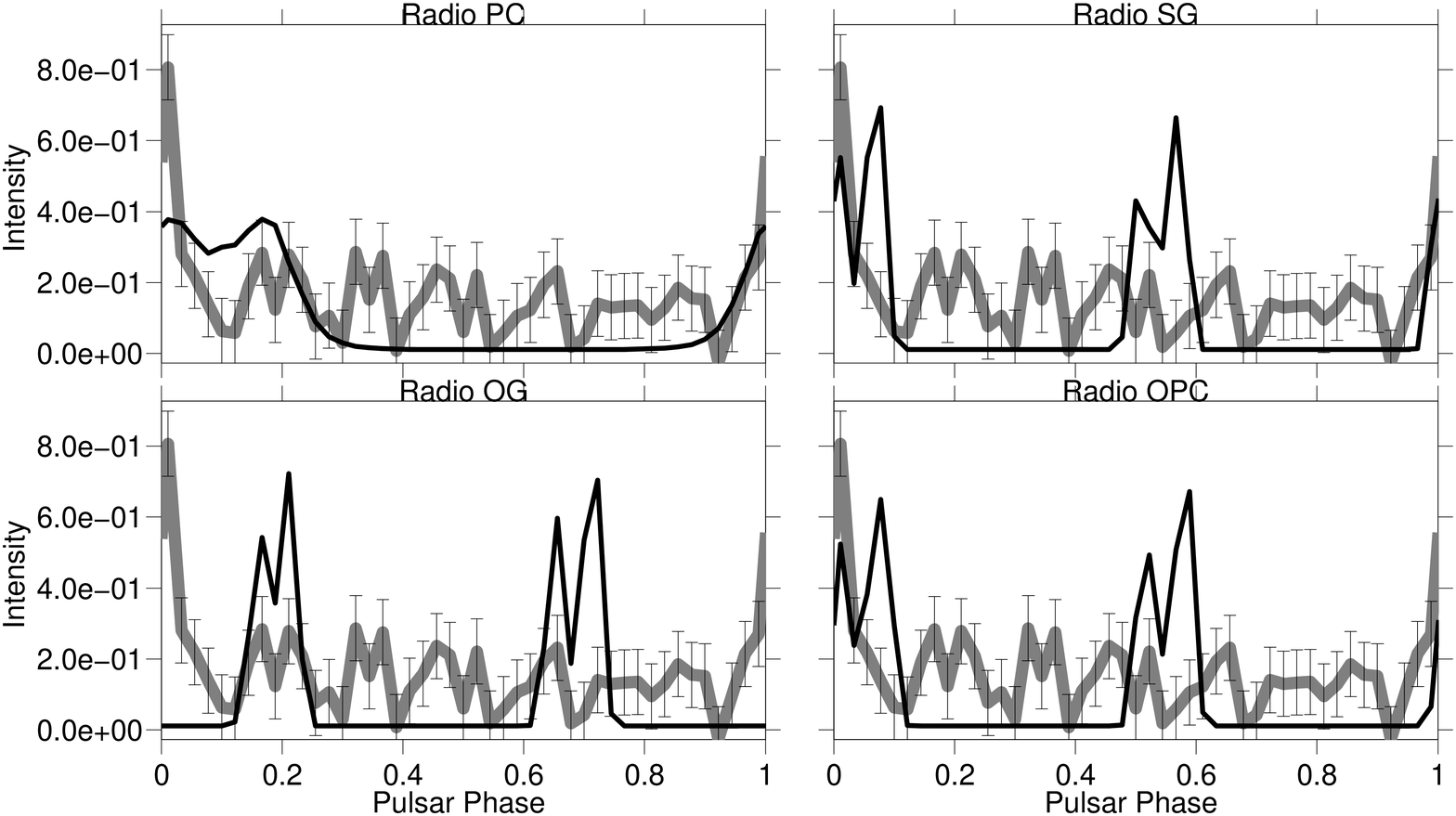}
\caption{PSR J0106$+$4855.  \emph{Top}: for each model the best joint fit solution $\gamma$-ray light-curve (thick black line) is superimposed on the LAT pulsar $\gamma$-ray light-curve (shaded histogram). The estimated background is indicated by the dash-dot line. \emph{Bottom}: for each model the best joint fit solution radio light-curve (black line) is  is superimposed on the LAT pulsar radio light-curve (grey thick line).  The radio model is unique, but the $(\alpha,\zeta)$ solutions vary for each $\gamma$-ray model.}
\label{fitJoint_GmR_2_1}
\end{figure}

Figure \ref{fitJoint_GmR_2_1} shows the best-fit radio and $\gamma$-ray light curves for pulsar J0106$+$4855 while its best joint-fit 
parameters are given in Table \ref{Joint_RQ2RL_1}. The PC joint-fit solution is characterised by lower $\alpha$ and $\zeta$ angles 
and similar $|\alpha-\zeta|$ and $f_{\Omega}$ values when compared with the $\gamma$-ray only fit solution, while the SG model 
joint-fit solution is overall consistent with the $\gamma$-ray-only fit results. 
For  OG and OPC models, the joint-fit $\alpha$ values are larger than the values obtained through $\gamma$-ray only fit. This implies a 
lower joint-fit $|\alpha-\zeta|$ value that favours simultaneous $\gamma$-ray and radio emission. Moreover the 
OG and OPC joint-fit values of $f_{\Omega}$ are larger than the values obtained with the $\gamma$-ray only fit and this 
favours the overlapping of $\gamma$-ray and radio beam to give a RL pulsar. 

Concerning the best fit radio and $\gamma$-ray light curves, the largest PC likelihood value shown in Table \ref{Joint_RQ2RL_1} is 
fictitious since the PC $\gamma$-ray fit shown in Figure \ref{fitJoint_GmR_2_1} explains just one of the two $\gamma$-ray peaks. 
In agreement with the $\gamma$-ray only fit that predicts a two peaks $\gamma$-ray light curve just for PC and SG models (Figure \ref{fitGm1}),
the SG is the model that best explains simultaneous $\gamma$-ray and radio emission from pulsar J0106$+$4855.
\begin{table*}[htbp!]
\def\arraystretch{1.5}
\centering
\begin{tabular}{| c || c | c | c | c | c || c | c | c | c | c |}
\hline
J0106$+$4855& $ PC $ & $ SG$  & $ OG$  & $ OPC$ \\
\hline
\hline
$\ln$ L & $ -115 $  & $ -157 $  & $ -235 $  & $ -209 $  \\
\hline
$\alpha~[^{\circ}]$& $ 18^{2}_{2} $  & $ 88^{2}_{6} $  & $ 89^{2}_{2} $  & $ 90^{2}_{2} $  \\
\hline
$\zeta~[^{\circ}]$&   $ 10^{2}_{2} $  & $ 90^{2}_{4} $  & $ 86^{2}_{2} $  & $ 90^{2}_{2} $  \\
\hline
$f_{\Omega}$ &  0.14   & 0.93  & 0.38  & 0.94  \\
\hline
L$_{\gamma}$ [W]&  $2.99\times10^{26}$   & $1.95\times10^{27}$  & $7.93\times10^{26}$  & $1.96\times10^{27}$  \\
\hline
\end{tabular}
\caption{Best fit parameters resulting from the joint fit of radio and $\gamma$-ray light curves of pulsar J0106$+$4855.  From 
top to bottom are listed, for each model, best fit log-likelihood value, magnetic obliquity $\alpha$, observer line of sight $\zeta$,  
$\gamma$-ray beaming factor $f_{\Omega}$, and  $\gamma$-ray Luminosity. The errors on $\alpha$ and $\zeta$ bigger than 
2 correspond to 3$\sigma$ statistical error.}
\label{Joint_RQ2RL_1}
\end{table*}

\subsection{ J1907$+$0602}
\begin{figure}[htbp!]
\centering
\includegraphics[width=0.9\textwidth]{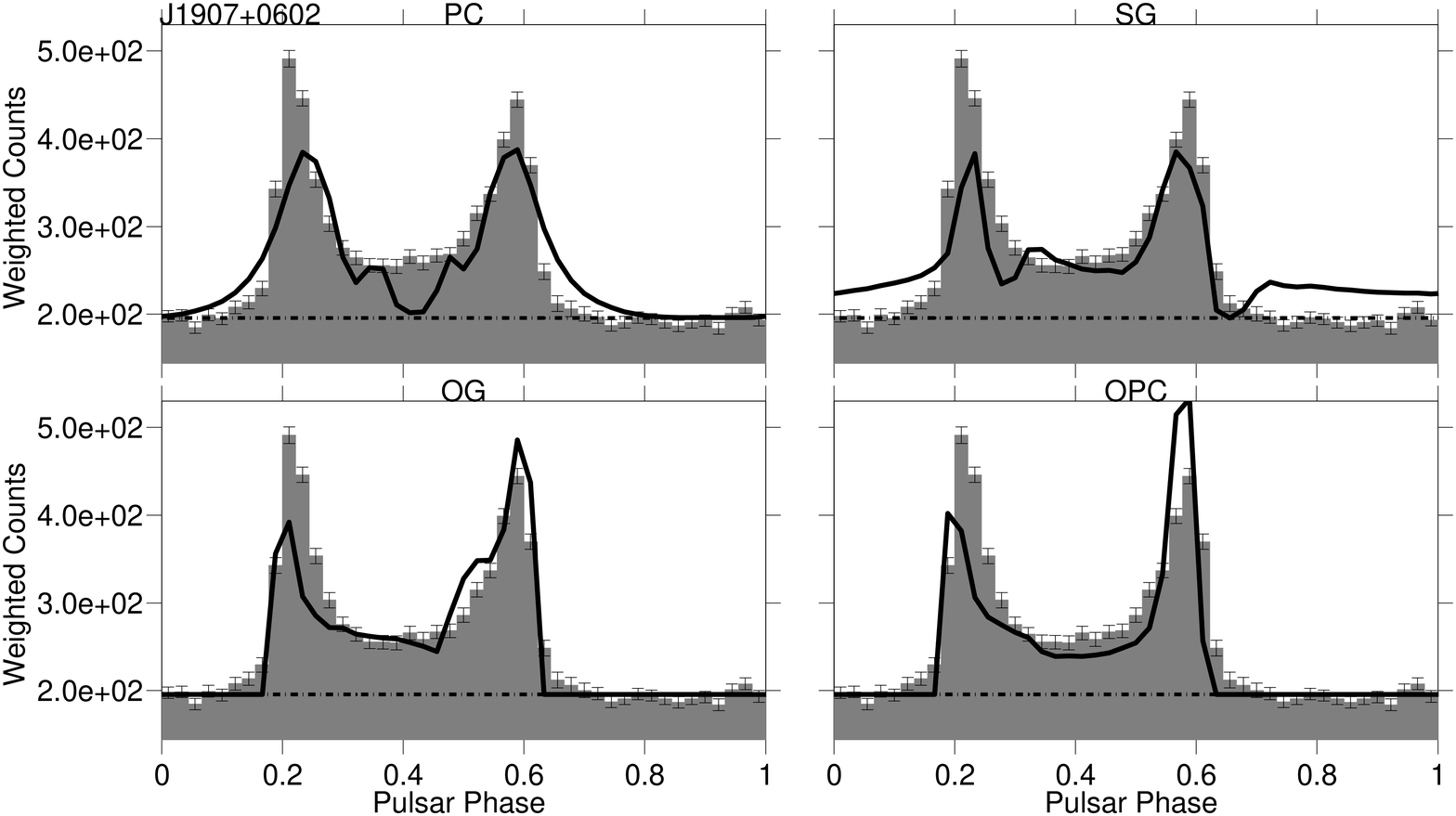}
\includegraphics[width=0.9\textwidth]{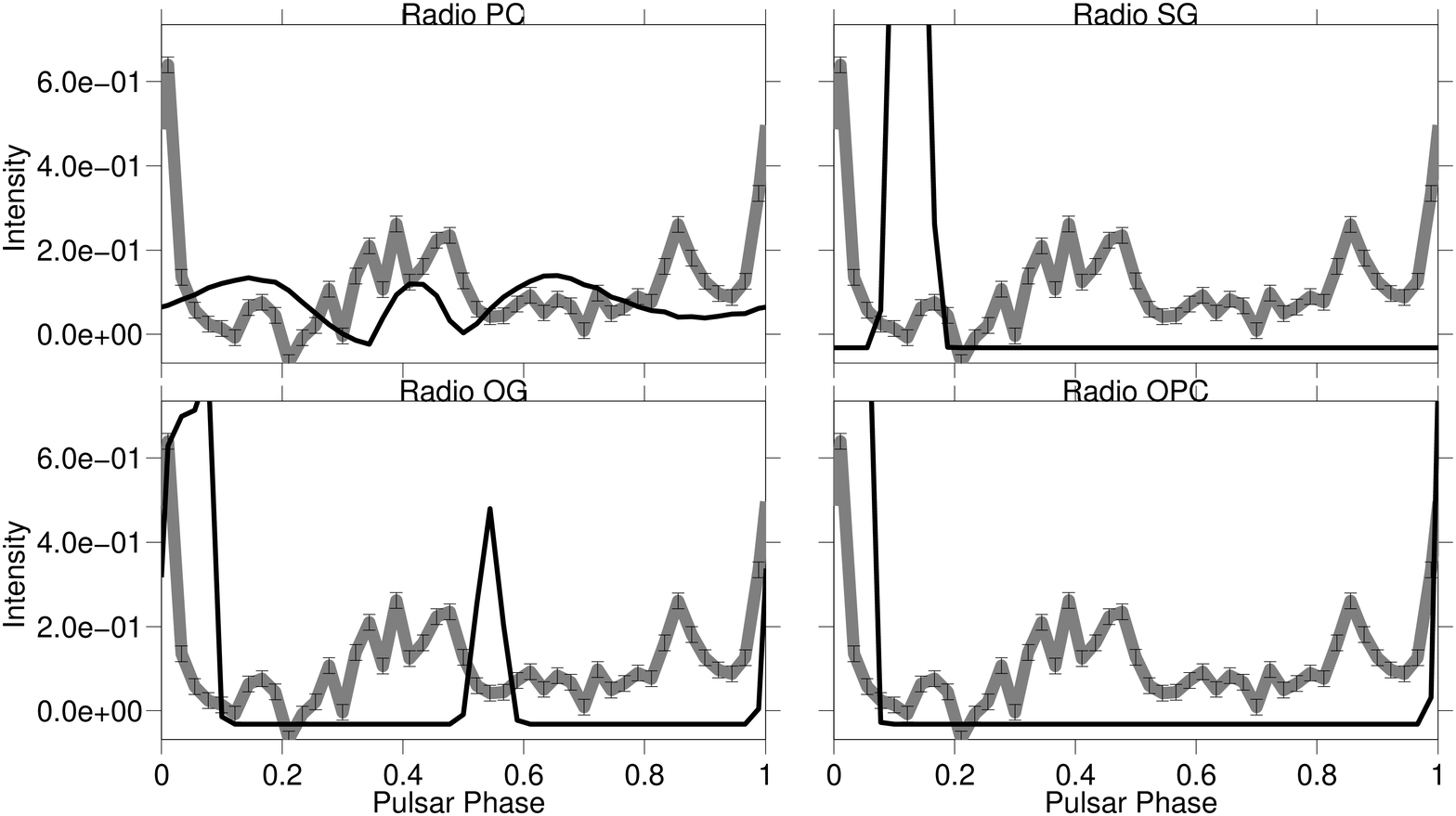}
\caption{PSR J1907$+$0602. \ \emph{Top}: for each model the best joint fit solution $\gamma$-ray light-curve (thick black line) is superimposed on the LAT pulsar $\gamma$-ray light-curve (shaded histogram). The estimated background is indicated by the dash-dot line. \emph{Bottom}: for each model the best joint fit solution radio light-curve (black line) is  is superimposed on the LAT pulsar radio light-curve (grey thick line).  The radio model is unique, but the $(\alpha,\zeta)$ solutions vary for each $\gamma$-ray model.}
\label{fitJoint_GmR_2_2}
\end{figure}

Figure \ref{fitJoint_GmR_2_2} shows the best-fit radio and $\gamma$-ray light curves for pulsar J1907$+$0602 while its 
best joint-fit parameters are given in Table \ref{Joint_RQ2RL_2}. Both PC and SG model best joint-fit parameters are consistent with the
$\gamma$-ray-only fit results. 
As for pulsar J0106$+$4855, the OG and OPC models best-fit results predict $\alpha$ values larger than the values obtained through $\gamma$-ray 
only fit and larger values of $f_{\Omega}$. The lower $|\alpha-\zeta|$ joint-fit values and the larger $f_{\Omega}$ joint-fit values 
favour the overlapping of $\gamma$-ray and radio beam to give a RL pulsar.

In agreement with the $\gamma$-ray only fit that predicts a $\gamma$-ray light curves with two peaks connected by a high bridge just for OG and OPC 
models (Figure \ref{fitGm25}), the OG is the model that best explains simultaneous $\gamma$-ray and radio emission from pulsar J1907$+$0602.
\begin{table*}[htbp!]
\def\arraystretch{1.5}
\centering
\begin{tabular}{| c || c | c | c | c | c || c | c | c | c | c |}
\hline
J1907$+$0602& $ PC $ & $ SG$  & $ OG$  & $ OPC$ \\
\hline
\hline
$\ln$ L & $ -368 $  & $ -957 $  & $ -353 $  & $ -580 $  \\
\hline
$\alpha~[^{\circ}]$& $ 7^{2}_{2} $  & $ 61^{2}_{6} $  & $ 87^{2}_{2} $  & $ 63^{2}_{2} $  \\
\hline
$\zeta~[^{\circ}]$ &   $ 9^{2}_{2} $  & $ 54^{2}_{4} $  & $ 79^{2}_{2} $  & $ 72^{2}_{2} $  \\
\hline
$f_{\Omega}$ &  0.03   & 0.97  & 0.78  & 0.78  \\
\hline
L$_{\gamma}$ [W]&  $8.52\times10^{26}$   & $3.06\times10^{28}$  & $2.45\times10^{28}$  & $2.44\times10^{28}$  \\
\hline
\end{tabular}
\caption{Best fit parameters resulting from the joint fit of radio and $\gamma$-ray light curves of pulsar J1907$+$0602.  From 
top to bottom are listed, for each model, best fit log-likelihood value, magnetic obliquity $\alpha$, observer line of sight $\zeta$,  
$\gamma$-ray beaming factor $f_{\Omega}$, and  $\gamma$-ray Luminosity. The errors on $\alpha$ and $\zeta$ bigger than 
2 correspond to 3$\sigma$ statistical error.}
\label{Joint_RQ2RL_2}
\end{table*}

 \clearpage

\clearpage

